\documentclass[1p,final,dvipsnames]{elsarticle}

\usepackage{graphicx}

\usepackage[T1]{fontenc}
\usepackage{dsfont}               
\usepackage{mathrsfs}             
\usepackage{slashed}              
\usepackage{amsmath}
\usepackage{amssymb, bbm}
\usepackage{amsbsy}
\usepackage{amsfonts}
\usepackage{booktabs} 
\usepackage{array} 

\usepackage{subcaption}
\renewcommand\vec[1]{\ifstrequal{#1}{0}{\ensuremath{\mathbf{0}}}{\ensuremath{\boldsymbol{#1}}}}

\usepackage{overpic}
\usepackage{slashed} 
\usepackage{xspace} 
\usepackage{tikz}
\usepackage{diagbox}

\numberwithin{equation}{section}
\numberwithin{table}{section}
\numberwithin{figure}{section}

\journal{Progress in Particle and Nuclear Physics}

\usepackage{titlesec}
\usepackage{sectsty}
\titleformat{\section}{\normalfont\Large\bfseries}{\thesection}{1em}{}
\titleformat{\subsection}{\normalfont\large\bfseries}{\thesubsection}{1em}{}
\titleformat{\subsubsection}{\normalfont\normalsize\bfseries}{\thesubsubsection}{1em}{}

\newcommand{\C}{C_\alpha}
\newcommand{\La}{\Lambda}
\newcommand{\Si}{\Sigma}
\newcommand{\Kb}{\bar K}

\newcommand{\Ps}{P^{(\sigma)}}
\newcommand{\Pt}{P^{(\tau)}}
\newcommand{\Pp}{P^{(p)}}
\newcommand{\ir}[1]{\mathbf{#1}}
\newcommand{\extfield}{(\nabla^i \phi)}
\protected\def\lc{C}
\protected\def\ld{D}
\protected\def\lC{H}
\newcommand{\trace}[1]{\ensuremath{\langle #1\rangle}}
\newcommand\numberthis{\addtocounter{equation}{1}\tag{\theequation}}

\makeatletter
\newcommand{\Biggg}{\bBigg@{4}}
\makeatother

\newcommand{\sdot}[2]{\left( #1\cdot #2 \right)}
\newcommand{\order}[1]{\mathcal{O}\left( #1 \right)}

\newcommand{\B}{\mathcal{B}}

\newlength{\feynwidth} 
\newlength{\feynwidthb} 

\protected\def\fig{\@ifstar\@fig\@@fig} \def\@fig#1{\ref{#1}} \def\@@fig#1{Fig.~\ref{#1}}
\protected\def\tab{\@ifstar\@tab\@@tab} \def\@tab#1{\ref{#1}} \def\@@tab#1{Tab.~\ref{#1}}
\protected\def\eq{\@ifstar\@eq\@@eq} \def\@eq#1{\eqref{#1}}\def\@@eq#1{Eq.~\eqref{#1}}
\protected\def\ch{\@ifstar\@ch\@@ch} \def\@ch#1{\ref{#1}} \def\@@ch#1{Ch.~\ref{#1}}
\protected\def\sect{\@ifstar\@sect\@@sect} \def\@sect#1{\ref{#1}} \def\@@sect#1{Sec.~\ref{#1}}
\protected\def\ssect{\@ifstar\@ssect\@@ssect} \def\@ssect#1{\ref{#1}} \def\@@ssect#1{Subsec.~\ref{#1}}
\protected\def\app{\@ifstar\@app\@@app} \def\@app#1{\cite{#1}} \def\@@app#1{App.~\ref{#1}}
\protected\def\ct{\@ifstar\@ct\@@ct} \def\@ct#1{\cite{#1}} \def\@@ct#1{Ref.~\cite{#1}}
\protected\def\figs{\@ifstar\@figs\@@figs} \def\@figs#1{\ref{#1}} \def\@@figs#1{Figs.~\ref{#1}}
\protected\def\tabs{\@ifstar\@tabs\@@tabs} \def\@tabs#1{\ref{#1}} \def\@@tabs#1{Tabs.~\ref{#1}}
\protected\def\eqs{\@ifstar\@eqs\@@eqs} \def\@eqs#1{\eqref{#1}} \def\@@eqs#1{Eqs.~\eqref{#1}}
\protected\def\chs{\@ifstar\@chs\@@chs} \def\@chs#1{\ref{#1}} \def\@@chs#1{Chs.~\ref{#1}}
\protected\def\sects{\@ifstar\@sects\@@sects} \def\@sects#1{\ref{#1}} \def\@@sects#1{Secs.~\ref{#1}}
\protected\def\ssects{\@ifstar\@ssects\@@ssects} \def\@ssects#1{\ref{#1}} \def\@@ssects#1{Subsecs.~\ref{#1}}
\protected\def\apps{\@ifstar\@apps\@@apps} \def\@apps#1{\cite{#1}} \def\@@apps#1{Apps.~\ref{#1}}
\protected\def\cts{\@ifstar\@cts\@@cts} \def\@cts#1{\cite{#1}} \def\@@cts#1{Refs.~\cite{#1}}

\newcommand{\eg}{e.g., }

\newcommand{\cheft}{\(\chi\)EFT\xspace}

\newlength{\usualunit} 
\unitlength = \usualunit

\newcommand{\defeq}{\ensuremath{:=}}

\usepackage[colorlinks,citecolor=blue,linktoc=all,linkcolor=cyan]{hyperref}
\usepackage{orcidlink}

\begin{document}
	
	\begin{frontmatter}
		
		\title{{\em Ab initio} description of hypernuclei}
        \author[a]{Johann Haidenbauer\orcidlink{0000-0002-0923-8053}}\ead{j.haidenbauer@fz-juelich.de}
        \author[b,a,c,d]{Ulf-G. Mei{\ss}ner\orcidlink{0000-0003-1254-442X}\corref{mycorrespondingauthor}}\ead{meissner@hiskp.uni-bonn.de}
        \author[a,c]{Andreas Nogga\orcidlink{0000-0003-2156-748X}}
  		\cortext[mycorrespondingauthor]{Corresponding author}
		\ead{a.nogga@fz-juelich.de}
		
		\address[a]{Institute for Advanced Simulation (IAS-4), Forschungszentrum J\"ulich, 52428 J\"ulich, Germany}
		\address[b]{Helmholtz-Institut~f\"{u}r~Strahlen-~und~Kernphysik~and~Bethe~Center~for~Theoretical~Physics, Universit\"{a}t~Bonn,~53115~Bonn,~Germany}
		\address[c]{Centre for Advanced Simulation and Analytics (CASA), Forschungszentrum J\"ulich, 52428 J\"ulich, Germany}
        \address[d]{Peng Huanwu Collaborative Center for Research and Education, International Institute for Interdisciplinary and Frontiers,\\
        Beihang University, Beijing 100191, China}
		
		\begin{abstract}
			Hypernuclei are bound states of neutrons, protons and one or two hyperons, thus extending the nuclear landscape to 
            a third dimension. They also encode information about the baryon-baryon and three-baryon interactions. Here, we review recent work
            on chiral effective field theory for two- and three-baryon interactions and their application in nuclei based
            on {\em ab initio} methods. These include the Faddeev-Yakubovsky equations, the no-core-shell-model (NCSM) and nuclear lattice effective field theory (NLEFT). 
Besides of providing an overview of the formalisms explicit results for the separation energies
of light $\Lambda$ hypernuclei are provided. Two-body
and three-body forces are included consistently, in line with the underlying power counting.
Calculations of $\Lambda$ hypernuclei within the NCSM, performed up to A=7 so far, 
suggest that agreement with the experimental binding energies can be achieved once 
appropriate three-body forces are taken into account. 
Similar conclusions are drawn from the study based on NLEFT, 
where even hypernuclei up to A=16 can be computed.
            Additionally, 
            applications of {\it ab initio} approaches in calculations of $\Lambda\Lambda$ 
            and $\Xi$ hypernuclei are discussed and possible
candidates for the lightest systems that could be bound are identified, 
namely $^{\ \ 5}_{\Lambda\Lambda}$He and $^4_\Xi$H.
            
		\end{abstract}
		
		
	\end{frontmatter}
	
	\newpage
	
	\thispagestyle{empty}
	\tableofcontents
	

	\newpage
	
    \section{Introduction}\label{intro}

\begin{figure}[h]
\centering
\includegraphics[scale=0.75]{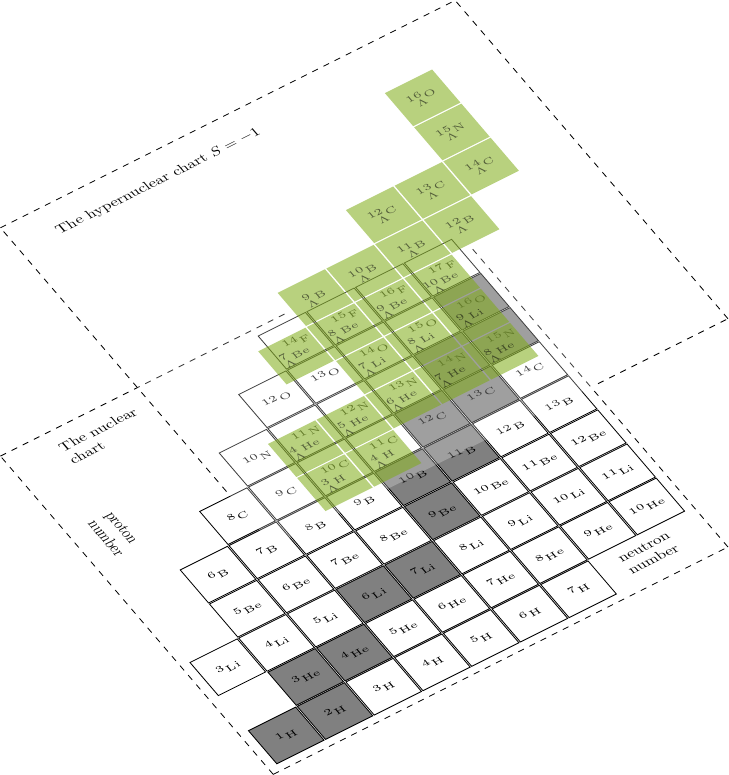}
\caption{
Sketch of the hypernuclear landscape for hypernuclei with strangeness $S=-1$ up to 
the oxygen region. Figure courtesy of Fabian Hildenbrand.
\label{fig:HLS}
}
\end{figure}

Conventional atomic nuclei that make up the visible matter in the Universe are
made of the light quarks $u$ and $d$, generating the two-dimensional nuclear
landscape that is investigated at radioactive beam  and other accelerator facilities worldwide. With the
strange quark, that is considerably heavier than the light quarks, a new dimension
in the nuclear chart opens. While conventional nuclei have strangeness $S=0$,
adding one, two, ... hyperons ($\Lambda, \Sigma, \Xi$ baryons) to a nucleus generates 
hypernuclei with strangeness $S=-1,-2, ..\,$. To understand the formation and
the properties of these ``strange nuclei'' is the main goal of strangeness 
nuclear physics, for an early review see e.g.~\cite{Povh:1978mx}. A sketch
of the hypernuclear landscape for nuclei with atomic number $A\leq 16$ is given in Fig.~\ref{fig:HLS}.

Hypernuclei constitute an important source of information on the 
hyperon-nucleon (YN) interaction. 
Indeed, from the first discovery of a $\Lambda$ hypernucleus in 
1952~\cite{Danysz:1953} up to 1964 the binding (or separation) energies of the
observed hypernuclei \cite{LEVI-SETTI:1958mcn}
were practically the only experimental information
that could be used to estimate the strength of the 
YN interaction and to establish appropriate potential models.
General conclusions like that the $\Lambda$N interaction in the
spin-singlet state ($^1S_0$) has to be more attractive than in the triplet
state ($^3S_1$) were already drawn in those days, based on the observations
that the hypertriton has angular momentum $J=1/2$ and that
the ground state of $^4_\Lambda$H has $J=0$ rather than $J=1$
\cite{Dalitz:1958zza}.
 
From 1964 onward, with the arrival of a more extended set of $\Lambda p$ 
cross sections for low momenta~\cite{SechiZorn:1964,Alexander:1964},
potentials could be constructed by fitting them directly to scattering 
data. Subsequently also low-momentum data for $\Sigma^+p$ and
$\Sigma^-p$ became available~\cite{Engelmann:1966npz,Eisele:1971mk}
and paved the way for a more refined
theoretical treatment of the YN interaction. In particular, exploiting 
the approximate SU(3) flavor symmetry \cite{deSwart:1963pdg} of the quark model 
(and of QCD) allowed to interconnect the interaction between octet baryons
(N, $\Lambda$, $\Sigma$, $\Xi$) and by building on the meson-exchange
picture a realistic and physically sound description of the YN interaction 
could be achieved. The driving force of that development was certainly 
the Nijmegen group which published various potentials over the years
\cite{DeSwart:1971dr,Nagels:1973rq,Nagels:1976xq,Maessen:1989sx,Rijken:1998yy,Nagels:2015lfa},
later in competition with the J\"ulich-Bonn group 
\cite{Holzenkamp:1989tq,Reuber:1993ip,Haidenbauer:2005zh}
and some others~\cite{Greenberg:1992fp,Tominaga:2001ra,Fujiwara:2006yh,Garcilazo:2007ss}. 

Nonetheless, the insight into the properties of the $\Lambda N$ interaction 
that one can gain from the presently available scattering experiments 
\cite{Alexander:1968acu,Sechi-Zorn:1968mao,Kadyk:1971tc,Hauptman:1977hr,CLAS:2021gur,BESIII:2023trh}
is still somewhat limited. In particular, essential features like the
already mentioned spin dependence cannot be deduced from those data.
Therefore, hypernuclei continue to play an essential role in testing and 
improving our understanding of the YN interaction.
Of course, for that aim one needs reliable and efficient tools for evaluating
the properties of hypernuclei such as separation energies directly for
the underlying YN interaction. 
Indeed, it took until 1993 for the first microscopic calculation 
of hypernuclei which utilized realistic meson-exchange YN potentials in 
their full complexity, i.e. with the tensor coupling between the $^3S_1$ and
$^3D_1$ partial waves and with the coupling between the $\Lambda$N and 
$\Sigma$N systems. First results for the hypertriton 
based on the Faddeev approach~\cite{Miyagawa:1993rd,Miyagawa:1995sf}
were followed by calculations of $^4_\Lambda$H and $^4_\Lambda$He
by applying the Faddeev-Yakubovsky formalism~\cite{Nogga:2001ef} and
variational methods~\cite{Hiyama:2001zt,Nemura:2002fu}. 
In fact within a cluster model approach the calculations could be even pushed 
up to A=10-13 $\Lambda$ hypernuclei \cite{Hiyama:2000jd,Hiyama:2012sq}. 
However, this was only possible at the expense of having to use simplified 
phenomenological effective potentials as input. 

Fortunately, in the past decade the theoretical and computational tools have 
been significantly improved. Specifically, new {\it ab initio} methods like
the no-core shell model (NCSM) have been developed 
\cite{Wirth:2014apa,Wirth:2017bpw,Le:2020zdu} that allow one to
compute binding energies for hypernuclei well beyond the $s$-shell, 
retaining the full complexity of the YN interaction. 
So far, studies of hypernuclei up to $^{13}_{\ \Lambda}$C have been
reported \cite{Wirth:2014apa}.  
Moreover, YN interactions based on modern approaches like chiral effective 
field theory (EFT) have been employed. Originally suggested for the 
nucleon-nucleon (NN) interaction, chiral EFT can be easily extended to baryons 
with strangeness, again by exploiting the approximate SU(3) flavor symmetry of QCD
\cite{Polinder:2006zh,Haidenbauer:2013oca,Haidenbauer:2019boi,Haidenbauer:2023qhf}.  
One essential merit of this approach is that two-body forces 
and three-body forces (3BFs) can be treated/derived consistently
\cite{Petschauer:2015elq}. 
Hypernuclei have been also studied in the framework
of nuclear lattice effective field theory (NLEFT) \cite{Hildenbrand:2024ypw}.  
This framework is
a powerful quantum many-body method that combines aspects of effective
field theories with lattice (stochastic) methods. Based on it, results for hypernuclei
up to $^{16}_{\ \Lambda}$O have been obtained. Other approaches like the 
auxiliary field diffusion Monte Carlo algorithm allowed to extend the
investigations to much heavier hypernuclei \cite{Lonardoni:2013rm,Lonardoni:2013gta}, though 
only for simplified representations of the YN interaction.  Recently, the properties of light hypernuclei have been studied within a variational Monte Carlo method based
on  neural network quantum states within a pionless EFT framework, showing  
good agreement with binding energies up to $^{16}_\Lambda$O~\cite{DiDonna:2025oqf}.

Recent calculations of light hypernuclei within {\it ab initio} 
approaches and with realistic $\Lambda N$-$\Sigma N$ potentials as 
input show that the binding energies and mass spectra can be qualitatively well reproduced \cite{Haidenbauer:2019boi,Le:2023bfj}. 
But they also revealed noticeable deviations from the experimental values. Whether those point already to shortcomings in the employed
YN interactions or rather are a sign for the need to include 3BFs 
remains unclear at present. But at least a partial answer has been
given by investigations that employ NN and YN interactions derived within
chiral EFT. In this case the inherent power counting not only determines
the order at which 3BFs start to contribute and, thereby, enables conclusions 
on their relative importance, but it allows also an explicit estimation for 
their magnitude \cite{Le:2023bfj}.
Calculations of $\Lambda$ hypernuclei up to A=7, guided
by such principles, suggest that 3BFs alone could be sufficient to achieve
agreement with the experimental binding energies \cite{Le:2024rkd}.
Similar conclusions 
can be also drawn from the work based on NLEFT, where even hypernuclei
up to A=16 were considered \cite{Hildenbrand:2024ypw}.

Another aspect where measured binding energies of hypernuclei play an
essential role is charge-symmetry breaking (CSB). Up to now there are 
no data on $\Lambda n$ scattering, which one could compare with the 
ones for $\Lambda p$, or for $\Sigma^- n$, which one could compare with
$\Sigma^+p$, for establishing a possible CSB in the $\Lambda N$ and/or
$\Sigma N$ systems. However, and already for a long time, there is clear 
experimental evidence for CSB from the separation energies of the hypernuclei $^4_\Lambda$H and $^4_\Lambda$He, 
in the ground state ($0^+$) as well as in the excited state ($1^+$)
\cite{HypernuclearDataBase}. 
The differences
in the separation energies are in the order of $100$ to $200$~keV and, 
thus, much 
larger than the ones one knows from ordinary nuclei, say
 $^3$H and $^3$He, 
when disregarding the trivial effect due to the Coulomb interaction. 
There is also a noticeable CSB in heavier systems like in the 
isospin triplet $^7_\Lambda$He, $^7_\Lambda$Li$^*$, $^7_\Lambda$Be, 
and the $I=1/2$ doublet $^8_\Lambda$Li, $^8_\Lambda$Be \cite{Botta:2016kqd}.   
CSB effects in light $\Lambda$ hypernuclei have been studied in recent
times based on a heuristic treatment of CSB~\cite{Gazda:2015qyt,Gazda:2016qva}
but also by employing a CSB interaction
derived consistently within chiral EFT \cite{Haidenbauer:2021wld}.  
In the latter case the CSB in the energy
levels can be well described and leads to the conclusion that the $\Lambda n$
interaction in the $^1S_0$ state must be noticeably more attractive than the
one in $\Lambda p$, whereas the difference in the $^3S_1$ is much smaller
and also of opposite sign. 
An exploration has been made to examine whether
the CSB effects established from the $^4_\Lambda$H/$^4_\Lambda$He systems are consistent 
with those in the mentioned $A=7,8$ hypernuclei \cite{Le:2022ikc}.  
But firm conclusions are
difficult to draw because the present experimental uncertainty is large
for the latter. New measurements with improved precision would be highly
desirable for those systems but also for other light hypernuclei for which 
it is still feasible to perform calculations within {\it ab initio}
approaches. 

Finally, it should be mentioned that {\it ab initio} approaches have been 
also successfully applied in calculations of $\Lambda\Lambda$ and $\Xi$ hypernuclei 
\cite{Hiyama:2019kpw,Contessi:2019csf,Le:2021wwz,Le:2021gxa}. 
Candidates for the possible lightest bound systems have been identified, 
for example $^{\ \ 5}_{\Lambda\Lambda}$He \cite{Contessi:2019csf,Le:2021wwz}
or $^4_\Xi$H \cite{Le:2021gxa}. They are waiting for
an experimental confirmation - or disproof - though it is clear that
corresponding experiments are rather challenging. 

We should also mention the femtoscopic approach, that recently has been
used to get information on various YN and YY (as well as other hadronic) interactions and also promises
to pin down three-body forces with unprecedented precision, see e.g.~\cite{ALICE:2021njx,ALICE:2020mfd,ALICE:2019gcn,ALICE:2019eol,ALICE:2019buq,ALICE:2023eyl,ALICE:2023bny}.
While this approach is certainly capable of getting some  information on
interactions that are otherwise difficult to access, in view of the recent
criticism related mostly to the model-dependence of the hadronic source
\cite{Epelbaum:2025aan} and the frequent use of simplified interaction models \cite{Albaladejo:2025lhn},
we do not consider this approach here in detail but just mention it whenever appropriate.

In this review we summarize the present status of {\it ab initio}
calculations of hypernuclei. Several previous review articles exist focusing more on 
experimental data  \cite{Gibson:1995an,Davis:2005mb,Hashimoto:2006aw,Gal:2016boi}
or describing the historical development of hypernuclear physics \cite{Hiyama:2022htx}.
Therefore, in our review we focus specifically on recent developments related to the
application of effective field theories and also we include an introduction
of the pertinent computational approaches.
In Sect.~\ref{second}, we introduce the chiral effective
field theory (EFT) approach to baryon-baryon interactions. The three-baryon forces
from chiral EFT are discussed in Sect.~\ref{sec:BBB}. Then, in Sect.~\ref{sec:third}
we discuss the Faddeev-Yakubovsky approach that allows for {\em ab initio} studies
for hypernuclei with atomic number $A=3,4$. To tackle heavier hypernuclei,
we first consider the No-Core Shell Model (NCSM) approach in Sect.~\ref{sec:fourth}, 
in particular the Jacobi NCSM, and we present results for $p$-shell hypernuclei.
A different {\em ab initio} approach is presented in Sect.~\ref{sec:fifth},
namely Nuclear Lattice EFT (NLEFT). We outline how it can be applied to hypernuclei
and present result for hypernuclei up-to-and including $A=16$. We end with a
summary and outlook in Sect.~\ref{sec:sum}.


    \section{Baryon-baryon interaction in chiral 
 effective field theory}\label{second}

The baryon-baryon (BB) interactions considered in the
present review are all constructed 
within SU(3) chiral effective field theory (EFT), an
approach which exploits the symmetries of the underlying fundamental theory of strong
interactions, quantum chromodynamics (QCD), 
and the breaking of those symmetries.
In that approach a potential is established
via an expansion in terms of small momenta, subject to an appropriate power counting, so
that the results can be improved systematically by going to higher orders, while at the same
time theoretical uncertainties can be estimated
\cite{Epelbaum:2008ga,Machleidt:2011zz}. 
Furthermore, two- and three-baryon
forces can be constructed in a consistent way. The resulting interaction potentials can be
readily employed in standard two- and few-body calculations. They consist of contributions
from an increasing number of pseudoscalar-meson exchanges, determined by the underlying
chiral symmetry, and of contact terms which encode the unresolved short-distance dynamics
and whose strengths are parameterized by a priori unknown low-energy constants (LECs). The latter need to be determined by 
a fit to experimental data. 

In the following we describe the construction of the
potential within the Weinberg power counting.
So far the J\"ulich-Bonn (and partly J\"ulich-Bonn-TUM) group
has established YN potentials at leading order (LO)
\cite{Polinder:2006zh}, up to next-to-leading order (NLO)
\cite{Haidenbauer:2013oca,Haidenbauer:2019boi}, and, recently
even up to next-to-next-to-leading order (N2LO) 
\cite{Haidenbauer:2023qhf} in the chiral expansion. 
We note that alternative approaches have been pursued in \cite{Korpa:2001au},
where the so-called Kaplan-Savage-Weise (KSW) resummation 
scheme \cite{Kaplan:1998we} has been applied, and in \cite{Li:2016paq,Ren:2019qow,Song:2021yab} where calculations
are performed in covariant chiral EFT, but so far only at LO. 
The $\Lambda$N and $\Lambda\Lambda$ interactions have been also treated
within pion-less EFT 
\cite{Hammer:2001ng,Hildenbrand:2019sgp,Contessi:2018qnz,Contessi:2019csf}. 

We would like to mention that there are also ongoing efforts to
compute the YN interaction within lattice QCD, spearheaded
by the NPLQCD \cite{Beane:2010em} and HAL QCD \cite{Aoki:2012tk} collaborations. 
In fact, over the past few years it has become possible to perform such lattice simulations for
quark (pion) masses close to the physical point and
the HAL QCD collaboration has published corresponding results
for $\La N$ \cite{Nemura:2022wcs} and $\Si N$ 
\cite{Nemura:2017vjc}, and also for $S=-2$ systems
\cite{HALQCD:2019wsz}.

 \subsection{Formalism}
\label{sec:form}
\begin{figure}[t]
 \centering
\includegraphics[width=\feynwidth]{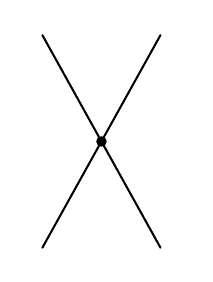}
 \includegraphics[width=\feynwidth]{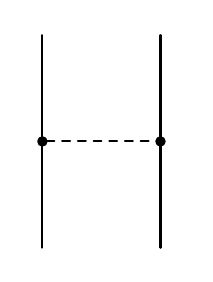}
 
\includegraphics[width=\feynwidth]{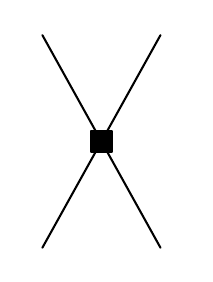}
 \includegraphics[width=\feynwidth]{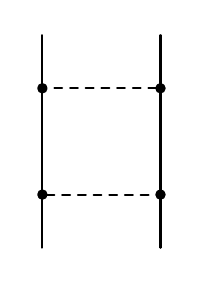}
 \includegraphics[width=\feynwidth]{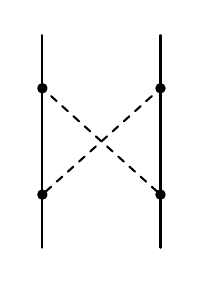}
 \includegraphics[width=\feynwidth]{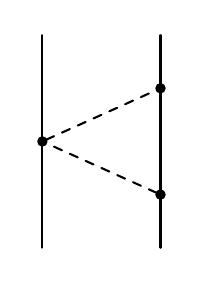}
 \includegraphics[width=\feynwidth]{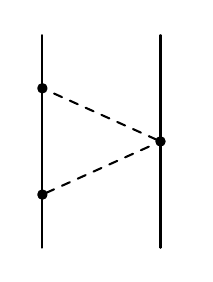}
 \includegraphics[width=\feynwidth]{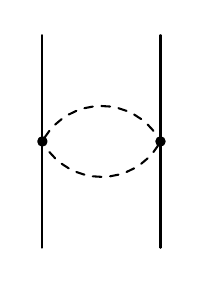}

\includegraphics[width=\feynwidthb]{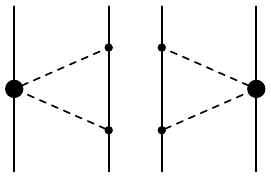}{\quad}
 \includegraphics[width=\feynwidthb]{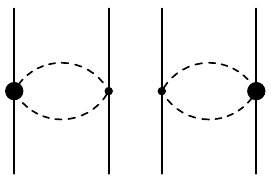}
 \caption{Diagrams contributing at LO (top), NLO (center), and N2LO (bottom) in chiral EFT. 
 Solid and dashed lines
 denote octet baryons and pseudoscalar mesons, respectively. For the meson-baryon interaction,
 the small (fat) circle denotes a one-(two-)derivative vertex, whereas for the four-baryon contact
 terms the small circle (big square) represents a zero-(two-)derivative interaction.
} 
\label{fig:chEFT}
\end{figure}

The application of chiral effective field theory to the NN system 
is thoroughly documented in various reviews, see e.g. 
\cite{Epelbaum:2008ga,Machleidt:2011zz}. 
Details on the derivation of the chiral baryon-baryon potentials for the strangeness sector 
at LO using the Weinberg power counting can be found in Refs.~\cite{Polinder:2006zh}
and \cite{Haidenbauer:2007ra}. 
The LO potential consists of four-baryon contact terms without derivatives and of one-pseudoscalar-meson exchanges, see Fig.~\ref{fig:chEFT} (top). 
 
The LO SU(3) invariant contact terms for the octet baryon-baryon interactions that are Hermitian 
and invariant under Lorentz transformations follow from the Lagrangians
\begin{eqnarray}
{\mathcal L}^1 &=& C^1_i \left<\bar{B}_a\bar{B}_b\left(\Gamma_i B\right)_b\left(\Gamma_i B\right)_a\right>\ , \nonumber \\
{\mathcal L}^2 &=& C^2_i \left<\bar{B}_a\left(\Gamma_i B\right)_a\bar{B}_b\left(\Gamma_i B\right)_b\right>\ , \nonumber \\
{\mathcal L}^3 &=& C^3_i \left<\bar{B}_a\left(\Gamma_i B\right)_a\right>\left<\bar{B}_b\left(\Gamma_i B\right)_b\right>\  .
\label{jeq:2.1}
\end{eqnarray}
Here $a$ and $b$ denote the Dirac indices of the particles, $B$ is the irreducible octet representation of ${\rm SU(3)}_{\rm f}$, in form of
a traceless \(3\times3\) matrix, 
\begin{equation}
  B=
 \begin{pmatrix}
  \frac{\Sigma^0}{\sqrt 2} + \frac{\Lambda}{\sqrt 6} & \Sigma^+ & p \\
  \Sigma^- & -\frac{\Sigma^0}{\sqrt 2} + \frac{\Lambda}{\sqrt 6} & n \\
  -\Xi^- & \Xi^0 & -\frac{2\Lambda}{\sqrt 6}
 \end{pmatrix}\,.
\end{equation}
and the $\Gamma_i$ are the usual elements of the 
Clifford algebra \cite{Polinder:2006zh}. 
 
As described in Ref.~\cite{Polinder:2006zh}, 
to LO the Lagrangians in Eq.~(\ref{jeq:2.1}) give rise to only six independent 
low-energy coefficients (LECs) -- $C^1_S$, $C^1_T$, $C^2_S$, $C^2_T$, $C^3_S$ and $C^3_T$ --
due to SU(3) constraints. 
The subscripts $S$ and $T$ refer to the central and spin-spin parts of the 
potential, respectively. 
The spin- and momentum structure of the potentials to LO is given by 
\begin{eqnarray}
V^{(0),\, CT} &=& C_S + C_T 
(\mbox{\boldmath $\sigma$}_1\cdot\mbox{\boldmath $\sigma$}_2) .
\end{eqnarray} 
For concrete applications it is convenient to re-express the 
LECs $C^i_S$ and $C^i_T$ in terms of LECs that correspond to the 
SU(3) irreducible representations,
$C^{27}$, $C^{10}$, $C^{10^*}$, $C^{8_s}$, $C^{8_a}$, $C^1$, 
according to the decomposition 
of the tensor product for two octet baryons, 
$\ir8 \otimes \ir8 =
 \ir{27} \oplus \ir{10} \oplus \ir{\overline{10}} \oplus \ir{8}_s
\oplus \ir{8}_a \oplus \ir{1} $, 
see e.g. \cite{deSwart:1963pdg,Dover:1991sh}. 
Then the contact interaction can be written in the practical form  
\begin{equation}
V^{(0),\, CT}=
\frac{1}{4}(1-\mbox{\boldmath $\sigma$}_1\cdot \mbox{\boldmath $\sigma$}_2) \, C_{1S0}
+ \frac{1}{4}(3+\mbox{\boldmath $\sigma$}_1 \cdot\mbox{\boldmath $\sigma$}_2) \, C_{3S1} \ ,
\label{jeq:contact}
\end{equation}
where the LECs for the $^1S_0$ and $^3S_1$ partial waves 
in the various baryon-baryon channels are linear combinations of 
$C^{27}$, $C^{8_s}$, $C^{1}$ and $C^{10}$, $C^{10^*}$,  $C^{8_a}$, respectively. 
The explicit relations which express 
the constraints imposed by the assumed SU(3) flavor symmetry are
summarized in Table~\ref{tab:SU3}.
 
The lowest order SU(3) invariant pseudoscalar-meson--baryon
interaction Lagrangian with the appropriate symmetries was discussed in \cite{Polinder:2006zh,Haidenbauer:2013oca}. 
Introducing the matrix $\phi$ representing the pseudoscalar mesons,
\begin{equation}
 \phi =
 \begin{pmatrix}
  \frac{\pi^0}{\sqrt 2} + \frac{\eta}{\sqrt 6} & \pi^+ & K^+ \\
  \pi^- & -\frac{\pi^0}{\sqrt 2} + \frac{\eta}{\sqrt 6} & K^0 \\
   K^- & \Kb^0 & -\frac{2\eta}{\sqrt 6}
 \end{pmatrix}\,,
\end{equation}
it can be written in the form 
\begin{equation} \label{eq:pseudovector}
  \mathcal{L}_{BB\phi} = -\frac{\sqrt2}{2F_0}\left<D \bar B \gamma^\mu\gamma_5\left\{\partial_\mu \phi,B\right\} + F \bar B \gamma^\mu\gamma_5\left[\partial_\mu \phi,B\right]\right> \,,\\
 \end{equation}
which leads to a vertex between two baryons and one meson.
Here, $F_0$ is the pseudoscalar-meson decay constant in the chiral limit. 
$F$ and $D$ are coupling constants which satisfy the relation $F+D=g_A\simeq 1.26$, where
$g_A$ is the axial-vector strength measured in neutron $\beta$--decay. Further, $\langle ...\rangle$
denotes the trace in flavor space.

In the isospin basis the Lagrangian reads
\begin{eqnarray}
{\mathcal L}&=&-f_{NN\pi}\bar{N}\gamma^\mu\gamma_5\mbox{\boldmath $\tau$}N\cdot\partial_\mu\mbox{\boldmath $\pi$} 
+if_{\Sigma\Sigma\pi}\bar{\mbox{\boldmath $ \Sigma$}}\gamma^\mu\gamma_5\times{\mbox{\boldmath $ \Sigma$}}\cdot\partial_\mu\mbox{\boldmath $\pi$} \nonumber \\
&&-f_{\Lambda\Sigma\pi}\left[\bar{\Lambda}\gamma^\mu\gamma_5{\mbox{\boldmath $ \Sigma$}}+\bar{\mbox{\boldmath $\Sigma$}}\gamma^\mu\gamma_5
\Lambda\right]\cdot\partial_\mu\mbox{\boldmath $\pi$}-f_{\Xi\Xi\pi}\bar{\Xi}\gamma^\mu\gamma_5\mbox{\boldmath $\tau$}\Xi\cdot\partial_\mu\mbox{\boldmath $\pi$} \nonumber \\
&&-f_{\Lambda NK}\left[\bar{N}\gamma^\mu\gamma_5\Lambda\partial_\mu K+\bar{\Lambda}\gamma^\mu\gamma_5N\partial_\mu K^\dagger\right]
\nonumber \\&&
-f_{\Xi\Lambda K}\left[\bar{\Xi}\gamma^\mu\gamma_5\Lambda\partial_\mu K_c+\bar{\Lambda}\gamma^\mu\gamma_5\Xi\partial_\mu K_c^\dagger\right]
\nonumber \\&&
-f_{\Sigma NK}\left[\bar{\mbox{\boldmath $ \Sigma$}}\cdot\gamma^\mu\gamma_5\partial_\mu K^\dagger\mbox{\boldmath $\tau$}N+\bar{N}\gamma^\mu\gamma_5\mbox{\boldmath $\tau$}\partial_\mu K\cdot{\mbox{\boldmath $ \Sigma$}}\right]
\nonumber \\&&
-f_{\Xi \Sigma K}\left[\bar{\mbox{\boldmath $ \Sigma$}}\cdot\gamma^\mu\gamma_5\partial_\mu K_c^\dagger\mbox{\boldmath $\tau$}\Xi+\bar{\Xi}
\gamma^\mu\gamma_5\mbox{\boldmath $\tau$}\partial_\mu K_c\cdot{\mbox{\boldmath $ \Sigma$}}\right]
\nonumber \\&&
-f_{NN\eta_8}\bar{N}\gamma^\mu\gamma_5N\partial_\mu\eta
\nonumber \\&&
-f_{\Lambda\Lambda\eta_8}\bar{\Lambda}\gamma^\mu\gamma_5\Lambda\partial_\mu\eta-f_{\Sigma\Sigma\eta_8}\bar{\mbox{\boldmath $ \Sigma$}}\cdot\gamma^\mu\gamma_5{\mbox{\boldmath $ \Sigma$}}\partial_\mu\eta
\nonumber \\&& 
-f_{\Xi\Xi\eta_8}\bar{\Xi}\gamma^\mu\gamma_5\Xi\partial_\mu\eta \ .
\label{jeq:2.3}
\end{eqnarray}

The interaction Lagrangian in Eq.~(\ref{jeq:2.3}) is invariant under SU(3) transformations if the various coupling constants fulfill specific relations which can be expressed in terms of the coupling constant $f$ and the $F/(F+D)$-ratio $\alpha$ as \cite{deSwart:1963pdg},
\begin{equation}
\begin{array}{rlrl}
f_{NN\pi}  = & f, & f_{NN\eta_8}  = & \frac{1}{\sqrt{3}}(4\alpha -1)f, \\
f_{\Lambda NK} = & -\frac{1}{\sqrt{3}}(1+2\alpha)f, & f_{\Xi\Xi\pi}  = & -(1-2\alpha)f, \\
f_{\Xi\Xi\eta_8}  = & -\frac{1}{\sqrt{3}}(1+2\alpha )f, & f_{\Xi\Lambda K} = & \frac{1}{\sqrt{3}}(4\alpha-1)f, \\
f_{\Lambda\Sigma\pi}  = & \frac{2}{\sqrt{3}}(1-\alpha)f, & f_{\Sigma\Sigma\eta_8}  = & \frac{2}{\sqrt{3}}(1-\alpha )f,\\
f_{\Sigma NK} = & (1-2\alpha)f, & f_{\Sigma\Sigma\pi}  = & 2\alpha f, \\
f_{\Lambda\Lambda\eta_8}  = & -\frac{2}{\sqrt{3}}(1-\alpha )f, & f_{\Xi\Sigma K} = & -f \, ,
\end{array}
\label{jeq:2.5}
\end{equation}
where $f\equiv g_A/2F_0$. In the actual calculations $F_0$ was fixed to
the weak pion decay constant, $F_\pi =  92.4$ MeV (LO and NLO). 
In the N2LO potential the explicit SU(3) breaking of the decay constant
was taken into account by using the empirical values for $F_\pi$, $F_K$, 
and $F_\eta$ \cite{Haidenbauer:2023qhf}. 
For the $F/(F+D)$-ratio the SU(6) value ($\alpha=0.4$) was adopted in 
the LO, NLO, and N2LO studies.

The spin-space part of the LO one-pseudoscalar-meson-exchange potential is similar to the 
static one-pion-exchange potential in \cite{Epelbaum:1998ka} (recoil and relativistic corrections give 
higher order contributions),
\begin{eqnarray}
V^{(0),\,OBE}_{B_1B_2\to B_3B_4}
&=&-f_{B_1B_3\phi}f_{B_2B_4\phi}\frac{\left(\mbox{\boldmath $\sigma$}_1\cdot{\bf q}\right)\left(\mbox{\boldmath $\sigma$}_2\cdot{\bf q}\right)}{{\bf q}^2+M^2_\phi}\ ,
\label{jeq:14}
\end{eqnarray}
where $M_\phi$ is the mass of the exchanged pseudoscalar meson. The transferred 
and average momentum, ${\bf q}$ and ${\bf k}$, are defined in terms of the final and initial 
center-of-mass (c.m.) momenta of the baryons, ${\bf p}'$ and ${\bf p}$, as 
${\bf q}={\bf p}'-{\bf p}$ and ${\bf k}=({\bf p}'+{\bf p})/2$. 
Note that we use the physical masses of the exchanged pseudoscalar mesons. 
Thus, the explicit ${\rm SU(3)}$ breaking reflected in the mass splitting between the 
pseudoscalar mesons is taken into account. 
The $\eta$ meson was identified with the octet $\eta$ ($\eta_8$) and its physical 
mass was used.


In next-to-leading order (NLO) contributions from (non-iterated) 
two-pseudoscalar-meson exchange diagrams arise  \cite{Epelbaum:2004fk,Haidenbauer:2013oca}. 
Those involve the leading meson-baryon Lagrangian 
\begin{equation}
  \mathcal{L}_{PB} = \frac1{4F_0^2}\left<\mathrm i \bar B \gamma^\mu \left[\left[\phi,\partial_\mu \phi\right],B\right]\right>\,,\\
\end{equation}
which describes a (Weinberg-Tomozawa) vertex between two baryons and two mesons.
Explicit expressions for the resulting NLO potential can be found in the appendix
of Ref.~\cite{Haidenbauer:2013oca}. A graphical representation of the 
contributions is provided in the middle line of Fig.~\ref{fig:chEFT} and
consists of (irreducible) planar box, crossed box, triangle, and 
football diagrams. 
In addition there are four-baryon contact terms with two derivatives,
which can be constructed analogous to Eq.~(\ref{jeq:2.1}).
The spin- and momentum structure of the latter is given by 
\begin{eqnarray}
V^{(2),\,CT} &=& C_1 {\bf q}^{\,2}+ C_2 {\bf k}^{\,2} + (C_3 {\bf q}^{\,2}+ C_4 {\bf k}^{\,2})
(\mbox{\boldmath $\sigma$}_1\cdot\mbox{\boldmath $\sigma$}_2) \nonumber \\
&+& i C_5 (\mbox{\boldmath $\sigma$}_1+\mbox{\boldmath $\sigma$}_2)\cdot ({\bf q} \times {\bf k})
+ C_6 ({\bf q} \cdot \mbox{\boldmath $\sigma$}_1) ({\bf q} \cdot \mbox{\boldmath $\sigma$}_2)
\nonumber \\
&+& C_7 ({\bf k} \cdot \mbox{\boldmath $\sigma$}_1) ({\bf k} \cdot \mbox{\boldmath $\sigma$}_2)
+ i C_8 (\mbox{\boldmath $\sigma$}_1-\mbox{\boldmath $\sigma$}_2)\cdot ({\bf q} \times {\bf k}). 
\label{jeq:15}
\end{eqnarray}
The $C_i$'s are additional LECs. Performing a partial wave projection and
imposing again SU(3) symmetry one finds that in case of the 
$Y$N interaction 
there are ten new coefficients entering the $S$ waves and $S$-$D$ transitions, 
respectively, and eight coefficients in the $P$ waves. The potential up to NLO
can then be written in the form
\begin{eqnarray}
\label{VCT}
V^{CT}_{B_1B_2\to B_3B_4}
&=& \tilde{C}_{\alpha} + {C}_{\alpha} \, ({p}^2+{p}'^2)~
\quad (\alpha =\,^1S_0,\, ^3S_1), \nonumber \\
&=& {C}_{\beta}\, {p}^2~, \ {C}_{\beta}\, {p'}^2~ 
\quad \quad \quad (\beta =\,^3S_1-^3D_1,\,^3D_1-^3S_1), \\
&=& {C}_{\gamma}\, {p}\, {p}'~
\qquad \qquad \qquad \  (\gamma = \,^3P_0,\,^1P_1,\,^3P_1,\,^3P_2), \nonumber
\end{eqnarray}
with $p = |{\bf p}\,|$ and ${p}' = |{\bf p}\,'|$. 
$\tilde C_\alpha$ and $C_\alpha$, etc. are appropriate combinations of the
$C_i$'s appearing in Eqs.~(\ref{jeq:contact}) and (\ref{jeq:15}), see Ref.~\cite{Haidenbauer:2013oca}. As before, for each partial wave the 
contact interactions in the various BB channels are interrelated with each
other by the imposed SU(3) symmetry, see Table~\ref{tab:SU3}.
Note that at NLO additional contact terms arise from an insertion of the
external field $\chi$ \cite{Petschauer:2013uua}, 
\begin{equation}
 \chi = 2B_0\begin{pmatrix} m_u & 0 & 0 \\ 0 & m_d & 0 \\ 0 & 0 & m_s \end{pmatrix}
 \approx \begin{pmatrix} M_\pi^2 & 0 & 0 \\ 0 & M_\pi^2 & 0 \\ 0 & 0 & 2M_K^2-M_\pi^2 \end{pmatrix} \, ,
 \label{VCTSB1}
\end{equation}
into the Lagrangians in Eq.~(\ref{jeq:2.1}). 
The parameter $B_0$ is related to the vacuum quark condensate. Those introduce an 
explicit SU(3) symmetry breaking~\cite{Haidenbauer:2013oca,Petschauer:2013uua} 
and lead to a modification of the $S$-wave contact potentials given in 
Eq.~(\ref{VCT}):
\begin{eqnarray}
\label{VCTSB2}
V^{CT}_{B_1B_2\to B_3B_4}
&=& \tilde{C}_{\alpha} + {C}_{\alpha} \, ({p}^2+{p}'^2) + 
{C}_{\alpha,\,\chi} \, (M^2_K - M^2_\pi)~.
\end{eqnarray}
The pertinent LECs ${C}_{\alpha,\,\chi}$ are likewise constrained by SU(3) 
symmetry and 
the relations are given in the last two columns of Table~\ref{tab:SU3}.

Finally, at next-to-next-to-leading order (N2LO), contributions involving the
subleading meson-baryon Lagrangian \cite{Haidenbauer:2023qhf} arise 
and are depicted at the bottom of Fig.~\ref{fig:chEFT}. Explicit expressions
for those contributions can be found in Ref.~\cite{Haidenbauer:2023qhf}.

\begin{table}[ht]
\caption{SU(3) relations for the various contact potentials in the 
isospin basis 
for strangeness $S=0,\;-1$, and $-2$. 
$C^{27}_{\alpha}$ etc. refers to the corresponding irreducible SU(3) representation
for a particular partial wave ${\alpha}$, see text and Ref.~\cite{Haidenbauer:2013oca}.
The isospin is denoted by $I$. The actual potential still needs to be multiplied 
by pertinent powers of the momenta $p$ and $p'$.
}
\label{tab:SU3}
\vskip 0.1cm
\renewcommand{\arraystretch}{1.3}
\centering
\resizebox*{.7\textheight}{!}{
\begin{tabular}{|l|c|c||c|c||c|c|}
\hline
S&Channel &I &\multicolumn{4}{|c|}{$V({\alpha})$} \\
\hline
\hline
&        &  &$\alpha = \,^1S_0, \, ^3P_0, \, ^3P_1, \, ^3P_2 $
& $\alpha= \,^3S_1, \, ^3S_1$-$^3D_1, \, ^1P_1$ & $^1S_0$ $\chi$ & $^3S_1$ $\chi$ \\
\hline
${\phantom{-}0}$&$NN\rightarrow NN$ &$0$ & \ \ -- & $\C^{10^*}$ &  -- &  $\frac{1}{2}C^7_\chi$ \\
                       &$NN\rightarrow NN$ &$1$ & $\C^{27}$ &  -- &  $\frac{1}{2}C^1_\chi$ &  --\\
\hline
\hline
${-1}$&$\La N \rightarrow \La N$ &$\frac{1}{2}$ &$\frac{1}{10}\left(9C^{27}_{\alpha}+C^{8_s}_{\alpha}\right)$
& $\frac{1}{2}\left(C^{8_a}_{\alpha}+C^{10^*}_{\alpha}\right)$   & $C^{2}_{\chi}$ &  $C^{8}_{\chi}$ \\
&$\La N \rightarrow \Si N$ &$\frac{1}{2}$        &$\frac{3}{10}\left(-C^{27}_{\alpha}+C^{8_s}_{\alpha}\right)$
& $\frac{1}{2}\left(-C^{8_a}_{\alpha}+C^{10^*}_{\alpha}\right)$   & $-C^{3}_{\chi}$& $ -C^{9}_{\chi}$ \\
&$\Si N \rightarrow \Si N$  &$\frac{1}{2}$        &$\frac{1}{10}\left(C^{27}_{\alpha}+9C^{8_s}_{\alpha}\right)$
& $\frac{1}{2}\left(C^{8_a}_{\alpha}+C^{10^*}_{\alpha}\right)$   & $C^{4}_{\chi} $ & $C^{10}_{\chi}$\\
&$\Si N \rightarrow \Si N$  &$\frac{3}{2}$        &$C^{27}_{\alpha}$
& $C^{10}_{\alpha}$ & $\frac{1}{4}C^1_\chi$ & $-\frac{1}{4}C^7_\chi$ \\
\hline
\hline
${-2}$&$\La\La \rightarrow \La\La$ &$0$   &$\frac{1}{40}\left(27C^{27}_{\alpha}+8C^{8_s}_{\alpha}+5{C^{1}_{\alpha}}\right)$ & --
 & $\frac{1}{2}C^5_\chi$ & --\\
&$\La\La\rightarrow \Xi N$ &$0$  &$\frac{-1}{40}\left(18C^{27}_{\alpha}-8C^{8_s}_{\alpha}-10{C^{1}_{\alpha}}\right)$ & --
 & $\frac{3}{4}C^1_\chi-{3}C^2_\chi -C^3_\chi + \frac{3}{4}C^5_\chi$ & -- \\
&$\La\La\rightarrow \Si\Si$ &$0$ &$\frac{\sqrt{3}}{40}\left(-3C^{27}_{\alpha}+8C^{8_s}_{\alpha}-5{C^{1}_{\alpha}}\right)$ & --
 & --  & --\\
&$\Xi N \rightarrow \Xi N$ &$0$  &$\frac{1}{40}\left(12C^{27}_{\alpha}+8C^{8_s}_{\alpha}+20{C^{1}_{\alpha}}\right)$ & $C^{8_a}_{\alpha}$ 
 &$\frac{2}{3}C^1_\chi-{3}C^2_\chi +\frac{1}{3}C^4_\chi + \frac{9}{8}C^5_\chi$ & $C^{11}_{\chi}$ \\
&$\Xi N \rightarrow \Si\Si$ &$0$ &$\frac{\sqrt{3}}{40}\left(2C^{27}_{\alpha}+8C^{8_s}_{\alpha}-10{C^{1}_{\alpha}}\right)$ & -- 
 &$-\frac{1}{4\sqrt{3}}C^1_\chi+{\sqrt{3}}C^3_\chi +\frac{1}{\sqrt{3}}C^4_\chi$ & --\\
&$\Si\Si\rightarrow \Si\Si$ &$0$ &$\frac{1}{40}\left(C^{27}_{\alpha}+24C^{8_s}_{\alpha}+15{C^{1}_{\alpha}}\right)$ & --
 & --& --\\
&$\Xi N \rightarrow \Xi N$ &$1$  &$\frac{1}{5}\left(2C^{27}_{\alpha}+3C^{8_s}_{\alpha}\right)$ & $\frac{1}{3}\left(C^{10}_{\alpha}+C^{10^*}_{\alpha}+C^{8_a}_{\alpha}\right)$ 
 & $C^{6}_{\chi}$ &  $C^{12}_{\chi}$ \\
&$\Xi N \rightarrow \Si\La$ &$1$ &$\frac{\sqrt{6}}{5}\left(C^{27}_{\alpha}-C^{8_s}_{\alpha}\right)$ & $\frac{\sqrt{6}}{6}\left(C^{10}_{\alpha}-C^{10^*}_{\alpha}\right)$ 
 &$-\frac{\sqrt{2}}{3\sqrt{3}}C^{1}_\chi+\sqrt{\frac{3}{2}}C^2_\chi-\frac{1}{3\sqrt{6}}C^4_\chi -\sqrt{\frac{2}{3}}C^6_\chi$ 
 &$-\frac{1}{\sqrt{6}}C^{10}_\chi+\sqrt{\frac{2}{3}}C^{12}_\chi+\frac{1}{2\sqrt{6}}C^7_\chi 
   -\sqrt{\frac{3}{2}}C^8_\chi +\sqrt{\frac{2}{3}}C^9_\chi$ \\
&$\Xi N \rightarrow \Si\Si$ &$1$ & --& $\frac{\sqrt{2}}{6}\left(C^{10}_{\alpha}+C^{10^*}_{\alpha}-2C^{8_a}_{\alpha}\right)$ 
 & -- & $\sqrt{2}C^{10}_\chi-\frac{1}{2\sqrt{2}}C^7_\chi-{\sqrt{2}}C^9_\chi$ \\
&$\Si\La\rightarrow \Si\La$ &$1$ &$\frac{1}{5}\left(3C^{27}_{\alpha}+2C^{8_s}_{\alpha}\right)$ & $\frac{1}{2}\left(C^{10}_{\alpha}+C^{10^*}_{\alpha}\right)$ 
  &$-\frac{1}{9}C^1_\chi+\frac{4}{3}C^3_\chi +\frac{4}{9}C^4_\chi + \frac{2}{3}C^6_\chi$ 
  &$\frac{4}{3}C^{10}_\chi+\frac{2}{3}C^{12}_\chi -\frac{1}{3}C^7_\chi - \frac{4}{3}C^9_\chi$ \\
&$\Si\La\rightarrow \Si\Si$ &$1$ & --&$\frac{\sqrt{3}}{6}\left(C^{10}_{\alpha}-C^{10^*}_{\alpha}\right)$ 
 & -- & --\\
&$\Si\Si\rightarrow \Si\Si$ &$1$ & --&$\frac{1}{6}\left(C^{10}_{\alpha}+C^{10^*}_{\alpha}+4C^{8_a}_{\alpha}\right)$ 
 & -- & --\\
&$\Si\Si\rightarrow \Si\Si$ &$2$   &$C^{27}_{\alpha}$ & --
 & -- & --\\
\hline
\hline
\end{tabular}
}
\renewcommand{\arraystretch}{1.0}
\end{table}


The presently available YN scattering data do not allow
any conclusions on possible effects from charge symmetry
breaking (CSB), say regarding $\La p$ versus $\La n$ or
$\Si^+p$ versus $\Si^-n$. 
However, the experimentally established difference  
of the $\Lambda$ separation energies
in the mirror nuclei ${^4_\Lambda \rm He}$ and ${^4_\Lambda \rm  H}$
provides clear evidence for a CSB between the
$\La p$ and $\La n$ interactions. 
CSB effects can be included in the YN interaction
\cite{Haidenbauer:2021wld} following the arguments and the 
power counting outlined for the NN system
in Ref.~\cite{Epelbaum:2004fk}. 
According to that work (and notation) at L{\O}
CSB forces are due to the Coulomb interaction and due to the
mass differences of the exchanged pseudoscalar mesons. 
However, in case of $\La$N, the former is absent while
the latter is rather small, since the mass difference 
between the $K^0$ and $K^\pm$ mesons is small compared to their
averaged mass. More important are
CSB contributions to the $\La N$ potential at NL{\O}
which are given by \cite{Haidenbauer:2021wld} 
\begin{eqnarray}
V^{CSB}_{\La N\to \La N}
&=& \Bigg[ -f^{(\La-\Si^0)}_{\La\La\pi} f_{NN\pi}\frac{\left(\mbox{\boldmath $\sigma$}_1\cdot{\bf q}\right)
\left(\mbox{\boldmath $\sigma$}_2\cdot{\bf q}\right)}{{\bf q}^2+M^2_{\pi^0}} \nonumber\\
&-&f^{(\eta-\pi^0)}_{\La\La\pi} f_{NN\pi}\left(\mbox{\boldmath $\sigma$}_1\cdot{\bf q}\right)
\left(\mbox{\boldmath $\sigma$}_2\cdot{\bf q}\right)
\left( \frac{1}{{\bf q}^2+M^2_{\pi^0}}-\frac{1}{{\bf q}^2+M^2_{\eta}} \right) \nonumber\\
&+&\frac{1}{4}(1-\mbox{\boldmath $\sigma$}_1\cdot\mbox{\boldmath $\sigma$}_2)\, C^{CSB}_{^1S_0}
\,+\,\frac{1}{4}(3+\mbox{\boldmath $\sigma$}_1\cdot\mbox{\boldmath $\sigma$}_2)\, C^{CSB}_{^3S_1}
\Bigg] \  \tau_N \ .
\label{VCSB}
\end{eqnarray}
These CSB contributions arise from a non-zero $\La\La\pi$
coupling constant which can be estimated from
$\Lambda-\Sigma^0$ ($f^{(\La-\Si^0)}_{\La\La\pi}$) and $\eta-\pi^0$ 
($f^{(\eta-\pi^0)}_{\La\La\pi}$) mixing \cite{Dalitz:1964es}, respectively,
and from two contact terms, $C^{CSB}_{^1S_0}$ and $C^{CSB}_{^3S_1}$,
that represent short-ranged CSB forces. Note that 
$\tau_p = 1$ and $\tau_n =-1$.

The reaction amplitudes are obtained from the solution of a coupled-channels Lippmann-Schwinger (LS) 
equation for the interaction potentials: 
\begin{eqnarray}
T_{\rho''\rho'}^{\nu''\nu',J}(p'',p';E)=V_{\rho''\rho'}^{\nu''\nu',J}(p'',p')+
\sum_{\rho,\nu}\int_0^\infty \frac{dpp^2}{(2\pi)^3} \, V_{\rho''\rho}^{\nu''\nu,J}(p'',p)
\frac{2\mu_{\nu}}{q_{\nu}^2-p^2+i\eta}T_{\rho\rho'}^{\nu\nu',J}(p,p';E)\ .
\label{jeq:LS}
\end{eqnarray}
The label $\nu$ indicates the particle channels and the label $\rho$ the partial wave. $\mu_\nu$ is the pertinent reduced mass. The on-shell momentum in the intermediate state, $q_{\nu}$, is defined by 
$E=E_{B_{1,\nu}} + E_{B_{2,\nu}} =
\sqrt{m^2_{B_{1,\nu}}+q_{\nu}^2}+\sqrt{m^2_{B_{2,\nu}}+q_{\nu}^2}$. Relativistic kinematics is used for relating the laboratory energy $T_{{\rm lab}}$ of the hyperons to the c.m. momentum.
We solve the LS equation in the particle basis, in order to incorporate the correct physical
thresholds. Depending on the specific values of strangeness and charge up to six baryon-baryon
channels can couple. For the $S=-1$ sector, where a detailed comparison with scattering 
data is 
possible, the Coulomb interaction is taken into account appropriately. 

When the chiral potentials are inserted into the LS equation a regulator has to 
be introduced in order to remove high-momentum components of the baryon 
and pseudoscalar meson fields \cite{Epelbaum:2004fk}. 
In the chiral YN potentials of the J\"ulich-Bonn group up 
to 2019 
\cite{Polinder:2006zh,Haidenbauer:2013oca,Haidenbauer:2019boi}
a non-local regulator function of the form
$\exp\left[-\left(p'^4+p^4\right)/\Lambda^4\right]$ 
has been applied, analogous to the procedure used for chiral NN potentials
\cite{Epelbaum:2004fk}. 
In particular, values for the cut-off mass $\La$ 
in a comparable range, i.e. 550, ..., 700 MeV, were considered. 
Please note that there are two variants of the NLO
interaction, which are denoted according to the year of
publication, i.e. NLO13 \cite{Haidenbauer:2013oca}
and NLO19 \cite{Haidenbauer:2019boi}, respectively. These yield practically equivalent results for $\La$N and $\Si$N
scattering observables, but they differ in ``non-observable''
properties like the strength of the $\La$N$\to$$\Si$N transition potential~\cite{Haidenbauer:2019boi}. 
The impact of this difference on predictions for light
hypernuclei is instructive and will be discussed below.

In the extension of the YN potentials to N2LO in
\cite{Haidenbauer:2023qhf} a novel regularization scheme
has been adopted, following the development in the
NN sector \cite{Reinert:2017usi}.
Specifically, now a local regulator is applied
to the meson-exchange contributions and only the contact terms,
being non-local by themselves, are regularized with a 
non-local function. Following Reinert et al.~\cite{Reinert:2017usi}
we call the resulting interactions 
semilocal momentum-space (SMS) regularized potentials. 
Here the considered cut-off masses $\La$ 
are in the range of 500, ..., 600 MeV, i.e. slightly
larger than those used in the NN sector. This choice reflects
the requirement that the cutoff mass has to be larger than the
mass of the exchanged meson and that in SU(3) chiral EFT 
not only pions are exchanged but also $K$- and $\eta$ 
mesons. Regarding two-meson exchange only contributions from
$\pi\pi$ have been taken into account in the SMS potentials. 
For those involving a $K$ and/or $\eta$ 
($\pi K$, $KK$, etc.) the combined masses exceed the 
cutoff value so that they should be strongly suppressed, see the pertinent discussion in Ref.~\cite{Haidenbauer:2023qhf}. 
Note that in the SMS potentials deviations of the meson-baryon coupling constants 
from the SU(3) values are taken into account. 
Specifically, the explicit 
SU(3) symmetry breaking in the empirical values of the decay constants \cite{ParticleDataGroup:2012pjm},
\begin{equation}
F_\pi  =   92.4~{\rm MeV}, \quad
F_K  =  (1.19\pm 0.01) F_\pi, \quad
F_\eta  = (1.30\pm 0.05) F_\pi \ .
\end{equation}
is implemented. 
Furthermore, following the practice in chiral NN 
potentials \cite{Reinert:2017usi}, 
we use $g_A = 1.29$, which is slightly larger than the experimental 
value, in order to account for the Goldberger-Treiman discrepancy~\cite{Fettes:1998ud}.
 

\begin{figure}[htbp] 
\vskip -0.4cm
\begin{center}
\includegraphics[width=0.65\linewidth]{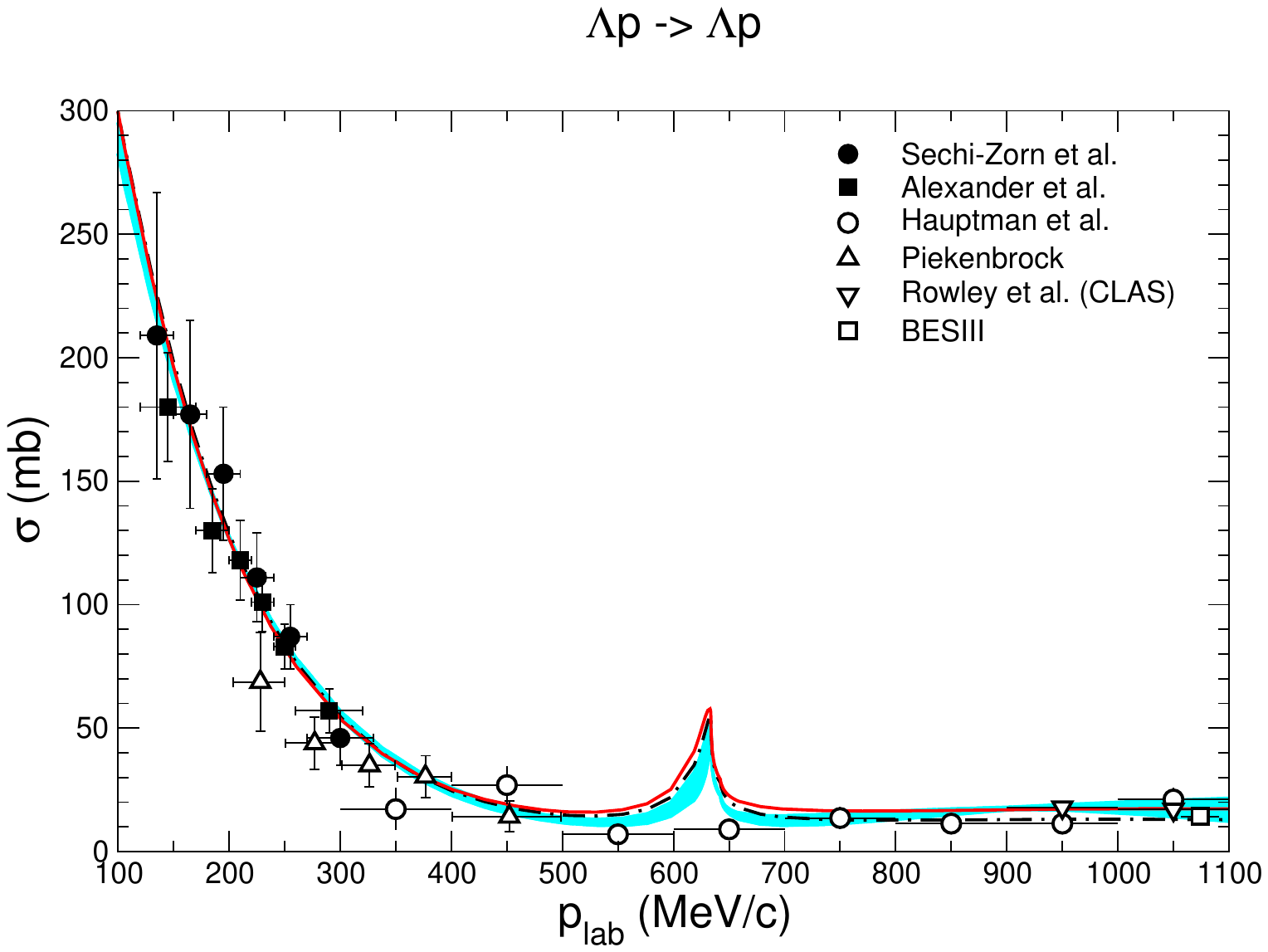}
\end{center}
\vskip -0.5cm 
\caption{$\La p$ cross section. 
Results are shown for the chiral YN potentials NLO19 
\cite{Haidenbauer:2019boi} (cyan band) and 
the SMS potentials NLO (550) (black dash-dotted line)
and N2LO (550) (red solid line)
\cite{Haidenbauer:2023qhf}.
Data shown as open symbols have not been included in
the fitting procedure. 
}
\label{fig:YNcross}
\end{figure}

\subsection{Results}
\label{jsec:LN}

In order to establish the potentials the involved LECs
have to be fixed by a fit to data. For that the J\"ulich-Bonn
group considered the “standard” set of 35 $\La$N and $\Si$N cross sections at 
low momenta that have been also used in studies based on the conventional
meson-exchange framework in the past
\cite{DeSwart:1971dr,Nagels:1973rq,Nagels:1976xq,Maessen:1989sx,Rijken:1998yy,Nagels:2015lfa,
Holzenkamp:1989tq,Reuber:1993ip,Haidenbauer:2005zh}. 
These data consists of two sets 
of cross sections for $\La p$ \cite{Alexander:1968acu,Sechi-Zorn:1968mao}
(6 data points each), and single sets for  
$\Si^+ p$ \cite{Eisele:1971mk} (4 data points), 
$\Si^-p$ \cite{Eisele:1971mk} (7 data points),
and for the transitions $\Si^-p \to \La n$ \cite{Engelmann:1966npz} (6 data points)
and $\Si^-p \to \Si^0 n$ \cite{Engelmann:1966npz} (6 data points). 
In addition the inelastic capture ratio at rest
\cite{Hepp:1968zza,Stephen:1970} is taken into account. 
Besides these YN data the empirical binding energy of the hypertriton $^3_\La$H
was used as a further constraint. Otherwise it would not be possible to fix 
the relative strength of the spin-singlet and spin-triplet $S$-wave contributions 
to the $\Lambda p$ interaction. See also the discussion in Sect.~\ref{sec:third} 
below.

In practice, the LECs in the $S$-waves can be fixed fairly well from the data.
Those in the $P$-waves could be only determined qualitatively, namely by 
imposing more general restrictions like that the partial-wave cross sections
for $\La$p remain small at higher energies~\cite{Haidenbauer:2013oca}, 
so that the total cross section remains small as well, 
as indicated by the measurements.
Luckily, when the extension to N2LO was in progress, the first extended
measurements of differential cross sections for $\Si^+p$ and $\Si^-p$ away
from the threshold region became available from the E40 experiment at
J-PARC \cite{J-PARCE40:2021bgw,J-PARCE40:2021qxa,J-PARCE40:2022nvq} 
which allowed to pin down the $P$-wave LECs much better than before~\cite{Haidenbauer:2023qhf}.

As expected, at LO only the bulk properties of the YN interaction can be 
reproduced~\cite{Polinder:2006zh}. However, already   
at NLO the description of the YN system achieved is on
the same level of quality as the one by the most advanced meson-exchange 
YN interactions \cite{Rijken:1998yy}. 
In terms of the $\chi^2$ its values are around 15-16
for the total of 36 $\La$N and $\Si$N data included.
It amounts to $\approx$~30 for the LO interactions.

Interestingly, with the NLO interactions NLO13
and NLO19 it was possible to achieve a combined 
description of the $\La$p and $\Si$N systems without any explicit 
SU(3) symmetry breaking in the contact interactions. 
Of course, 
SU(3) symmetry is broken by the used physical masses of
the involved pseudoscalar mesons in the potential and 
of the hyperons when solving the LS equation. 
On the other hand, a unified description of other BB interactions,
for example, in the NN system or of $\La\La$ and $\Xi$N, 
with contact terms fulfilling strict SU(3) symmetry 
turned out to be not possible \cite{Haidenbauer:2014rna}. 
In particular, the strength of the contact interaction in the 
{$\bf 27$} representation of SU(3), see Table~\ref{tab:SU3}, 
that is needed to reproduce the $pp$ (or $np$) $^1S_0$ phase shifts 
is simply not compatible with what is required 
for the description of the empirical $\Si^+$p cross section
\cite{Haidenbauer:2015zqb}.
 
\subsubsection{\texorpdfstring{$\La$N}{Lambda-N} and \texorpdfstring{$\Si$N}{Sigma-N} systems}
In the following we focus on a comparison with cross section 
data at low energies. Here the results are dominated by the 
$S$-wave interactions, which are also the most relevant 
ones for the predictions of hypernuclei that will be discussed below. 

\begin{figure}[htbp] 
\begin{center}
\includegraphics[width=0.42\linewidth]{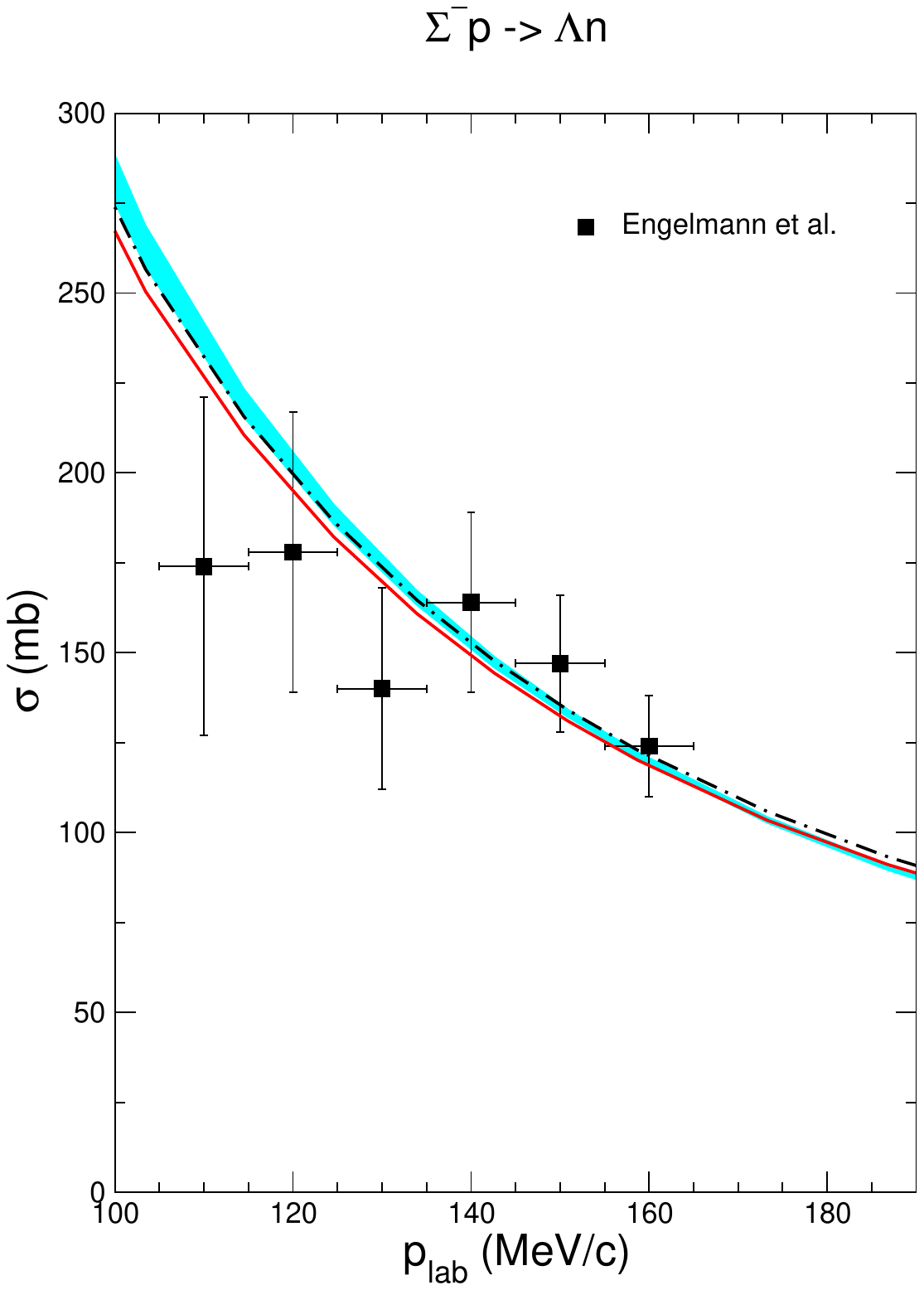}
\includegraphics[width=0.42\linewidth]{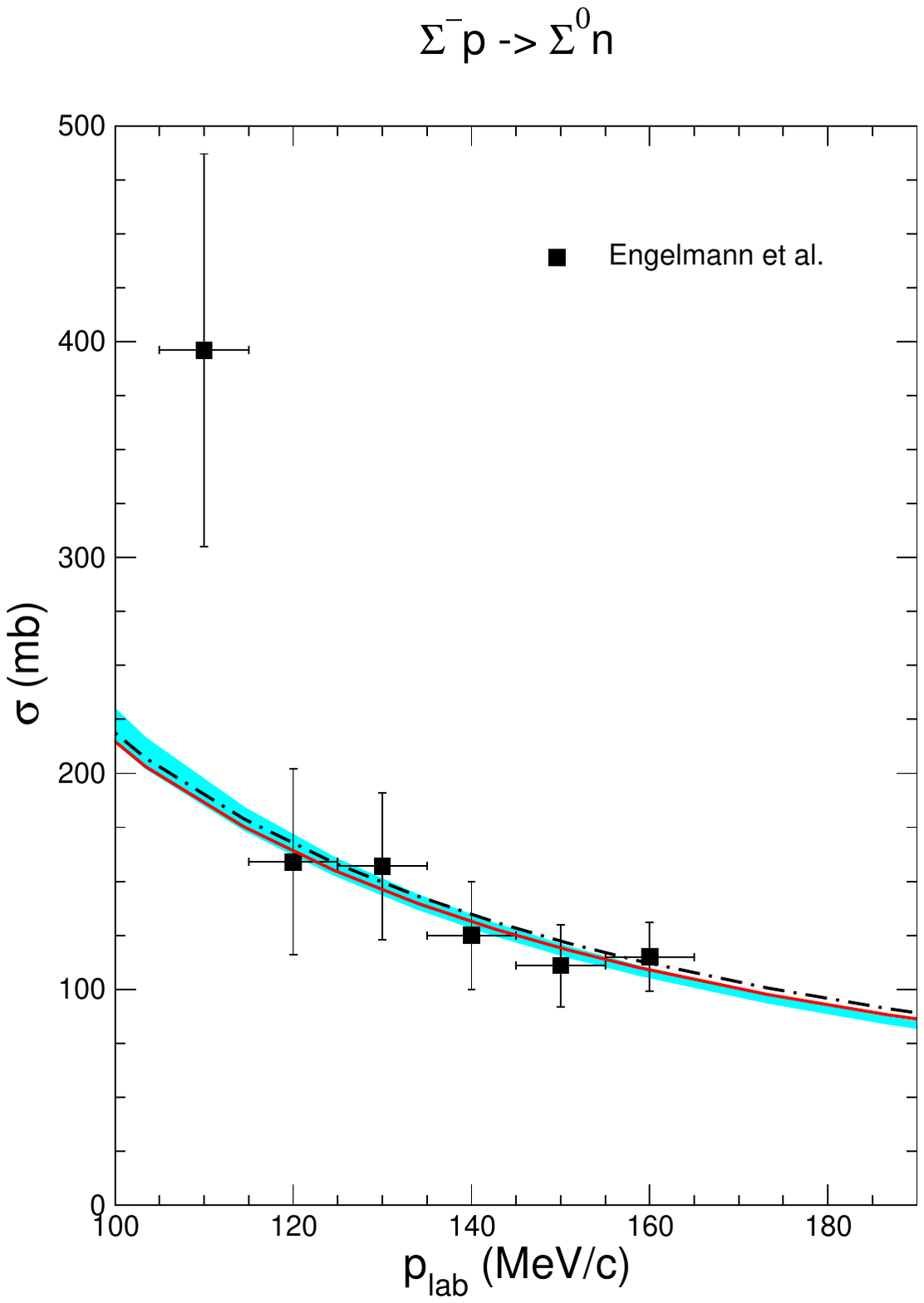}

\includegraphics[width=0.42\linewidth]{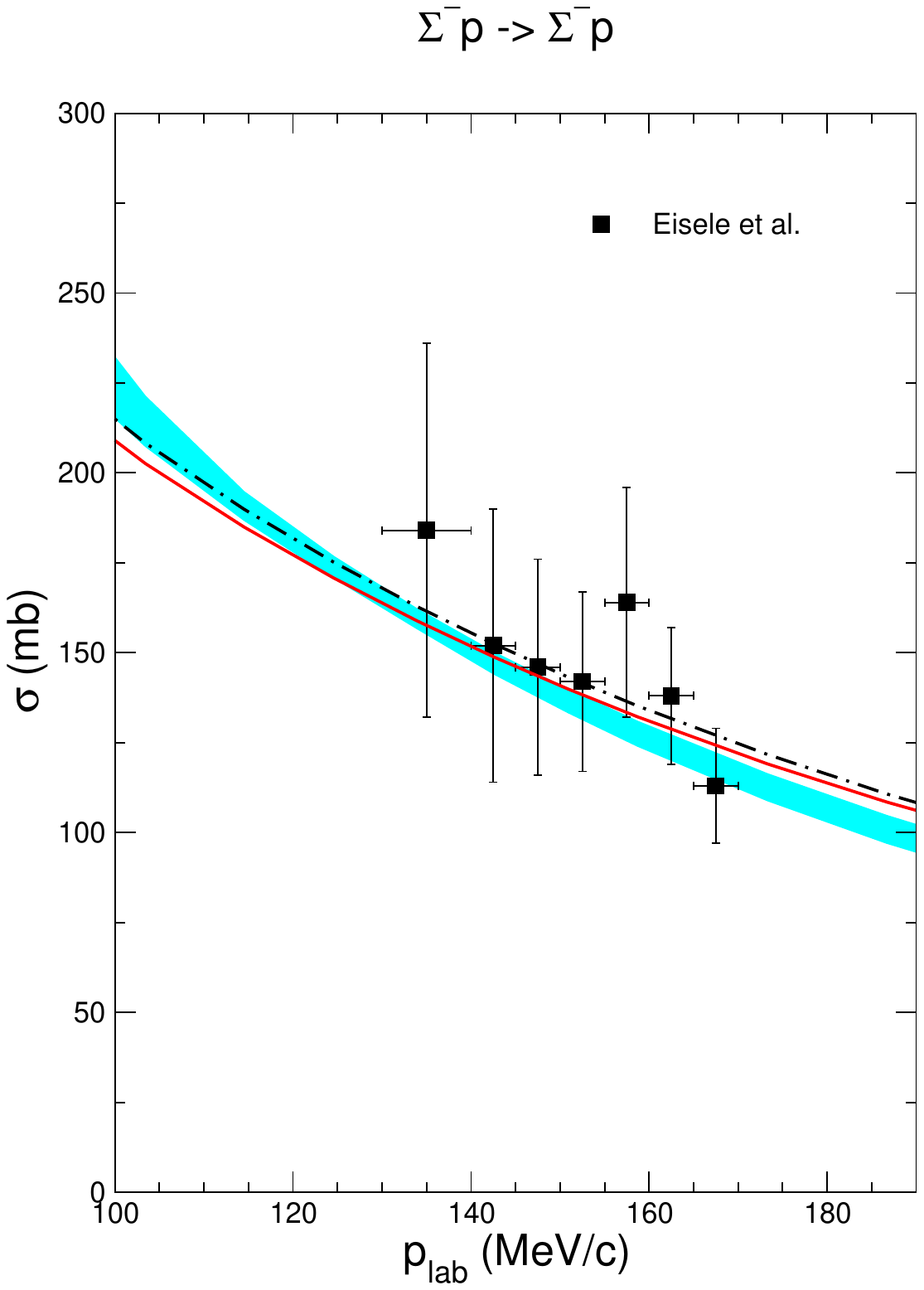}
\includegraphics[width=0.42\linewidth]{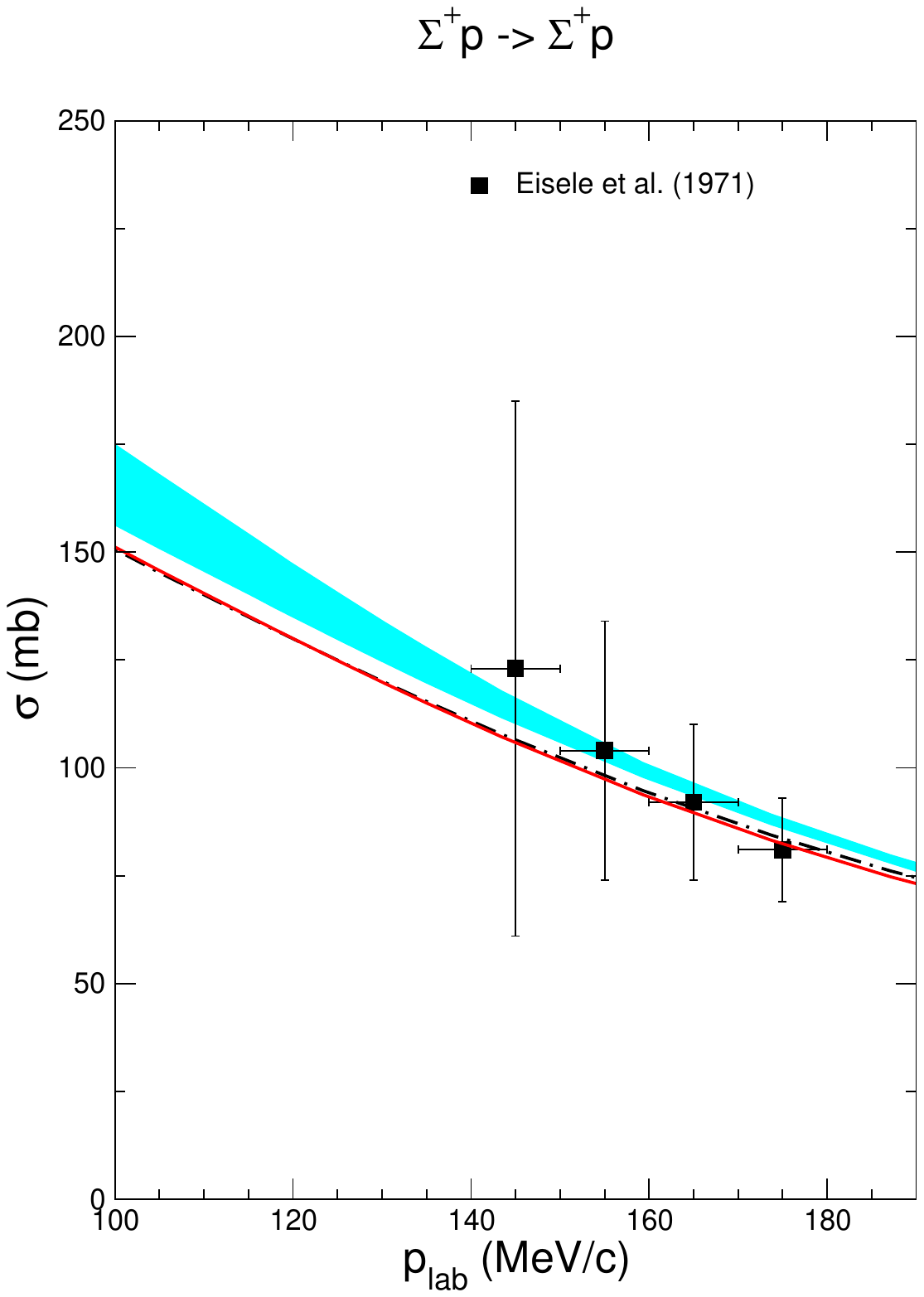}
\end{center}
\vskip -0.5cm 
\caption{$\Si^- p$ and $\Si^+ p$ cross sections. 
Results are shown for the chiral YN potentials NLO19 
\cite{Haidenbauer:2019boi} (cyan band) and SMS \cite{Haidenbauer:2023qhf}
NLO (550) (dash-dotted/black line) and N2LO (550) (solid/red line).
}
\label{fig:YNcrossS}
\end{figure}

Cross sections for $\La p$ are shown in Fig.~\ref{fig:YNcross}, while those 
for $\Si^-p$ and $\Si^+p$ can be found in Fig.~\ref{fig:YNcrossS}.
We include results for the NLO19 YN potential
(the variation with the cutoff is indicated by the
band) and those for the SMS potentials up to
NLO and N2LO, respectively. In the latter case
the results correspond to the cutoff of $550$~MeV. 
Data that have been included in the fitting 
procedure are shown as filled symbols. 
Note that the results for the NLO13 and NLO19 potentials 
for $\La p$ and $\Si$N scattering are practically identical
\cite{Haidenbauer:2019boi} and, therefore, the former are 
not included/discussed here.

The results for $\La p$ are shown over an extended momentum range so that 
we can also include the very recent data from CLAS~\cite{CLAS:2021gur} and 
from BESIII~\cite{BESIII:2023trh}. 
Interestingly, the predictions from the chiral potentials are quite
well in line with those measurements. Nonetheless,
one should keep in mind that, strictly speaking, the potentials are only 
valid for momenta below the $\pi\La$N threshold, i.e. for
$p_{\rm lab} < 890$~MeV/c. 

We refrain from showing more results here. Further
observables, specifically $\Si^+$p and $\Si^-$p cross sections at higher momenta, 
differential cross sections \cite{J-PARCE40:2021bgw,J-PARCE40:2021qxa,J-PARCE40:2022nvq},  
and $S$- and $P$-wave phase shifts in the $\La$N and $\Si$N channels can be found in 
Refs.~\cite{Haidenbauer:2013oca,Haidenbauer:2019boi,Haidenbauer:2023qhf}. 
However, we list results for the
effective range parameters, see Table~\ref{tab:ereF}.
Those parameters provide a simple and illuminating view on
the strength of the $S$-wave interactions in the various 
$\La$N and $\Si$N channels. 

\begin{table*}[t]
\caption{Scattering lengths ($a$) and effective ranges ($r$) 
for singlet (s) and triplet (t) $S$-waves (in fm), 
for $\Lambda N$, $\Si N$ with isospin $I=1/2,\, 3/2$, and for 
$\Si^+ p$ with inclusion of the Coulomb interaction. 
  }
\label{tab:ereF}
\vskip 0.1cm
\renewcommand{\arraystretch}{1.4}
\begin{center}
\begin{tabular}{|l||rrr|rrr||r|r|}
\hline
\hline
& \multicolumn{3}{c|}{SMS NLO} & \multicolumn{3}{c||}{SMS N2LO} & NLO13 & NLO19  \\
\hline
${\Lambda}$ [MeV] 
& 500     & 550     & 600     & 500     & 550     & 600     & 600     & 600  \\
\hline
\hline
$a^{\La N}_s$ 
& $-2.80$ & $-2.79$ & $-2.79$ & $-2.80$ & $-2.79$ & $-2.80$  & $-2.91$ & $-2.91$  \\
$r^{\La N}_s$ 
& $ 2.87$ & $ 2.72$ & $ 2.63$ & $ 2.82$ & $ 2.89$ & $ 2.68$  & $ 2.78$ & $ 2.78$  \\
\hline
$a^{\La N}_t$ 
& $-1.59$ & $-1.57$ & $-1.56$ & $-1.56$ & $-1.58$ & $-1.56$  & $-1.54$ & $-1.41$ \\
$r^{\La N}_t$ 
& $ 3.10$ & $ 2.99$ & $ 3.00$ & $ 3.16$ & $ 3.09$ & $ 3.17$  & $ 2.72$  & $ 2.53$ \\
\hline
\hline
Re\,$a^{\Si N \ (I=1/2)}_{s}$
& $ 1.14$ & $ 1.15$ & $ 1.10$ & $ 1.03$ & $ 1.12$ & $ 1.06$  & $ 0.90$ & $ 0.90$ \\
Im\,$a^{\Si N}_{s}$ 
& $ 0.00$ & $ 0.00$ & $ 0.00$ & $ 0.00$ & $ 0.00$ & $ 0.00$  & $ 0.00$ & $ 0.00$ \\
\hline
Re\,$a^{\Si N \ (I=1/2)}_{t}$
& $ 2.58$ & $ 2.42$ & $ 2.31$ & $ 2.60$ & $ 2.38$ & $ 2.53$  & $ 2.27$ & $ 2.29$  \\
Im\,$a^{\Si N}_{t}$ 
& $-2.60$ & $-2.95$ & $-3.09$ & $-2.56$ & $-3.26$ & $-2.64$  & $-3.29$ & $-3.39$  \\
\hline
$a^{\Si N \ (I=3/2)}_s$
& $-4.21$ & $-4.05$ & $-4.11$ & $-4.37$ & $-4.19$ & $-4.03$  & $-4.45$ & $-4.55$  \\
$r^{\Si N}_s$ 
& $ 3.93$ & $ 3.89$ & $ 3.75$ & $ 3.73$ & $ 3.89$ & $ 3.74$  & $ 3.68$ & $ 3.65$  \\
\hline
$a^{\Si N \ (I=3/2)}_t$
& $ 0.46$ & $ 0.47$ & $ 0.47$ & $ 0.38$ & $ 0.44$ & $ 0.41$  & $ 0.44$ & $ 0.43$  \\
$r^{\Si N}_t$ 
& $-5.08$ & $-4.74$ & $-4.82$ & $-5.70$ & $-4.96$ & $-5.72$  & $-4.59$ & $-5.27$  \\
\hline
\hline
$a^{\Si^+ p}_s$ 
& $-3.41$ & $-3.30$ & $-3.44$ & $-3.47$ & $-3.39$ & $-3.25$  & $-3.56$ & $-3.62$  \\
$r^{\Si^+ p}_s$ 
& $ 3.75$ & $ 3.73$ & $ 3.59$ & $ 3.61$ & $ 3.73$ & $ 3.65$  & $ 3.54$ & $ 3.50$  \\
\hline
$a^{\Si^+ p}_t$ 
& $ 0.51$ & $ 0.52$ & $ 0.52$ & $ 0.41$ & $ 0.48$ & $ 0.45$  & $ 0.49$ & $ 0.47$  \\
$r^{\Si^+ p}_t$ 
& $-5.46$ & $-5.12$ & $-5.19$ & $-6.74$ & $-5.50$ & $-6.41$  & $-5.08$ & $-5.77$  \\
\hline
\hline
\end{tabular}
\end{center}
\renewcommand{\arraystretch}{1.0}
\end{table*}

\subsubsection{\texorpdfstring{$\La\La$}{Lambda-Lambda} and \texorpdfstring{$\Xi$N}{Cascade-N} systems}
\label{subsec:LLXiN}

Chiral potentials up to NLO for the BB interaction in the strangeness $S=-2$ systems $\La\La$ and 
$\Xi N$ have already been established in 2016 \cite{Haidenbauer:2015zqb}. 
Thereby constraints from the $\La\La$ scattering length in the $^1S_0$ state
together with experimental upper bounds on the cross sections for $\Xi N$
scattering and for the transition $\Xi N \to \La\La$ have been exploited.
This allowed us to fix the additional LECs that arise in the ${\bf 1}$
irreducible representation of SU(3)~\cite{Haidenbauer:2015zqb},
see Table~\ref{tab:SU3}.
Furthermore, the consideration of those empirical constraints necessitated to add
SU(3) symmetry breaking contact terms (cf. Eq.~(\ref{VCTSB1})) in other irreps
(${\bf 27}$, ${\bf 10}$, ${\bf\overline{10}}$, ${\bf 8_s}$, ${\bf 8_a}$),
with regard to those determined from the $\La N$ and $\Si N$ data,
as already mentioned above. 
In 2019 a modified version has been suggested \cite{Haidenbauer:2018gvg} 
which only differs from the one in Ref.~\cite{Haidenbauer:2015zqb} by a more attractive 
interaction in the $^3{\rm S}_1$ partial wave with isospin $I=1$.
That potential yields a moderately attractive (in-medium) 
$\Xi$-nuclear interaction
\cite{Haidenbauer:2018gvg} and supports the existence of bound $\Xi$-hypernuclei 
as will be discussed below. 
In both variants a non-local regulator is applied throughout. 
An extension of the S=-2 interaction up to N2LO based on the SMS scheme
has not been attempted yet. This would be only meaningful once more extensive
and more quantitative experimental information becomes available.

\begin{figure}
\begin{center}
\includegraphics[height=4.5cm,keepaspectratio]{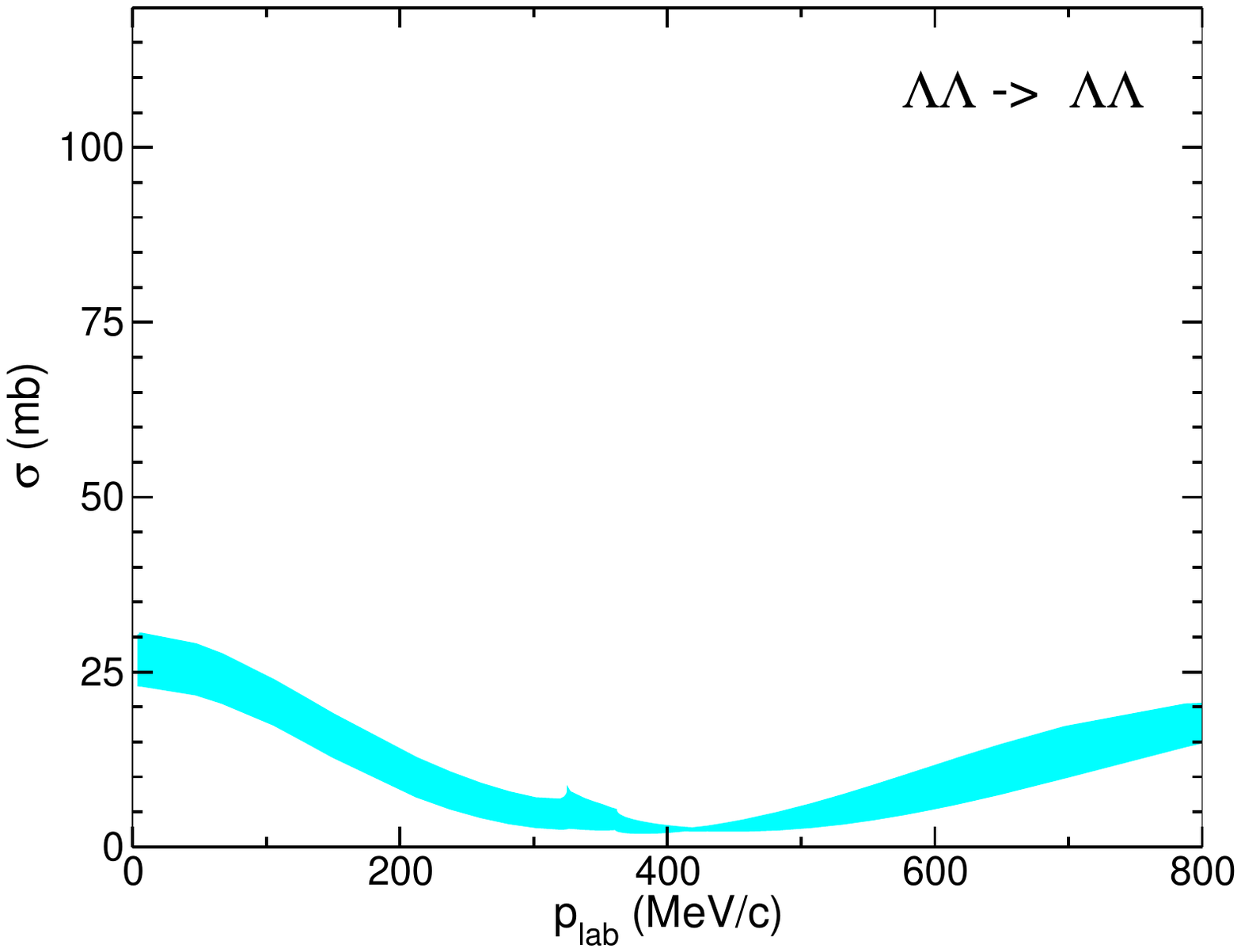}
\includegraphics[height=4.5cm,keepaspectratio]{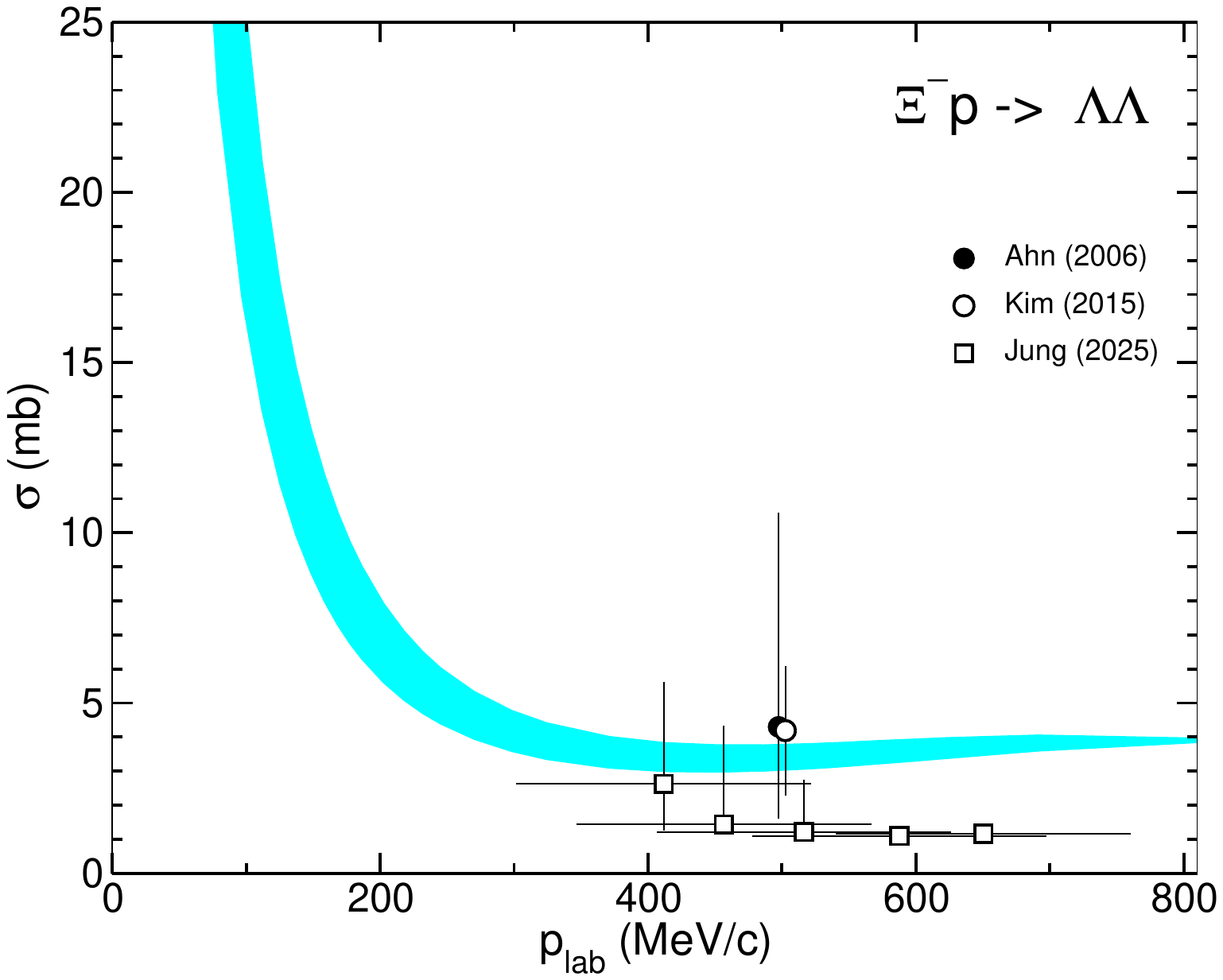}
\end{center}
\vspace*{-0.3cm}
\caption{Results for the $\La\La$ cross sections of the NLO potential 
from Ref.~\cite{Haidenbauer:2015zqb}.
Data are from Refs.~\cite{Ahn:2005jz,Jung:2025hrx}.
\vspace*{-0.1cm}
} 
\label{fig:LL}
\end{figure}

\begin{figure}
\begin{center}
\includegraphics[height=4.3cm,keepaspectratio]{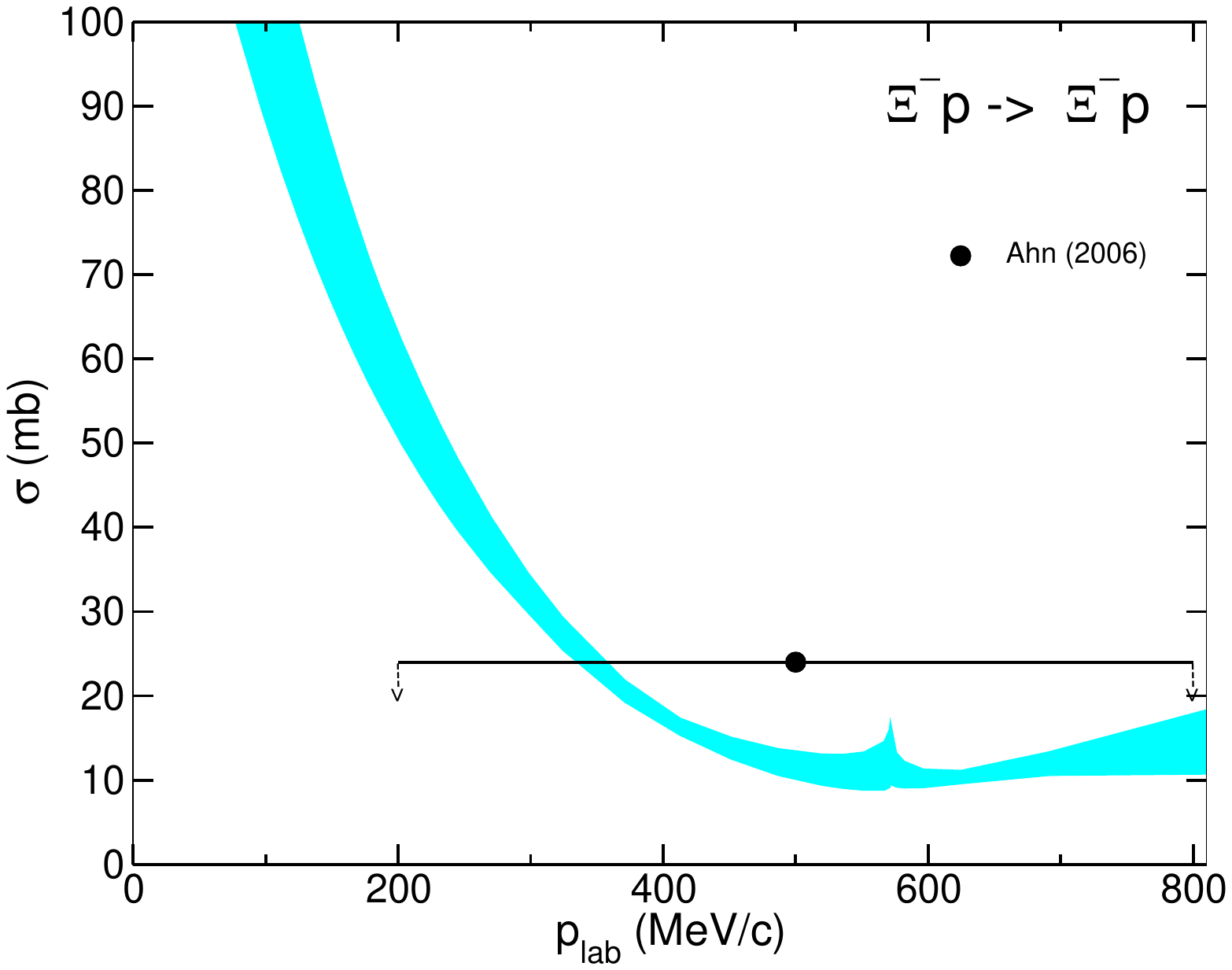}
\includegraphics[height=4.3cm,keepaspectratio]{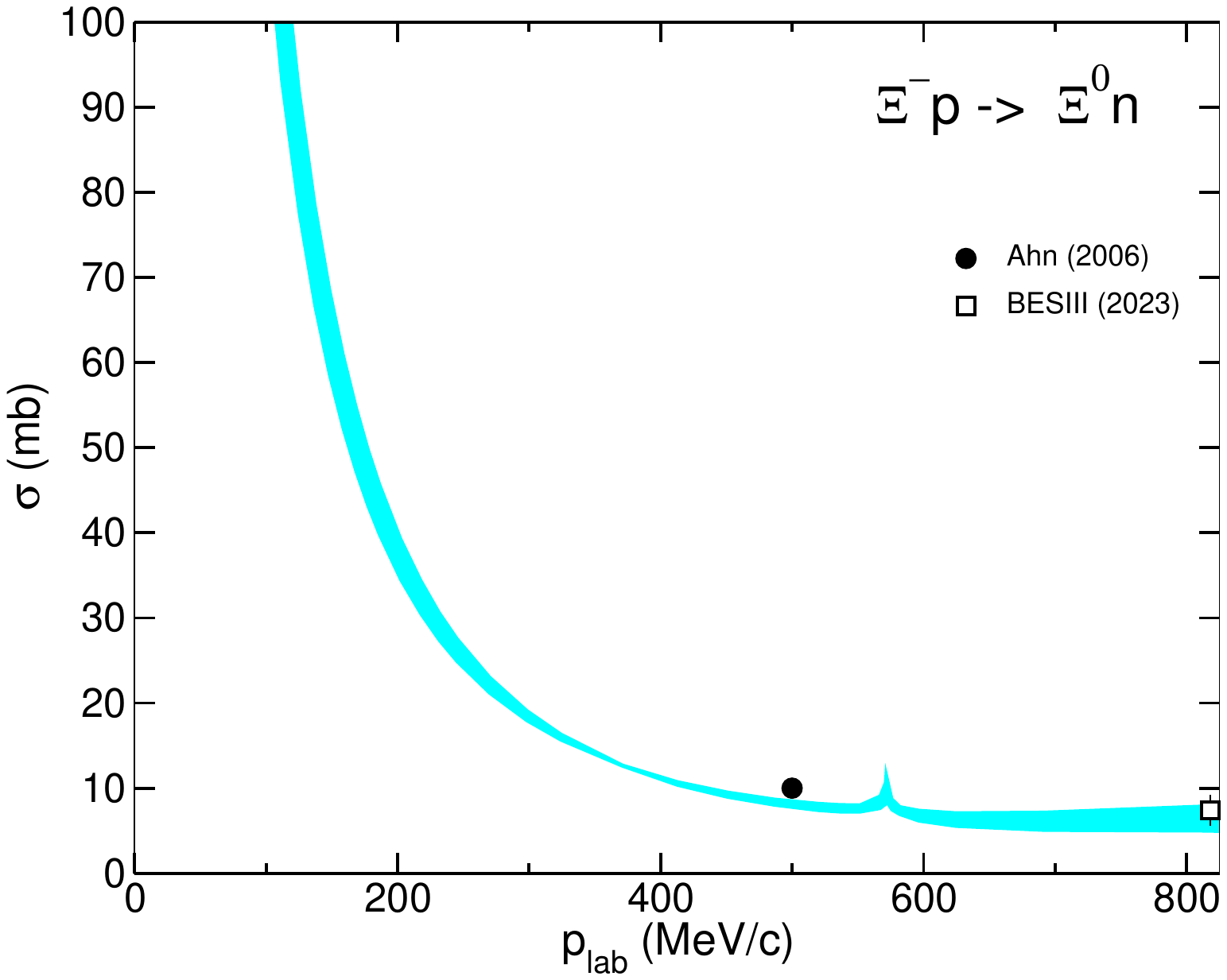}
\includegraphics[height=4.3cm,keepaspectratio]{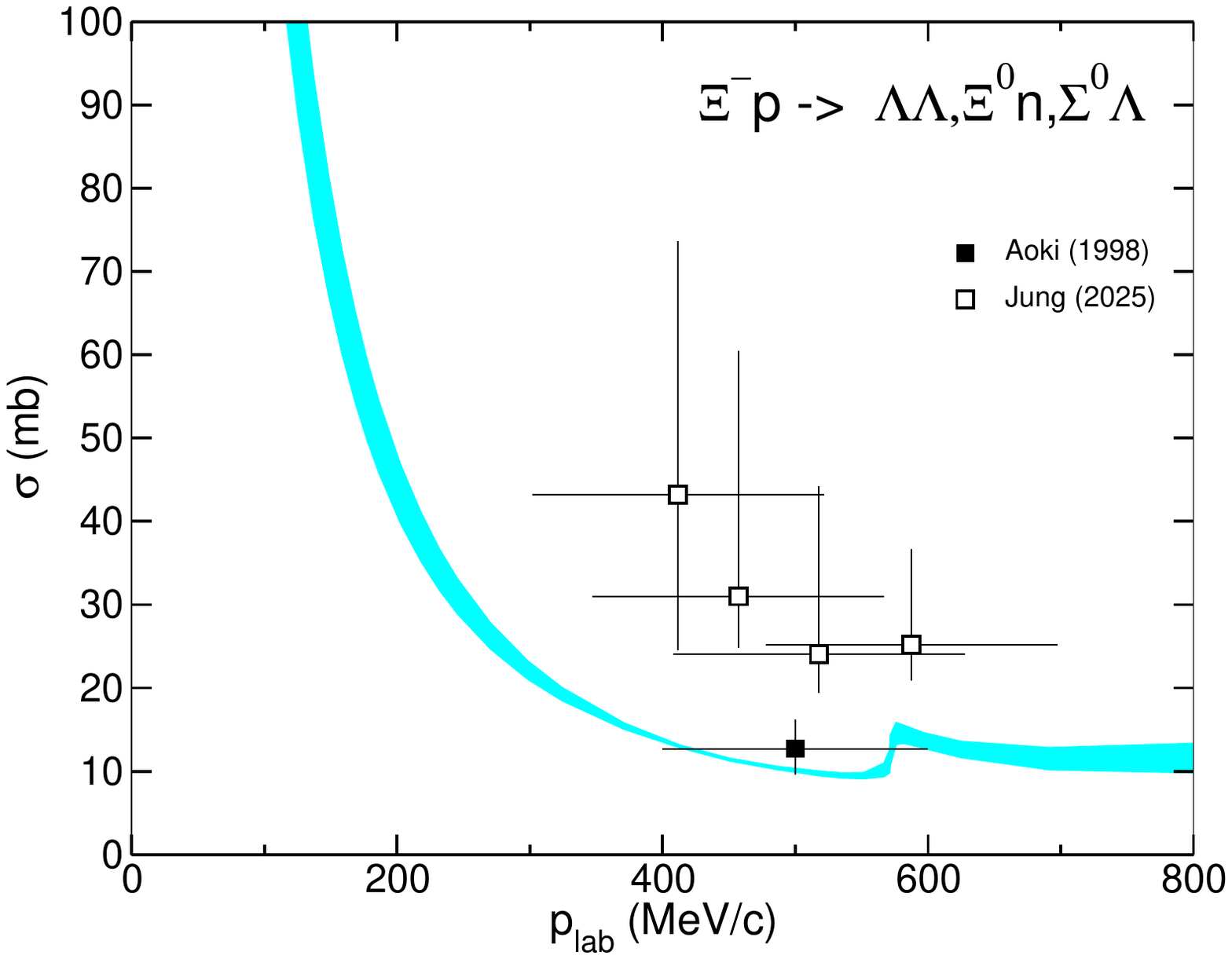}
\end{center}
\vspace*{-0.3cm}
\caption{Results for the $\Xi N$ cross sections of the NLO potential 
from Ref.~\cite{Haidenbauer:2018gvg}. Data are from 
Refs.~\cite{Aoki:1998sv,Ahn:2005jz,BESIII:2023clq,Jung:2025hrx}.
} 
\label{fig:xmpX}
\end{figure}

Results for the $\La\La$ cross sections are presented in Fig.~\ref{fig:LL}  
while those for the $\Xi N$ cross sections (based on the $\Xi N$ potential 
from 2019 \cite{Haidenbauer:2018gvg}) are presented in Fig.~\ref{fig:xmpX}. 
As one can see there is no experimental information for low momenta.
Moreover, in some cases only upper limits for the cross sections
are available.
We include here two recent additions to the data base. 
One is from the BESIII Collaboration~\cite{BESIII:2023clq} 
who deduced the $\Xi^0 n\to \Xi^-p$ cross section from the 
$\Xi^0 + ^9$Be reaction.
The others are new $\Xi^-$p cross sections from the J-PARC 
E42 experiment~\cite{Jung:2025hrx}, estimated from 
the production of $\Xi^-$ and $\La\La$ in the reaction 
$^{12}C(K^-,K^+)$ at an incident beam momentum of 
$1.8$~GeV/c using the eikonal approximation. As one
can see in Fig.~\ref{fig:LL} the cross section for $\Xi^-p\to\La\La$ 
is well in line with older empirical information \cite{Ahn:2005jz} 
and with results from chiral EFT. The same can be said about the
$\Xi^0 n\to \Xi^-p$ cross section, cf. Fig.~\ref{fig:xmpX}. 
On the other hand there is a 
striking discrepancy in the $\Xi^-p$ inelastic cross section in 
comparison to previous experimental evidence~\cite{Aoki:1998sv} 
and also to theory. 
Note that such large values as suggested by the new estimate  
would require a much stronger attractive $\Xi$N
interaction, likely even a bound state in one of
the $^3S_1$ partial waves, see Ref.~\cite{Haidenbauer:2015zqb}, 
for which, however, there is no empirical evidence. 

An interesting and independent test for the $\Xi N$ interaction is provided by 
two-particle momentum correlation functions. The ones for $\Xi^- p$
have been measured recently by the ALICE Collaboration in $p$-Pb collisions 
at $5.02$ TeV \cite{ALICE:2019hdt} 
and in $pp$ collisions at $13$ TeV \cite{ALICE:2020mfd}.
As has been shown in Refs.~\cite{Haidenbauer:2022esw,Doring:2025sgb}, those data 
are nicely reproduced by the chiral $\Xi$N interaction from 2019 \cite{Haidenbauer:2018gvg}. 

\subsection{Large-\texorpdfstring{$N_C$}{NC} constraints}

We end this section by discussing constraints from large-$N_C$ QCD. It was introduced by 't Hooft \cite{tHooft:1973alw} as a means
of studying QCD amplitudes in a systematic way using the inverse number of colors, $1/N_c$, as the expansion parameter.
Witten \cite{Witten:1979kh} developed  a Hartree-like picture of large-$N_c$ baryons. Shortly after this, not only
the connection to the Skyrme model \cite{Skyrme:1961vq,Skyrme:1962vh} could be uncovered \cite{Witten:1983tx,Gervais:1984rc},
but also the fact that baryons with an $\text{SU}(N_f)\times\text{SU}(2)_\text{spin}$ symmetry come with an exact
contracted $\mathrm{SU}(2N_f)$ spin-flavor symmetry in the large-$N_c$ limit leading to a tower of degenerate
$\mathrm{SU}(N_f)$ baryon multiplets \cite{Gervais:1983wq,Dashen:1993as,Dashen:1994qi}. The investigation of the NN interaction
was pioneered in Refs.~\cite{Kaplan:1995yg,Kaplan:1996rk} and the SU(3) BB interaction was studied first in this framework focusing on the leading order chiral contact interactions in~\cite{Liu:2017otd}. Note that the basis of the BB interaction in the  large-$N_C$ scenario was already
spelled out in Ref.~\cite{Kaplan:1996rk}. Here, we follow the more recent and more detailed work of Ref.~\cite{Vonk:2024lce} and refer to
that paper for calculational details.

First, we summarize the large-$N_C$ scaling of the various ingredients entering the BB potential. These are
\begin{equation}
m_B  \sim N_c, ~~ |\mathbf{q}|^2  \sim 1,~~ q_0  \sim N_c^{-1}, ~~
\Delta m_B \sim N_c^{-1}, ~~ |\mathbf{k}|^2  \sim 1, ~~ \sdot{\mathbf{k}}{\mathbf{q}} \sim 1.
\label{eq:momentascalings}
\end{equation}
Here, $m_B$ denotes the baryon mass, $\Delta m_B$ the splitting within a given multiplet, $\bf{q}=\bf{p}'-\bf{p}$ is the three-momentum transfer,
$\bf{k}=\bf{p}'+\bf{p}$ the momentum sum, and $q_0 = \Delta m_B + \sdot{\mathbf{k}}{\mathbf{q}} /(2m_B)$ the energy transfer in the
non-relativistic limit. Moreover, expanding the
BB potential in a Taylor series of the above momenta leads to the second source of
$1/N_c$ suppressions due to factors of $1/m_B$. As argued in \cite{Kaplan:1996rk}, this
suppression follows the general rule that terms proportional to $\mathbf{q}^m\,\mathbf{k}^n$
are suppressed by
\begin{equation}\label{eq:qksuppression}
1/N_c^{\operatorname{min}(m,n)} .
\end{equation}
Restricting ourselves to the pure octet baryon sector, the resulting 
BB potential takes the form
\begin{align}\label{eq:largeNCPot}
\begin{split}
V&_{\B^\alpha\B^\beta\to \B^\gamma\B^\delta} =  N_c  \Biggg\{ v_{0,0} + v_{0,1}^{(T)}
\frac{\sdot{\hat{\mathcal{T}}_1}{\hat{\mathcal{T}}_2}}{2 N_c^2} + v_{0,1}^{(S)}
\frac{\sdot{\hat{\mathcal{S}}_1}{\hat{\mathcal{S}}_2}}{3 N_c^2} + 2 v_{0,1}^{(G)} \frac{\sdot{\hat{\mathcal{G}}_1}{\hat{\mathcal{G}}_2}}{N_c^2} 
\\ & + \left[ \left(v_{1,0} + v_{1,1}^{(T)}
\frac{\sdot{\hat{\mathcal{T}}_1}{\hat{\mathcal{T}}_2}}{2N_c^2} \right) \frac{\left(\hat{\mathcal{S}}^i_1 + \hat{\mathcal{S}}_2^i\right)}{\sqrt{3}N_c} + v_{2,0} \frac{\left(\hat{\mathcal{G}}_2^{ia}\hat{\mathcal{T}}^a_1 + \hat{\mathcal{G}}_1^{ia}\hat{\mathcal{T}}_2^a
\right)}{N_c^2} \right] \left(\mathbf{q}\times\mathbf{k}\right)^i \\
& + \Biggl[ v_{4,0} \left( \mathbf{q}^i\mathbf{q}^j-\frac{1}{3}\left|\mathbf{q}\right|^2\delta^{ij}\right)
  + v_{5,0} \left( \mathbf{k}^i\mathbf{k}^j-\frac{1}{3}\left|\mathbf{k}\right|^2\delta^{ij}\right)\Biggr] 
\frac{2\left(\hat{\mathcal{G}}_1^{ia}\hat{\mathcal{G}}_2^{ja}\right)}{N_c^2}\Biggg\} + \order{1/N_c^3}~,
\end{split}
\end{align}
in terms of the generators of the contracted SU(6) spin-flavor symmetry.
\begin{equation}
\hat{\mathcal{S}}^i = q^\dagger\left(\frac{\sigma^i}{2} \otimes \mathbbm{1} \right) q~, \quad
\hat{\mathcal{T}}^a  = q^\dagger\left(\mathbbm{1} \otimes \frac{\lambda^a}{2} \right) q~, \quad
\hat{\mathcal{G}}^{ai}  = q^\dagger\left(\frac{\sigma^i}{2} \otimes \frac{\lambda^a}{2} \right)q~.
\end{equation}
Here, $q = (u,d,s)$ represents a three flavor bosonic quark operator that carries no color, the $\sigma_i$'s are the three Pauli spin matrices and the $\lambda_a$'s are the eight Gell-Mann matrices. Furthermore, the coefficients $v_{n,k}$ are scalar functions of $|\mathbf{q}|^2$ and
$|\mathbf{k}|^2$. It is instructive to compare this to the generic formulation of the SU(3) BB potential with flavor labels $a\dots d$, which can be written as
\begin{equation}
\label{eq:symbolicPot}
V_{B^aB^b\to B^cB^d} = V_0^0 + V_\sigma^0 \sdot{\boldsymbol{\sigma}_1}{\boldsymbol{\sigma}_2}+ V_\text{LS}^0  \sdot{\mathbf{L}}{\mathbf{S}} + V_\text{T}^0 S_{12}  + V_0^1 \rho_0^{abcd} + V_\sigma^1 \sdot{\boldsymbol{\sigma}_1}{\boldsymbol{\sigma}_2} \rho_\sigma^{abcd} + V_\text{LS}^1  \sdot{\mathbf{L}}{\mathbf{S}} \rho_\text{LS}^{abcd} + V_\text{T}^1  S_{12} \rho_\text{T}^{abcd} , 
\end{equation}
where
$S_{12}(\mathbf{\hat{r}}) = 3 \sdot{\mathbf{\hat{r}}}{\boldsymbol{\sigma}_1}\sdot{\mathbf{\hat{r}}}{\boldsymbol{\sigma}_2} - \sdot{\boldsymbol{\sigma}_1}{\boldsymbol{\sigma}_2}$,
with $\mathbf{\hat{r}} = \mathbf{r}/|\mathbf{r}|$, and the $\rho^{abcd}_{\left\{0,\sigma,\text{LS},\text{T}\right\}}$ represent some appropriate structure in accordance with SU(3) flavor symmetry not important at this stage. Here, we have deliberately mimicked the generic NN potential given in Ref.~\cite{Kaplan:1996rk} in order to faciliate the comparision. For the NN interaction, the $\rho^{abcd}_{\left\{0,\sigma,\text{LS},\text{T}\right\}}$ are simply given by $\sdot{\boldsymbol{\tau}_1}{\boldsymbol{\tau}_2}$, with $\boldsymbol{\tau}$ being the isospin operator. What the authors of Ref.~\cite{Kaplan:1996rk} have shown is that in this case only $V_0^0$, $V_\sigma^1$, and $V_\text{T}^1$ are of leading $\order{N_c}$, while all other contributions are of $\order{1/N_c}$. Similarly, one finds for the SU(3) BB interaction considering baryons of strangeness of $\order{1}$
\begin{equation}\label{eq:Vscalings}
V_0^0  \sim V_0^1 \sim V_\sigma^1 \sim V_\text{T}^1\sim N_c , \quad
V_\sigma^0  \sim V_\text{LS}^0 \sim V_\text{LS}^1 \sim V_\text{T}^0 \sim 1/N_c ,
\end{equation}
which is basically the same as for the NN case except for the lifting of $V_0^1$, which is related to the terms $\sim\sdot{\hat{\mathcal{T}}_1}{\hat{\mathcal{T}}_2}$ which in the corresponding NN potential are suppressed by a relative factor of $1/N_c^2$ but in
general are not suppressed in the BB case. {Note that in the most general case the BB potential Eq.~\eqref{eq:symbolicPot} can also have an antisymmetric spin-orbit term $\sim \mathbf{L}\cdot\left(\boldsymbol{\sigma}_1-\boldsymbol{\sigma}_2\right)$ \cite{Haidenbauer:2013oca}. This force describing spin singlet-triplet transitions is absent in isospin-symmetric NN
potentials but is in accordance with SU(3) symmetry. However, in the large-$N_c$ case this contribution comes with the same suppression that also showed up in the $V_\text{LS}^i$ case above due to Eq.~\eqref{eq:qksuppression}. As none of the contributions at LO and NLO in $1/N_C$  does actually generate such antisymmetric spin-orbit interactions, this term is excluded from the analysis and from Eq.~\eqref{eq:symbolicPot}. We further note that the large-$N_c$ results for the potential are not RG-invariant and that there is a preferred scale, see e.g. Ref.~\cite{Lee:2020esp} (and references therein).
However, the extraction of this preferred scale as discussed in the NN case 
\cite{Lee:2020esp} can not be answered at present as corresponding data are either absent or too imprecise.

The contact terms of leading order in chiral perturbation theory, see Sec.~\ref{sec:form},
generate a potential that includes central, spin-spin, spin-orbit, and tensorial parts. However, only the central
and spin-spin parts of this potential are indeed of $\order{N_c}$, while all
other contributions are suppressed by a factor $1/m_B^2$. The contact terms alone hence do not generate the full
leading $\order{N_c}$ potential, but only terms corresponding to $V_0^0$, $V_0^1$, and $V_\sigma^1$ in
Eq.~\eqref{eq:symbolicPot}, while an $\order{N_c}$ tensorial part is missing. Moreover, the spin-orbit part
 is of subleading $\order{1/N_c}$ as expected. What these contact terms also add is a partial expansion
of the large-$N_c$ coefficients in Eq.~\eqref{eq:largeNCPot} in the momenta, which can not be determined from the
large-$N_c$ Hartree scenario.  The leading $\order{N_c}$ contact contributions $\sim C_S^{abcd}$ and $\sim C_T^{abcd}$,
where $a,b,c,d$ are flavor indices, consist of linear combinations of six of the original 15 low-energy constants of the contact Lagrangian. 
One can derive sum rules valid at leading order in $1/N_c$ allowing to reduce the
number of independent parameters to three. These read
\begin{eqnarray}
\label{eq:YNlargeNC}
C_{1S0}^{\Sigma\Sigma} & \approx & \frac{1}{9}\left(20\, C_{1S0}^{\Lambda\Lambda} - 11\, C_{3S1}^{\Lambda\Lambda}-7\, C_{3S1}^{\Lambda\Sigma}  \right)~, \nonumber\\
C_{3S1}^{\Sigma\Sigma} & \approx & -12\, C_{1S0}^{\Lambda\Lambda} +13\, C_{3S1}^{\Lambda\Lambda}+9\, C_{3S1}^{\Lambda\Sigma}~.
\end{eqnarray}
These large-$N_c$ sum rules of the leading order contact terms are indeed fulfilled to a good accuracy as can be seen from Table~\ref{tab:YN}. Especially for small cutoff masses, the agreement is formidable with deviations just within what is expected from $1/N_c$ corrections\footnote{Note that these sum rules differ from the ones given in~\cite{Liu:2017otd}, from the details given in that paper we were not able to arrive at their results.}.
\begin{table}[h]
\renewcommand{\arraystretch}{1.3}
\centering
\begin{tabular}{|l||c|c|c||c|c||c|c|}
\hline
Cutoff & $C_{1S0}^{\Lambda\Lambda}$ & $C_{3S1}^{\Lambda\Lambda}$ & $C_{3S1}^{\Lambda\Sigma}$ & \multicolumn{2}{c||}{$C_{1S0}^{\Sigma\Sigma}$} &  \multicolumn{2}{c|}{$C_{3S1}^{\Sigma\Sigma}$} \\ \hline
$550$\,MeV & $-0.0466$ & $-0.0222$ & $-0.0016$ & $-0.0766$ & $\mathbf{-0.0751}$ & $0.2336$ & $\mathbf{0.2562}$ \\
$600$\,MeV & $-0.0403$ & $-0.0163$ & $-0.0019$ & $-0.0763$ & $\mathbf{-0.0682}$ & $0.2391$ & $\mathbf{0.2546}$ \\
$650$\,MeV & $-0.0322$ & $-0.0097$ & $\phantom{-}0.0000$ & $-0.0757$ & $\mathbf{-0.0597}$ & $0.2392$ & $\mathbf{0.2603}$ \\
$700$\,MeV & $-0.0304$ & $-0.0022$ & $\phantom{-}0.0035$ & $-0.0744$ & $\mathbf{-0.0676}$ & $0.2501$ & $\mathbf{0.3677}$ \\
\hline
\end{tabular}
\renewcommand{\arraystretch}{1.0}
\caption{Comparing best fit YN potentials from Ref.~\cite{Polinder:2006zh} and corresponding large-$N_c$ predictions (in units of $10^4$ GeV$^{-2}$). The bold values of $C_{1S0}^{\Sigma\Sigma}$ and $C_{3S1}^{\Sigma\Sigma}$ are obtained using the large-$N_c$ sum rules Eq.~\eqref{eq:YNlargeNC}.}
\label{tab:YN}
\end{table}

A BB potential derived from SU(3) Chiral Perturbation Theory (CHPT) must include one-meson exchange
contributions in order to fully reproduce the leading order large-$N_c$ potential, as the tensorial part $V_T^1$
of $\order{N_c}$ can not be generated by the contact terms alone, which only generate a tensorial part of
$\order{1/N_c}$. This is in accordance with chiral power counting. 
Matching the one-meson exchange contributions with the large-$N_c$ potential yields the already known
ratio $F/D = 2/3 \left( 1 + \order{1/N_c^{2}}\right)$, see e.\,g. \cite{Dashen:1993jt}. One can further derive an
effective coupling $g_{BB\Phi}$ in terms of $g_A=F+D$ that is valid at leading order in $1/N_c$. In the literature it
is common to use YN and YY couplings $f_{BB\Phi}$ expressed in terms of $f_{NN\pi}=g_A/(2F_0)$
and $\alpha=F/(F+D)$ based on Ref.~\cite{deSwart:1963pdg}. The effective large-$N_c$ coupling $g_{BB\Phi}$ just
reproduces these $f_{BB\Phi}$ after forming approriate isospin combinations and setting $\alpha=2/5$.
It is also of relevance that the full large-$N_c$ scaling of $\order{N_c}$ in the one-meson exchange
case is only achieved by exchanging pions, while exchanging kaons are of $\order{1}$ and exchanging $\eta$'s are
even more suppressed and of $\order{1/N_c}$ which is a consequence of the choice to match real-world baryons with
those large-$N_c$ baryons that have strangeness of $\order{1}$. At the level of
quarks and gluons, this is just a result of combinatorics, as with this choice there are about $N_c$ choices
to pick up an up or down quark, but only $\order{1}$ choices to find a strange quark. 

Finally, we note that the large-$N_c$ scalings of many-meson exchange contributions can not be assessed by means of a naive
power counting of the involved meson-baryon couplings alone, as this might lead to results that contradict the
assumption that the BB potential is of $\order{N_c}$. However, imposing spin-flavor symmetry and
considering all diagrams of a given type including the full baryon tower retains consistency. Summing over all
$n$-meson exchange diagrams of a given type yields a contribution that at most scales as $\order{N_c^{2-n}}$.
For the two-meson exchange contributions in SU(3) CHPT, this is this explicitly shown in~\cite{Vonk:2024lce}. In this case, the inclusion of
the decuplet baryons is mandatory, and a cancellation between the deceptive $\order{N_c^2}$ contributions of the box and
crossed box diagrams appears if the large-$N_c$ ratio $C/D = 2$ in addition to the ratio $F/D= 2/3$, where $C$
is the leading baryon octet-baryon decuplet-Goldstone boson coupling constant (often also denoted as $H$). To leading
order it is thus possible to describe one-meson and two-meson exchange diagrams by a single parameter, e.\,g.
by setting $D=3/5\,g_A$,  $F=2/5\,g_A$, and $C=6/5\,g_A$.
    \section{Three-baryon forces in chiral effective
field theory}  \label{sec:BBB}

One of the merits of chiral EFT is that two-body forces
and three-body forces (3BFs) can be derived in a consistent way. 
Given the importance of three-nucleon forces (3NFs) for
an accurate description of nuclei one expects that
3BFs play likewise an important role in nuclear systems with strangeness. Specifically, the $\La$NN interaction
could play a decisive role for reproducing the separation
energies of $\La$ hypernuclei, which are experimentally well
established. 

In this section we review the derivation of the leading irreducible three-baryon interactions from SU(3) chiral effective field theory following Ref.~\cite{Petschauer:2015elq}.
We introduce the minimal effective Lagrangian required for the pertinent vertices.
Furthermore the estimation of the corresponding LECs through decuplet saturation
will be covered \cite{Petschauer:2016pbn}.
According to the power counting the 3BFs arise formally at N2LO in the chiral expansion.
Three types of diagrams contribute: three-baryon contact terms, one-meson and two-meson exchange diagrams, cf. Fig.~\ref{fig:3BF}.
Note that a two-meson exchange diagram, such as on the left of Fig.~\ref{fig:3BF}, with a (LO) Weinberg-Tomozawa vertex in the middle, would formally be an NLO contribution.
However, as in the nucleonic sector, this contribution is kinematically suppressed due to the fact that the involved meson energies are differences of baryon kinetic energies.
Anyway, parts of these N2LO contributions get promoted to NLO by the introduction of intermediate decuplet baryons.

\begin{figure}[ht]
\centering
\includegraphics[scale=0.5]{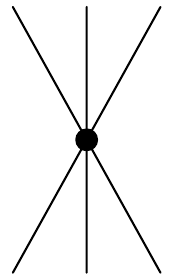}\qquad
\includegraphics[scale=0.5]{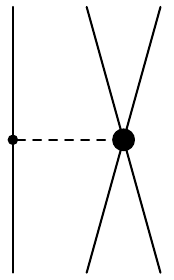}\qquad
\includegraphics[scale=0.5]{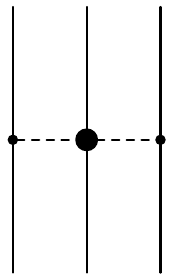}
\caption{
Leading three-baryon interactions: contact term, one-meson exchange and two-meson exchange.
Filled circles and solid dots denote vertices with \(\Delta_i=1\) and \(\Delta_i=0\), respectively.
\label{fig:3BF}
}
\end{figure}

\subsection{Contact interaction} \label{subsec:BBBct}

In this subsection, we consider the leading three-baryon contact interaction.
The corresponding Lagrangian has been constructed in Ref.~\cite{Petschauer:2015elq},
where the following possible structures in flavor space have been considered
\begin{eqnarray}\label{eq:BBBflavor}
 &\trace{\bar B\bar B\bar B BBB}\,,
 \trace{\bar B\bar B B \bar BBB}\,,
 \trace{\bar B\bar B B B \bar BB}\,, \
 \trace{\bar B B\bar B B \bar BB}\,, \\&
 \trace{\bar B \bar BBB}\trace{\bar B B}\,,
 \trace{\bar B B \bar BB}\trace{\bar B B}\,, 
 \trace{\bar B \bar B \bar BB}\trace{B B}\,, \\&
 \trace{\bar B\bar B\bar B}\trace{BBB}\,,
 \trace{\bar B\bar BB}\trace{B\bar BB}\,, \\ \label{eq:BBBflavorE}&
 \trace{\bar BB}\trace{\bar BB}\trace{\bar BB}\,,
 \trace{\bar B\bar B}\trace{\bar BB}\trace{BB}\,. 
\end{eqnarray}
The possible Dirac structures are
\begin{eqnarray}
 &\mathbbm1\otimes\mathbbm1\otimes\mathbbm1\,,\
 \mathbbm1 \otimes \gamma_5\gamma^\mu \otimes \gamma_5\gamma_\mu\,,\
 \gamma_5\gamma^\mu \otimes \mathbbm1 \otimes \gamma_5\gamma_\mu\,,\\
 &\gamma_5\gamma^\mu \otimes \gamma_5\gamma_\mu \otimes \mathbbm1\,,\
 \gamma_5\gamma_\mu \otimes \mathrm i\;\sigma^{\mu\nu} \otimes \gamma_5\gamma_\nu\,, 
\end{eqnarray}
which lead to the following operators in the three-body spin space
\begin{equation} \label{eq:BBBbasis}
\mathbbm1\,,\
\vec\sigma_1\cdot\vec\sigma_2\,,\
\vec\sigma_1\cdot\vec\sigma_3\,,\
\vec\sigma_2\cdot\vec\sigma_3\,,\
\mathrm i\;\vec\sigma_1\cdot(\vec\sigma_2\times\vec\sigma_3) \,.
\end{equation}
All combinations of these possibilities leads to a (largely overcomplete) set of terms for the leading covariant Lagrangian.
Note that in \cite{Petschauer:2015elq} the starting point is a covariant Lagrangian, but the goal is the minimal non-relativistic Lagrangian.
Therefore, only Dirac structures leading to independent (non-relativistic) spin operators are relevant.

Let us consider the process \(B_1 B_2 B_3 \rightarrow B_4 B_5 B_6\), where the \(B_i\) are baryons in the particle basis, \\ \(B_i\in\{n,p,\Lambda,\Sigma^+,\Sigma^0,\Sigma^-,\Xi^0,\Xi^-\}\).
The contact potential \(V\) has to be derived within a threefold spin space for this process.
The operators in spin-space 1 is defined to act between the two-component Pauli spinors of \(B_1\) and \(B_4\).
In the same way, spin-space 2 belongs to \(B_2\) and \(B_5\), and spin-space 3 to \(B_3\) and \(B_6\).
For a fixed spin configuration the potential can be calculated from
\begin{equation} \label{eq:spinpot}
{\chi_{B_4}^{(1)}}^\dagger {\chi_{B_5}^{(2)}}^\dagger {\chi_{B_6}^{(3)}}^\dagger \, V \, \chi_{B_1}^{(1)}\chi_{B_2}^{(2)}\chi_{B_3}^{(3)} \,,
\end{equation} 
where the superscript of a spinor denotes the spin space and the subscript denotes the baryon to which the spinor belongs.
The potential is obtained as
\(
V = -\langle B_4 B_5 B_6\vert \ \mathscr L \ \vert B_1 B_2 B_3\rangle
\),
where the contact Lagrangian \(\mathscr L\) has to be inserted, and the 36 Wick contractions need to be performed.
The number 36 corresponds to the \(3!\times3!\) possibilities to arrange the three initial and three final baryons into Dirac bilinears.
One obtains six direct terms, where the baryon bilinears combine the baryon pairs 1--4, 2--5 and 3--6, as shown in 
Eq.~(\ref{eq:spinpot}).
For the other 30 Wick contractions, the resulting potential is not fitting to the form of Eq.~(\ref{eq:spinpot}), because the wrong baryon pairs are connected in a separate spin space.
Hence, an appropriate exchange of the spin wave functions in the final state  has to be performed.
This is achieved by multiplying the potential with the well-known spin-exchange operators \(\Ps_{ij}=\frac12(\mathbbm1+\vec\sigma_i\cdot\vec\sigma_j)\).
Furthermore additional minus signs arise from the interchange of anticommuting baryon fields.
The full potential is then obtained by adding up all 36 contributions to the potential.
One obtains a potential that fulfills automatically the generalized Pauli principle and that is fully antisymmetrized.

In order to obtain a minimal set of Lagrangian terms of the final potential matrix,
redundant terms have been eliminated until the rank of the final potential matrix
(consisting of multiple flavor structures, see Eqs. (\ref{eq:BBBflavor})
-(\ref{eq:BBBflavorE}), 
and the spin structures in Eq.~(\ref{eq:BBBbasis})) 
matches the number of terms in the Lagrangian.
The minimal non-relativistic six-baryon contact Lagrangian is \cite{Petschauer:2015elq}
\begin{align*} \label{eq:minct}
 \mathscr L =
 -\,&\lc_1 \trace{\bar B_a\bar B_b\bar B_c B_a B_b B_c} 
 +\,\lc_2 \trace{\bar B_a\bar B_b B_a\bar B_c B_b B_c} \displaybreak[0]\\
 -\,&\lc_3 \trace{\bar B_a\bar B_b B_a B_b\bar B_c B_c} \displaybreak[0]
 +\,\lc_4 \trace{\bar B_a B_a\bar B_b B_b\bar B_c B_c} \displaybreak[0]\\
 -\,&\lc_5 \trace{\bar B_a\bar B_b B_a B_b}\; \trace{\bar B_c B_c} \displaybreak[0]\\
 -\,&\lc_6 \Big(\trace{\bar B_a\bar B_b\bar B_c B_a(\sigma^i B)_b(\sigma^i B)_c}
 + \trace{\bar B_c\bar B_b\bar B_a(\sigma^i B)_c(\sigma^i B)_b B_a}\Big)\displaybreak[0]\\
 +\,&\lc_7 \Big(\trace{\bar B_a\bar B_b B_a\bar B_c(\sigma^i B)_b(\sigma^i B)_c}
 + \trace{\bar B_c\bar B_b(\sigma^i B)_c\bar B_a(\sigma^i B)_b B_a}\Big) \displaybreak[0]\\
 -\,&\lc_8 \Big(\trace{\bar B_a\bar B_b B_a(\sigma^i B)_b\bar B_c(\sigma^i B)_c}
 + \trace{\bar B_b\bar B_a(\sigma^i B)_b B_a\bar B_c(\sigma^i B)_c}\Big) \displaybreak[0]\\
 +\,&\lc_9 \trace{\bar B_a B_a\bar B_b(\sigma^i B)_b\bar B_c(\sigma^i B)_c} \displaybreak[0]\\
 -\,&\lc_{10} \Big(\trace{\bar B_a\bar B_b B_a(\sigma^i B)_b}\; \trace{\bar B_c(\sigma^i B)_c}
 + \trace{\bar B_b\bar B_a(\sigma^i B)_b B_a}\; \trace{\bar B_c(\sigma^i B)_c}\Big) \displaybreak[0]\\
 -\,&\lc_{11} \trace{\bar B_a\bar B_b\bar B_c(\sigma^i B)_a B_b(\sigma^i B)_c} \displaybreak[0]
 +\,\lc_{12} \trace{\bar B_a\bar B_b(\sigma^i B)_a\bar B_c B_b(\sigma^i B)_c} \displaybreak[0]\\
 -\,&\lc_{13} \trace{\bar B_a\bar B_b(\sigma^i B)_a(\sigma^i B)_b\bar B_c B_c} \displaybreak[0]
 -\,\lc_{14} \trace{\bar B_a\bar B_b(\sigma^i B)_a(\sigma^i B)_b}\; \trace{\bar B_c B_c} \displaybreak[0]\\
 -\,&\mathrm i\, \epsilon^{ijk}\lc_{15}\trace{\bar B_a\bar B_b\bar B_c(\sigma^i B)_a(\sigma^j B)_b(\sigma^k B)_c} \displaybreak[0]
 +\,\mathrm i\, \epsilon^{ijk}\lc_{16}\trace{\bar B_a\bar B_b(\sigma^i B)_a\bar B_c(\sigma^j B)_b(\sigma^k B)_c} \displaybreak[0]\\
 -\,&\mathrm i\, \epsilon^{ijk}\lc_{17}\trace{\bar B_a\bar B_b(\sigma^i B)_a(\sigma^j B)_b\bar B_c(\sigma^k B)_c}
 +\,\mathrm i\, \epsilon^{ijk}\lc_{18}\trace{\bar B_a(\sigma^i B)_a\bar B_b(\sigma^j B)_b\bar B_c(\sigma^k B)_c} \,, \numberthis
\end{align*}
with vector indices \(i,j,k\) and two-component spinor indices \(a,b,c\).
In total 18 low-energy constants \(\lc_1\dots \lc_{18}\) are present.
The low-energy constant \(E\) of the six-nucleon contact term (see \cite{Epelbaum:2002vt}) can be expressed through these LECs by \(E=2(\lc_4-\lc_9)\).

\begin{table*}
\centering
\vspace{.3\baselineskip}

\begin{tabular}{>{$}c<{$}>{$}c<{$}>{$}c<{$}>{$}c<{$}}
\hline
\text{states} & (S,I) & {}^2S_{1/2} & {}^4S_{3/2} \\
\cmidrule(lr){1-2} \cmidrule(lr){3-4}
NNN & (0,\frac12) & \ir{\overline{35}}\\
\cmidrule(lr){1-2} \cmidrule(lr){3-4}
\Lambda NN,\Sigma NN & (-1,0) & \ir{\overline{10}},\ir{\overline{35}} & \ir{\overline{10}}_a \\
\Lambda NN,\Sigma NN & (-1,1) & \ir{27},\ir{\overline{35}} & \ir{27}_a \\
\Sigma NN & (-1,2) & \ir{35}\\
\cmidrule(lr){1-2} \cmidrule(lr){3-4}
\Lambda\Lambda N,\Sigma\Lambda N,\Sigma\Sigma N,\Xi NN & (-2,\frac12) & \ir{8},\ir{\overline{10}},\ir{27},\ir{\overline{35}} & \ir{8}_a,\ir{\overline{10}}_a,\ir{27}_a\\
\Sigma\Lambda N,\Sigma\Sigma N,\Xi NN & (-2,\frac32) & \ir{10},\ir{27},\ir{35},\ir{\overline{35}} & \ir{10}_a,\ir{27}_a\\
\Sigma\Sigma N & (-2,\frac52) & \ir{35} \\
\cmidrule(lr){1-2} \cmidrule(lr){3-4}
\Lambda\Lambda\Lambda,\Sigma\Sigma\Lambda,\Sigma\Sigma\Sigma,\Xi\Lambda N,\Xi\Sigma N & (-3,0) & \ir{8},\ir{27} & \ir{1}_a,\ir{8}_a,\ir{27}_a\\
\Sigma\Lambda\Lambda,\Sigma\Sigma\Lambda,\Sigma\Sigma\Sigma,\Xi\Lambda N,\Xi\Sigma N & (-3,1) & \ir{8},\ir{10},\ir{\overline{10}},\ir{27},\ir{35},\ir{\overline{35}} & \ir{8}_a,\ir{10}_a,\ir{\overline{10}}_a,\ir{27}_a\\
\Sigma\Sigma\Lambda,\Sigma\Sigma\Sigma,\Xi\Sigma N & (-3,2) & \ir{27},\ir{35},\ir{\overline{35}} & \ir{27}_a \\
\cmidrule(lr){1-2} \cmidrule(lr){3-4}
\Xi\Lambda\Lambda,\Xi\Sigma\Lambda,\Xi\Sigma\Sigma,\Xi\Xi N & (-4,\frac12) & \ir{8},\ir{10},\ir{27},\ir{35} & \ir{8}_a,\ir{10}_a,\ir{27}_a\\
\Xi\Sigma\Lambda,\Xi\Sigma\Sigma,\Xi\Xi N & (-4,\frac32) & \ir{\overline{10}},\ir{27},\ir{35},\ir{\overline{35}} & \ir{\overline{10}}_a,\ir{27}_a \\
\Xi\Sigma\Sigma & (-4,\frac52) & \ir{\overline{35}} \\
\cmidrule(lr){1-2} \cmidrule(lr){3-4}
\Xi\Xi\Lambda,\Xi\Xi\Sigma & (-5,0) & \ir{10},\ir{35} & \ir{10}_a\\
\Xi\Xi\Lambda,\Xi\Xi\Sigma & (-5,1) & \ir{27},\ir{35} & \ir{27}_a \\
\Xi\Xi\Sigma & (-5,2) & \ir{\overline{35}}\\
\cmidrule(lr){1-2} \cmidrule(lr){3-4}
\Xi\Xi\Xi & (-6,\frac12) & \ir{35}\\
\bottomrule
\end{tabular}
\caption{Irreducible representations for three-baryon states with strangeness \(S\) and isospin \(I\) in partial waves \(\vert {}^{2S+1}L_J \rangle\),
with the total spin \(S=\frac12,\frac32\), the angular momentum \(L=0\) and the total angular momentum \(J=\frac12,\frac32\) \cite{Petschauer:2015elq}.
\label{tab:isoBBB}
}
\end{table*}

As in the two-body sector, group theoretical considerations can deliver valuable constrains on the resulting potentials.
In flavor space, the three octet baryons form the 512-dimensional tensor product \(\ir8 \otimes \ir8 \otimes \ir8\), which decomposes into the following irreducible SU(3) representations
\begin{align*} \label{eq:irrBBB}
 \ir8 \otimes \ir8 \otimes \ir8 =
 \ir{64} \oplus (\ir{35} \oplus \ir{\overline{35}})_2 \oplus \ir{27}_6
\oplus (\ir{10} \oplus \ir{\overline{10}})_4 \oplus \ir{8}_8 \oplus \ir{1}_2\,, \numberthis
\end{align*}
where the multiplicity of an irreducible representations is denoted by subscripts.
In spin space, one obtains for the product of three doublets
\begin{equation}
\ir2 \otimes \ir2 \otimes \ir2 = \ir{2}_2 \oplus \ir{4}\,.
\end{equation}
Transitions are only allowed between irreducible representations of the same type.
Analogous to \cite{Dover:1990id} for the two-baryon sector, the contributions of different irreducible representations to three-baryon multiplets can be established,
which is summarized in Table \ref{tab:isoBBB}. 
At LO only transitions between \(S\)-waves are possible, since the potentials are momentum-independent.
Due to the Pauli principle, the totally symmetric spin-quartet \(\mathbf4\) must combine with the totally antisymmetric part of \(\ir8 \otimes \ir8 \otimes \ir8\) in flavor space,%
\begin{equation} \label{eq:alt38}
 \text{Ant}_3(\mathbf8) = \ir{56}_a = \ir{27}_a+\ir{10}_a+\ir{\overline{10}}_a+\ir{8}_a+\ir{1}_a \,.
\end{equation}
It follows, that these totally antisymmetric irreducible representations are present only in states with total spin 3/2.
The totally symmetric part of \(\ir8 \otimes \ir8 \otimes \ir8\) leads to
\begin{equation}
\text{Sym}_3(\mathbf8) = \ir{120}_s = \ir{64}_s+\ir{27}_s+\ir{10}_s+\ir{\overline{10}}_s+\ir{8}_s+\ir{1}_s \,.
\end{equation}
However, the totally symmetric flavor part has no totally antisymmetric counterpart in spin space, hence these representations do not contribute to the potential.
In Table \ref{tab:isoBBB}, these restrictions obtained by the generalized Pauli principle have already be incorporated.
The potentials of \cite{Petschauer:2015elq} (decomposed in isospin basis and partial waves) fulfill the restrictions of Table \ref{tab:isoBBB}.
For example the combination of LECs related to the representation \(\ir{\overline{35}}\) is present in the \(NNN\) interaction as well as in  the \(\Xi\Xi\Sigma\) $({\rm strangeness,isospin})=(S,I)=(-5,2)$  interaction.

\subsection{One-meson exchange component} \label{subsec:BBBome}

The meson-baryon couplings in the one-meson exchange diagram of Fig.~\ref{fig:3BF} emerge from the LO chiral Lagrangian 
\(\mathscr{L}_\mathrm{BBP}\) given in Eq.~(\ref{eq:pseudovector}). It specifies 
the $BB\phi$ vertices where $\phi =\{\pi^0,\pi^+,\pi^-,K^+,K^-,K^0,\bar K^0,\eta\}$. 
The other vertex involves four baryon fields and one pseudo\-scalar-meson field.
In \cite{Petschauer:2015elq} first an overcomplete set of terms for the 
corresponding Lagrangian has been constructed. Subsequently, 
in order to obtain the complete minimal Lagrangian from the overcomplete set of terms, the matrix elements of the process \(B_1B_2\to B_3B_4\phi_1\) have been 
evaluated. The corresponding spin operators in the potential are
\begin{equation}
\vec\sigma_1\cdot\vec q\,,\quad
\vec\sigma_2\cdot\vec q\,,\quad
\mathrm i\,(\vec\sigma_1\times\vec\sigma_2)\cdot\vec q \,,
\end{equation}
where \(\vec q\) denotes the momentum of the emitted meson.
Redundant term are removed until the rank of the potential matrix formed by all transitions and spin operators matches the number of terms in the Lagrangian.
One ends up with the minimal non-relativistic chiral Lagrangian
\begin{align*} \label{eq:LBBMBBmin}
 \mathscr L ={}
 &\ld_1/F_0 \trace{\bar B_a \extfield B_a\bar B_b(\sigma^i B)_b}\displaybreak[0]\\
 &+\ld_2/F_0 \Big( \trace{\bar B_a  B_a \extfield\bar B_b(\sigma^i B)_b} 
 + \trace{\bar B_a  B_a\bar B_b(\sigma^i B)_b \extfield}\Big)\displaybreak[0]\\
 &+\ld_3/F_0 \trace{\bar B_b \extfield(\sigma^i B)_b\bar B_a B_a}\displaybreak[0]\\
 &-\ld_4/F_0 \Big( \trace{\bar B_a\extfield\bar B_b B_a(\sigma^i B)_b}
 + \trace{\bar B_b \bar B_a (\sigma^i B)_b\extfield B_a}\Big)\displaybreak[0]\\
 &-\ld_5/F_0 \Big( \trace{\bar B_a\bar B_b\extfield B_a(\sigma^i B)_b}
 + \trace{\bar B_b \bar B_a \extfield (\sigma^i B)_b B_a}\Big)\displaybreak[0]\\
 &-\ld_6/F_0 \Big( \trace{\bar B_b\extfield\bar B_a(\sigma^i B)_b B_a}
 + \trace{\bar B_a \bar B_b B_a\extfield (\sigma^i B)_b}\Big)\displaybreak[0]\\
 &-\ld_7/F_0 \Big( \trace{\bar B_a\bar B_b B_a(\sigma^i B)_b\extfield} 
 + \trace{\bar B_b \bar B_a (\sigma^i B)_b B_a\extfield}\Big)\displaybreak[0]\\
 &+\ld_8/F_0 \trace{\bar B_a\extfield B_a}\trace{\bar B_b(\sigma^i B)_b}\displaybreak[0]
 +\ld_9/F_0 \trace{\bar B_a B_a\extfield}\trace{\bar B_b(\sigma^i B)_b}\displaybreak[0]\\
 &+\ld_{10}/F_0 \trace{\bar B_b\extfield(\sigma^i B)_b}\trace{\bar B_a B_a}\displaybreak[0]\\
 &+\mathrm i\,\epsilon^{ijk}\ld_{11}/F_0 \trace{\bar B_a (\sigma^i B)_a (\nabla^k \phi)\bar B_b(\sigma^j B)_b}\displaybreak[0]\\
 &-\mathrm i\,\epsilon^{ijk}\ld_{12}/F_0 \Big( \trace{\bar B_a(\nabla^k \phi)\bar B_b(\sigma^i B)_a(\sigma^j B)_b}
 - \trace{\bar B_b \bar B_a (\sigma^j B)_b(\nabla^k \phi) (\sigma^i B)_a}\Big)\displaybreak[0]\\
 &-\mathrm i\,\epsilon^{ijk}\ld_{13}/F_0 \trace{\bar B_a\bar B_b(\nabla^k \phi)(\sigma^i B)_a(\sigma^j B)_b}\displaybreak[0]
 -\mathrm i\,\epsilon^{ijk}\ld_{14}/F_0 \trace{\bar B_a\bar B_b(\sigma^i B)_a(\sigma^j B)_b(\nabla^k \phi)}\,, \numberthis
\end{align*}
with two-component spinor indices  \(a\) and \(b\) and 3-vector indices \(i\), \(j\) and \(k\).
For all possible strangeness sectors \(S=-4\ldots0\) one obtains in total 14 low-energy constants \(\ld_1\dots \ld_{14}\) .
The low-energy constant of the corresponding vertex in the nucleonic sector \(D\) 
\cite{Epelbaum:2002vt} is related to the LECs above 
by \(D=4 (\ld_1 - \ld_3 + \ld_8 - \ld_{10})\).

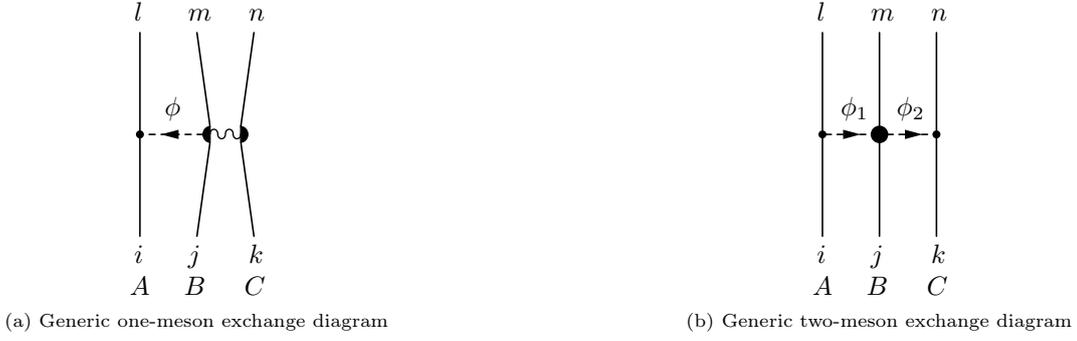
\begin{figure*}
\centering
\hfill
\begin{subfigure}[t]{.45\textwidth}
\centering
\vspace{.3\baselineskip}
\begin{overpic}[scale=.6]{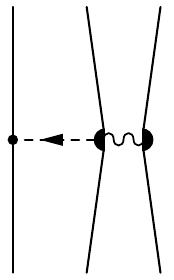}
\put(2,103){$l$}\put(27,103){$m$}\put(55,103){$n$}
\put(2,-10){$i$}\put(27,-10){$j$}\put(55,-10){$k$}
\put(0,-25){$A$}\put(25,-25){$B$}\put(53,-25){$C$}
\put(16,59){$\phi$}
\end{overpic}
\vspace{1.7\baselineskip}
\caption{
Generic one-meson exchange diagram
\label{fig:ome-gen}
}
\end{subfigure}
\hfill
\begin{subfigure}[t]{.45\textwidth}
\centering
\vspace{.3\baselineskip}
\begin{overpic}[scale=.6]{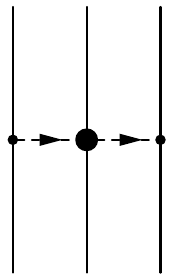}
\put(2,103){$l$}\put(27,103){$m$}\put(55,103){$n$}
\put(2,-10){$i$}\put(27,-10){$j$}\put(55,-10){$k$}
\put(0,-25){$A$}\put(25,-25){$B$}\put(53,-25){$C$}
\put(13,59){$\phi_1$} \put(39,59){$\phi_2$}
\end{overpic}
\vspace{1.7\baselineskip}
\caption{
Generic two-meson exchange diagram
\label{fig:tme-gen}
}
\end{subfigure}
\hfill\mbox{}
\caption{Generic meson-exchange diagrams.
The wiggly line symbolized the four-baryon contact vertex, to illustrate the baryon bilinears.}
\end{figure*}

To obtain the 3BF one-meson-exchange diagram, the generic one-meson-exchange diagram in Fig.~\ref{fig:ome-gen} has to be evaluated.
It involves the baryons \(i,j,k\) in the initial state, the baryons \(l,m,n\) in the final state and an exchanged meson \(\phi\).
The contact vertex on the right is pictorially separated into two parts to indicate that baryon j--m and k--n are in the same bilinear.
The spin spaces corresponding to the baryon bilinears are denoted by \(A,B,C\).

One obtains a generic potential of the form
\begin{equation}
V = \frac{1}{2F_0^2} \frac{\vec\sigma_A\cdot\vec q_{li}}{\vec q_{li}^{\,2}+M_{\phi}^2}\, \Big(
N_1 \vec\sigma_C\cdot\vec q_{li}
+N_2 \mathrm i\,(\vec\sigma_B\times\vec\sigma_C)\cdot\vec q_{li}
\Big)\,,
\end{equation}
with the momentum transfer \(\vec q_{li} = \vec p_l-\vec p_i\) carried by the exchanged meson.
The constants \(N_1\) and \(N_2\) are linear combinations of low-energy constants.

The complete one-meson exchange three-baryon potential for the process \(B_1B_2B_3\to B_4B_5B_6\) is finally obtained by summing up the 36 permutations of initial-state and final-state baryons for a fixed meson and by summing over all mesons \(\phi \in \left\{\pi^0,\pi^+,\pi^-,K^+,K^-,K^0,\bar K^0,\eta\right\}\).
%
%
As defined before, the baryons \(B_1\), \(B_2\) and \(B_3\) belong to the spin-spaces 1, 2 and 3, respectively.

\subsection{Two-meson exchange component} \label{subsec:2MEtme}

The two-meson exchange diagram of Fig.~\ref{fig:3BF} includes the vertices 
arising from the $BB\phi$ Lagrangian. Furthermore, we need the well-known
\(\mathcal O(q^2)\) meson-baryon Lagrangian \cite{Krause:1990xc}.
The relevant terms are \cite{Oller:2006yh}
\begin{align*} \label{eq:MBMBLagr}
\mathscr L ={}
& b_D\langle\bar B\{\chi_+,B\}\rangle
+ b_F\langle\bar B[\chi_+,B]\rangle
+ b_0\langle\bar BB\rangle\,\langle\chi_+\rangle \\
&+ b_1\langle\bar B[u^\mu,[u_\mu,B]]\rangle
+ b_2\langle\bar B\{u^\mu,\{u_\mu,B\}\}\rangle
+ b_3\langle\bar B\{u^\mu,[u_\mu,B]\}\rangle
+ b_4\langle\bar BB\rangle\,\langle u^\mu u_\mu\rangle \\
&+\mathrm i d_1\langle\bar B\{[u^\mu,u^\nu],\sigma_{\mu\nu}B\}\rangle
+\mathrm i d_2\langle\bar B[[u^\mu,u^\nu],\sigma_{\mu\nu}B]\rangle 
+\mathrm i d_3\langle\bar Bu^\mu\rangle\langle u^\nu\sigma_{\mu\nu}B\rangle \,, \numberthis
\end{align*}
with \(u_\mu = -\frac1{F_0}\partial_\mu\phi+\mathcal O(\phi^3)\) and \(\chi_+ = 2\chi-\frac1{4F_0^2}\{\phi,\{\phi,\chi\}\}\linebreak[0]+\mathcal O(\phi^4)\), 
and $\chi$ given in Eq.~(\ref{VCTSB1}). 
The terms proportional to \(b_D,b_F,b_0\) break explicitly SU(3) flavor symmetry, because of different meson masses \(M_K\neq M_\pi\).
The LECs of Eq.~\eqref{eq:MBMBLagr} are related to the conventional LECs of the nucleonic sector by \cite{Frink:2004ic,Mai:2009ce}
\begin{align*} \label{eq:LECc134}
c_1 &= \frac{1}{2} (2 b_0+b_D+b_F)\,,\\
c_3 &= b_1+b_2+b_3+2 b_4\,,\\
c_4 &= 4 (d_1+d_2)\,. \numberthis
\end{align*}
Note, however, that these are not the matching relations when going from SU(3) 
to SU(2) but rather the relations that
are generated when one compares the dimension-two meson-baryon operators in the two- and the three-flavor representations.
The matching relations are discussed in Refs.~\cite{Frink:2004ic,Mai:2009ce}.

To obtain the potential of the two-meson exchange diagram of Fig.~\ref{fig:3BF}, the generic diagram of 
Fig.~\ref{fig:tme-gen} can be considered.
It includes the baryons \(i,j,k\) in the initial state, the baryons \(l,m,n\) in the final state, and two exchanged mesons \(\phi_1\) and \(\phi_2\).
The spin spaces corresponding to the baryon bilinears are denoted by \(A,B,C\) and they are aligned with the three initial baryons.
The momentum transfers carried by the virtual mesons are \(\vec q_{li} = \vec p_l-\vec p_i\) and \(\vec q_{nk} = \vec p_n-\vec p_k\).
One obtains the generic transition amplitude
\begin{align*}
V ={} & -\frac{1}{4F_0^4} \frac{\vec\sigma_A\cdot\vec q_{li}\ \vec\sigma_C\cdot\vec q_{nk}}{(\vec q_{li}^{\,2}+M_{\phi_1}^2)(\vec q_{nk}^{\,2}+M_{\phi_2}^2)}\,
\Big(N'_1 + N'_2\,\vec q_{li}\cdot\vec q_{nk} +N'_3\,\mathrm i\,(\vec q_{li}\times\vec q_{nk})\cdot\vec\sigma_B\Big) \,, \numberthis
\end{align*}
with \(N^\prime_i\) linear combinations of the low-energy constants of the three involved vertices.
The complete three-body potential for a transition \(B_1B_2B_3\rightarrow B_4B_5B_6\) can be calculated by summing up the contributions of all 18 distinguable Feynman diagrams and by summing over all possible exchanged mesons.
If the baryon lines are not in the configuration 1--4, 2--5 and 3--6 additional (negative) spin-exchange operators have to be included.

\subsection{\texorpdfstring{$\La$}NNN and NNN three-baryon potentials} \label{subsec:BBBpotex}

In order to give concrete examples, the explicit expressions for the \(\Lambda NN\) three-body potentials in spin-, isospin- and momentum-space are presented
for the contact interaction and one- and two-pion exchange contributions \cite{Petschauer:2015elq}. In addition we provide those for the corresponding 
NNN forces for the ease of comparison. 

\subsubsection{Contact interaction}

The \(\Lambda NN\) contact interaction is described by the following potential
\begin{align*}
\label{eq:lnncontact}
V_{\Lambda NN}^\mathrm{CT} ={}
& \phantom{{}+{}} \lc'_1\ (\mathbbm1 - \vec\sigma_2\cdot\vec\sigma_3 ) ( 3 + \vec\tau_2\cdot\vec\tau_3 ) \\
& + \lc'_2\ \vec\sigma_1\cdot(\vec\sigma_2+\vec\sigma_3)\,(\mathbbm1 - \vec\tau_2\cdot\vec\tau_3) \\
& + \lc'_3\ (3 + \vec\sigma_2\cdot\vec\sigma_3 ) ( \mathbbm1 - \vec\tau_2\cdot\vec\tau_3 ) \,, \numberthis
\end{align*}
where the primed constants are linear combinations of \(\lc_1\dots \lc_{18}\) of Eq.~(\ref{eq:minct}).
The symbols \(\vec\sigma\) and \(\vec\tau\) denote the usual Pauli matrices in spin and isospin space.
The constant \(\lc'_1\) appears only in the transition with total isospin \(I=1\).
The constants \(\lc'_2\) and \(\lc'_3\) contribute for total isospin \(I=0\).

For comparison, the NNN contact potential of \cite{Epelbaum:2002vt} in its antisymmetrized form is
\begin{equation}
V_{NNN}^\mathrm{CT} = \frac12 E\, \mathcal A\, \sum_{j\neq k}\vec\tau_j\cdot\vec\tau_k \,,
\end{equation}
where \(\mathcal A\) denotes the three-body antisymmetrization operator, \(\mathcal A = (\mathbbm1-\mathcal P_{12}) (\mathbbm1-\mathcal P_{13}-\mathcal P_{23})\).
Each two-particle exchange operator
\(\mathcal P_{ij} = \Ps_{ij}\Pt_{ij}\Pp_{ij}\)
is the product of an exchange operator in spin space \(\Ps_{ij}=\frac12(\mathbbm1+\vec\sigma_i\cdot\vec\sigma_j)\), in isospin space \(\Pt_{ij}=\frac12(\mathbbm1+\vec\tau_i\cdot\vec\tau_j)\) and in momentum space \(\Pp_{ij}\).
However, the latter has no effect here, since the LO NNNN contact 
potential is momentum-independent.

Note that the NNN force is usually written in terms of the dimensionless coupling
$c_E = E F_\pi^4 \Lambda_\chi$, where $c_E$ is then in the order of one. 
Here $\Lambda_\chi$ is the chiral symmetry breaking scale assumed to be
of the order of the $\rho$ meson mass, i.e. $\Lambda_\chi \approx 700$~MeV \cite{Epelbaum:2002vt}. 

\subsubsection{One-pion exchange}
Isospin conservation implies that the \(\Lambda\Lambda\pi\) coupling constant
is zero. This reduces the number of possible diagrams that can contribute
to the $\Lambda$NN one-pion exchange three-body potentials.
One obtains the following potential
\begin{align*} \label{eq:LNNope}
V_{\Lambda NN}^\mathrm{OPE} =&{} -\frac{g_A}{2F_0^2} \,\bigg(
\frac{\vec\sigma_2\cdot\vec q_{52}}{\vec q_{52}^{\,2}+M_\pi^2} \vec\tau_2\cdot\vec\tau_3
\Big[  (\ld'_1\vec\sigma_1+\ld'_2\vec\sigma_3)\cdot\vec q_{52} \Big] 
+\frac{\vec\sigma_3\cdot\vec q_{63}}{\vec q_{63}^{\,2}+M_\pi^2} \vec\tau_2\cdot\vec\tau_3
\Big[  (\ld'_1\vec\sigma_1+\ld'_2\vec\sigma_2)\cdot\vec q_{63} \Big] \\
&+\Ps_{23}\Pt_{23} \Ps_{13}\frac{\vec\sigma_2\cdot\vec q_{62}}{\vec q_{62}^{\,2}+M_\pi^2} \vec\tau_2\cdot\vec\tau_3 
\Big[  -\frac{\ld'_1+\ld'_2}2 (\vec\sigma_1+\vec\sigma_3)\cdot\vec q_{62}
+ \frac{\ld'_1-\ld'_2}2\,\mathrm i\,(\vec\sigma_3\times\vec\sigma_1)\cdot\vec q_{62}  \Big] \\
&+\Ps_{23}\Pt_{23} \Ps_{12}\frac{\vec\sigma_3\cdot\vec q_{53}}{\vec q_{53}^{\,2}+M_\pi^2} \vec\tau_2\cdot\vec\tau_3 
\Big[  -\frac{\ld'_1+\ld'_2}2 (\vec\sigma_1+\vec\sigma_2)\cdot\vec q_{53} 
- \frac{\ld'_1-\ld'_2}2\,\mathrm i\,(\vec\sigma_1\times\vec\sigma_2)\cdot\vec q_{53}  \Big]
\bigg)\,, \numberthis
\end{align*}
with only two constants \(\ld'_1\) and \(\ld'_2\), which are linear combinations of the constants \(\ld_1\dots\ld_{14}\).
Exchange operators in spin space \(\Ps_{ij}=\frac12(\mathbbm1+\vec\sigma_i\cdot\vec\sigma_j)\) and in isospin space \(\Pt_{ij}=\frac12(\mathbbm1+\vec\tau_i\cdot\vec\tau_j)\) have been introduced.

The corresponding result for the NNN potential is equal to the antisymmetrization 
of the expression given in \cite{Epelbaum:2002vt}, \pagebreak[0]
\begin{equation}
V_{NNN}^\mathrm{OPE} =-\frac{g_A}{8F_\pi^2} D \mathcal{A} \sum_{i\neq j\neq k}\frac{\vec\sigma_j\cdot\vec q_j}{\vec q_j^{\,2}+m_\pi^2} \vec\tau_i\cdot\vec\tau_j\ \vec\sigma_i\cdot\vec q_j \,,
\end{equation}
inserting the momentum transfers \(\vec q_1 = \vec q_{41} = \vec p_4-\vec p_1\), \(\vec q_2 = \vec q_{52} = \vec p_5-\vec p_2\), \(\vec q_3 = \vec q_{63} = \vec p_6-\vec p_3\).
In this case the momentum part of each two-body exchange operator, \(\Pp_{ij}\), exchanges also the momenta in the final state.%
\footnote{For example, \(\Pp_{23}\) leads to the replacements \(q_{41},q_{52},q_{63}\to q_{41},q_{62},q_{53}\) and \(\Pp_{12}\Pp_{13}\) to \(q_{41},q_{52},q_{63}\to q_{61},q_{42},q_{53}\).}

Also here it is standard to rewrite the NNN
forces in terms of dimensionless couplings, namely in the form 
$c_D = D F_\pi^2 \Lambda_\chi$~\cite{Epelbaum:2002vt}.

\subsubsection{Two-pion exchange}

The $\Lambda$NN three-body interaction generated by two-pion exchange is given by
\begin{align*} \label{eq:LNNtpe}
&V_{\Lambda NN}^\mathrm{TPE} ={}
\frac{g_A^2}{3F_0^4}
\frac{\vec\sigma_3\cdot\vec q_{63}\ \vec\sigma_2\cdot\vec q_{52}}{(\vec q_{63}^{\,2}+M_{\pi}^2)(\vec q_{52}^{\,2}+M_{\pi}^2)} \vec\tau_2\cdot\vec\tau_3 
 \,  \Big( -(3 b_0 + b_D) M_\pi^2      +      (2 b_2 + 3 b_4)      \,\vec q_{63}\cdot\vec q_{52}\Big) \\
&\quad- \Ps_{23}\Pt	_{23} \frac{g_A^2}{3F_0^4}
\frac{\vec\sigma_3\cdot\vec q_{53}\ \vec\sigma_2\cdot\vec q_{62}}{(\vec q_{53}^{\,2}+M_{\pi}^2)(\vec q_{62}^{\,2}+M_{\pi}^2)} \vec\tau_2\cdot\vec\tau_3
 \, \Big( -(3 b_0 + b_D) M_\pi^2      +      (2 b_2 + 3 b_4)      \,\vec q_{53}\cdot\vec q_{62}\Big) \,. \numberthis
\end{align*}
Due to the vanishing of the \(\Lambda \Lambda \pi\) vertex, only those two diagrams contribute, where the (final and initial) \(\Lambda\) hyperon are attached to the central baryon line, see Fig.~\ref{fig:tme-gen}. 

Regarding the NNN case the result is equal to the antisymmetrization of the expression given in \cite{Epelbaum:2002vt}:
\begin{equation}
V_\mathrm{NNN}^\mathrm{TPE} = \frac{g_A^2}{8F_\pi^2} \mathcal{A} \sum_{i\neq j\neq k} \frac{\vec\sigma_i\cdot\vec q_i\ \vec\sigma_j\cdot\vec q_j}{ (\vec q_i^{\,2}+M_\pi^2)(\vec q_j^{\,2}+M_\pi^2)} F^{\alpha\beta}_{ijk}\tau^\alpha_i\tau^\beta_j \,,
\end{equation}
with
\begin{align*}
& F^{\alpha\beta}_{ijk} =
 \frac{\delta^{\alpha\beta}}{F_\pi^2} (-4 c_1 M_\pi^2+2c_3\vec q_i\cdot\vec q_j)
 +\sum_\gamma \frac{c_4}{F_\pi^2}\epsilon^{\alpha\beta\gamma}\tau_k^\gamma\ \vec\sigma_k\cdot(\vec q_i\times\vec q_j) \,, \numberthis
\end{align*}
where the $c_i$ are given in Eq.~(\ref{eq:LECc134}). 


\subsection{Three-baryon force through decuplet saturation}  \label{subsec:BBBDec}

As should be clear from the discussion above, the number of LECs of the 
leading YNN 3BF is by far too large as compared to the available 
experimental constraints from light hypernuclei which could be used to
pin them down. Of course, the rather large number of unknown LECs presented in the
previous subsections is related to the whole multitude of three-baryon multiplets, 
with strangeness S ranging from \(0\) to \(-6\). 
But the number is not that much reduced if we restrict ourselves to the S=-1 sector, 
i.e. to $\La$NN and $\Si$NN. Even if one considers the $\La$NN 3BF alone 
(and disregards $K$ and $\eta$ exchanges), as done in Sect.~\ref{subsec:BBBpotex},
one would have to deal with 5 LECs (3 for the contact terms and 2
from the one-pion exchange 3BF) to be compared with only 2 LECs in the NNN case. 

Varying all LECs or setting some of them arbitrarily to zero is neither practical
nor a sound procedure for concrete few-body calculations. 
Therefore, in a first application of the chiral 3BFs in studying hypernuclei
\cite{Le:2024rkd} a different strategy was followed. In that work 
the mechanism of resonance saturation
via decuplet baryons ($\Delta$, $\Sigma^*$, $\Xi^*$) was exploited 
to estimate the strengths of chiral 3BFs 
\cite{Petschauer:2016pbn,Petschauer:2016tee,Petschauer:2020urh}.
Moreover, meson-exchange contributions were restricted to those
from pion-exchange, in line with the SMS YN potential~\cite{Haidenbauer:2023qhf} 
where two-meson contributions involving the $K$ and/or $\eta$ were neglected.  

The utilization of decuplet saturation is well motivated by corresponding observations/studies
in the purely nucleonic sector~\cite{Bernard:1996gq}. There, some of the a priori unknown LECs turned out to 
be fairly large compared to their order of magnitude as expected from the hierarchy 
of nuclear forces in chiral EFT, when fitted to, say, NN scattering data 
and 3N observables such as 3-body binding energies \cite{Epelbaum:2002vt}.
The physical origin of this behavior is obviously the strong coupling of the 
$\pi$-system to the low-lying \(\Delta(1232)\)-resonance.
It is therefore natural to include the \(\Delta(1232)\)-isobar as an explicit degree 
of freedom in the chiral Lagrangian 
(see \cite{Bernard:1996gq,Hemmert:1997ye,Kaiser:1998wa,Krebs:2007rh}).
The small mass difference between nucleons and $\Delta$ (\(293\ \mathrm{MeV}\)) introduces a small scale, which can be included consistently in the chiral power counting scheme and the hierarchy of nuclear forces.
The dominant parts of the three-nucleon interaction mediated by two-pion exchange at N2LO are then promoted to NLO through the $\Delta$ contributions.
The appearance of the inverse mass splitting explains the large numerical values of the corresponding LECs \cite{Bernard:1996gq,Epelbaum:2008ga,Epelbaum:2007sq}.

In SU(3) \cheft the situation is similar.
In systems with strangeness \(S=-1\) like $\Lambda$NN, resonances such as the \(\Sigma^*\)(1385) play a comparable role to that of the \(\Delta\) in the NNN system.
The small decuplet-octet mass splitting (in the chiral limit), \(\Delta\defeq m_{10}-m_{8}\), is counted together with external momenta and meson masses as \(\mathcal O(q)\) and, thus, parts of the N2LO three-baryon interaction are promoted to NLO by the explicit inclusion of the baryon decuplet, as illustrated in Fig.~\ref{fig:hierdec}. 
Therefore, in principle, one could incorporate those contributions to the three-baryon interaction 
already together with the NLO YN interaction.
Note that in the nucleonic sector, only the two-pion exchange diagram with an intermediate \(\Delta\)-isobar is allowed.
Other diagrams are forbidden due to the Pauli principle, as we will show later.
For three flavors more particles are involved and, in general, also the other diagrams (contact and one-meson exchange) with intermediate decuplet baryons in Fig.~\ref{fig:hierdec} appear.
In the following we review the estimation of these LECs by resonance saturation which
has been worked out in \cite{Petschauer:2016pbn}.

\begin{figure*}
\centering
\newcommand{\dechierscale}{.5}
\setlength{\tabcolsep}{8pt}
\begin{tabular}{lcc}
\toprule \addlinespace[1ex]
& \multicolumn{2}{c}{three-baryon force} \\ \addlinespace[1ex]
& decuplet-less EFT & decuplet contribution \\ \addlinespace[1ex] \midrule \addlinespace[2ex]
LO & $\vcenter{\hbox{\rule{0cm}{2cm}}}$ & \\ \addlinespace[2ex]
NLO & &
$\vcenter{\hbox{
\includegraphics[scale=\dechierscale]{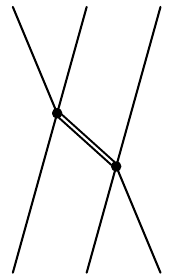}\ \
\includegraphics[scale=\dechierscale]{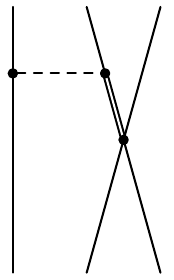}\ \
\includegraphics[scale=\dechierscale]{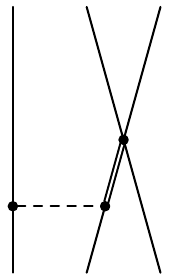}\ \
\includegraphics[scale=\dechierscale]{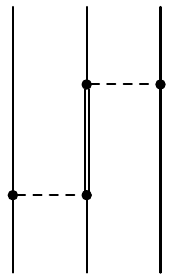}}}$ \\ \addlinespace[2ex]
N2LO &
$\vcenter{\hbox{
\includegraphics[scale=\dechierscale]{figures/FBBBcont}\ \
\includegraphics[scale=\dechierscale]{figures/FBBB1ME}\ \
\includegraphics[scale=\dechierscale]{figures/FBBB2ME2}}}$ &
$\cdots$ \\ \addlinespace[2ex]
\bottomrule
\end{tabular}
\caption{Hierarchy of three-baryon forces with explicit introduction of the baryon decuplet (represented by double lines) \cite{Meissner:2008zza,Petschauer:2016pbn}.} \label{fig:hierdec}
\setlength{\tabcolsep}{6pt}
\end{figure*}

The LO non-relativistic interaction Lagrangian between octet and decuplet baryons (see, \eg \cite{Sasaki:2006cx}) is
\begin{align*} \label{eq:LMBD}
\mathscr{L} = \frac C{F_0} \sum_{a,b,c,d,e=1}^3 \epsilon_{abc} \Bigg(& \bar T_{ade} \vec S^{\,\dagger} \cdot \left(\vec\nabla \phi_{db}\right)B_{ec}
+ \bar B_{ce} \vec S \cdot \left(\vec\nabla\phi_{bd}\right)T_{ade} \Bigg)\,, \numberthis
\end{align*}
where the decuplet baryons are represented by the totally symmetric three-index tensor \(T\), see \cite{Petschauer:2016pbn}.
At this order only a single LEC \(C\) appears.
Typically the (large-\(N_c\)) value \(C=\frac34g_A\approx 1\) is used, as it leads to a decay width \(\Gamma(\Delta\to\pi N)=110.6\ \mathrm{MeV}\) that is in good agreement with the empirical value of \(\Gamma(\Delta\to\pi N)=(115\pm 5)\ \mathrm{MeV}\) \cite{Kaiser:1998wa}.
The spin \(\frac12\) to \(\frac32\) transition operators \(\vec S\) connect the two-component spinors of octet baryons with the four-component spinors of decuplet baryons (see \eg \cite{Ericson:1988gk}).
In their explicit form they are given as \(2\times4\) transition matrices
\begin{align*}
S_1 &= \begin{pmatrix}
-\frac{1}{\sqrt{2}} & 0 & \frac{1}{\sqrt{6}} & 0 \\
0 & -\frac{1}{\sqrt{6}} & 0 & \frac{1}{\sqrt{2}}
\end{pmatrix} ,\\
S_2 &= \begin{pmatrix}
-\frac{i}{\sqrt{2}} & 0 & -\frac{i}{\sqrt{6}} & 0 \\
0 & -\frac{i}{\sqrt{6}} & 0 & -\frac{i}{\sqrt{2}}
\end{pmatrix} ,\\
S_3 &= \begin{pmatrix}
0 & \sqrt{\frac{2}{3}} & 0 & 0 \\
0 & 0 & \sqrt{\frac{2}{3}} & 0
\end{pmatrix}. \numberthis
\end{align*}
These operators fulfill the relation \( S_i {S_j}^\dagger = \frac13 ( 2\delta_{ij}-\mathrm i\epsilon_{ijk} \sigma_k )\).

A non-relativistic \(B^*BBB\) Lagrangian with a minimal set of terms is given by \cite{Petschauer:2016pbn}:
\begin{align*} \label{eq:LDBBBmin}
\mathscr{L} = &\quad\,
\lC_1\sum_{\substack{a,b,c,\\d,e,f=1}}^3 \epsilon_{abc}
\big[
\left(\bar T_{ade}\vec S^\dagger B_{db}\right)\cdot\left(\bar B_{fc}\vec\sigma B_{ef}\right)
 +\left(\bar B_{bd}\vec S\, T_{ade}\right)\cdot\left(\bar B_{fe}\vec\sigma B_{cf}\right)
\big]\\
&+\lC_2
\sum_{\substack{a,b,c,\\d,e,f=1}}^3 \epsilon_{abc}
\big[
\left(\bar T_{ade}\vec S^\dagger B_{fb}\right)\cdot\left(\bar B_{dc}\vec\sigma B_{ef}\right) 
+\left(\bar B_{bf}\vec S\, T_{ade}\right)\cdot\left(\bar B_{fe}\vec\sigma B_{cd}\right)
\big] \,, \numberthis
\end{align*}
with the LECs \(\lC_1\) and \(\lC_2\).
Again one can employ group theory to justify the number of two constants for a transition \(BB\to B^*B\).
In flavor space the two initial octet baryons form the tensor product \(\mathbf8\otimes\mathbf8\), and in spin space they form the product \(\mathbf{2} \otimes \mathbf{2}\).
These tensor products can be decomposed into irreducible representations:
\begin{align*} \label{eq:BBgroup}
\mathbf8\otimes\mathbf8 &= \underbrace{{\mathbf{27}}\oplus{\mathbf{8}_s}\oplus{\mathbf1}}_\text{symmetric}\oplus\underbrace{\mathbf{10}\oplus
\mathbf{\overline{10}}\oplus\mathbf{8}_a}_\text{antisymmetric}\,,\\
\mathbf{2} \otimes \mathbf{2} &= \mathbf{1}_a \oplus \mathbf{3}_s \,. \numberthis
\end{align*}
In the final state, having a decuplet and an octet baryon, the situation is similar:
\begin{align*} \label{eq:DBgroup}
\mathbf{10}\otimes\mathbf8 &= \mathbf{35}\oplus\mathbf{27}\oplus\mathbf{10}\oplus\mathbf{8}\,,\\
\mathbf{4} \otimes \mathbf{2} &= \mathbf{3} \oplus \mathbf{5} \,.\numberthis
\end{align*}
As seen in the previous sections, at LO only \(S\)-wave transitions occur, as no momenta are involved.
Transitions are only allowed between the same types of irreducible (flavor and spin) representations.
Therefore, in spin space the representation \(\mathbf{3}\) has to be chosen.
Because of the Pauli principle in the initial state, the symmetric \(\mathbf{3}\) in spin space combines with the antisymmetric representations \(\mathbf{10},
\mathbf{\overline{10}},\mathbf{8}_a\) in flavor space.
But only \(\mathbf{10}\) and \(\mathbf{8}_a\) have a counterpart in the final state flavor space.
This number of two allowed transitions matches the number of two LECs in the minimal Lagrangian.
Another interesting observation can be made from Eqs.~(\ref{eq:BBgroup}) and (\ref{eq:DBgroup}).
For NN states only the representations \(\mathbf{27}\) and \(\mathbf{\overline{10}}\) can contribute.
But these representations combine either with the wrong spin, or have no counterpart in the final state.
Therefore, NN$\to$$\Delta$N transitions in \(S\)-waves are not allowed because of the Pauli principle.

Having the above two interaction types at hand, one can estimate the low-energy constants of the leading three-baryon interaction by decuplet saturation using the diagrams shown in 
Fig.~\ref{fig:hierdec}.
At this order, where no loops are involved, one just needs to evaluate the diagrams with an intermediate decuplet baryon and the diagrams without decuplet baryons and compare them with each other.

\begin{figure*}
\centering
\hfill
\begin{subfigure}[b]{.25\textwidth}
\centering
\(
\vcenter{\hbox{\includegraphics[scale=.45]{figures/FBBBcont}}}
\ =\
\vcenter{\hbox{\includegraphics[scale=.45]{figures/FBBBdeccont}}}
\)
\caption{Saturation of the six-baryon contact interaction}
\label{fig:3BFctDec}
\end{subfigure}
\hfill
\begin{subfigure}[b]{.3\textwidth}
\centering
\(
\vcenter{\hbox{\includegraphics[scale=.62]{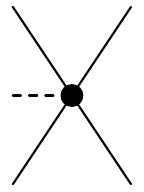}}} =
\vcenter{\hbox{\includegraphics[scale=.62]{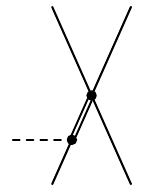}}} +
\vcenter{\hbox{\includegraphics[scale=.62]{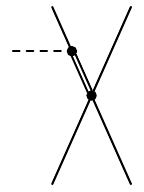}}}
\)
\vspace{.3\baselineskip}
\caption{Saturation of the \(BB\to BB\phi\) vertex \\
\qquad $\phantom{x}$
}
\label{fig:3BF1MEDec}
\end{subfigure}
\hfill
\begin{subfigure}[b]{.4\textwidth}
\centering
\(
\vcenter{\hbox{\includegraphics[scale=.5]{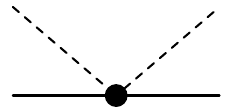}}}
\ = \
\vcenter{\hbox{\includegraphics[scale=.5]{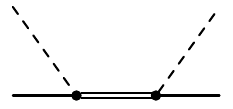}}}
\ + \
\vcenter{\hbox{\includegraphics[scale=.5]{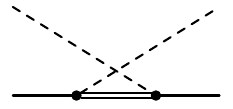}}}
\)
\caption{Saturation of the NLO baryon-meson vertex \\
\qquad $\phantom{x}$}
\label{fig:3BF2MEDec}
\end{subfigure}
\hfill\mbox{}
\caption{Saturation via decuplet resonances.}
\end{figure*}
In order to estimate the LECs of the six-baryon contact Lagrangian of Eq.~(\ref{eq:minct}), one can consider the process \(B_1B_2B_3\to B_4B_5B_6\) as depicted in 
Fig.~\ref{fig:3BFctDec}.
The left side of Fig.~\ref{fig:3BFctDec} has already been introduced in the previous subsection and can be obtained by performing all 36 Wick contractions. For the 
diagrams on the right side of Fig.~\ref{fig:3BFctDec} the procedure is similar.
After summing over all intermediate decuplet baryons \(B^*\), the full three-body potential of all possible combinations of baryons on the left side of Fig.~\ref{fig:3BFctDec} can be compared with the ones on the right side.
In the end the 18 LECs of the six-baryon contact Lagrangian \(\lc_1,\dots,\lc_{18}\) of Eq.~(\ref{eq:minct}) can be expressed as linear combinations of the combinations \(\lC_1^2\), \(\lC_2^2\) and \(\lC_1\lC_2\)
and are proportional to the inverse average decuplet-octet baryon mass splitting \(1/\Delta\) \cite{Petschauer:2016pbn}.

Since we are at LO only tree-level diagrams are involved and we can estimate the LECs of the one-meson-exchange part of the 3BFs already on the level of the vertices, as depicted in Fig.~\ref{fig:3BF1MEDec}.
We consider the transition matrix elements of the process \(B_1B_2\to B_3B_4\phi\) and start with the left side of Fig.~\ref{fig:3BF1MEDec}.
After doing all possible Wick contractions, summing over all intermediate decuplet baryons, and comparing the left side of Fig.~\ref{fig:3BF1MEDec} with the right hand side for all combinations of baryons and mesons, the LECs can be estimated.
The LECs of the minimal non-relativistic chiral Lagrangian for the four-baryon vertex including one meson of Eq.~(\ref{eq:LBBMBBmin}) \(\ld_1,\dots,\ld_{14}\) are then proportional to \(C/\Delta\) and to linear combinations of \(\lC_1\) and \(\lC_2\) \cite{Petschauer:2016pbn}.

The last class of diagrams is the three-body interaction with two-meson exchange.
As done for the one-meson exchange, the unknown LECs can be saturated directly on the level of the vertex and one can consider the process \(B_1\phi_1\to B_2\phi_2\) as shown in Fig.~\ref{fig:3BF2MEDec}. 
A direct comparison of the transition matrix elements for all combinations of baryons and mesons after summing over all intermediate decuplet baryons \(B^*\) leads to the following contributions to the LECs of the meson-baryon Lagrangian in Eq.~(\eqref{eq:MBMBLagr}):
\begin{align*}
b_D  &= 0 \,,\
b_F  = 0 \,,\
b_0  = 0 \,,\\
b_1  &= \frac{7 C^2}{36 \Delta } \,,\
b_2  = \frac{C^2}{4 \Delta } \,,\
b_3  = -\frac{C^2}{3 \Delta } \,,\
b_4  = -\frac{C^2}{2 \Delta } \,,\\
d_1  &= \frac{C^2}{12 \Delta } \,,\
d_2  = \frac{C^2}{36 \Delta } \,,\
d_3  = -\frac{C^2}{6 \Delta } \,, \numberthis
\end{align*}
These findings are consistent with the \(\Delta\)(1232) contribution to the LECs \(c_1,c_3,c_4\) (see Eq.~(\ref{eq:LECc134})) in the nucleonic sector \cite{Bernard:1996gq}:
\begin{equation}
c_1=0\,,\qquad c_3=-2c_4=-\frac{g_A^2}{2 \Delta } \,.
\end{equation}

Applying decuplet saturation to the $\Lambda$NN interaction (contact interaction,
one-pion and two-pion exchange) of Section \ref{subsec:BBBpotex} one obtains:
\begin{align*}
\label{eq:decsatlnn}
&\lc^\prime_1  = \lc^\prime_3 = {} \frac{(\lC_1+3 \lC_2)^2}{72 \Delta } \,, \quad
\lc^\prime_2  ={} 0 \,, \\
&\ld'_1  ={} 0 \,, \quad
\ld'_2  ={} \frac{2 C (\lC_1+3 \lC_2)}{9 \Delta } \,, \\
&3b_0+b_D ={} 0 \,, \quad
2b_2+3b_4 ={} -\frac{C^2}{\Delta } \,. \numberthis
\end{align*}
Evidently, there is only one unknown constant here, namely the combination 
\(\lC_1+3 \lC_2\).
It is also interesting to see that the sign of the constants \(\lc^\prime_i\) 
for the contact interaction is already fixed, independently of the values of the 
two LECs \(\lC_1\) and \(\lC_2\). The fact that it is positive implies that the 
$\La$N 3BF from the contact interaction is always repulsive. 
Estimates for the order of magnitude of the parameters $H_1$ and $H_2$, based 
on dimensional scaling arguments, have been given in Ref.~\cite{Petschauer:2016pbn}.
These suggest $H_1 \approx H_2 \approx \pm 1/F_0^2$.
Those estimates can be compared with the ones for the NNN forces,
$E = 2(C_4-C_9) = c_E /( F_\pi^4 \Lambda_\chi)$ 
and $D=4(D_1-D_3+D_8-D_{10}) = c_D / ( F_\pi^2 \Lambda_\chi)$
where $c_E$ and $c_D$ are in the order of one.
Obviously, the decuplet saturation mechanism promotes the constants $C_i$ 
and $D_i$ to ${\cal O}(1/( F_\pi^4 \Delta))$ and ${\cal O}(1/( F_\pi^2 \Delta))$,
respectively. 

	\section{Faddeev-Yakubovsky approach}\label{sec:third}
	
In order to test and improve the two- and three-baryon interactions, solutions of the Schr\"odinger equation 
for few-baryon systems are helpful. Properties like the separation energies of hypernuclear systems do not only reflect 
the overall strength of the interactions but they are also sensitive to their angular and spin dependence 
and even to their isospin or charge dependence. Various methods exist that allow one to solve 
the Schr\"odinger equation for realistic BB interactions. The probably most direct way consists
in solving Faddeev or Yakubovsky equations in momentum space. This approach is very flexible 
with respect to the interaction used and gives high accuracy solutions even for interactions that 
induce strong correlations between the baryons. The $A=3$ Faddeev equations for bound states of the hypertriton 
have been formulated in \cite{Miyagawa:1993rd,Miyagawa:1995sf} taking $\Lambda$-$\Sigma$ conversion into account.
This method has been applied to the phenomenological interactions of the Nijmegen \cite{Maessen:1989sx,Rijken:1998yy} and 
J\"ulich \cite{Reuber:1993ip,Haidenbauer:2005zh} groups and also to the chiral interactions discussed before. 
Importantly, it was realized early on \cite{Herndon:1967zza,Dalitz:1972vzj} that the hypertriton binding energy depends very 
sensitively on the $\Lambda$N interaction in the $^1{\rm S}_0$ channel. Therefore, the spin dependence of 
the YN interactions is usually adjusted to reproduce the hypertriton binding energy. 
The Faddeev equations of the $\Lambda$NN system have also been implemented for 
scattering states. They were applied to $\Lambda$-d scattering by Kohno and 
Kamada~\cite{Kohno:2024tkq} using realistic YN and NN forces. 
The possibly earliest example of a $\Lambda$-d calculation
within a Faddeev-type approach has been reported already in 1965~\cite{Hetherington:1965zz}.
For references to further works see \cite{Haidenbauer:2020uew}.

Extending to $A=4$, the Schr\"odinger equation can be rewritten in a set of Yakubovsky equations. For the 
strangeness $S=-1$ hypernuclei $^{4}_{\Lambda}$He and $^{4}_{\Lambda}$H, one finds a set of 
five Yakubovsky equations for five independent Yakubovsky components (see below). The equations were 
likewise solved 
for the phenomenological interactions of the Nijmegen und J\"ulich groups \cite{Nogga:2001ef} and for 
chiral interactions in LO \cite{Nogga:2008zz}, NLO \cite{Haidenbauer:2013oca,Haidenbauer:2019boi} 
and N2LO \cite{Haidenbauer:2023qhf}. It turns out that the predictions strongly depend on the 
interaction and even the regulator used. The older models of the J\"ulich group result in the wrong 
ordering of the $0^+$ and $1^+$ states. Some of the Nijmegen interactions lead to an almost unbound $1^+$ state. 
The separation energies for these hypernuclei are significantly influenced by off-shell properties 
of the YN interactions, and specifically by the strength of the $\Lambda$-$\Sigma$ conversion. As has been discussed in 
\cite{Haidenbauer:2019boi}, such properties of the YN interaction are linked to corresponding contributions 
of 3BFs that have usually not been taken into account. Therefore, the deviations of the different 
interactions can be explained by the incompleteness of the underlying Hamiltonian. We discuss results including 
such 3BFs later in Sec.~\ref{sec:ncsmynn}. 

An issue practically unaffected by
the problem of missing 3BFs are predictions for the charge-symmetry breaking (CSB) of the $A=4$
hypernuclei. Experiments suggest that the difference of the $\Lambda$-separation energies is quite large, of the order 
of 200-300~keV and strongly depend on the spin of the nucleus. These values could not be explained 
by the phenomenological models \cite{Nogga:2001ef}. We will come back to this issue in the subsection of CSB 
in the context of chiral interactions that provide much for freedom to determine the CSB contributions. 

In the following, we will start with a brief derivation of the pertinent equations.

\subsection{Derivation of the Yakubovsky equations}

The aim is to obtain solution of the Schr\"odinger equation for a four-baryon bound state 
in momentum space. Using momentum space naturally allows for the application of non-local 
interactions. For a bound state, it is also simple to use long-ranged interactions like the 
Coulomb force. Therefore, this method has been used regularly to obtain high accuracy 
solutions for few-baryon problems \cite{Miyagawa:1993rd,Nogga:2001ef,Nogga:2001cz,Nogga:2002qp}. 
Practical calculations show that the few-baryon wave functions converge only slowly with 
respect to a partial wave decomposition although the interactions themselves usually 
only require small angular momenta for an accurate representation. This is due to the need to also represent 
correlations of the baryon pairs in Jacobi coordinates that to do single out these pairs. 
This problem can be tamed by rewriting the Schr\"odinger equation into Faddeev or Yakubovsky 
equations. Different Faddeev and Yakubovsky components then represent the various 
correlations efficiently and lead to an improved convergence with respect to partial waves. 

We start with the Schr\"odinger equation for a system of three nucleons and a hyperon 
with pair interactions $V_{ij}$, 3BFs $V_{ijk}$ and, to be most general, with the four-baryon force (4BF) $V_{1234}$
\begin{equation}
    \label{eq:schroedynnn}
    \left( E- H_0 \right) \left| \Psi \right\rangle 
    = \sum_{i<j} V_{ij} \ \left| \Psi \right\rangle +\sum_{i<j<k} V_{ijk} \ \left| \Psi \right\rangle  +V_{1234} \ \left| \Psi \right\rangle \ .
\end{equation}
The indices $i,j,k=1,\ldots,3$ thereby refer to the nucleons and $4$ to the hyperon ($\Lambda$ or $\Sigma$).
$H_0$ is the intrinsic kinetic energy of the four-baryon system and $E$ the energy of the system. 
Note that $H_0$ also includes the rest masses of the baryons. Usually an overall shift 
of the energy is applied so that $E=0$ corresponds to the mass of the baryons in the lightest particle 
channel evolved. When $\Lambda$-$\Sigma$ conversion is included, the kinetic energy of the $\Sigma$ 
channel includes the mass difference of the $\Lambda$ and $\Sigma$. 
In order to define Faddeev components in presence of the three-baryon interactions, we split each of them 
into three different parts
\begin{equation}
    V_{ijk} = V_{ijk}^{(i)} +V_{ijk}^{(j)}  +V_{ijk}^{(k)}  
\end{equation}
such that the interaction is symmetrical under the exchange of two nucleons $ij$ in $V_{ijk}^{(k)}$. 
For three-nucleon forces (3NFs), such a decomposition is usually already done when generating the 
interactions. For the YNN forces, each of the parts can be defined as $V_{ijk}^{(k)} = 1/3 \ V_{ijk}$.
Similarly, we decompose the 4BF into 12 components $V_{1234}^{(ij)k,l}$ which are also symmetric 
under the exchange of particles $i$ and $j$
\begin{equation}
    V_{1234} = V_{1234}^{(12)3,4}+V_{1234}^{(12)4,3}+V_{1234}^{(23)1,4}+V_{1234}^{(23)4,1}+V_{1234}^{(31)2,4}+V_{1234}^{(31)4,2}+V_{1234}^{(14)2,3}+V_{1234}^{(14)3,2}+V_{1234}^{(24)3,1}+V_{1234}^{(24)1,3}+V_{1234}^{(34)1,2}+V_{1234}^{(34)2,1} \ .
\end{equation}
A possible decomposition is just $V_{1234}^{(ij)k,l} = \frac{1}{12} V_{1234}$. 
With this definition, one can define Faddeev components 
\begin{equation}
    \left| \psi_{ij} \right\rangle = G_0 \ \left( V_{ij} + V_{ijk}^{(k)} + V_{ijl}^{(l)}  + V_{1234}^{(ij)k,l} + V_{1234}^{(ij)l,k}
    \right) \, \left| \Psi \right\rangle 
\end{equation}
where $G_0 = \left( E- H_0 \right)^{-1}$. 
This definition ensures that 
\begin{equation}
    \label{eq:faddecomp}
    \left| \Psi \right\rangle = \sum_{i<j} \left| \psi_{ij} \right\rangle 
\end{equation}
and that the Faddeev components fulfill the Faddeev equations 
\begin{equation}
    \left| \psi_{ij} \right\rangle  = G_0 \, t_{ij} \left(  \left| \Psi \right\rangle -   \left| \psi_{ij} \right\rangle \right)
    + \left( 1 + G_0 \, t_{ij} \right) \, G_0 \, \left( V_{ijk}^{(k)} + V_{ijl}^{(l)} +V_{1234}^{(ij)k,l} + V_{1234}^{(ij)l,k}\right) \left| \Psi \right\rangle \ . 
\end{equation}
For the Faddeev equations, we introduced the two-baryon off-shell $t$-matrices $t_{ij}$ that are solutions of the
Lippmann-Schwinger equations
\begin{eqnarray}
    t_{ij} = V_{ij} + V_{ij} \, G_0 \, t_{ij} \ .
\end{eqnarray}

Now we are ready to further decompose the Faddeev components into Yakubovsky components 
\begin{equation}
    \label{eq:yakdecomp}
    \left| \psi_{ij} \right\rangle  = \left| \psi_{(ij)k,l} \right\rangle + \left| \psi_{(ij)l,k} \right\rangle  + \left| \psi_{(ij)kl} \right\rangle 
\end{equation}
defined by 
\begin{eqnarray}
  \left| \psi_{(ij)k,l} \right\rangle & = & G_0 \, t_{ij} \left(  \left| \psi_{ik} \right\rangle +  \left| \psi_{jk} \right\rangle  \right)   
     +   \left( 1 + G_0 \, t_{ij} \right) \, G_0 \, \left( V_{ijk}^{(k)} + V_{1234}^{(ij)k,l}  \right) \left| \Psi \right\rangle \nonumber \\
    \left| \psi_{(ij)kl} \right\rangle & = &   G_0 \, t_{ij}  \, \left| \psi_{kl} \right\rangle \ . 
\end{eqnarray}
Inserting the decomposition of the Faddeev components leads to the set of Yakubovsky equations that we use for bound state 
calculations\footnote{Note that these equations are still not connected when iterated. For a scattering problem, a resummation of 
parts involving connected clusters is necessary.}
\begin{eqnarray}
     \left| \psi_{(ij)k,l} \right\rangle 
        & = & G_0 \, t_{ij} \left( \left| \psi_{(ik)j,l} \right\rangle 
                                 + \left| \psi_{(ik)l,j} \right\rangle 
                                 + \left| \psi_{(ik)jl} \right\rangle 
                                 + \left| \psi_{(jk)i,l} \right\rangle 
                                 + \left| \psi_{(jk)l,i} \right\rangle 
                                 + \left| \psi_{(jk)il} \right\rangle 
                                 \right)    \nonumber \\
        & &    + \left( 1 + G_0 \, t_{ij} \right) \, G_0 \, \left( V_{ijk}^{(k)} + V_{1234}^{(ij)k,l}\right) \left| \Psi \right\rangle \nonumber \\
     \left| \psi_{(ij)kl} \right\rangle & = &   G_0 \, t_{ij} \left( \left| \psi_{(kl)i,j} \right\rangle 
                                                                   + \left| \psi_{(kl)j,i} \right\rangle 
                                                                   + \left| \psi_{(kl)ij} \right\rangle  \right)  \ . 
\end{eqnarray}

\begin{table}
\begin{center}
\begin{tabular}{l|lllll}
& $\psi_{1A}=\psi_{(12)3,4}$ & $\psi_{1B}=\psi_{(12)4,3}$ & $\psi_{1C}=\psi_{(14)2,3}$ &
$\psi_{2A}=\psi_{(12)34}$ & $\psi_{2B}=\psi_{(34)12}$ \cr
\hline
$\psi_{(12)3,4}$&  $1$ &  - &  - &
 - &  -  \cr
$\psi_{(23)1,4}$&  $P_{12}P_{23}$ &  - &  - &
 - &  -  \cr 
$\psi_{(31)2,4}$&  $P_{13}P_{23}$ &  - &  - &
 - &  -  \cr
$\psi_{(12)4,3}$&  - &  1  &  - &
 - &  -  \cr
$\psi_{(23)4,1}$&  - &  $P_{12}P_{23}$&  - &
 - &  -  \cr
$\psi_{(31)4,2}$&  - &  $P_{13}P_{23}$ &  - &
 - &  -  \cr
$\psi_{(14)2,3}$&  - &  - &  1 &
 - &  -  \cr
$\psi_{(24)3,1}$&  - &  - &  $P_{12}P_{23}$ &
 - &  -  \cr
$\psi_{(34)1,2}$&  - &  - &  $P_{12}P_{13}$ &
 - &  -  \cr
$\psi_{(14)3,2}$&  - &  - &  $-P_{23}$ &
 - &  -  \cr
$\psi_{(34)2,1}$&  - &  - &  $-P_{13}$ &
 - &  -  \cr
$\psi_{(24)1,3}$&  - &  - &  $-P_{12}$ &
 - &  -  \cr
$\psi_{(12)34}$&  - &  - &  - &
 1  &  -  \cr
$\psi_{(23)14}$&  - &  - &  - &
 $P_{12}P_{23}$&  -  \cr
$\psi_{(31)24}$&  - &  - &  - &
 $P_{13}P_{23}$ &  -  \cr
$\psi_{(14)23}$&  - &  - &  - &
 - &  $P_{12}P_{23}$  \cr
$\psi_{(24)31}$&  - &  - &  - &
 - &  $P_{13}P_{23}$  \cr
$\psi_{(34)12}$&  - &  - &  - &
 - &  1   \cr
\end{tabular}
\end{center}
\caption{\label{tab:yakperm} Connection between the different Yakubovsky components.
The transpositions have to be applied to the Yakubovsky components in the first row.}
\end{table}

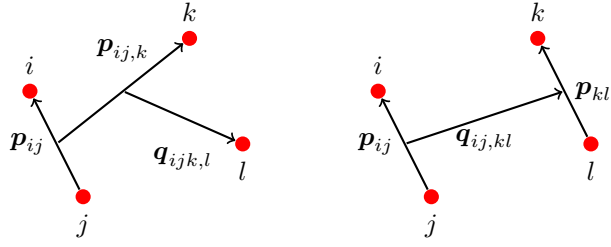
\begin{figure}
\begin{center}

\begin{tikzpicture}[scale=0.7]

    \node[circle, fill=red, inner sep=2pt, label=above:{$i$}] (P1) at (0, 2) {};
    \node[inner sep=0pt] (P12) at (0.5,1) {};
    \node[circle, fill=red, inner sep=2pt, label=below:{$j$}] (P2) at (1, 0) {};
    \node[circle, fill=red, inner sep=2pt, label=above:{$k$}] (P3) at (3, 3) {};
    \node[inner sep=0pt,label={[yshift=7pt]above:{$\vec p_{ij,k}$}}] (P123) at (1.75,2) {};
    
    \node[inner sep=0pt,label=left:{$\vec p_{ij}$}] (P12) at (0.5,1) {};
    \node[inner sep=0pt,label={[yshift=-7pt]below:{$\vec q_{ijk,l}$}}] (P1234) at (2.875,1.5) {};
    
    \node[circle, fill=red, inner sep=2pt, label=below:{$l$}] (P4) at (4, 1) {};

    \draw[thick, solid,<-] (P1) -- (P2);
    \draw[thick, solid,->] (P12) -- (P3);
    \draw[thick, solid,->] (P123) -- (P4);
\end{tikzpicture}
\hspace{1cm}
\begin{tikzpicture}[scale=0.7]

    \node[circle, fill=red, inner sep=2pt, label=above:{$i$}] (P1) at (0, 2) {};
    \node[inner sep=0pt,label=left:{$\vec p_{ij}$}] (P12) at (0.5,1) {};
    \node[circle, fill=red, inner sep=2pt, label=below:{$j$}] (P2) at (1, 0) {};
    \node[circle, fill=red, inner sep=2pt, label=above:{$k$}] (P3) at (3, 3) {};
    \node[inner sep=0pt,label=right:{$\vec p_{kl}$}] (P34) at (3.5,2) {};
    \node[circle, fill=red, inner sep=2pt, label=below:{$l$}] (P4) at (4, 1) {};
    \node[inner sep=0pt,label=below:{$\vec q_{ij,kl}$}] (P1234) at (2,1.5) {};

    \draw[thick, solid,<-] (P1) -- (P2);
    \draw[thick, solid,->] (P12) -- (P34);
    \draw[thick, solid,<-] (P3) -- (P4);
\end{tikzpicture}

\end{center}
    \caption{``3+1'' and ``2+2'' Jacobi coordinates to represent Yakubovsky components.}
    \label{fig:jacobimomenta}
\end{figure}

So far, we have not made use of the antisymmetry of the wave function under permutation of the three nucleons. 
Introducing the permutation operators $P_{ij}$ that commute the momenta and quantum numbers of the $i$-th and $j$-th nucleon,
we can relate the 18 different  Yakubovsky components to each other leading to a set of five independent  
Yakubovsky components fulfilling a set of five Yakubovsky equations. The relation of the different Yakubovsky 
components is shown in Table~\ref{tab:yakperm} and the resulting set of equations reads 
\begin{eqnarray}
\label{eq:yakeq}
    \left|  \psi_{1A}  \right\rangle & = &   \left| \psi_{(12)3,4} \right\rangle  
     =   G_0 \, t_{12} \, P \, \left(  
     \left|  \psi_{1A}  \right\rangle +  \left|  \psi_{1B}  \right\rangle +  \left|  \psi_{2A}  \right\rangle 
     \right) 
     + \left( 1 + G_0 \, t_{12} \right) \, G_0 \, \left( V_{123}^{(3)} + V_{1234}^{(12)3,4} \right) \left| \Psi \right\rangle   \nonumber  \\
    \left|  \psi_{1B}  \right\rangle & = &   \left| \psi_{(12)4,3} \right\rangle 
     =   G_0 \, t_{12} \, \left(   
       \left(1-P_{12} \right)  \left(1-P_{23} \right) \left|  \psi_{1C}  \right\rangle 
       +  P \, \left|  \psi_{2B}  \right\rangle  
       \right) 
       + \left( 1 + G_0 \, t_{12} \right) \, G_0 \, \left( V_{124}^{(4)} + V_{1234}^{(12)4,3} \right) \left| \Psi \right\rangle  \nonumber  \\  
    \left|  \psi_{1C}  \right\rangle & = &   \left| \psi_{(14)2,3} \right\rangle 
     =   G_0 \, t_{14} \, \left(   
         \left|  \psi_{1A}  \right\rangle 
       + \left|  \psi_{1B}  \right\rangle 
       +  P_{12} \left( P_{23}-1)\right)  \, \left|  \psi_{1C}  \right\rangle  
       +  \left|  \psi_{2A}  \right\rangle  
       +  P_{13}P_{23} \, \left|  \psi_{2B}  \right\rangle  
       \right) \nonumber \\ 
       & & \hspace{2cm}+ \left( 1 + G_0 \, t_{14} \right) \, G_0 \, \left( V_{124}^{(2)} 
       + V_{1234}^{(14)2,3} \right) \left| \Psi \right\rangle  \nonumber  \\  
    \left|  \psi_{2A}  \right\rangle & = &   \left| \psi_{(12)34} \right\rangle 
     =   G_0 \, t_{12} \, \left(   
         \left(P_{12}-1\right) P_{13} \, \left|  \psi_{1C}  \right\rangle 
       + \left|  \psi_{2B}  \right\rangle 
       \right) \nonumber \\
    \left|  \psi_{2B}  \right\rangle & = &   \left| \psi_{(34)12} \right\rangle 
     =   G_0 \, t_{34} \, \left(   
         \left|  \psi_{1A}  \right\rangle 
       + \left|  \psi_{1B}  \right\rangle 
       + \left|  \psi_{2A}  \right\rangle 
       \right) 
\end{eqnarray}
where the combination of permutation operators $P=P_{12} P_{23} + P_{13} P_{23} $ was introduced. 
Using Eqs.~(\ref{eq:faddecomp}) and (\ref{eq:yakdecomp}), the wave function is the sum of all Yakubovsky components
which can be simplified to 
\begin{equation}
    \label{eq:4bwf}
    \left| \Psi \right\rangle = \left(1+P\right) \left( \left|  \psi_{1A}  \right\rangle + \left|  \psi_{1B}  \right\rangle + \left|  \psi_{2A}  \right\rangle +\left|  \psi_{2B}  \right\rangle   \right)  + \left( 1-P_{12} \right) \left(1+P\right) \, \left|   \psi_{1C}  \right\rangle 
\end{equation}

In order to fully exploit that the Yakubovsky components converge faster than the 
wave functions, we need to define different sets of Jacobi coordinates 
for each Yakubovsky component. The natural coordinates for $ \left|  \psi_{1A}  \right\rangle $,
$ \left|  \psi_{1B}  \right\rangle $ and $ \left|  \psi_{1C}  \right\rangle $ are of the 
``3+1'' type shown in Fig.~\ref{fig:jacobimomenta}, the ones for $ \left|  \psi_{2A}  \right\rangle $ and $ \left|  \psi_{2B}  \right\rangle $ are of the ``2+2'' type. Thereby, the 
particles $i$, $j$, $k$ and $l$ are selected such that they single out the same pairs and triplets as the Yakubovsky component. The corresponding momenta are define as 
\begin{eqnarray}
    \label{eq:jmom}
    \vec p_{ij} & = &  \frac{1}{m_i+m_j}\left(m_j \, \vec k_i- m_i \, \vec k_j \right) \nonumber \\
    \vec p_{ij,k} & = &  \frac{1}{m_i+m_j+m_k}\left((m_i+m_j)\, \vec k_k- m_k\, \left( \vec k_i+\vec k_j\right)  \right) \nonumber \\
    \vec q_{ijk,l} & = &  \frac{1}{m_i+m_j+m_k+m_l}\left((m_i+m_j+m_k)\, \vec k_l- m_l\,\left( \vec k_i+\vec k_j +\vec k_k \right) \right) \nonumber \\
    \vec p_{kl} & = &  \frac{1}{m_k+m_l}\left(m_l \, \vec k_k- m_k \, \vec k_l \right) \nonumber \\
    \vec q_{ij,kl} & = &  \frac{1}{m_i+m_j+m_k+m_l}\left((m_k+m_l)\, \left( \vec k_i+\vec k_j\right) - (m_i+m_j)\, \left( \vec k_k+\vec k_l\right)  \right) \nonumber \\
\end{eqnarray}
For a more efficient representation, we expand the angular dependencies in orbital angular momenta 
which are then coupled with the spins to a fixed total angular momentum. The ``3+1'' and ``2+2'' basis states 
then read 
\begin{eqnarray}
    \left|  (ij)k,l \right\rangle & =&  
    \left| p_{ij}  p_{ij,k} q_{ijk,l}  
    \left[  \left( \left(l^{}_{ij}s^{}_{ij}\right) j^{}_{ij}  
                  \left( l^{}_{k} 1/2 \right) I^{}_{k} \right) j^{}_{k}
                  \left( l^{}_{l} 1/2 \right) I^{}_{l}  \right] J 
    \ \left[ \left( \left( t^{}_{i} t^{}_{j} \right) t^{}_{ij}  t^{}_{k}  \right) \tau_{k} \right] T M_T \right\rangle~, \nonumber \\
    \left|  (ij)kl \right\rangle & =&  
    \left| p_{ij}  p_{kl} q_{ij,kl}  \left[ \left( \left(l^{}_{ij}s^{}_{ij}\right) j^{}_{ij}  
    \lambda \right) I  \left( \left(l^{}_{kl}s^{}_{kl}\right) j^{}_{kl}  \right) \right] J 
    \ \left[ \left( \left( t^{}_{i} t^{}_{j} \right) t^{}_{ij}  \left( t^{}_{k} t^{}_{l} \right) t^{}_{kl} \right) \right] T M^{}_T \right\rangle \ .
\end{eqnarray}
$p_{ij}$, $p_{ij,k}$, $q_{ijk,l}$, $p_{kl}$ and  $q_{ij,kl}$ are the magnitude of the momenta. Their angular 
dependence is expanded in orbital angular momenta $l^{}_{ij}$, $l^{}_{k}$,  $l^{}_{l}$, $l^{}_{kl}$ and $\lambda$, respectively. The spin $1/2$ of the baryons couple to two-baryon spin $s_{ij}$ and $s_{kl}$ 
and these couple with the orbital angular momenta to intermediate angular momenta $j_{ij}$, $I^{}_{k}$, $I^{}_{l}$, $j_{k}$, $j_{kl}$, and $I$ to the total angular momentum $J$ 
as indicated. Since isospin symmetry is preserved rather accurately, the states are formulated 
in the isospin basis. The baryon isospins $t^{}_{i}$, $t^{}_{j}$, $t^{}_{k}$, and $t^{}_{l}$ 
are coupled to two-baryon isospins $t^{}_{ij}$ and $t^{}_{kl}$ and three-baryon isospins $\tau^{}_{k}$ and finally to the total isospin $T$ and its third component $M^{}_T$. Following the standard notation, we use here $t$ for isospins and the $I$ for a coupled angular momentum which differs 
from the notation of $A=2$ systems used in Sec.~\ref{second}. 
Note that the inclusion of the isospin quantum numbers allows one to distinguish N, $\Lambda$
and $\Sigma$ baryons by their isospin. Depending on the Yakubovsky component, 
states with different positions of the hyperon, the fourth particle in our convention, 
are used. E.g., for $\left|  \psi_{1A}  \right\rangle$, `3+1' states with $l=4$ are used 
and, for $\left|  \psi_{2B}  \right\rangle$, `2+2' states with $j=4$. 

Since we use  different sets of basis states for the different Yakubovsky components, 
coordinate transformations between the different states are necessary. These are not 
explicitly shown in the Eqs.~(\ref{eq:yakeq}) but are in fact 
equivalent to the permutation operators. 
To obtain energies converged up to the keV level, still a very large number of partial 
waves of the order or several thousand is necessary. Using approximately $40$ momentum 
grid points for each of the momenta, the discretized set of equations can become as large
as $10^8$ dimensional. Clearly, storing and applying 
a $10^8 \times 10^8$ matrix is impossible even on today's supercomputers. But fortunately, 
the Yakubovsky equations can be decomposed in a series of sparse matrices that can be 
handled on massively parallel systems. To achieve this form, we apply coordinate 
transformations and permutation operators such that either particle $i$ and $j$ are not affected,
or, for ``3+1'' coordinates,we leave the outer $l$-th particle unaffected, 
or, for ``2+2'' coordinates, we only change the subsystems. 
In these cases, the two-baryon subsystem(s) or $l$-th particle  quantum numbers and momenta 
are conserved and the transformation becomes sparse. The application of the two-baryon 
$t$-matrix is sparse since it only changes quantum numbers of the particles $i$ and $j$. 
Note however that the possibility of $\Lambda$-$\Sigma$ conversion changes the rest masses 
of the $ij$ subsystem. This induces a shift of the other momenta when 
using ``3+1'' coordinates. In order to end up with sparse matrices, we only apply the $t$-matrices  using ``2+2'' coordinates where such a shift of Jacobi momenta shows up. The application of the three-baryon interaction is implemented in ``3+1'' coordinates. In this case, all momenta and quantum numbers of particle $l$ are conserved. 

Including the different transformations and the different sets of coordinates, Eqs.~(\ref{eq:yakeq}) become more lengthy  
\begin{eqnarray}
 \left\langle (12)3,4 \right| \left.  \psi_{1A} \right\rangle \! \!  &  =  & \!\!
 \left\langle (12)3,4  \right| G_0 t_{12} \left| (12)3,4 ' \right\rangle 
     \left\langle (12)3,4'   \right| P \left| (12)3,4 '' \right\rangle \nonumber \\
     & & \quad \left[ \left\langle (12)3,4''   \right| \left.  \psi_{1A} \right\rangle
            +  \left\langle (12)3,4''   \right| \left. (12)4,3''' \right\rangle
                    \left\langle (12)4,3'''   \right| \left.  \psi_{1B} \right\rangle
            +  \left\langle (12)3,4''   \right| \left. (12)34''' \right\rangle
                    \left\langle (12)34'''   \right| \left.  \psi_{2A} \right\rangle      
        \right] \nonumber \\
    & & +  \left\langle (12)3,4  \right| 1+ G_0 t_{12} \left| (12)3,4 ' \right\rangle
           \left\langle (12)3,4'   \right| V_{123}^{(3)} + V_{1234}^{(12)3,4}\left| (12)3,4 '' \right\rangle
              \left\langle (12)3,4''   \right| \left.  \Psi  \right\rangle,\nonumber \\
 \left\langle (12)4,3 \right| \left.  \psi_{1B} \right\rangle \! \!  &  =  & \!\!
 2 \left\langle (12)4,3  \right| G_0 t_{12} \left| (12)4,3 ' \right\rangle 
          \left\langle (12)4,3'   \right| \left. (14)2,3'' \right\rangle
            \left[  \left\langle (14)2,3''   \right| 1 -P_{23} \left| (14)2,3''' \right\rangle
               \left\langle (14)2,3'''   \right| \left.  \psi_{1C} \right\rangle \right. \nonumber \\
        & & \left. \hspace{8cm}  +      \left\langle (34)1,2''   \right| \left. (34)12''' \right\rangle 
                   \left\langle (34)12'''   \right| \left.  \psi_{2B} \right\rangle 
            \right] \nonumber \\      
    & & +  \left\langle (12)4,3  \right| 1+ G_0 t_{12} \left| (12)4,3 ' \right\rangle
           \left\langle (12)4,3'   \right| V_{124}^{(4)} + V_{1234}^{(12)4,3} \left| (12)4,3 '' \right\rangle
              \left\langle (12)4,3''   \right| \left.  \Psi  \right\rangle, \nonumber \\
 \left\langle (14)2,3 \right| \left.  \psi_{1C} \right\rangle \! \!  &  =  & \!\!    
 \left\langle (14)2,3 \right| \left. (14)23 ' \right\rangle  
       \left\langle (14)23' \right|  G_0 t_{14}  \left| (14)23 '' \right\rangle  
       \left\langle (14)23'' \right| \left. (14)2,3 ''' \right\rangle  \nonumber \\
       & & \quad  \left[  \left\langle (14)2,3''' \right| \left. (12)4,3 ^{IV} \right\rangle 
       \left(  \left\langle (12)4,3^{IV} \right| \left. \psi_{1B} \right\rangle 
             + \left\langle (12)4,3^{IV} \right| \left. (12)3,4 ^{V} \right\rangle 
                   \left\langle (12)3,4^{V} \right|      \left. \psi_{1A} \right\rangle \right.\right.
                \nonumber \\   
       && \qquad  \left. \left.    + \left\langle (12)4,3^{IV} \right| \left. (12)34 ^{V} \right\rangle 
                         \left\langle (12)34^{V} \right| \left. \psi_{2A} \right\rangle 
       \right) \right. \nonumber \\
       & & \quad + \left. \left\langle (14)2,3''' \right| P_{12} \left| (14)2,3 ^{IV} \right\rangle 
         \left( - \left\langle (14)2,3^{IV} \right| \left. \psi_{1C} \right\rangle 
             + \left\langle (14)2,3^{IV} \right| P_{23} \left| (14)2,3 ^{V} \right\rangle 
                   \left\langle (14)2,3^{V} \right|      \left. \psi_{1C} \right\rangle 
               \right. \right. \nonumber \\    
             & & \qquad \left. \left. + \left\langle (34)2,1^{IV} \right| \left. (34)12 ^{V} \right\rangle 
                   \left\langle (34)12^{V} \right|      \left. \psi_{2B} \right\rangle          
         \right) 
       \right] \nonumber \\      
    & & +  \left\langle (14)2,3 \right| \left. (14)23 ' \right\rangle  
       \left\langle (14)23' \right|  1+ G_0 t_{14}   \left| (14)23 '' \right\rangle  
       \left\langle (14)23'' \right| \left. (14)2,3 ''' \right\rangle  \nonumber \\
    && \quad        \left\langle (14)2,3'''   \right| V_{124}^{(2)} + V_{1234}^{(14)2,3} \left| (14)2,3 ^{IV} \right\rangle
              \left\langle (14)2,3^{IV}   \right| \left.  \Psi  \right\rangle, \nonumber \\  
\left\langle (12)34 \right| \left.  \psi_{2A} \right\rangle \! \!  &  =  & \!\!  
     \left\langle (12)34  \right| G_0 t_{12} \left| (12)34 ' \right\rangle 
      \left\langle (12)34'  \right| \left. (34)12 '' \right\rangle 
      \left[ -2  \left\langle (34)12''  \right| \left. (34)2,1 ''' \right\rangle  
                      \left\langle (14)2,3''' \right|      \left. \psi_{1C} \right\rangle 
             + \left\langle (34)12''  \right| \left. \psi_{2B} \right\rangle 
     \right], \nonumber \\
\left\langle (34)12 \right| \left.  \psi_{2B} \right\rangle \! \!  &  =  & \!\!  
     \left\langle (34)12  \right| G_0 t_{34} \left| (34)12 ' \right\rangle 
      \left\langle (34)12'  \right| \left. (12)34 '' \right\rangle \nonumber \\
      & & \quad \left[ \left\langle (12)34''  \right| \left. (12)3,4 ''' \right\rangle  
                      \left\langle (12)3,4''' \right|      \left. \psi_{1A} \right\rangle 
             +\left\langle (12)34''  \right| \left. (12)4,3 ''' \right\rangle  
                      \left\langle (12)4,3''' \right|      \left. \psi_{1B} \right\rangle 
             + \left\langle (12)34''  \right| \left. \psi_{2A} \right\rangle 
     \right]. \nonumber \\
\end{eqnarray}

This form can be achieved using the antisymmetry of the basis states for two-nucleon 
subsystems and the fact that permutation operators  relate different sets of Jacobi coordinates 
with each other. Summation and integration  of the quantum numbers and coordinates 
of intermediate states are omitted to shorten the notation. 
Using the equivalence of permutation operators and coordinate transformations 
is also used to replace permutation operators by using different sets of coordinates 
for bra and ket vectors in some of the expressions above. For the application of the 
3BFs, we need the wave function defined in Eq.~(\ref{eq:4bwf}) in different Jacobi 
coordinates. Also these can be obtained by appropriate coordinate transformations.  
In the following, we do not consider the 4BFs anymore since they are, at this point, usually neglected. 

Note that that our requirements on the sparsity of the different matrices are fulfilled in this form and the YN $t$-matrix is always applied in ``2+2'' coordinates to avoid a momentum 
shift due to particle conversion. 

Due to the distribution of different parts of the 3BFs 
to different Yakubovsky components, the coordinates used are the ones naturally used 
for these interactions. The improved convergence is primarily enforced  by 
the suppression of high orbital angular momenta  due to $t_{ij}$ and $V_{ijk}^{(k)}$. 
Similarly, the expectation values  $\left\langle \Psi \right| V_{ij} \left| \Psi \right\rangle $ 
and $\left\langle \Psi \right| V_{ijk} \left| \Psi \right\rangle$ only require a limited 
range of partial waves if evaluated in the appropriate Jacobi coordinates. 
The kinetic energy can be easily expressed in arbitrary Jacobi coordinates. 
In this case, it is advantageous to use the antisymmetry under exchange 
of the nucleons to rewrite the expectation value using Eq.~(\ref{eq:4bwf}) as 
\begin{equation}
    \label{eq:ekin}
    \left\langle \Psi \right| H_0 \left. \Psi \right\rangle = 
       3 \left\langle \Psi \right| H_0 \left| \psi_{1A} \right\rangle 
       + 3 \left\langle \Psi \right| H_0 \left| \psi_{1B} \right\rangle 
       + 6 \left\langle \Psi \right| H_0 \left| \psi_{1C} \right\rangle 
       + 3 \left\langle \Psi \right| H_0 \left| \psi_{2A} \right\rangle 
       + 3 \left\langle \Psi \right| H_0 \left| \psi_{2B} \right\rangle \ .
\end{equation}
When different parts are then calculated using the appropriate Jacobi coordinate 
so that again, due to the Yakubovsky components, the contributing number of 
partial waves is restricted. 
In order to check the numerical accuracy of our calculations, we compare  
the energy values $E$ entering the Yakubovsky equations to the value of the expectation 
value calculated as outlined above. The presence of the wave function still 
limits the accuracy of the expectation value. But we find that our energies 
for the most recent calculations are converged to 20~keV and the expectation values
to 50~keV. Both results generally agree within these uncertainties. 

\begin{table}[]
    \centering
    \begin{tabular}{lll|rrr|rrr|rrr}
YN   & NN  & 3NF & \multicolumn{3}{c|}{$^3_\Lambda$H}& \multicolumn{3}{c|}{$^4_\Lambda$He($0^+$)}& \multicolumn{3}{c}{$^4_\Lambda$He($1^+$)} \cr
         & & 
          & $E_\Lambda^{}$ & $E$    & $\langle H \rangle$   
          & $E_\Lambda^{}$ & $E$    & $\langle H \rangle$   
          & $E_\Lambda^{}$ & $E$    & $\langle H \rangle$   \\
          \hline 
LO(700) & N4LO$^+$(450) & N2LO & 0.135 & -2.359 & -2.360 & 3.088 & -10.827 & -10.803 & 2.275 & -10.015 & -9.992 \\
NLO13(500) & N4LO$^+$(450) & N2LO & 0.134 & -2.357 & -2.358 & 1.654 & -9.394 & -9.367 & 0.778 & -8.518 & -8.495 \\
NLO19(500) & N4LO$^+$(450) & N2LO & 0.100 & -2.323 & -2.324 & 1.630 & -9.370 & -9.349 & 1.221 & -8.960 & -8.940 \\
NLO19(550) & N4LO$^+$(450) & N2LO & 0.093 & -2.317 & -2.317 & 1.534 & -9.274 & -9.243 & 1.238 & -8.978 & -8.949 \\
NLO19(600) & N4LO$^+$(450) & N2LO & 0.091 & -2.314 & -2.314 & 1.444 & -9.183 & -9.136 & 1.055 & -8.795 & -8.755 \\
NLO19(650) & N4LO$^+$(400) & N2LO & 0.098 & -2.321 & -2.322 & 1.553 & -9.289 & -9.232 & 0.929 & -8.664 & -8.622 \\
NLO19(650) & N4LO$^+$(450) & N2LO & 0.095 & -2.318 & -2.319 & 1.525 & -9.265 & -9.201 & 0.917 & -8.657 & -8.608 \\
NLO19(650) & N4LO$^+$(500) & N2LO & 0.090 & -2.313 & -2.314 & 1.480 & -9.223 & -9.153 & 0.891 & -8.635 & -8.581 \\
NLO19(650) & N4LO$^+$(550) & N2LO & 0.085 & -2.308 & -2.308 & 1.429 & -9.176 & -9.102 & 0.856 & -8.604 & -8.547 \\
NLO(500) & N4LO$^+$(450) & N2LO & 0.127 & -2.350 & -2.351 & 2.009 & -9.749 & -9.701 & 1.041 & -8.781 & -8.746 \\
NLO(550) & N4LO$^+$(450) & N2LO & 0.124 & -2.347 & -2.348 & 2.102 & -9.841 & -9.788 & 1.102 & -8.842 & -8.800 \\
NLO(600) & N4LO$^+$(450) & N2LO & 0.122 & -2.345 & -2.346 & 2.021 & -9.761 & -9.681 & 0.927 & -8.667 & -8.602 \\
NLONM(550) & N4LO$^+$(550) & N2LO & 0.105 & -2.328 & -2.328 & 1.910 & -9.657 & -9.595 & 1.017 & -8.765 & -8.718 \\
NNLO(500) & N4LO$^+$(450) & N2LO & 0.147 & -2.371 & -2.371 & 2.001 & -9.741 & -9.686 & 1.002 & -8.741 & -8.706 \\
NNLO(550) & N4LO$^+$(400) & N2LO & 0.137 & -2.360 & -2.361 & 1.995 & -9.731 & -9.673 & 1.249 & -8.985 & -8.944 \\
NNLO(550) & N4LO$^+$(450) & N2LO & 0.139 & -2.362 & -2.363 & 2.004 & -9.743 & -9.677 & 1.251 & -8.990 & -8.943 \\
NNLO(550) & N4LO$^+$(500) & N2LO & 0.138 & -2.362 & -2.362 & 1.990 & -9.734 & -9.661 & 1.230 & -8.974 & -8.924 \\
NNLO(550) & N4LO$^+$(550) & N2LO & 0.136 & -2.359 & -2.360 & 1.962 & -9.710 & -9.633 & 1.205 & -8.953 & -8.901 \\
NNLO(600) & N4LO$^+$(450) & N2LO & 0.172 & -2.395 & -2.396 & 2.263 & -10.002 & -9.933 & 1.181 & -8.921 & -8.872 \\
NNLO(E)(550) & N4LO$^+$(450) & N2LO & 0.125 & -2.348 & -2.349 & 1.969 & -9.709 & -9.643 & 1.188 & -8.928 & -8.881 \\
\hline 
J\"ulich $\tilde A$ & Nijm 93 & -- & --    & --    & --    & 0.430 & -7.400 &      & 0.480 & -7.450 &     \\
J\"ulich '04 & Nijm 93       & --   & 0.13  &        &        &  1.9     &        &     &  2.3     &        &        \\
SC89 & Nijm 93 & -- & 0.143 & -2.367 & -2.366 & 2.140 & -9.140 &  & 0.020 & -6.990 &   \\
SC97d & Nijm 93  --  & --   & --    &   --   &   --   &  1.3     &        &     &  0.8     &        &        \\
SC97e & Nijm 93 & -- & 0.023 & -2.247 & -2.246 & 1.540 & -8.550 &  & 0.720 & -7.690 &  \\
SC97e & Nijm 93 & TM & 0.023 & -2.247 & -2.246 & 1.560 & -9.320 &  & 0.700 & -8.350 &  \\
\hline 
Experiment   &  &  & $0.164(43)$& & & 
\multicolumn{3}{c|}{\parbox{3cm}{$2.347(36) \ \left(\phantom{.}^4_\Lambda {\rm He}\right) $ 
            \\ $2.169(42) \ \left(\phantom{.}^4_\Lambda {\rm H}\right) $
             \\ $2.258(55) \ \left({\rm average} \right)$ }}    &
\multicolumn{3}{c}{\parbox{3cm}{$0.942(36) \ \left(\phantom{.}^4_\Lambda {\rm He}\right) $ \\ $1.081(46) \ \left(\phantom{.}^4_\Lambda {\rm H}\right) $  \\ 
$1.012(58) \ \left({\rm average} \right)$ }}
\end{tabular}
    \caption{$\Lambda$-separation energies and binding for $^3_\Lambda$H and the $0^+$ and $1^+$ states of $^4_\Lambda$He compared to the experimental values 
    \cite{HypernuclearDataBase}. Besides the chiral interactions, results are shown for selected combinations 
    of phenomenological YN interactions: SC89~\cite{Maessen:1989sx}, SC97d-e~\cite{Rijken:1998yy}, 
    J\"ulich \~A \cite{Reuber:1993ip} and J\"ulich '04~\cite{Haidenbauer:2005zh}. 
    As NN interaction also Nijm~93~\cite{Stoks:1994wp} was used.
    Additionally, the corresponding binding energies    
    and the expectation values $\langle H \rangle$ are given.  
    Energies are given in MeV. }
    \label{tab:fylamsepener}
\end{table}

\subsection{Application of chiral baryon-baryon interactions}

\begin{figure}
    \centering
    \includegraphics[width=0.8\linewidth]{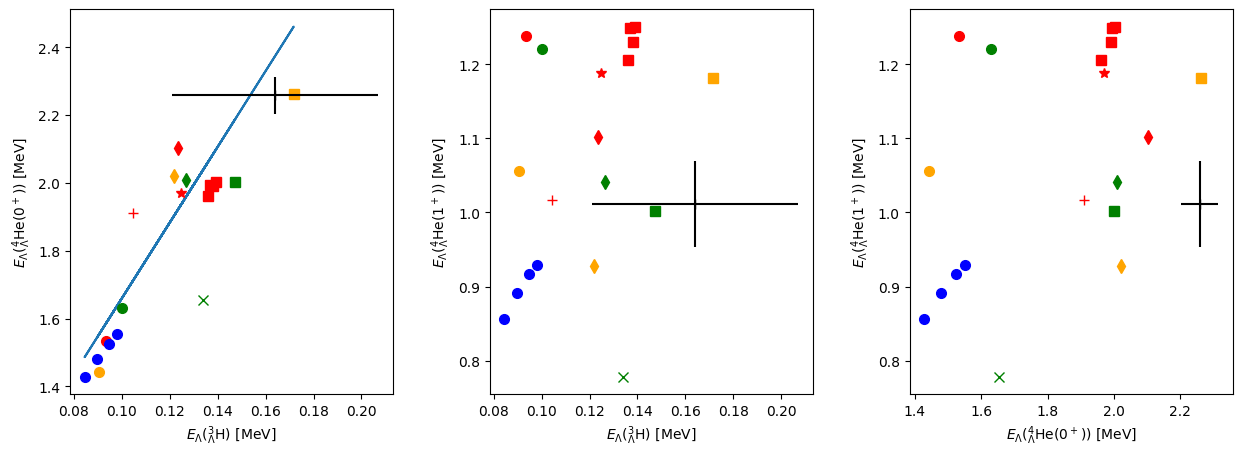}
    \caption{Correlation of the separation energies of the $^3_\Lambda$H, 
    $^4_\Lambda$He$(0^+)$ and $^4_\Lambda$He$(1^+)$ states for different orders 
    and cutoffs. The symbols distinguish the different orders or versions 
    of the chiral interactions: NLO13 (stars), NLO19 (crosses), 
    and the SMS YN potentials NLONM (diamonds), NLO (squares), N2LO (circles), and N2LO(550)$^b$(+). 
    Colors distinguish different 
    cutoffs of the YN interaction: 500~MeV (green), 550~MeV (red), 600~MeV (orange), 650~MeV (blue). 
    The different cutoffs of the NN and 3N interaction are not distinguished.
    The black cross indicates the experimental values 
    and their uncertainty.}
    \label{fig:energycorrFY} 
\end{figure}

Results for the light hypernuclei for the different chiral interactions 
and some selected phenomenological interactions are compiled in 
Table~\ref{tab:fylamsepener} and summarized in Fig.~\ref{fig:energycorrFY}. 

Before starting a discussion on the physical implications, let us 
note that it has been checked that the separation energies are numerically 
converged up to 1~keV for $^3_\Lambda$H and up to 20~keV for $^4_\Lambda$He
except for the older calculations based on phenomenological interactions.
In those calculations, the accuracy is about 50~keV. 
The expectation values require the calculation of the wave functions. Therefore,
their uncertainty is slightly larger, 2~keV for $^3_\Lambda$H and up to 100~keV for $^4_\Lambda$He. 
It is reassuring to see that the energy and the expectation value agree within these uncertainties in all case. In order to get 
the most accurate expectation values, the kinetic energy has been evaluated 
using Eq.~(\ref{eq:ekin}). The expectation value for the NN and 3N interactions 
was calculated using the coordinates $\left| (12)3,4 \right\rangle$.
For the evaluation of the YN interaction, it turned out that 
$\left| (34)12 \right\rangle$ coordinates lead to the fastest convergence 
with respect to partial waves. 

The results for $^4_\Lambda$H are not given in the table since the YN 
interactions used here do not include charge-symmetry breaking (CSB). 
The separation energies therefore only show a negligible difference 
between $^4_\Lambda$He and $^4_\Lambda$H. CSB will be discussed in the next section 
in detail. 

Light hypernuclei are an important source of information on the YN and YNN 
interactions. Thereby, the $^3_\Lambda$H is of similar importance for the YN 
interaction as the deuteron for the NN interaction. First of all, it is the 
lightest bound hypernucleus and therefore the simplest bound system that can 
be studied. Second, it has been realized very early that the angular momentum $J=1/2$ 
and isospin $T=0$ of the only bound state in this system 
implies that the $\Lambda$N interaction is more attractive in the 
$^1{\rm S}_0$  than in the $^3{\rm S}_1$  channel \cite{Dalitz:1958zza}. 
In absence of any polarization data, this information is a key element to fix 
the spin dependence of the $\Lambda$N interaction. 
In order to better constrain this spin dependence, one can exploit a series of YN interactions 
provided by the Nijmegen group (SC97a-f) that essentially differ by their 
prediction for the $\Lambda$N $^1{\rm S}_0$ scattering length while keeping the description of the unpolarized cross sections unaltered~\cite{Rijken:1998yy}. It was found that the hypertriton is only bound by the last two variants of these interactions \cite{Nogga:2001cz}. This information suggests that the $\Lambda$N 
$^1{\rm S}_0$ scattering length should be larger (in magnitude) than $-2.10$~fm.  

\begin{figure}
    \centering
    \includegraphics[width=0.6\linewidth]{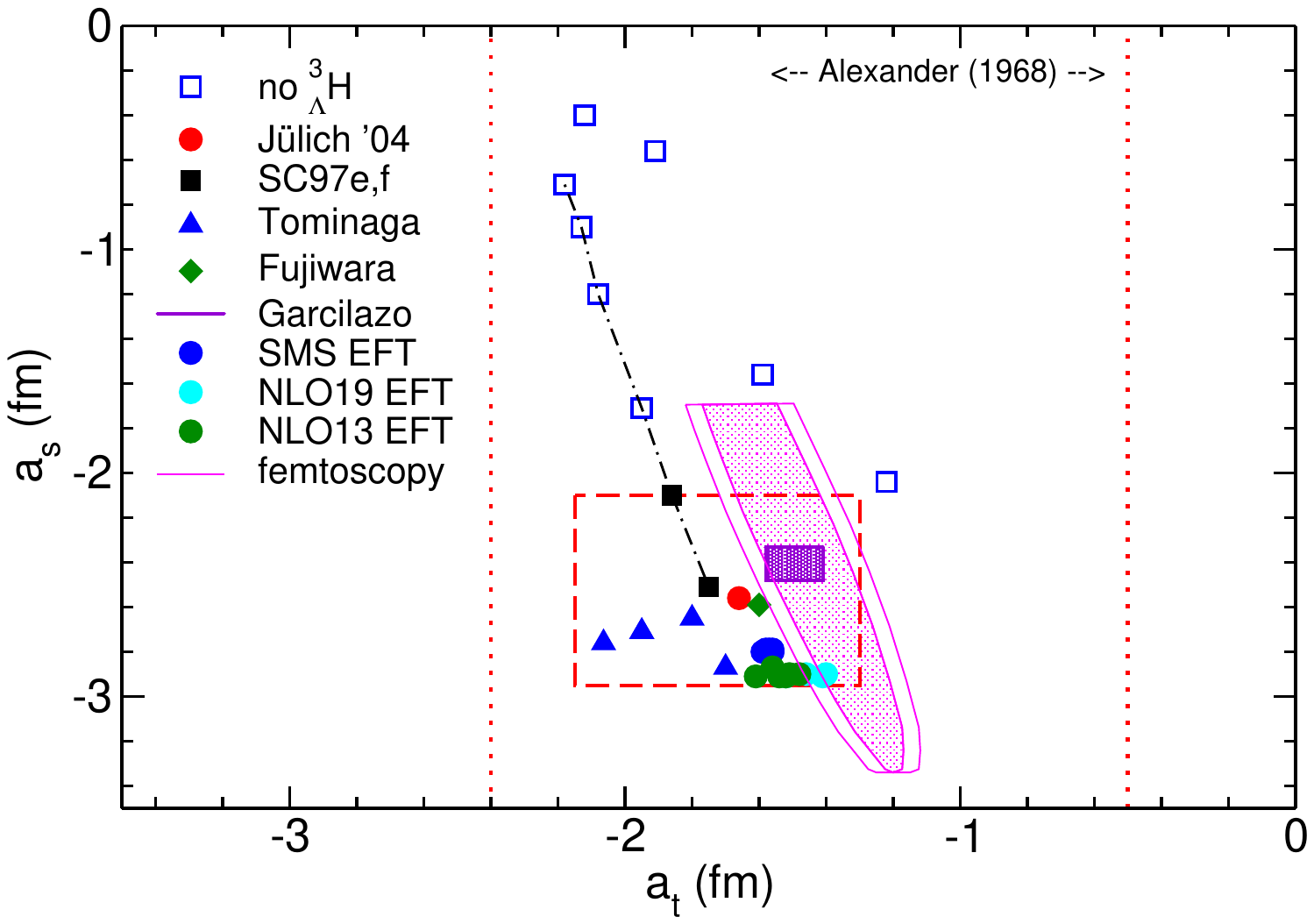}
    \caption{Correlation between the $\Lambda$N
    $^1S_0$ and $^3S_1$ scattering lengths ($a_s$ and $a_t$) and the
    existence of a bound $^3_\Lambda$H. Results are
    shown for various YN potentials from the literature where bound-state calculations have
    been performed. The region where the $^3_\Lambda$H is bound is indicated by a dashed
    rectangle. Open symbols correspond to results of potentials for which the 
    $^3_\Lambda$H is not bound. 
The squares connected by a dash-dotted line correspond to the series of 
YN interactions by the Nijmegen group~\cite{Rijken:1998yy}, see text. 
    In addition estimates for the scattering lengths from the measured $\Lambda$p cross
    section \cite{Alexander:1968acu} (dotted lines)
    and from femtoscopic data \cite{Mihaylov:2023ahn} 
    (hatched area) are indicated. }
    \label{fig:asat}
\end{figure}

The world average of the $\Lambda$ separation energy of  $^3_\Lambda$H is 
$0.164\pm0.043$~MeV~\cite{HypernuclearDataBase} making this nucleus the  
lightest halo nucleus and effectively suppressing contributions 
of YNN forces. Based on power counting arguments \cite{Le:2023bfj} and 
explicit calculation of the YNN contribution \cite{Le:2024rkd}, one can 
expect that the YNN force contribution is smaller than the experimental 
uncertainty of the separation energy. This motivates using the separation 
energy as an additional datum for constraining YN interactions. This 
approach has been used for fitting all chiral YN interactions 
of the J\"ulich-Bonn group \cite{Haidenbauer:2005zh,Polinder:2006zh, Haidenbauer:2019boi, Haidenbauer:2023qhf}. Without using charge-symmetry breaking (CSB), all interactions lead to singlet scattering lengths  
between -2.5 and -2.9~fm except the result of the LO chiral interaction \cite{Polinder:2006zh} which predicted -1.90~fm because its momentum 
dependence is too weak to get a reasonable 
description of the $\Lambda$N cross section for all momenta below the opening 
of the $\Sigma$ channel.  
 An overview of the correlation between the $\Lambda$p scattering lengths and the existence
 of a bound $^3_\Lambda$H is provided in 
 Fig.~\ref{fig:asat}, based on the YN potentials of the J\"ulich-Bonn group
 and for several YN potentials from the literature where bound-state calculations
 have been performed
\cite{Tominaga:2001ra,Fujiwara:2006yh,Fujiwara:2007en,Garcilazo:2007ss}. 
Besides the $(a_s,a_t)$ region where a bound hypertriton has been found for the
corresponding YN potentials one can also see the clear correlation between 
$a_s$ and $a_t$ that result from fitting to the $\Lambda$p cross-section 
data \cite{Alexander:1968acu,Sechi-Zorn:1968mao}, and a very similar though slightly
shifted correlation when considering recent femtoscopic data by the ALICE
Collaboration~\cite{ALICE:2021njx,Mihaylov:2023ahn}.
Indeed, also femtoscopic measurements of the $\Lambda$p 
 interaction do not allow to disentangle the spin dependence. However, the analysis
 of such experiments in Ref.~\cite{Mihaylov:2023ahn} indicates that the $\Lambda$N 
 interaction could be overall slightly weaker than what follows from the
 available $\Lambda$p cross sections.
 Of course, one should keep in mind that the systematic uncertainty due to
 the employed theoretical tools (Koonin-Pratt formalism, source function)
 \cite{Epelbaum:2025aan} has not yet been accounted for.

The resulting separation energies can be seen in Table~\ref{tab:fylamsepener}. Note that the $^3_\Lambda$H energies are not exactly 
the same even if the values were used to fit the interaction. 
First of all, the separation energy has recently changed 
due to new data leading to a slightly larger world average \cite{STAR:2019wjm,ALICE:2022sco}. 
However, and more important, because of the rather large experimental 
uncertainty of this datum, it was only used as guideline and not as a 
stringent fitting constraint. Moreover, the fit was usually only done for 
one of the cutoffs of the interaction and the other cases where 
adjusted to the resulting scattering lengths leading to small 
variations of the separation energy for the different orders and 
cutoffs. 

The results for the $A=4$ hypernuclei show much more dependence on the interaction. Especially, the 
separation energies for the LO interaction are unrealistically large which can be already expected 
from the overprediction of the $\Lambda$p cross sections around the $\Sigma$N threshold. 
In the following, we will not consider the LO results since they are affected by 
an unrealistic description of YN data. 

For the rest of the interactions, the available YN data is generally well described. 
Nevertheless, one observes significant variations of the $A=4$ energies. Some insight can 
be gained by considering the energies from the chiral interactions, summarized in Fig.~\ref{fig:energycorrFY}. The plots show that there is a correlation 
of the $^3_\Lambda$H and the $^4_\Lambda$He$~0^+$ energies. The energy of the $1^+$ state is less 
related to the other two. Assuming only $S$-wave
interactions and neglecting possible contributions from the $\Lambda$-$\Sigma$ transitions, 
it can be shown that the $1^+$ state 
is driven by the $^3{\rm S}_1$ interaction \cite{Herndon:1967zza,Dalitz:1972vzj,Haidenbauer:2019boi} 
whereas the $J^\pi=1/2^+$ state of $^3_\Lambda$H
is mostly affected by the $^1{\rm S}_0$ state and the $J^\pi=0^+$ state of $^4_\Lambda$He gets approximately equal contributions from both partial waves. 
Those observations taken together suggest that contributions of the 
$^3{\rm S}_1$ state to the separation energies are much more dependent on the detailed  
realization of the YN interactions. 
This assessment is in line with the observed effects from the difference of the $\Lambda$-$\Sigma$  conversion strength documented in \cite{Haidenbauer:2019boi}, by 
comparing the NLO13 and NLO19 interactions. 
Since both interactions are phase-shift equivalent to a high degree, 
it is clear that the actual $\Lambda$-$\Sigma$ conversion strength in the potential, which 
is basically an ``off-shell'' property of the YN interaction,  
cannot be fixed uniquely by YN data. 
Different realizations of the YN interaction will need different additional YNN interactions (see Section~\ref{sec:ncsmynn}
for first results including such interactions). The variation of the separation energies seen in Fig.~\ref{fig:energycorrFY}
is therefore also a measure of the size of the YNN forces. Note that none of the interactions can describe the $0^+$ and $1^+$ state simultaneously indicating that 
YNN forces, that are able to resolve this discrepancy, need to be spin dependent. 

Recently, 
it has been discussed that also differences in the NN interactions could affect the separation 
energies of $\Lambda$ hypernuclei \cite{Htun:2021jnu,Gazda:2022fte}. 
In these works, a family of NN interactions and corresponding 
3NFs were used to investigate this aspect in more 
detail. The results show  a significant dependence of the order of several hundred keV for 
$A=4$ hypernuclei and still 100~keV for $^3_\Lambda$H. This is in contradiction to earlier observations that the NN interaction 
dependence is minor \cite{Nogga:2001cz}.     
It is however also seen that the dependence is systematically linked to 
the cutoff of the interaction (40~keV for $^3_\Lambda$H \cite{Htun:2021jnu}) and also to the maximal kinetic energy of the 
NN data that was used in the fit of the employed NN interaction (60~keV for $^3_\Lambda$H \cite{Htun:2021jnu}). 
The numbers are based on LO chiral interactions \cite{Polinder:2006zh}. 
The two contributions to the NN-force dependence have clearly a different theoretical origin. 
The dependence on the NN data used is related to an inaccuracy of the NN interaction that is irrelevant 
when high order and high accuracy NN interactions are used as is by now standard in any calculations for 
hypernuclei. The cutoff dependence indicates a sensitivity to the realization of the interaction that should 
be absorbed by higher order contributions, i.e. 3BFs. Therefore, the results of  
Refs.~\cite{Htun:2021jnu,Gazda:2022fte} indicate a contribution of YNN interactions of the order of 20~keV 
for $^3_\Lambda$H and 100~keV for $^4_\Lambda$He which is in better agreement with previous assessments. 
Tab.~\ref{tab:fylamsepener} and Fig.~\ref{fig:energycorrFY} include a series of calculations 
for NLO19(650) and N$^2$LO(550) that show that the dependence on the NN interactions is even lower for 
$^3_\Lambda$H than estimated above. This can be traced back to using higher order YN interactions 
instead of LO, which seems to strongly reduce the sensitivity to the 
short range properties of the NN interaction. For these higher order interactions, 
the estimate for the YNN contribution necessary is less than 10~keV (30~keV) for $^3_\Lambda$H
($^4_\Lambda$He) which is smaller (comparable) to the experimental uncertainty. 
Certainly, the dependence on the realization of the YN interaction is much larger. 

One of the remarkable things one can learn from Table~\ref{tab:fylamsepener} is 
that some of the phenomenological interactions predict vast deviations 
from the experimental value. For example, the J\"ulich '04 interaction leads to reversed 
ordering of the $0^+$ and $1^+$ states although the $\Lambda$N scattering lengths and
the $^3_\Lambda$H energy
are quite comparable to the other interactions. The origin of this failure is most likely 
related to the rather weak $\Lambda$N-$\Sigma$N transition potential that characterizes this 
YN model. Actually, it is so weak that even the $\Sigma^-$p$\to$$\Lambda$n transition
cross section is somewhat underestimated, see the results and the discussion in 
Ref.~\cite{Haidenbauer:2019boi}. One would need
large YNN force effects to reproduce the experimental separation energies
which, however, cannot be systematically obtained due to the 
phenomenological character of the interaction. For the Nijmegen SC97 interactions, 
the $^3_\Lambda$H and  $^4_\Lambda$He energies are generally too small. 
The SC89 is quite different 
since it describes the $^3_\Lambda$H and the $0^+$ state well but barely binds the excited 
state. Also these deviations are probably mostly due to missing YNN forces that cannot 
be systematically obtained in the framework of these models. 

\subsection{Charge-symmetry breaking of \texorpdfstring{$A=4$}{A=4} hypernuclei}
\label{sec:csbyn}

\begin{table}
\renewcommand{\arraystretch}{1.4}
    \caption{\label{tab:csb}CSB of the $A=4$ hypernuclear separation energies in MeV, based on 
    the scenario CSB1 considered in Ref.~\cite{Haidenbauer:2021wld}. 
    For completeness also the $^3_\Lambda$H separation energy is given.  
    The predicted  singlet  $a_{s}$ and triplet $a_{t}$ are given in fm. In the last row the experimental values for 
    the separation energy from the Mainz database \cite{HypernuclearDataBase} are given (the fit used slightly different values).
    For the scattering length the averaged scattering length from non-CSB versions of NLO13 and NLO19 is added. 
    }
\begin{center}    
\begin{tabular}{|l|r|rr|rr|rr|rr|}
\hline
      Interaction &  $E_\Lambda(^3_\Lambda{\rm H})$ &  \multicolumn{2}{c|}{$E_\Lambda(^4_\Lambda{\rm He})$ } &  \multicolumn{2}{c|}{$E_\Lambda(^4_\Lambda{\rm H})$ }&  \multicolumn{2}{c|}{$a^{}_{s}$}&  \multicolumn{2}{c|}{ $a^{}_{t}$}\\
                     &     &   $J^\pi=0^+$ & $J^\pi=1^+$ & $J^\pi=0^+$ & $J^\pi=1^+$ &
$\Lambda p \ $ & $\Lambda n \ $ & $\Lambda p \ $ & $\Lambda n \ $ \\                                      
\hline
NLO13(500)           & $0.14$ &  $1.82$ &  $0.76$ &  $1.56$ & $0.82$  & $-2.604$ & $-3.267$ & $-1.647$ & $-1.561$ \\
NLO13(550)           & $0.10$ &  $1.62$ &  $0.56$ &  $1.36$ & $0.61$  & $-2.586$ & $-3.291$ & $-1.551$ & $-1.469$ \\
NLO13(600)           & $0.09$ &  $1.59$ &  $0.55$ &  $1.34$ & $0.60$  & $-2.588$ & $-3.291$ & $-1.573$ & $-1.487$ \\
NLO13(650)           & $0.09$ &  $1.61$ &  $0.59$ &  $1.36$ & $0.64$  & $-2.592$ & $-3.271$ & $-1.538$ & $-1.452$ \\
\hline
NLO19(500)           & $0.10$ &  $1.77$ &  $1.19$ &  $1.52$ & $1.27$  & $-2.649$ & $-3.202$ & $-1.580$ & $-1.467$ \\
NLO19(550)           & $0.10$ &  $1.67$ &  $1.21$ &  $1.42$ & $1.28$  & $-2.640$ & $-3.205$ & $-1.524$ & $-1.407$ \\
NLO19(600)           & $0.09$ &  $1.58$ &  $1.03$ &  $1.34$ & $1.09$  & $-2.632$ & $-3.227$ & $-1.473$ & $-1.362$ \\
NLO19(650)           & $0.10$ &  $1.65$ &  $0.89$ &  $1.40$ & $0.96$  & $-2.620$ & $-3.225$ & $-1.464$ & $-1.365$ \\
\hline 
Expt.                & $0.164(43)$  & $2.347(36)$ & $0.942(36)$ & $2.169(42)$ & $1.081(46)$ & \multicolumn{2}{c|}{$-2.906(5)$} & \multicolumn{2}{c|}{$-1.496(65)$} \\
      \hline
\end{tabular}
\end{center}
\end{table}

One important motivation to study hypernuclear interactions is the possible 
contribution of hyperons to the equation of state (EOS) of neutron matter. 
It is discussed quite extensively that a possible softening of the EOS due to the 
appearance of hyperons would be in contradiction to the observation of neutron stars 
with masses around or larger than two solar masses 
\cite{Ghosh:2022lam,Chatterjee:2015pua,Weissenborn:2011kb,Weissenborn:2011ut,Djapo:2008au}. 
Since the $\Lambda$ is 
expected to be the most relevant hyperon admixture to neutron matter, 
good knowledge of the $\Lambda$-neutron ($\Lambda$n) interaction is highly desired. 
YN data is only available for the $\Lambda$-proton ($\Lambda$p) system (except for $\Lambda$N-$\Sigma$N
transition cross sections). Therefore, usually, isospin symmetry is assumed leading 
to practically equal $\Lambda$n  and $\Lambda$p  interactions. 

However, CSB of the YN interaction has already been noted and discussed in the 
1960's \cite{Dalitz:1964es,Raymund:1964an}. The data at the time showed that the 
separation energies of $^4_\Lambda$He and $^4_\Lambda$H are significantly different. 
Since there is no direct 
contribution to the $\Lambda$N interaction from the Coulomb interaction much smaller 
differences were expected. Dalitz and von~Hippel realized that there is definitely  
a pion-ranged CSB interaction due to the $\Lambda$-$\Sigma^0$ mixing. Their estimate of 
this contribution is still part of many YN  interaction models and also of the 
chiral interactions. More details can be found around Eq.~(\ref{VCSB}) above. 

Since then many investigations have been performed to better understand this 
mechanism and to explain the data. Three main contributions have been identified. 
First, as mentioned above, the Coulomb interaction has been 
shown to contribute only a small fraction of the difference and this contribution 
even goes into the wrong direction \cite{Bodmer:1985km,Nogga:2001ef}. 
In addition there is an effect associated with the $\Lambda$-$\Sigma$ conversion which 
generates a CSB contribution to the kinetic energy due to the mass difference 
of $\Sigma^+$ and $\Sigma^-$. The main part of this contribution is just related 
to the probabilities to find the various charge states of $\Sigma$'s in 
the wave function \cite{Nogga:2001ef}. Since this is a non-observable property 
of the wave function that can be changed by unitary transformations, it should in principle 
be linked to other contributions that accommodate any changes in the $\Sigma$ probabilities. 
In practical calculations, for YN potentials that predict the $A=4$ state
realistically, this contribution 
is of the order of 50~keV \cite{Nogga:2013pwa}. The last contribution is due to the 
CSB interaction which is usually driven by the mechanism of Dalitz and von~Hippel. 

Unfortunately, the actual comparison of calculations to experiment is hindered by the 
large uncertainty of the data. In fact, the central values of the currently 
accepted values \cite{HypernuclearDataBase} of the difference of the 
separation energies of the $0^+$ and $1^+$ states
\begin{eqnarray}
    \label{eq:csba4}
    \Delta E_\Lambda\left( 0^+\right) 
       & = & E_\Lambda\left( 0^+,  ^4_\Lambda \! {\rm He}\right) - E_\Lambda\left( 0^+,  ^4_\Lambda \! {\rm H}\right) = 178\pm 55 \; {\rm keV}\nonumber \\
    \Delta E_\Lambda\left( 1^+\right) 
       & = &  E_\Lambda\left( 1^+, ^4_\Lambda \! {\rm He}\right) - E_\Lambda\left( 1^+,  ^4_\Lambda \! {\rm H}\right) = -139\pm 58 \; {\rm keV} \nonumber \\       
\end{eqnarray}
are quite different from the long-term accepted CSB 
based on Refs.~\cite{Juric:1973zq,CERN-Lyon-Warsaw:1979ifx} of 340~keV (240~keV)  for the $0^+$ ($1^+$) states (see experimental values in Table~\ref{tab:csb}) used up to 2015. 
Not only the overall size changed 
but also the spin dependence is now much more significant. 
Therefore, many studies were hampered by the goal to explain partly misleading data. 

The mechanism of Dalitz and von~Hippel promises a straight forward explanation of 
the CSB. However, many models implementing this contribution 
did not properly describe the data \cite{Nogga:2001ef,Nogga:2013pwa} and studies based on 
different realizations of the LO chiral YN interaction showed a large regulator 
dependence \cite{Gazda:2015qyt,Gazda:2016qva} which could also not be resolved 
by using the complete $\Lambda$-$\Sigma$ conversion potential to fix the 
CSB interaction \cite{Gazda:2016qva}. The reason for this confusing results was 
that all these interactions were missing the two LO CSB contact interactions that are 
necessary to properly renormalize the 1$\pi$ exchange (see Eq.~(\ref{VCSB})).  
Indeed, a proper application of
effective field theory requires these two contact interaction that have to be 
determined from hypernuclear data. In absence of any $\Lambda$n data, 
the two constants can only be determined using hypernuclei  and the CSB of 
the $^4_\Lambda$He/$^4_\Lambda$H nuclei are probably the best observables to fit the 
interactions since the experimental uncertainties for other hypernuclear 
isospin multiplets are even larger and lattice QCD predictions for 
differences of $\Lambda$p and $\Lambda$n scattering will not be 
available in the foreseeable future. Using these two hypernuclei for the 
determination, the interactions can be used to predict the $\Lambda$n
interaction. This was performed in Ref.~\cite{Haidenbauer:2021wld} in 2021 based 
on the NLO13 and NLO19 interactions. Knowing that the experimentally accepted values 
are still changing, the fit was performed for the central values of three 
scenarios. The first one, labeled as CSB1, assumes that the $A=4$ hypernuclei 
have splitting of $233\pm 92$~keV ($-83\pm 94$~keV) for the $0^+$ ($1^+$) states. 
These numbers reflect the experimental situation after new data from J-PARC \cite{J-PARCE13:2015uwb}
and Mainz \cite{A1:2016nfu} became available and was the experimental status 
when the study was performed. The second one, CSB2, is based on the old 
data mentioned above  $340$~keV ($240$~keV) \cite{Juric:1973zq,CERN-Lyon-Warsaw:1979ifx}
used up to 2015. 
This scenario requires a very different spin dependence than CSB1. The third one, CSB3, 
is based on the experimental situation before the Mainz experiment was finished but 
taking the J-PARC result into account: $350\pm 50$~keV ($30\pm 50$~keV).  
The scenario is similar to CSB1 since there is only a spin independent 
shift between them. The overall size is still comparable to the older values. 
Note that the most recent measurement, from the STAR Collaboration  \cite{STAR:2022zrf}, 
suggests again different values for the CSB splittings, 
$160\pm 140 ({\rm stat}) \pm 100 ({\rm syst}) $~keV 
($-160\pm 140 ({\rm stat}) \pm 100 ({\rm syst})$~keV),
albeit with rather large uncertainties. 
The results for the scenario CSB1 are summarized in Table~\ref{tab:csb}. 

The study in Ref.~\cite{Haidenbauer:2021wld}
only took the central values into account and aimed at investigating 
whether predictions for the $\Lambda$n scattering length can be made 
once reliable data for the CSB of the $A=4$ hypernuclei is available. Therefore, the 
experimental uncertainties of the difference scenarios were not considered. 
For CSB1 and CSB3, the determination 
of the LECs of the CSB contact interactions are straight forward. The adjustment to the 
$A=4$ energies only leads to small variations of the predictions. For CSB2, the changes 
were more significant so that a refit of the YN interaction was necessary. In this case,
the final values were obtained in a multi-step process. For CSB1, the fit was done 
for all cutoffs and both interactions, NLO13 and NLO19. Since both interactions differ 
by a significantly different strength of $\Lambda$-$\Sigma$ conversion, one can expect
that the choice of interactions covers a wider range of possible predictions. 
As can be seen in the table, the results for singlet $\Lambda$n scattering length 
only vary by 3\% and the one for the triplet by 15\%. The larger variation for the triplet 
is not surprising since there is more dependence on the realization of the interaction 
even for the standard non-CSB interactions. For CSB2 and CSB3, fits have only been 
performed for a cutoff of $\Lambda=600$~MeV but still for both versions, NLO13 and NLO19. 
As can be seen in the table, also for the other two scenarios the dependence on the realization 
is small compared to the shifts of the $\Lambda$p and  $\Lambda$n scattering 
length. CSB2 is quite different to the other two. First of all, due to the necessity of 
refitting the interaction, the shift compared to the original NLO interactions is unsymmetrical
and even more interesting, the shift is opposite to the other two and also affects the triplet.
CSB2 was mainly included to exemplify that accurate $A=4$ separation energies are relevant 
to determine the $\Lambda$n interaction. Since the experimental values used for this 
scenario are updated, we will not consider it further. CSB1 and CSB3 are qualitatively 
similar to each other. The CSB of the triplet is smaller than the singlet one and the 
$\Lambda$n interaction becomes more attractive in the singlet. The larger size 
of the CSB of the separation energies in CSB3 shows up in a more visible shift of 
the triplet. It has been argued based on pionless EFT calculations and SU(3)$_f$ 
that the CSB in the singlet and triplet can be related to each other \cite{Schafer:2022une}. 
Such considerations could provide arguments that the effect is more pronounced in 
the singlet. Under these assumptions, only one CSB datum is required to determine 
the interaction. Since the excitation energies of the hypernuclei can be measured 
with high accuracy, the determinations are not so much limited by the experimental 
uncertainty anymore. However, this approach requires model assumptions and an EFT that 
has a lower break down scale than chiral EFT. 

As mentioned above, these results do not take the experimental uncertainty 
into account. Also the averaged experimental results have shifted again since this study was completed. Fortunately, at the same time, the uncertainty has been reduced. Therefore, in 
near future, the work of Ref.~\cite{Haidenbauer:2021wld} should be repeated 
using the new data, the updated interactions \cite{Haidenbauer:2023qhf} and taking 
experimental uncertainties into account. One can expect that model-independent 
determinations of the strength of the $\Lambda$n interaction can be obtained 
using the $A=4$ data. 

The application of CSB interactions to $p$-shell hypernuclei will be discussed in 
Section~\ref{sec:csbpshell}. 

\subsection{Estimate of YNN interactions}
\label{subsec:ynnestimate}

Separation energies, at least for light hypernuclei, can be calculated with 
high numerical accuracy. Besides these purely numerical discretization errors or 
uncertainties due to the truncation of the partial wave representation, one can 
expect also uncertainties due to our incomplete knowledge of the hypernuclear interactions. 
In order to better understand these uncertainties, the chiral EFT approach to 
(hyper-)nuclear interactions is very useful. By power counting, higher 
orders of the interaction are expected to contribute less to the result. 
An arbitrary observable $X$ can therefore be expanded as 
\begin{equation}
    X=X_{\rm ref}\ \sum_{k=0}^{\infty} c_k Q^k  \ .
\end{equation}
$X_{\rm ref}$ is dimensionful and sets a scale for the observable that is considered. Often this value 
is obtained based on the LO result, experiment or any other estimate. $c_k$ are dimensionless expansion coefficients. 
Assuming a valid power countering, $c_k$ should be of the order of $1$. $Q$ is the expansion parameter. For nuclear 
bound states and chiral EFT, it is usually given by $M_\pi^{\rm eff}/\Lambda_b$. $\Lambda_b$ is the breakdown scale of 
chiral EFT of the order of $600$~MeV and $M_\pi^{\rm eff}\approx 200$~MeV captures typical momenta in 
nuclei and the expansion in the pion mass \cite{Epelbaum:2019wvf}. The actual value of $Q$ can be determined using the sizes 
of the coefficients $c_k$. From a number of calculations using different orders of the interaction, 
the first few coefficients $c_k$ ($k \le K$) can be determined. 
Assuming that the probability distribution for all other $c_k$, $k>K$ is equal, one can obtain a probability 
distribution for
\begin{equation}
    \delta X_K  = X_{\rm ref}  \ \sum_{k=K+1}^{\infty} c_k Q^k  \ \mbox{ with } \ X = X_K + \delta X_K \mbox{ where } X_K  = X_{\rm ref}  \ \sum_{k=0}^{K} c_k Q^k \ . 
\end{equation}
Based on this probability, it is straightforward to obtain an 
uncertainty estimate for the $K$-th order prediction $X_K$
of the observable within a Bayesian analysis \cite{Melendez:2017phj,Melendez:2019izc}.  

\begin{figure}
    \centering
    \includegraphics[width=0.7\linewidth]{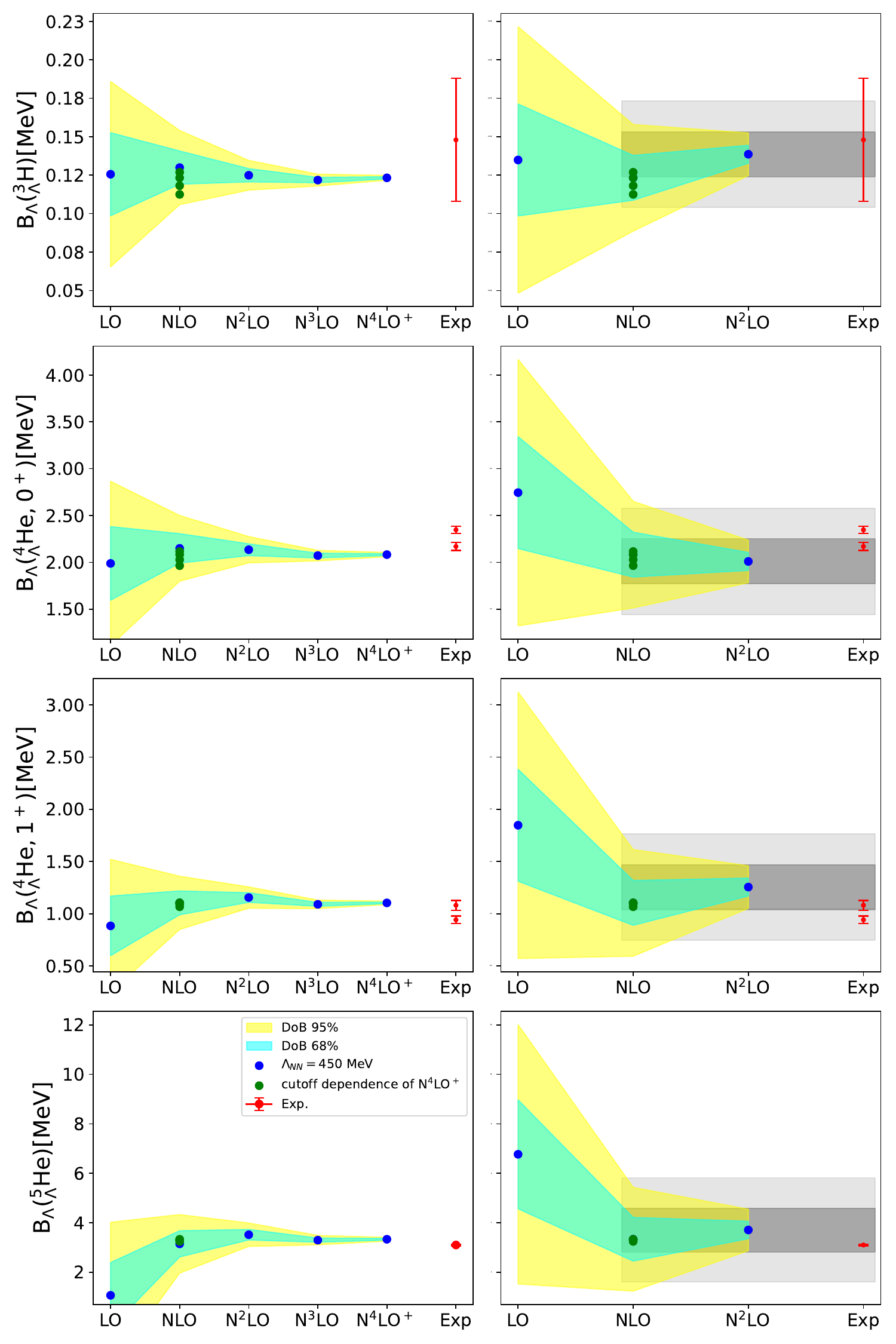}
    \caption{Predictions for the separation energies of $^3_\Lambda$H, $^4_\Lambda$He($0^+$), $^4_\Lambda$He($1^+$), and  $^5_\Lambda$He (from top to bottom) based of different orders of the SMS YN interactions for $\Lambda_{YN}=550$~MeV and using the SMS NN interactions for $\Lambda_{NN}=450$~MeV (blue circles). On the left hand side, the order of the NN interaction is increased
    using the YN interaction at NLO. On the right hand side, the order of the YN interaction is increased keeping the order of the NN interaction at N4LO$^+$. Green circles show results 
    for $\Lambda_{NN} = 400$-$550$~MeV. The red crosses show the experimental values including the error bars. For $A=4$, the two experimental values correspond to 
    $^4_\Lambda$H and $^4_\Lambda$He. The yellow and cyan bands show the 95\% and 68\% degree of believe intervals (DoBs) from the Bayesian analysis. The light and dark 
    grey band are the NLO DoBs plotted around the N2LO result. Energies are in MeV.  }
    \label{fig:uncertainty}
\end{figure}

In Ref.~\cite{Le:2023bfj}, this was done for hypernuclei with $A\le5$. 
The calculations used the SMS YN interactions up to N2LO omitting 
YNN interactions. Although the highest order calculations 
were not complete because of the missing YNN forces, it was possible 
to reliably estimate the size of the N2LO contribution which 
represents the uncertainty at order NLO. 
Also the expected uncertainty 
once the calculations include a complete set of the leading 
YNN forces has been estimated. Fig.~\ref{fig:uncertainty} shows 
the results at different order of the SMS YN interactions 
for a high order NN interaction. Obviously, the uncertainty shrinks for 
higher orders. The grey bands show the NLO uncertainty of the N2LO result where the YNN forces are still omitted. The experimental values 
are within the estimated uncertainties. The NLO uncertainty, 
e.g. given by the $68$\% degree of believe (DoB) interval, 
also provides an estimate of the YNN force contribution
which is 20~keV, 240~keV, and 900~keV for $^3_\Lambda$H, 
$^4_\Lambda$He/$^4_\Lambda$H, and $^5_\Lambda$He, respectively. 
This clearly shows the importance to include YNN interactions 
in the calculations for $A\ge 4$. First calculations will be discussed in Sec.~\ref{sec:ncsmynn}. 

The analysis is also instructive to better understand how different realizations 
of the NN interaction change the $\Lambda$ separation energies. Based on phenomenological 
calculations, the NN force dependence was estimated  to be 120~keV for $A=4$ \cite{Nogga:2001ef}. Similar calculations for $^3_\Lambda$H indicated that the 
NN force dependence is only $10$~keV for this hypernucleus. This estimate has recently 
been criticized \cite{Htun:2021jnu,Gazda:2022fte}. Based on a series of N2LO NN
interactions called NNLO$_{\rm sim}$ \cite{Carlsson:2015vda}, 
the $\Lambda$ separation energies for $A=3-5$ hypernuclei were calculated. 
The results suggest that the NN force dependence is much larger than previously 
estimated: $100$~keV for $^3_\Lambda$H and $250$~keV $^4_\Lambda$He/$^4_\Lambda$H. 
The dependence is systematically related to the NN data taken into account for 
fitting the NN interaction and to the cutoff. Therefore, a statistical interpretation 
that reduces the nominal values of the uncertainty is not possible. 
On the other hand, standard NN interactions (also chiral ones at high order, say up to
N3LO or N4LO) 
describe the NN data up to the pion production threshold with high accuracy. 
This is not the case for some variants of the NNLO$_{\rm sim}$ interactions. 
It turns out that restricting the interactions to the ones that describe the NN 
data up to higher energies, the NN force dependence shrinks to $50$ and $130$~keV. 
For $A=4$, this is in line with previous estimates. Still, the uncertainty 
for $A=3$ is uncomfortably large since it becomes comparable to the experimental 
uncertainty and this would therefore cast doubt on the use of the $^3_\Lambda$H binding 
energy for constraining the singlet $\Lambda$N scattering length, as usually done by
the J\"ulich-Bonn group.
In order to obtain a similar estimate of the uncertainty due to the NN 
interaction, we also studied explicitly the convergence with respect to the 
order of the NN interactions and due to cutoff variations of the NN interactions 
on the left hand side of Fig.~\ref{fig:uncertainty}. The uncertainties due to missing 
higher orders of the NN interaction are generally smaller than the ones due to YN 
interactions as can be expected because part of the variation will be similar 
in the core nucleus and the hypernucleus and will drop out for the separation 
energies. Due to the larger order, the contributions of missing higher 
orders are tiny as can be seen for the DoB intervals at order 
N4LO$^+$. Interestingly, the cutoff variation is much larger. Clearly, the missing 
YNN forces need to be adjusted consistently with the NN and YN force. This also 
leads to an absorption of cutoff dependencies. Since YNN forces are still missing 
in these calculations, the cutoff dependence can be expected to be of order N2LO
and the results are consistent with this expectation as can be seen when comparing the 
NLO DoB intervals with the results for the different cutoffs. 
The NN interaction dependence is therefore linked  to interactions 
involving the hyperon in this case. It is therefore already included in the uncertainty 
estimates shown on the right hand side of Fig.~\ref{fig:uncertainty}. In the future,
a similar analysis of the uncertainty needs to be done including YNN interactions. 
This is still work in progress.

    
    \newcommand{\stateAmIIINNY}{\begin{tikzpicture}[baseline={([yshift=-0.2ex]current bounding box.center)},scale=0.65]
                \filldraw[color=black, ultra thick, fill=gray ]  (0.,0.) circle(0.19cm);
                \filldraw[red]  (-0.4,-0.45)  circle (0.7mm) ;
                \filldraw[black] (-0.74,0.24)   circle(0.55mm); 
                \filldraw[black] (-0.74,-0.24)  circle(0.55mm); 
                \draw[baseline,thick, -]  (-0.74,-0.24) -- (-0.74,0.24);
                \draw[baseline,thick, -]  (-0.74,0.) -- (-0.2,0.0);
                \draw[baseline,thick, -]  (-0.4,-0.38) -- (-0.4,0.0);
                \end{tikzpicture}}

\newcommand{\stateAmIY}{\begin{tikzpicture}[baseline={([yshift=-.5ex]current bounding box.center)},scale=0.6]
                \filldraw[color=black, ultra thick, fill=gray ]  (0.,0.) circle(0.22cm);
                \filldraw[red]  (0.6,0.) circle (0.7mm) ;
                \draw[baseline,thick, -]  (0.22,0.) -- (0.53,0.);
                \end{tikzpicture}}
                
\newcommand{\stateAmI}{\begin{tikzpicture}[baseline={([yshift=-.5ex]current bounding box.center)},scale=0.6]
                \filldraw[color=black, ultra thick, fill=gray ]  (0.,0.) circle(0.22cm);
                \end{tikzpicture}}

\newcommand{\stateAmIIINN}{\begin{tikzpicture}[baseline={([yshift=-0.5ex]current bounding box.center)},scale=0.6]
                \filldraw[color=black, ultra thick, fill=gray ]  (0.,0.) circle(0.19cm);
                \filldraw[black] (-0.74,0.24)   circle(0.55mm); 
                \filldraw[black] (-0.74,-0.24)  circle(0.55mm); 
                \draw[baseline,thick, -]  (-0.74,-0.24) -- (-0.74,0.24);
                \draw[baseline,thick, -]  (-0.74,0.) -- (-0.2,0.0);
                \end{tikzpicture}}

\newcommand{\stateAmIINY}{\begin{tikzpicture}[baseline={([yshift=-0.5ex]current bounding box.center)},scale=0.6]
                \filldraw[color=black, ultra thick, fill=gray ]  (0.,0.) circle(0.20cm);
                \filldraw[red] (-0.7,0.24)   circle(0.7mm); 
                \filldraw[black] (-0.7,-0.24)  circle(0.55mm); 
                \draw[baseline,thick, -]  (-0.7,-0.24) -- (-0.7,0.17);
                \draw[baseline,thick, -]  (-0.7,0.) -- (-0.2,0.0);
                \end{tikzpicture}}

\newcommand{\stateAmIIINNN}{\begin{tikzpicture}[baseline={([yshift=-0.5ex]current bounding box.center)},scale=0.6]
                \filldraw[color=black, ultra thick, fill=gray ]  (0.,0.) circle(0.19cm);
                \filldraw[black] (-0.72,0.31)   circle(0.55mm); 
                \filldraw[black] (-0.82,0.28)   circle(0.55mm); 
                \filldraw[black] (-0.75,0.20)  circle(0.55mm); 
                \draw[baseline,thick, -]  (-0.74,0.24) -- (-0.2,0.06);
                \end{tikzpicture}}

\newcommand{\stateAmIVNNNY}{\begin{tikzpicture}[baseline={([yshift=-0.5ex]current bounding box.center)},scale=0.6]
                \filldraw[color=black, ultra thick, fill=gray ]  (0.,0.) circle(0.19cm);
                \filldraw[black] (-0.72,0.31)   circle(0.55mm); 
                \filldraw[black] (-0.82,0.28)   circle(0.55mm); 
                \filldraw[black] (-0.75,0.20)  circle(0.55mm); 
                \filldraw[red] (-0.6,-0.2)  circle(0.55mm); 
                \draw[baseline,thick, -]  (-0.74,0.24) -- (-0.2,0.06);
                \draw[baseline,thick, -]  (-0.58,-0.165)  -- (-0.5,0.15);
                \end{tikzpicture} }               

\newcommand{\stateNNYAmIII}{\begin{tikzpicture}[baseline={([yshift=-0.5ex]current bounding box.center)},scale=0.6]
                \filldraw[color=black, ultra thick, fill=gray ]  (0.,0.) circle(0.19cm);
                \filldraw[black] (-0.6,0.30)   circle(0.55mm); 
                \filldraw[black] (-1.0,0.20)  circle(0.55mm); 
                \filldraw[red] (-0.8,-0.1)  circle(0.55mm); 
                \draw[baseline,thick, -]  (-0.6,0.30) -- (-1.0,0.20);
                \draw[baseline,thick, -]  (-0.8,0.15) -- (-0.2,0.04);
                \draw[baseline,thick, -]  (-0.8,0.25)  -- (-0.8,-0.05);
                \end{tikzpicture}}

\newcommand{\stateAmIIYY}{\begin{tikzpicture}[baseline={([yshift=-0.5ex]current bounding box.center)},scale=0.6]
                \filldraw[color=black, ultra thick, fill=gray ]  (0.,0.) circle(0.20cm);
                \filldraw[red] (-0.7,0.24)   circle(0.7mm) node[anchor=base, above=0.07em, text=black] {\scriptsize{$Y_1$}} ;
                \filldraw[red] (-0.7,-0.24)  circle(0.7mm) node[anchor=base, below=0.07em, text=black] {\scriptsize{$Y_2$}} ;
                \draw[baseline,thick, -]  (-0.7,-0.19) -- (-0.7,0.17);
                \draw[baseline,thick, -]  (-0.7,0.) -- (-0.2,0.0);
                \end{tikzpicture}}                

\newcommand{\stateAmIXi}{\begin{tikzpicture}[baseline={([yshift=-.5ex]current bounding box.center)},scale=0.6]
                \filldraw[color=black, ultra thick, fill=gray ]  (0.,0.) circle(0.22cm);
                \filldraw[red]  (0.6,0.) circle (0.7mm)  node[anchor=base, above=0.02em, text=black] {\scriptsize{$\Xi$}} ;
                \filldraw[red]  (0.6,0.) circle (0.7mm)  node[anchor=base, below=0.02em, text=white] {\scriptsize{$\Xi$}} ;
                \draw[baseline,thick, -]  (0.22,0.) -- (0.53,0.);
                \end{tikzpicture}}

\newcommand{\stateAmIIINYIYII}{\begin{tikzpicture}[baseline={([yshift=-0.5ex]current bounding box.center)},scale=0.55]
                \filldraw[color=black, ultra thick, fill=gray ]  (0.,0.) circle(0.20cm);
                 \filldraw[red] (-0.8,0.24)   circle(0.7mm) node[anchor=base, above=-0.1em, text=black] {\scriptsize{$Y_1$}} ;
                \filldraw[black] (-0.8,-0.24)  circle(0.55mm); 
                  \filldraw[red] (-0.37,-0.5)   circle(0.7mm) node[anchor=base, below=-0.1em, text=black] {\scriptsize{$Y_2$}} ;
                \draw[baseline,thick, -]  (-0.8,-0.24) -- (-0.8,0.17);
                \draw[baseline,thick, -]  (-0.8,0.) -- (-0.2,0.0);
                  \draw[baseline,thick, -]  (-0.37,-0.45) -- (-0.37,0.);
                \end{tikzpicture} }             

\newcommand{\stateAmIIINNXi}{\begin{tikzpicture}[baseline={([yshift=0.6ex]current bounding box.center)},scale=0.65]
                \filldraw[color=black, ultra thick, fill=gray ]  (0.,0.) circle(0.19cm);
                \filldraw[black] (-0.74,0.24)   circle(0.55mm); 
                \filldraw[black] (-0.74,-0.24)  circle(0.55mm); 
                \draw[baseline,thick, -]  (-0.74,-0.24) -- (-0.74,0.24);
                \draw[baseline,thick, -]  (-0.74,0.) -- (-0.2,0.0);
                \draw[baseline,thick, -]  (-0.4,-0.38) -- (-0.4,0.0);
                \filldraw[red]  (-0.4,-0.45) circle (0.7mm)  node[anchor=base, right=0.15em, yshift=-0.2em, text=black] {\scriptsize{$\Xi$}} ;
                \end{tikzpicture}}                

\newcommand{\stateAmIVNNYY}{\begin{tikzpicture}[baseline={([yshift=0.6ex]current bounding box.center)},scale=0.65]
                \filldraw[color=black, ultra thick, fill=gray ]  (0.,0.) circle(0.19cm);
                \filldraw[black] (-0.74,0.24)   circle(0.55mm); 
                \filldraw[black] (-0.74,-0.24)  circle(0.55mm); 
                \draw[baseline,thick, -]  (-0.74,-0.24) -- (-0.74,0.24);
                \draw[baseline,thick, -]  (-0.74,0.) -- (-0.2,0.0);
                \draw[baseline,thick, -]  (-0.4,-0.45) -- (-0.4,0.0);
                \draw[baseline,thick, -]  (-0.23,-0.45) -- (-0.57,-0.45);
                \filldraw[red]  (-0.64,-0.45) circle (0.7mm)  node[anchor=base, right=0.7em, yshift=-0.2em, text=black] {\scriptsize{$Y_2$}} ;
                \filldraw[red]  (-0.16,-0.45) circle (0.7mm)  node[anchor=base, left=0.6em, yshift=-0.2em, text=black] {\scriptsize{$Y_1$}} ;
                \end{tikzpicture}}

\newcommand{\stateYIAmIIINYII}{\begin{tikzpicture}[baseline={([yshift=-0.5ex]current bounding box.center)},scale=0.65]
                \filldraw[color=black, ultra thick, fill=gray ]  (0.,0.) circle(0.20cm);
                 \filldraw[red] (-0.8,0.24)   circle(0.7mm) node[anchor=base, above=0.07em, text=black] {\scriptsize{$Y_2$}} ;
                \filldraw[black] (-0.8,-0.24)  circle(0.55mm); 
                  \filldraw[red] (-0.37,-0.5)   circle(0.7mm) node[anchor=base, below=0.07em, text=black] {\scriptsize{$Y_1$}} ;
                      \filldraw[white] (-0.37,0.5)   circle(0.7mm) node[anchor=base, above=0.07em, text=white] {\scriptsize{$Y_1$}} ;
                \draw[baseline,thick, -]  (-0.8,-0.24) -- (-0.8,0.17);
                \draw[baseline,thick, -]  (-0.8,0.) -- (-0.2,0.0);
                  \draw[baseline,thick, -]  (-0.37,-0.45) -- (-0.37,0.);
                \end{tikzpicture}}               

\newcommand{\stateYIIAmIIINYI}{\begin{tikzpicture}[baseline={([yshift=-0.5ex]current bounding box.center)},scale=0.65]
                \filldraw[color=black, ultra thick, fill=gray ]  (0.,0.) circle(0.20cm);
                 \filldraw[red] (-0.8,0.24)   circle(0.7mm) node[anchor=base, above=0.07em, text=black] {\scriptsize{$Y_1$}} ;
                \filldraw[black] (-0.8,-0.24)  circle(0.55mm); 
                  \filldraw[red] (-0.37,-0.5)   circle(0.7mm) node[anchor=base, below=0.07em, text=black] {\scriptsize{$Y_2$}} ;
                   \filldraw[white] (-0.37,0.5)   circle(0.7mm) node[anchor=base, above=0.07em, text=white] {\scriptsize{$Y_2$}} ;
                \draw[baseline,thick, -]  (-0.8,-0.24) -- (-0.8,0.17);
                \draw[baseline,thick, -]  (-0.8,0.) -- (-0.2,0.0);
                  \draw[baseline,thick, -]  (-0.37,-0.45) -- (-0.37,0.);
                \end{tikzpicture}}             

\section{No-core shell model approach}\label{sec:fourth}
The no-core shell model (NCSM) (see \cite{Barrett:2013nh}) is one of the ab-initio approaches that is suitable 
for predicting properties of nuclei and hypernuclei beyond $A=4$. 
For this method, the Schr\"odinger equation is solved using a harmonic oscillator (HO) basis. In contrast to the shell-model, the approach does not single out an inert core and valence nucleons but assumes that all nucleons and hyperons are active. Over the years 
several variants have been developed. Most calculations are performed within the so-called 
$m$-scheme where the basis states are build from single particle HO states. This allows one to easily antisymmetrize the states. However, the model spaces used for the basis states 
are usually large since the center-of-mass (CM) coordinate cannot be separated and because 
the basis states are not eigenstates of total angular momentum and isospin. The original 
NCSM uses a complete set of basis states up to a total HO excitation $N_{\rm tot}$. 
The completeness of the model spaces insures that the CM motion can be exactly separated.
The importance truncated NCSM (IT-NCSM) \cite{Roth:2007sv} does not included 
a complete set of states up to a definite HO excitation.  Instead, the states are preselected using a perturbative estimate of their importance. This reduces the number of basis states significantly and allows to extend NCSM calculations to medium mass nuclei.  

Most of the calculations that will be presented here, have been obtained using the Jacobi NCSM (J-NCSM). In this approach, the basis states are given in terms of relative Jacobi coordinates. By construction, this separates out the CM coordinate and also allows one 
to define basis states with well-defined total angular momentum and isospin. Therefore, 
the number of basis states for a given maximal HO excitation is much smaller than 
in the $m$-scheme. The drawback of this variant is that it is difficult to define antisymmetric states. This has only been achieved using numerical 
diagonalization as described below. The J-NCSM was already applied in \cite{Navratil:1999pw} to $s$-shell nuclei but has only recently been extended to $p$-shell nuclei \cite{Liebig:2015kwa}. 

The direct application of the NCSM is only suitable for bound state calculations because of its basis states
are square-integrable. Nevertheless, several extensions have been devised to gain insight into
scattering processes \cite{Baroni:2012su,Shirokov:2016thl,Kruppa:2025ijb}. Such approaches can also help to improve the numerical accuracy for calculations of bound states close to threshold.  
The first application of the NCSM to hypernuclei  already compared 
results of the the J-NCSM to the $m$-scheme \cite{Wirth:2014apa}. The first applications 
of the J-NCSM concentrated on the CSB of the $A=4$ system  \cite{Gazda:2015qyt,Gazda:2016qva}. 
With increasing model space sizes, it also became possible to 
study the barely bound $^3_\Lambda$H leading to a study of the lifetime of this hypernucleus 
\cite{Perez-Obiol:2020qjy}. Finally, the J-NCSM has been applied to $s$-shell hypernuclei  
for an analysis of theoretical uncertainties \cite{Htun:2021jnu,Gazda:2022fte}. 
In the $m$-scheme, the NCSM extended calculations to $p$-shell hypernuclei \cite{Wirth:2017bpw}. 

Since the HO states are not well suited for representing correlations of baryons induced 
by the short-range repulsion which is typical for baryonic interactions, usually 
interactions are unitarily transformed to soften the interactions. The most commonly used 
approach for such a transformation is the similarity renormalization group (SRG) \cite{Bogner:2006pc}. The unitary transformation is obtained by an evolution 
of the interaction that starts from the bare BB or 3B interaction. 
State-of-the-art nuclear and hypernuclear calculations 
perform the transformation up to the three-baryon level. The size of the 
missing higher-body interactions is then estimated using the dependence on the 
SRG evolution parameter. It became quickly clear that the SRG evolution up to the two-baryon level 
leads to an unexpectedly large SRG parameter dependence for hypernuclei \cite{Wirth:2017bpw}. 
Therefore, realistic calculations can only be done using special choices for the 
SRG parameter \cite{Le:2020zdu}. Fortunately, extending the calculations to the 3B level 
reduced the SRG parameter dependence drastically
and leads to results that agree with bare calculations \cite{Wirth:2019cpp,Le:2022ikc} 
for a wider range of the SRG parameters. 

The J-NCSM was first extended to $p$-shell hypernuclei in Refs.~\cite{Le:2019gjp,Le:2020zdu} and also used for predictions 
of $S=-2$ hypernuclei \cite{Le:2021wwz,Le:2021gxa}. The details of the approach and the applications are discussed in the 
following sections.

 \subsection{Jacobi NCSM for hypernuclei}
 \label{sec:jncsm}

We follow here the formulation of the J-NCSM of Refs.~\cite{Le:2020zdu,Le:2021wwz}. The Hamilton operator 
of an $S=-1$ hypernucleus consisting of $A-1$ nucleons and the $\Lambda$ or $\Sigma$ hyperon can be written in the form  
\begin{eqnarray} \label{eq:hamiltonian}
H &   = &   \sum_{i < j=1}^{A-1} \Big( \frac{2 {\vec p}^{2}_{ij}}{M(t_{\rm Y})} + V^{\rm NN}_{ij} \Big) 
  + \sum_{i=1}^{A-1} \Big( \frac{m_N + m(t_{\rm Y})}{M(t_{\rm Y})} \frac{{\vec p}^2_{i\rm Y}}{2\mu_{\rm NY}}  +  V^{\rm YN}_{iY}
  + \frac{1}{A-1} \big(m(t_{\rm Y}) - m_{\Lambda}\big) \Big) \nonumber \\
  & & + \sum_{i < j < k=1}^{A-1} V^{\rm 3N}_{ijk} 
  +  \sum_{i < j =1}^{A-1} V^{\rm YNN}_{ijY}  .
\end{eqnarray} 
$m_N$, $m(t_Y)$ and $\mu_{NY}$ are  nucleon-, hyperon-, and their reduced masses, respectively. $t_{\rm Y} =0$
and $t_{\rm Y} =1$ distinguish $\Lambda$ and $\Sigma$
masses. The relative kinetic energy is included in terms of pair terms. We include NN ($V^{\rm NN}$), 
3N ($V^{\rm 3N}$), YN ($V^{\rm YN}$)and YNN ($V^{\rm YNN}$) interactions and neglect higher-body forces. 
Because of $\Lambda$-$\Sigma$ conversion, the total rest mass of the system $M(t_Y)=(A-1)m_N + m(t_Y)$ also depends on the isospin of the hyperon. The relative pair momenta are defined as in Eq.~(\ref{eq:jmom})
\begin{equation}
    {\vec p}_{ij} = \frac{1}{2}({\vec k}_i - {\vec k}_j),\hspace{0.5cm}
    {\vec p}_{iY } = \frac{m(t_Y)}{m_N + m(t_Y)} {\vec k}_{i}  - \frac{m_N}{m_N + m(t_Y)} {\vec k}_Y .
\end{equation}  

\subsubsection{\texorpdfstring{$S=-1$}{S=-1} basis states}
The appropriate basis states for the $S=-1$ hypernucleus  are  denoted as   
 \begin{eqnarray} \label{eq:hbasis}
& &  \big |\alpha^{* (Y)}(\mathcal{N}J T)\big \rangle  =  | \mathcal{N}{J}{T}, \alpha_{A-1}
\mathcal{N}_{A-1} J_{A-1} T_{A-1} \, n_{Y} (l_Y s_Y) I_Y t_Y ;  (J_{A-1} I_Y){J}, (T_{A-1} t_Y) {T} \rangle
 \equiv | \stateAmIY  \rangle . 
\end{eqnarray}
We use $\alpha$ as a index for the $A$-body state. The basis state has a well defined 
total HO excitation $\mathcal{N}$, angular momentum $J$ and isospin $T$.
The $*(Y)$ indicates that the hyperon is coupled out of the $A$-body system. We assume 
here that we have already defined antisymmetrized states 
$|\alpha_{A-1}(\mathcal{N}_{A-1}J_{A-1} T_{A-1})\big \rangle$  of the $A-1$-nucleon  subsystem as described in \cite{Liebig:2015kwa} with total HO quantum 
number $\mathcal{N}_{A-1}$, angular momentum $J_{A-1}$, and isospin $ T_{A-1}$. 
The HO oscillator state describes the relative motion of the hyperon 
with respect to the $A-1$ cluster with HO quantum number $n_Y$, its orbital, spin  
and total angular momentum $l_Y$, $s_Y$ and $I_Y$ and its isospin $t_Y$. Note that $\mathcal{N}=\mathcal{N}_{A-1} + 2n_Y+l_Y$. The couplings of $J_{A-1}$, $I_Y$,  
$T_{A-1}$ and $t_Y$ to the total angular momentum and isospin is also indicated. 
We drop the third component of $J$ and $T$ since 
the basis states themselves do not depend on them. The graphical representation 
on the right hand side shows the coupled out hyperon in red and the antisymmetrized 
$A-1$-nucleon state as grey blob. To complete the definition 
of the states, we assume that the HO state of the hyperon refers to 
Jacobi coordinates that are directed to towards the hyperon. 

\subsubsection{Separation of two- and three-baryon clusters}

The basis states are not suitable for the implementation of the Hamilton operator. 
In Eq.~(\ref{eq:hamiltonian}), the terms are already arranged in a way that indicates 
the necessary steps for the application. The first part is the nuclear interaction 
and kinetic energy acting in a pair of nucleons. For this part, we require 
transition coefficients defined for the $A-1$ subsystem that couple out a pair of nucleons. 
The corresponding basis states are defined as 

\begin{eqnarray}\label{eq:NNsingleout}
& & \big |\big(\alpha^{*(2)}\big)^{*(Y)} ({\mathcal{{\tilde N}}} {\tilde J} {\tilde T})  \big \rangle  =  \big | {\mathcal{{\tilde N}}} {\tilde J} {\tilde T}, \alpha^{*(2)}_{A-1} \mathcal{N}^{*(2)}_{A-1} J^{*(2)}_{A-1} T^{*(2)}_{A-1} \, \, { \tilde n}_{Y} ({\tilde  l}_Y {\tilde  s}_Y) {\tilde  I}_Y {\tilde t}_Y ;  
({J}^{*(2)}_{A-1} {\tilde I}_Y){\tilde J}, ({T}^{*(2)}_{A-1} {\tilde t_Y}) {\tilde T} \big \rangle  \equiv \big| \,  \stateAmIIINNY  \big\rangle.
\end{eqnarray}
The hyperon motion is described by the same quantum numbers ${\tilde n}_{Y}$, ${ \tilde l}_Y$, ${ \tilde s}_Y$,${ \tilde I}_Y$, and ${\tilde t}_Y$ as in Eq.~(\ref{eq:hbasis}). For the $A-1$ subcluster, states are used 
that couple out two nucleons as defined in Ref.~\cite{Liebig:2015kwa}. Here, we identify 
the states by their total HO quantum number, angular momentum and isospin 
$\mathcal{N}^{*(2)}_{A-1}$, $J^{*(2)}_{A-1}$, $T^{*(2)}_{A-1}$ 
and an index $\alpha^{*(2)}_{A-1}$ that summarizes the quantum numbers of the NN subsystem 
and the $A-3$ subcluster. Eq.~(\ref{eq:NNsingleout}) also indicates the coupling 
of the different angular momenta and isospin. On the right hand side, we introduce a
graphical representation of the states. Important for the application of the Hamilton operator 
are the transition coefficients between the states of Eqs.~(\ref{eq:hbasis}) and (\ref{eq:NNsingleout})       
              \begin{eqnarray} \label{eq:overlapNN}
 \langle \alpha^{*(Y)} | \big(\alpha^{*(2)}\big)^{*(Y)}\rangle  & = & 
\langle \stateAmIY | \stateAmIIINNY \rangle 
  = \delta_{{ n}_{Y} {\tilde n}_{Y}} \delta_{l_Y {\tilde  l}_Y} \delta_{{ s}_Y{ \tilde s}_Y} \delta_{{ I}_Y{ \tilde I}_Y} \delta_{{t}_Y {\tilde t}_Y}  \delta_{J {\tilde J}} \delta_{T {\tilde T}} \ 
 \langle \stateAmI
| \stateAmIIINN  \big\rangle_{A-1} 
\end{eqnarray}
where  $ \langle \stateAmI
| \stateAmIIINN  \big\rangle_{A-1} $ represent the transition coefficients 
for coupling out an NN cluster from the $A-1$ nucleon system \cite{Liebig:2015kwa}. 

For the application of an operator involving an YN pair, like the second term in Eq.~(\ref{eq:hamiltonian}), we need to couple out a YN pair. The corresponding states 
are defined as 
\begin{eqnarray}
\label{eq:YNsingleout}
|\alpha^{*(YN)} (\mathcal{\tilde N} \tilde J \tilde T) \rangle  & = &  | \mathcal{\tilde N} \tilde J  \tilde T, \tilde n_{\rm YN}  ((\tilde l_{YN} S_{\rm YN}) 
 \tilde J_{\rm YN} (\tilde t_Y \tilde t_N) \tilde T_{YN} \  \tilde n_{\lambda} \tilde \lambda \  \alpha_{A-2} \mathcal{N}_{A-2}  J_{A-2} T_{A-2}  ;   (\tilde J_{YN}(\tilde \lambda J_{A-2})\tilde I_{\lambda})\tilde J, ( \tilde T_{\rm YN}T_{A-2}) \tilde T \rangle  
 \nonumber \\ & \equiv & \big| \, \stateAmIINY \big\rangle  .
\end{eqnarray}
The state of the YN pair is defined by its HO quantum number 
$\tilde n_{\rm YN}$, the orbital angular momentum and the spin of the baryons 
$\tilde l_{YN}$,  $S_{\rm YN}$ that couple to the 
total two-baryon angular momentum $\tilde J_{\rm YN}$. The isospin of the hyperon 
$\tilde t_Y$ and nucleon $\tilde t_N=1/2$ couple to the YN isospin $\tilde T_{YN}$. 
The relative motion of the YN pair and the $A-2$ subcluster is given by its HO quantum 
number $\tilde n_\lambda$ and orbital angular momentum $\tilde \lambda$. The $A-2$
state is again given by an index $\alpha_{A-2}$ labeling the state within a block 
for fixed ${\cal N}$, $J$, and $T$. As before, we also include the angular momentum and isospin couplings and a graphical representation in Eq.~(\ref{eq:YNsingleout}). 
Note that coordinates related to the HO state $\tilde n_\lambda$, $\tilde \lambda$ 
point towards the $A-2$ subcluster. The calculation of the transition coefficients 
$\langle \alpha^{*(Y)} | \alpha^{*(YN)}\rangle =  \langle \stateAmIY | \stateAmIINY \rangle$ was outlined in Ref.~\cite{Le:2020zdu}. The calculation of these  coefficients relies 
on Talmi-Moshinsky (TM) brackets \cite{Moshinsky:1959qbh,Trlifaj:1972zz,Kamuntavicius:2001pf}.
Because of the  independence of the TM brackets of the HO frequency, also the transition 
coefficients and CFPs are independent of the HO frequency. The matrix elements have to be 
evaluated numerically. The values are publicly available via a python package 
for downloading the matrix elements in form of HDF5 files \cite{cfpdownload}.

The third part of Eq.~(\ref{eq:hamiltonian}) involves the 3N interaction. For 
this part, we need to couple out a 3N cluster from our sets of states. In order to 
match the definition of the J-NCSM for ordinary nuclei \cite{Liebig:2015kwa}, 
we define the states as follows 
\begin{eqnarray}
\label{eq:3Nsingleout}
|{\alpha^{*(A-4)}}^{*(Y)} (\mathcal{\tilde N} \tilde J \tilde T) \rangle  & = &  | \mathcal{\tilde N} \tilde J  \tilde T,   \alpha^{*(A-4)}_{A-1} \mathcal{N}^{*(A-4)}_{A-1}   J^{*(A-4)}_{A-1}  T^{*(A-4)}_{A-1} 
\, { \tilde n}_{Y} ({\tilde  l}_Y {\tilde  s}_Y) {\tilde  I}_Y {\tilde t}_Y ;     (J^{*(A-4)}_{A-1} \tilde I_{\lambda})\tilde J, ( T^{*(A-4)}_{A-1}  {\tilde t}_Y ) \tilde T \rangle  
 \nonumber \\ & \equiv & \big| \, \stateAmIVNNNY \big\rangle  .
\end{eqnarray}
The notation $*(A-4)$ to couple out a cluster of $A-4$ nucleons leads to a 3N rest cluster and 
also defines that the $(3{\rm N})-(A-4)$ coordinate points towards the $A-4$ system.  As before the hyperon 
is the spectator particle. The states $\alpha^{*(A-4)}_{A-1}$ index the possible 
combinations of $A=3$ clusters in the $A-1$ nucleon system with a rest cluster with $A-4$ systems. 
The total angular momentum $ J^{*(A-4)}_{A-1} $ and isospin $ T^{*(A-4)}_{A-1}$ of the $A-1$ nucleon 
system couples with the spectator angular momentum and isospin as indicate to the total 
angular momentum and isospin. The overlap with the basis states is then obtained by 
\begin{eqnarray} \label{eq:overlap3N}
 \langle \alpha^{*(Y)} | \big(\alpha^{*(A-4)}\big)^{*(Y)}\rangle  & = & 
\langle \stateAmIY | \stateAmIVNNNY  \rangle 
  = \delta_{{ n}_{Y} {\tilde n}_{Y}} \delta_{l_Y {\tilde  l}_Y} \delta_{{ s}_Y{ \tilde s}_Y} \delta_{{ I}_Y{ \tilde I}_Y} \delta_{{t}_Y {\tilde t}_Y}  \delta_{J {\tilde J}} \delta_{T {\tilde T}} \ 
 \langle \stateAmI
| \stateAmIIINNN \big\rangle_{A-1} \ . 
\end{eqnarray}
The transition coefficients $ \langle \stateAmI | \stateAmIIINNN \big\rangle_{A-1} $ have been introduced in 
\cite{Liebig:2015kwa} and can also be obtained from \cite{cfpdownload}. 

Finally, for the last part of  Eq.~(\ref{eq:hamiltonian}), we require coefficients that couple 
out a YNN cluster. These are defined for the basis states 
\begin{eqnarray}
\label{eq:YNNsingleout}
|\alpha^{*(YNN)} (\mathcal{\tilde N} \tilde J \tilde T) \rangle  & = &  | \mathcal{\tilde N} \tilde J  \tilde T,  \tilde \alpha_{\rm YNN}  \tilde{\mathcal{N}}_{\rm YNN}  \tilde J_{\rm YNN} \tilde T_{\rm YNN}  \tilde n_{\lambda} \tilde \lambda \  \alpha_{A-3} \mathcal{N}_{A-3}  J_{A-3} T_{A-3}  ; (J_{\rm YNN}  (\tilde \lambda J_{A-3})\tilde I_{\lambda})\tilde J, ( \tilde T_{\rm YNN}T_{A-3}) \tilde T \rangle  
 \nonumber \\ & \equiv & \big| \, \stateNNYAmIII \big\rangle  .
\end{eqnarray}
For convenience, we define the YNN states using a single index $\tilde \alpha_{\rm YNN}$ and its 
total HO quantum number $\tilde{\mathcal{N}}_{\rm YNN}$, angular momentum $\tilde J_{\rm YNN}$, and 
isospin  $\tilde T_{\rm YNN}$. The index labels the different combinations of NN angular momentum 
and spin and hyperon angular momentum and spin as indicated in the graphical representation 
of the right hand side. Again the coupling of the different angular momentum and isospins is also 
given. The corresponding transition coefficients $   \langle \stateAmIY \big| \, \stateNNYAmIII \big\rangle $ need to be calculated numerically and are available 
for download at \cite{cfpdownload}. 

The CFP and transition coefficients are the basic ingredient of any J-NCSM calculations. Because 
the states are based on HO states, the coefficients conserve the total HO quantum number 
and are independent of the HO frequency which simplifies the calculations considerably. 
Unfortunately, HO states are not well suited for a direct solution of the Schr\"odinger 
equation for (hyper-)nuclear interactions. Their Gaussian long range behavior does not match the 
exponential tail of bound state wave functions. For the analysis, this difficulty can be 
partly overcome by matching the expected exponential behavior to the wave function 
results \cite{Sun:2025yfo}. But for the calculation itself, a sufficient number of states 
is necessary for the tail of the wavefunctions. At the same time, a large number of states 
is required to represent any short range repulsion that is typical for (hyper-)nuclear 
interactions. Moreover, different frequency ranges are necessary for an optimal representation 
of the long- and short-range part of the wave functions. Especially for $p$-shell (hyper-)nuclei 
a softening of the chiral interactions is necessary to obtain converged results. The state-of-the-art 
method to soften the interaction is the SRG evolution described in Section~\ref{sec:srgncsm}. 

\subsubsection{\texorpdfstring{$S=-2$}{S=-2} basis states}
\label{sec:s2states}

The extension to $S=-2$ requires to take into account two cases. First of all, 
two nucleons are replaced by $S=-1$ hyperons $\Lambda$ and/or $\Sigma$, and second, one nucleon 
is replaced by a $\Xi$. In the first case, we need to take the Pauli principle 
into account for $\Lambda \Lambda$ and $\Sigma \Sigma$ states. We also assume that particle 
states follow a well defined ordering as defined in the two-baryon systems for the standard
interactions \cite{Haidenbauer:2015zqb}. Therefore, we only require $\Lambda \Sigma$ but no 
$\Sigma \Lambda$ states.  The most natural coordinates that allow us to implement these constraints 
separate a hyperon-hyperon pair from $A-2$ nucleon system. The corresponding basis states read \cite{Le:2021wwz}
 \begin{eqnarray}\label{eq:basisS=Y1Y2}
 |\alpha^{*(Y_1 Y_2)} (\mathcal{N}J T) \rangle \hspace{-1ex} & = &  \hspace{-1ex}  \! \! | \mathcal{N}{J}{T}, \alpha_{A-2} \mathcal{N}_{A-2}{J}_{A-2}{T}_{A-2} \, n_{\rm Y_1 \! Y_2} (l_{\rm Y_1 \! Y_2} S_{\rm Y_1 \! Y_2}) J_{\rm Y_1 \!  Y_2}  (t_{\rm Y_1} t_{\rm Y_2})T_{\rm Y_1 \!  Y_2}  \, n_{\lambda} \lambda;  (J_{\rm Y_1 \! Y_2} 
(\lambda J_{A-2})I_{\lambda}) J, \nonumber \\
& & (T_{\rm Y_1 \! Y_2} T_{A-2}) T \rangle  
\equiv  \big| \, \stateAmIIYY   \big\rangle \ . 
\end{eqnarray}
The two hyperon labels  $Y_1, Y_2 =\Lambda, \Sigma$  distinguish the  three two-hyperon states 
$|\Lambda \Lambda\rangle$, $|\Lambda \Sigma \rangle$  and $|\Sigma \Sigma\rangle$. 
As before the index $\alpha_{A-2}$ labels the antisymmetrized $A-2$ nucleon states with total 
$\mathcal{N}_{A-2}$ HO excitation and angular momentum and isospin $J_{A-2}$ and $T_{A-2}$. 
The YY subsystem is defined by its HO quantum number $n_{\rm Y_1 \! Y_2}$, the orbital angular momentum 
$l_{\rm Y_1 \! Y_2}$ and its spin and total angular momentum $S_{\rm Y_1 \! Y_2})$ and $J_{\rm Y_1 \!  Y_2}$. 
The isospins of the first and second hyperon $t_{\rm Y_1}$ $t_{\rm Y_2}$ coupled to the YY isospin $T_{\rm Y_1 \!  Y_2}$. The relative motion of the two clusters is given by the HO quantum number 
$n_{\lambda}$ and its orbital angular momentum $\lambda$. The coupling to the total angular momentum 
of the $S=-2$ state $J$ and its isospin $T$ is also indicated. As usual, the total HO excitation 
is given by the sum of each part 
$\mathcal{N} = \mathcal{N}_{A-2}+2n_\lambda + \lambda + 2n_{\rm Y_1 \! Y_2}+l_{\rm Y_1 \! Y_2}$

The $\Xi$ states are defined in the same way as the $S=-1$ states of Eq.~(\ref{eq:hbasis})
  \begin{eqnarray}\label{eq:basisS=Xi}
    |\alpha^{* (\Xi)} (\mathcal{N}J T) \rangle & =&    | \mathcal{N}{J}{T}, \alpha_{A-1} \mathcal{N}_{A-1} J_{A-1} T_{A-1}  \, n_{\Xi} (l_{\Xi}\, s_{\Xi}) I_{\Xi}t_{\Xi} ;  (J_{A-1} I_{\Xi}){J}, (T_{A-1}\, t_{\Xi}) {T} \rangle
  \equiv \big| \stateAmIXi  \big\rangle
\end{eqnarray}
including analogous definitions of the quantum numbers involved. In both, the hyperon-hyperon and the $\Xi$
ones, the coordinate of the relative motion points towards the hyperon subcluster. 

For the solution of the Schr\"odinger equation, the Hamilton operator needs to be expressed 
differently for basis states   $\big| \, \stateAmIIYY   \big\rangle$ and  
$\big| \, \stateAmIXi   \big\rangle$ and for transitions between these two sets of states.
For the  $\big| \, \stateAmIIYY   \big\rangle$ states, the following terms contribute 
\begin{eqnarray}  \label{eq:hamiltonian2Y}
 \big\langle \stateAmIIYY \big|  H \big| \stateAmIIYY \big\rangle  
& = &  \sum_{i < j=1}^{A-2} \Big( \frac{2p^{2}_{ij}}{M(t_{Y_1}, t_{Y_2})} \,+ V^{\rm NN}_{ij} \Big)
+  \sum_{i=1}^{A-2}
\Big( \frac{m_N + m(t_{Y_1})}{M(t_{Y_1}, t_{Y_2})} \,\frac{p^2_{iY_1}}{2\mu_{iY_1}} \,+ V^{\rm YN}_{iY_1}
+   \frac{m_N + m(t_{Y_2})}{M(t_{Y_1}, t_{Y_2})} \,\frac{p^2_{iY_2}}{2\mu_{iY_2}} \,+ V^{\rm YN}_{iY_2} \Big)\nonumber\\[3pt]
& + & \frac{m(t_{Y_1}) + m(t_{Y_2})}{M(t_{Y_1}, {t_{Y2}})}\, \frac{p^{2}_{Y_1 Y_2}}{2\mu_{Y_1 Y_2}}
 \, +V^{\rm YY}_{Y_1 Y_2} 
  + \big(m(t_{Y_1}) + m({t_{Y_2}}) -  2m_{\Lambda}\big) \ .
  \end{eqnarray}
The kinetic energies are expressed in pair particle momenta of two nucleons ($\vec p_{ij}$), a hyperon and a nucleon ($\vec p_{iY_k}$) and two hyperons ($\vec p_{Y_1Y_2}$). The total rest mass of the hypernucleus 
depends on the isospins of the two hyperons $M(t_{Y_1},t_{Y_2})$ and the mass of 
$k$-th hyperon $m(t_{Y_k})$ only on is isospin. The kinetic energies also depend on the reduced masses 
of the  YN and YY system, $\mu_{i Y_k}$ and $\mu_{{\rm Y}_1{\rm Y}_2}$.   
In this part of the Hamilton operator, the NN, YN and YY interaction contributes. The terms of the interactions and the kinetic energy are already grouped so that they 
can be applied in coordinates that separate NN, YN and YY pairs. For simplicity, we neglect all 3BFs 
although 3N and YNN forces are regularly applied. So far, no YYN forces or $\Xi$NN have been included 
since the chiral ones have not been implemented and the SRG-induced YYN interactions are presumably small \cite{Le:2021wwz}.

The next part of the Hamilton operator  for the $\big| \, \stateAmIXi   \big\rangle$ states is 
\begin{eqnarray} \label{eq:hamiltonianXi}
 & &  \big\langle  \stateAmIXi  \big|  H \big| \, \stateAmIXi   \big\rangle  = \sum_{i < j=1}^{A-1} \Big( \frac{2p^{2}_{ij}}{M({\Xi})} \,+ V^{\rm NN}_{ij} \Big) +  \sum_{i=1}^{A-1}
\Big( \frac{m_N + m_{\Xi}}{M({\Xi})} \,\frac{p^2_{\Xi i}}{2\mu_{\Xi i}} \,+ V^{\Xi \rm N}_{\Xi i }\Big) + \big(m_{\Xi} + m_N -  2m_{\Lambda}\big) 
  \end{eqnarray}  
which includes terms that single out NN pairs and $\Xi$N pairs. Additionally to Eq.~(\ref{eq:hamiltonian2Y}),
the total mass including the cascade particle $M(\Xi)$ and 
the reduced mass $\mu_{\Xi i}$ of the $\Xi$N show up. In these coordinates, 
only the NN and $\Xi$N interactions appear. 

\begin{table}[t]
 \renewcommand{\arraystretch}{1.8}
 \vskip 1 cm 
    \setlength{\tabcolsep}{0.09cm}
\centering
\begin{tabular}{|*{5}{c|}}
\cline{1-5}  
\diagbox[height=0.87cm, innerwidth=2cm, innerrightsep=0.3cm,trim=l]{\phantom{x}\raisebox{-1.5mm}{transition}}{ \phantom{.}\hspace{-3mm} \raisebox{1mm}{YN}}&  $\tilde{V}_{N\Lambda,N\Lambda}$ &
$\tilde{V}_{N\Lambda, N\Sigma}$ & $\tilde{V}_{N\Sigma, N\Lambda}$ &
 $\tilde{V}_{N\Sigma, N\Sigma}$\\
\cline{1-5}
$\Lambda\Lambda \rightarrow \Lambda \Lambda$ & $2(A-2)$ & -  & - & \\
\cline{1-5}
$\Lambda\Lambda \rightarrow \Lambda \Sigma$ & - & $\sqrt2(A-2)$  & - & \\
\cline{1-5}
$\Lambda\Sigma \rightarrow \Lambda \Sigma$ & $A-2$ &  -  & - & $A-2$\\ 
\cline{1-5}
$\Lambda\Sigma \rightarrow \Sigma \Sigma$ &   - & $\sqrt2(A-2)$& - &- \\
\cline{1-5}
$\Sigma\Sigma \rightarrow \Lambda \Sigma$ & - &  -  & $ \sqrt2(A-2)$ & - \\
\cline{1-5}
$\Sigma\Sigma \rightarrow \Sigma \Sigma$ & - & -  & - & $2(A-2)$  \\
\cline{1-5}
\end{tabular}
\vspace{0.4 cm}
\caption{ Combinatorial factors of the two-body  YN  interactions embedded   in the
$A$-body space with strangeness  $S=-2$.}
\label{tab:yncombfact}
\end{table}

For transition matrix elements, only the interactions inducing the $\Xi$N-$\Lambda \Lambda$-$\Lambda \Sigma$-$\Sigma \Sigma$ transitions enter
  \begin{align} \label{eq:hamiltonian_transition}
\begin{split}
  \big\langle \stateAmIIYY \big|  H \big| \, \stateAmIXi   \big\rangle = \sum_{i=1}^{A-1}   V^{{\rm YY},\Xi {\rm N}}_{Y_{1} Y_{2} , \Xi i} \ . 
\end{split}
\end{align}

For the $S=-2$ systems, the different clusters have different symmetry properties. This leads to non-trivial 
combinatorial factors when reformulating the field theoretical interactions in second quantization 
into an ordinary Schr\"odinger equation. For a system of $A-2$ identical nucleons and YY or $\Xi$N pairs,
the factors for the YN interaction are summarized in Table~\ref{tab:yncombfact} and for the YY interaction 
in Table~\ref{tab:yycombfact} (see Appendix A of \cite{Le:2021wwz}). The factors also replace the sums in Eqs.~(\ref{eq:hamiltonian2Y}), (\ref{eq:hamiltonianXi}), and (\ref{eq:hamiltonian_transition}).

\begin{table}[t]
  \renewcommand{\arraystretch}{0.9}
 \vskip 1 cm 
    \setlength{\tabcolsep}{0.04cm}
\centering
\begin{tabular}{|*{11}{c|}}
\cline{1-11} 
\diagbox[height=1cm, innerwidth=2cm, innerrightsep=0.3cm,trim=l]{\phantom{x}\raisebox{-2mm}{transition}}{ \phantom{x}\raisebox{-2mm}{YY}}  &   { ${\tilde V}_{\Lambda \Lambda,
\Lambda \Lambda}$} & {$ {\tilde{V}}_{\Lambda \Lambda, \Lambda
\Sigma}$} &
{${\tilde V}_{\Lambda \Lambda, \Sigma \Sigma}$} &
{$ {\tilde  V}_{\Lambda
\Sigma, \Lambda \Sigma}$} &   {$ {\tilde  V}_{\Lambda \Sigma, \Sigma
\Sigma}$}  &
{${\tilde{V}}_{\Sigma \Sigma, \Sigma \Sigma}$} & 
{$ {\tilde V}_{\Lambda
\Lambda, N \Xi}$} &{$ { \tilde V}_{\Lambda \Sigma, N \Xi} $} & 
{${\tilde V}_{\Sigma \Sigma
, N \Xi}$} & { ${\tilde V}_{N\Xi, N \Xi }$} \\[3pt]
& & & & & & & & & & \\
\cline{1-11}
& & & & & & & & & & \\
$\Lambda \Lambda \rightarrow \Lambda \Lambda $ & 1 & - &- &- &- &- &-
& - &- &-\\[9pt]
\cline{1-11}
& & & & & & & & & & \\
$\Lambda \Lambda \rightarrow \Lambda \Sigma $ & - & 1 &- &- &- &- &
- & - &- &-\\[9pt]
\cline{1-11}
& & & & & & & & & & \\
$\Lambda \Lambda \rightarrow \Sigma \Sigma $ & - & - &1 &- &- &- &
- & - &- &-\\[9pt]
\cline{1-11}
& & & & & & & & & & \\
$\Lambda \Sigma \rightarrow \Lambda \Sigma $ & - & - &- &1 &- &- &
- & - &- &-\\[9pt]
\cline{1-11}
& & & & & & & & & & \\
$\Lambda \Sigma \rightarrow \Sigma \Sigma $ & - & - &- &- &1 &- &
- & - &- &-\\[9pt]
\cline{1-11}
& & & & & & & & & & \\
$\Sigma \Sigma \rightarrow \Sigma \Sigma $ & - & - &- &- &- &1 &
- & - &- &-\\[9pt]
\cline{1-11}
& & & & & & & & & & \\
$\Lambda \Lambda \rightarrow N \Xi$ & - & - &- &- &- & - & $\sqrt{A-1}$ &
 - &- &-\\[9pt]
\cline{1-11}
& & & & & & & & & & \\
$\Lambda \Sigma \rightarrow N \Xi$ & - & - &- &- &- & - & -&  $\sqrt{A-1}$ &
 - &-\\[9pt]
\cline{1-11}
& & & & & & & & & & \\
$\Sigma \Sigma \rightarrow N \Xi$ & - & - &- &- &- & - & -& -& $\sqrt{A-1}$ &
 -\\[9pt]
\cline{1-11}
& & & & & & & & & & \\
$N \Xi \rightarrow N \Xi$ & - & - &- &- &- & - & -& -&- & $ A-1$ \\[9pt]
\cline{1-11}
\end{tabular} 
\caption{ Combinatorial factors of the two-body  YY  interactions  embedded in the
$A$-body space with strangeness $S=-2$. }
\label{tab:yycombfact}
\end{table}

For the evaluation of the matrix elements, we again need transition coefficients to states 
that single out NN, YN, and YY pairs. Most simple are the transition to YY pairs. For the application 
of interactions to these states, the basis states Eq.~(\ref{eq:basisS=Y1Y2}) are already appropriate. 
For the NN pairs, we require transitions from both parts of the basis which can be related to the 
$A-2$ ($\langle \stateAmI
| \stateAmIIINN  \big\rangle_{A-2}$) and $A-1$ ($\langle \stateAmI
| \stateAmIIINN  \big\rangle_{A-1}$) transition matrix elements for ordinary nuclei
\begin{eqnarray}
     \big\langle \stateAmIIYY \big|  \, \stateAmIVNNYY \big\rangle & = &  \delta_{Y_1 {\tilde Y_1}} \  \delta_{Y_2 {\tilde Y_2}} \ 
 \langle \stateAmI
| \stateAmIIINN  \big\rangle_{A-2}  \nonumber \\
     \big\langle \stateAmIXi \big|  \, \stateAmIIINNXi \big\rangle & = &  \delta_{\Xi {\tilde \Xi}} \ 
 \langle \stateAmI
| \stateAmIIINN  \big\rangle_{A-1}  \ . 
\end{eqnarray}
Here, $\delta_{Y_1 {\tilde Y_1}} \  \delta_{Y_2 {\tilde Y_2}}$  and $\delta_{\Xi {\tilde \Xi}}$ indicate 
that the quantum numbers related to the relative motion of the $Y_1$ and $Y_2$ and $\Xi$ 
are conserved. The $A-1$ and $A-2$ transition coefficients can be calculated numerically as outlined 
in \cite{Liebig:2015kwa} and are available for download \cite{cfpdownload}. 

The transition for YN pairs is only necessary for $\stateAmIIYY$ states. We define the two sets of 
states that either single out ${\rm Y}_1$N or ${\rm Y}_2$N pairs 
\begin{eqnarray}
    \left| \big( \alpha^{*(Y_1 N)} \big)^{*(Y_2)} ( \tilde{\mathcal{N}} \tilde J \tilde T ) \right\rangle  
& = & | \tilde{\mathcal{N}} \tilde J \tilde T, 
\alpha^{*(Y_1 N)}_{A-1}  {\mathcal{N}}^{*(Y_1 N)}_{A-1} J^{*(Y_1 N)}_{A-1}  T^{*(Y_1 N)}_{A-1}
\, \tilde{n}_{Y_2}  (\tilde l_{Y_2} \tilde s_{Y_2} )  \tilde{I}_{Y_2} \tilde{t}_{Y_2}; 
  (J^{*(Y_1 N)}_{A-1} \tilde I_{Y_2}) \tilde J, (T^{*(Y_1 N)}_{A-1} \tilde{t}_{Y_2}) \tilde T \rangle \nonumber \\
& \equiv &  \Big| \stateYIIAmIIINYI \Big\rangle \nonumber \\ 
|\big( \alpha^{*{(Y_2 N)}}  \big)^{*(Y_1)} (\tilde{\mathcal{N}} \tilde J \tilde T) \rangle   
& = &  | \tilde {\mathcal{N}} \tilde J \tilde T, \alpha^{*(Y_2 N)}_{A-1} {\mathcal{N}}^{*(Y_2 N)}_{A-1} J^{*(Y_2 N)}_{A-1}  T^{*(Y_2 N)}_{A-1} \, 
\tilde{n}_{Y_1}  (\tilde{l}_{Y_1} s_{Y_1})\tilde{I}_{Y_1} \tilde{t}_{Y_1}; (J^{*(Y_2 N)}_{A-1}\tilde{I}_{Y_1}) \tilde J,  (T^{*(Y_2 N)}_{A-1} \tilde{t}_{Y_1}) \tilde T \rangle \nonumber \\
& \equiv & \Big| \stateYIAmIIINYII  \Big\rangle
\end{eqnarray}
keeping in mind that our ordering of states and the Pauli principal reduces the number of 
required YY pairs to ${\rm Y}_1{\rm Y}_2=\Lambda \Lambda$, $\Lambda \Sigma$, and $\Sigma \Sigma$. 
The YN pair quantum numbers are labeled by an index $\alpha^{*(Y_i N)}_{A-1}$ of 
states with total HO quantum number ${\mathcal{N}}^{*(Y_i N)}_{A-1}$, total angular momentum 
$J^{*(Y_i N)}_{A-1}$, and total isospin  $T^{*(Y_i N)}_{A-1}$. The spectator hyperon quantum 
numbers $\tilde{n}_{Y_i}$, $\tilde{l}_{Y_i}$, $s_{Y_i}=1/2$, and $\tilde{t}_{Y_i}$  couple 
as indicated to total angular momentum and isospin. As usual, the coordinates point towards the 
coupled out baryons. The transition coefficients $ \big\langle \stateAmIIYY  \big| \stateYIIAmIIINYI  \big\rangle$ and $ \big\langle \stateAmIIYY \big| \stateYIAmIIINYII  \big\rangle$ for these states are derived in Ref.~\cite{Le:2021wwz} and can also be downloaded from \cite{cfpdownload}. 
Based on these transition coefficients, Eqs.~(\ref{eq:hamiltonian2Y}), (\ref{eq:hamiltonianXi}), and (\ref{eq:hamiltonian_transition}) can be evaluated.   

\subsection{SRG evolution for chiral interactions}
\label{sec:srgncsm}

For the {\it ab-initio} studies of hypernuclei, we are mainly interested in results that are 
based on the modern chiral BB and 3B interactions described in Sections~\ref{second} and \ref{sec:BBB}. 
It turns out that it is not possible to get converged NCSM calculations based on such 
interactions since even on today's supercomputers the model spaces required 
cannot be handled. This is already true for ordinary nuclei for $A>4$ but becomes 
more pressing for hypernuclei due to the increased dimensions caused by particle 
conversions, like $\Lambda$-$\Sigma$ or $\Xi$N-YY. This issue can be resolved 
by performing unitary transformations of the underlying interactions that remove or 
mitigate the short range repulsion of typical nuclear interactions and reduce correlations 
of the baryons. In early NCSM calculations \cite{Navratil:1998uf}, the Lee-Suzuki \cite{Suzuki:1980yp} 
or Okubo \cite{Okubo:1954zz} transformations were used based on a combination of the NN and HO interactions. 
The disadvantage of these approaches was that the procedure was only applicable to NCSM calculations for 
a specific nucleus. Therefore, it was difficult to compare results of calculations even when 
both were performed within the NCSM. Directly using the unitarily transformed interactions in few-baryon 
systems to obtain, e.g., scattering results was not possible. Unitary transformations that can be directly 
applied to the interactions like the $V_{\rm lowk}$ \cite{Epelbaum:1998na,Bogner:2003wn} 
or SRG \cite{Bogner:2006pc} transformations are therefore preferable. Since all of these 
transformations induce many-baryon interactions, the SRG is nowadays the standard method for 
softening nuclear interactions for many-body calculations. For this approach, the calculation of 
three-baryon interactions is feasible and four-baryon interactions have been shown to be smaller 
than the expected uncertainty due to higher orders in the chiral expansion \cite{Maris:2023esu,Le:2023bfj}.
In the following, we give a brief introduction to SRG transformations. 

The unitary SRG transformations can be obtained by solving flow equations for the Hamiltonian \cite{Wegner:1994fdg,Bogner:2006pc}. Introducing the flow parameter $s$, the unitary transformations $U(s)$
and the transformed Hamilton operator $H(s)$ are defined by 
\begin{equation}\label{eq:SRGtrans}
H(s) = U(s) H_0 U^{\dagger}_{}(s) \equiv T_{rel}^{} + V(s).
\end{equation}  
Note that the potential $V(s)$ is defined such that the kinetic energy is not transformed. 
Starting point of the evolution is the bare Hamiltonian $H(0)$ for $s=0$. $s$ is then increased 
and therefore always positive. Following \cite{Bogner:2006pc}, we will utilize 
the more intuitive variable $\lambda=\left(\frac{4 \mu^2}{s}\right)^{1/4}$
with $\mu = {m_N}/{2}$ for all BB forces.  
$\lambda$ can be approximately identified with the width of the band for which the SRG-evolved momentum space 
matrix elements of the interaction are non-zero. The flow equation is obtained 
by differentiating the transformation Eq.~(\ref{eq:SRGtrans}) 
\begin{equation}
\label{eq:SRGflow}
   \frac{d H(s)}{ds} = \frac{d V(s)}{ds} = [\eta(s), H(s)]
\end{equation}
where the generator 
\begin{equation}
    \eta(s) = \frac{dU(s)}{ds} U^{\dagger}_{}(s) = -\eta^{\dagger}_{}(s)
\end{equation}
is an anti-hermitian operator. These equations can also be used to define the unitary transformation
by choosing an appropriate anti-hermitian operator $ \eta(s)$ and solving Eq.~\eqref{eq:SRGflow}.  
Usually, $ \eta(s)$ is  taken as a commutator of the relative kinetic energy 
(without the mass shift in case of particle conversion) with the Hamiltonian: 
$\eta(s) = [ \tilde T_{rel}^{}, H(s)]$. This choice leads to a suppression of off-diagonal 
matrix elements in momentum space and softens the interaction. Other  
choices of generator have been discussed, e.g., in 
Refs.~\cite{Wirth:2016iwn,Wirth:2018scw,Wegner:1994fdg} and lead to Hamilton operators 
that are diagonal with respect to different quantum numbers or basis states. 
Note that our choice of $\eta$ implies that $s$ has the dimension [length]$^2$. 

As usual, the unitary transformations generate higher-body interactions even when the 
bare interactions did not involve such many-body forces. Exact unitarity could only be achieved 
if all of these many-baryon interactions were generated. This is not feasible and would be 
even more complicated than directly solving the many-body problem using the bare interactions. 
The unitary transformations are nevertheless useful since it turns out that the most 
important contributions to the interactions generated by the transformation, so-called induced 
interactions, are two-baryon ones. For hypernuclei, the induced three-baryon interactions are 
quantitatively important. Four- and more-baryon interactions contribute much less than 
missing higher order terms in the chiral expansion and are therefore irrelevant, at least 
for light hypernuclei \cite{Maris:2023esu,Le:2023bfj}. 
The SRG is particularly convenient since the two-baryon, three-baryon and higher-body 
contributions can be obtained step wise by solving the individual equations \cite{Bogner:2006pc,Hebeler:2012pr}
\begin{eqnarray}
    \label{eq:srg2b}
    \frac{dV_{ij}(s)}{ds} & = &  \left[ \left[  T_{ij}  , V_{ij}(s) \right]   , T_{ij}+V_{ij}(s) \right]~,\\
    \label{eq:srg3b}
    \frac{dV_{ijk}(s)}{ds} & = &  \left[  \left[ T_{ij} , V_{jj}(s) \right]  , V_{ki}(s) + V_{jk}(s) + V_{ijk}(s) \right] \nonumber \\
    & & + \left[  \left[ T_{jk} , V_{jk}(s) \right]  , V_{ki}(s) + V_{ij}(s) + V_{ijk}(s) \right] \nonumber \\
    & & + \left[  \left[ T_{ki} , V_{ki}(s) \right]  , V_{ij}(s) + V_{jk}(s) + V_{ijk}(s) \right] \nonumber \\
    & & + \left[  \left[ T_{ij} + T_k , V_{ijk}(s) \right]  , T_{ij} + T_k  + V_{ij}(s) +  V_{ij}(s) + V_{ki}(s) + V_{ijk}(s) \right]~, 
\end{eqnarray}
where $i$,$j$,$k$ number the individual baryons of the pair interactions and triplets. These equations 
have been solved for nuclear and hypernuclear interactions either in a HO basis \cite{Wirth:2019cpp} or using momentum space \cite{Hebeler:2012pr,Le:2022ikc}. In both cases, the matrix elements need to be expressed 
in the HO basis of the NCSM calculation. In the former case, this generally requires a frequency shift, the latter case, a momentum integration over HO wave functions. 

\begin{figure}
    \centering
    \includegraphics[width=0.9\linewidth]{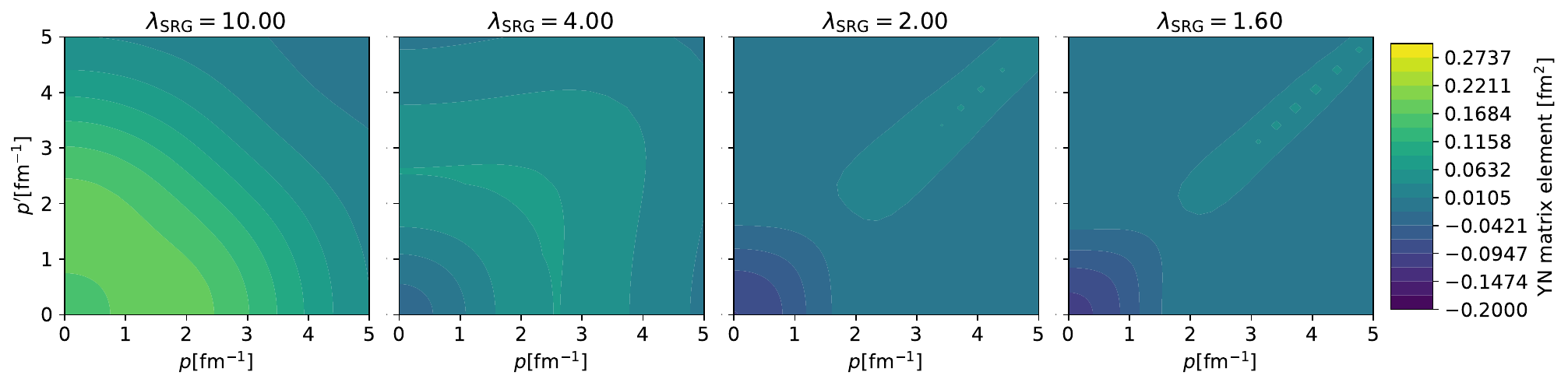}
    \caption{SRG evolution of the $\Lambda$N interaction in the $^3$S$_1$ partial wave for ($\lambda_{SRG}=10$, $4$, $2$ and $1.6$~fm$^{-1}$ (left to right)
     starting from the SMS N2LO(550) YN interaction. The matrix elements are given in fm$^2$ for 
     $\Lambda$N relative momenta in fm$^{-1}$.}
    \label{fig:LNsrg}
\end{figure}

\begin{figure}
    \centering
    \includegraphics[width=0.7\linewidth]{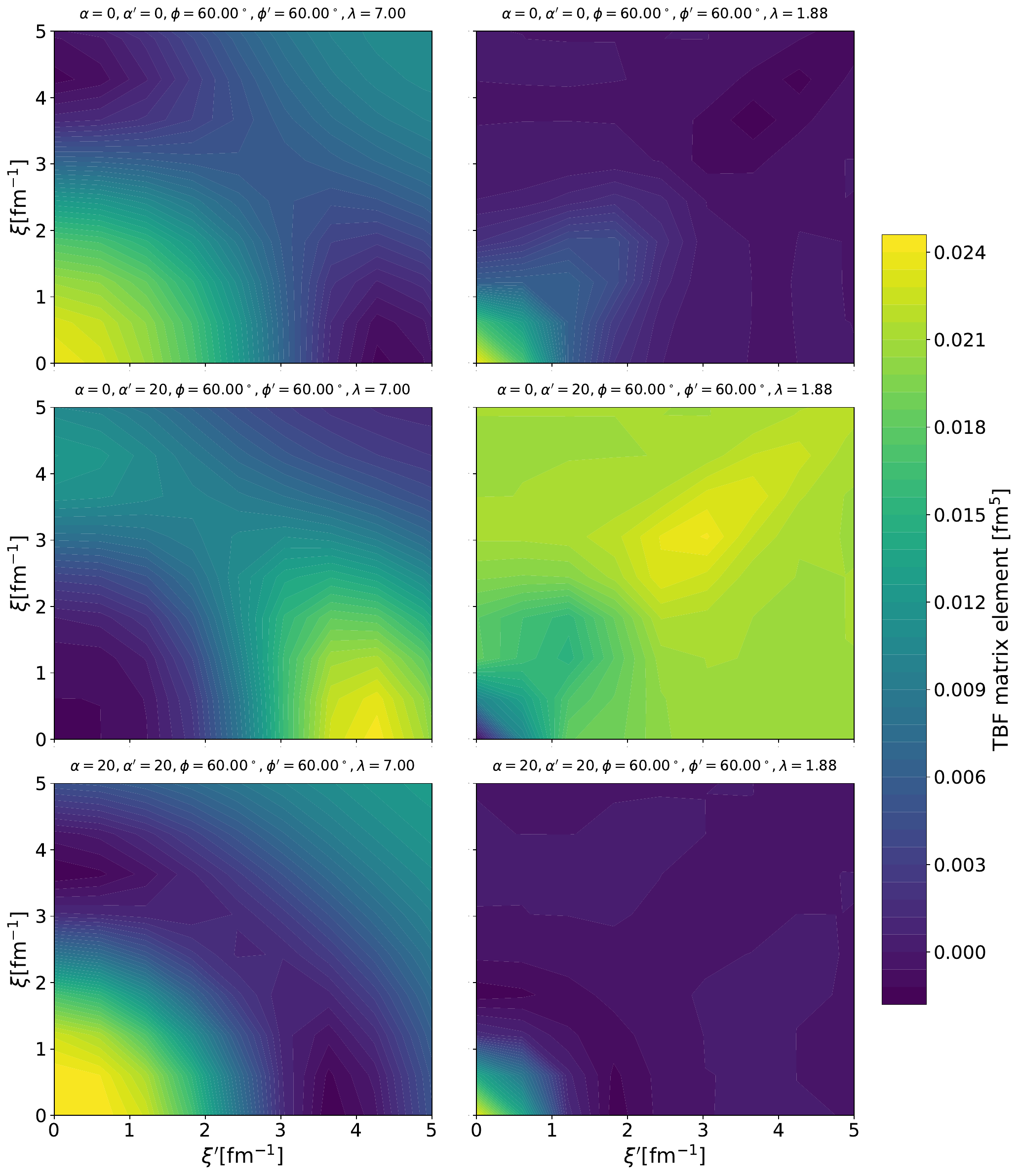}
    \caption{SRG evolution of the YNN interaction for 3B angular momentum, parity and isospin $J^\pi T=1/2^+ \, 0$ for ($\lambda_{SRG}=7$ and $1.88$~fm$^{-1}$ (left/right) starting from the SMS N2LO(550) YN and no YNN interaction. 
    The upper part are matrix elements for the 
    partial wave channel $\alpha=0$ to $\alpha=0$ ($S$-waves $\Lambda$NN channels), 
    the middle part transition $\alpha=0$ to $\alpha=20$ ($S$-waves $\Lambda$NN to $\Sigma$NN channels)
    and the lower part $\alpha=20$ to $\alpha=20$ ($S$-waves $\Sigma$NN channels). 
    The matrix elements are given in fm$^5$ for 
     hyper-momenta in fm$^{-1}$. The hyper-angles are set to $60^\circ$.}
    \label{fig:YNNsrg}
\end{figure}

Examples of the evolution in momentum space are shown in Figs.~\ref{fig:LNsrg} and \ref{fig:YNNsrg}. 
In both cases, the evolution leads to a suppression of transitions from low to high momenta. 
For the BB interactions this can be clearly seen in the dependence of the matrix elements 
on the relative momenta $p_{\rm YN}$ in the YN system. For the YNN interaction, we show here 
the dependence on the hyper-momentum 
$\xi = \sqrt{p_{\rm NN}^2 + \frac{m_{\rm Y}+2 m_{\rm N}}{4m_{\rm Y}} q_{\rm Y}^2 }$ (Y=$\Lambda,\Sigma)$). 
For this example, we fix the hyper-angle $\cos \Phi = p_{\rm NN} / \xi = 1/2$.  

\begin{figure}[tbp]
\begin{center}
    \includegraphics[scale=0.35]{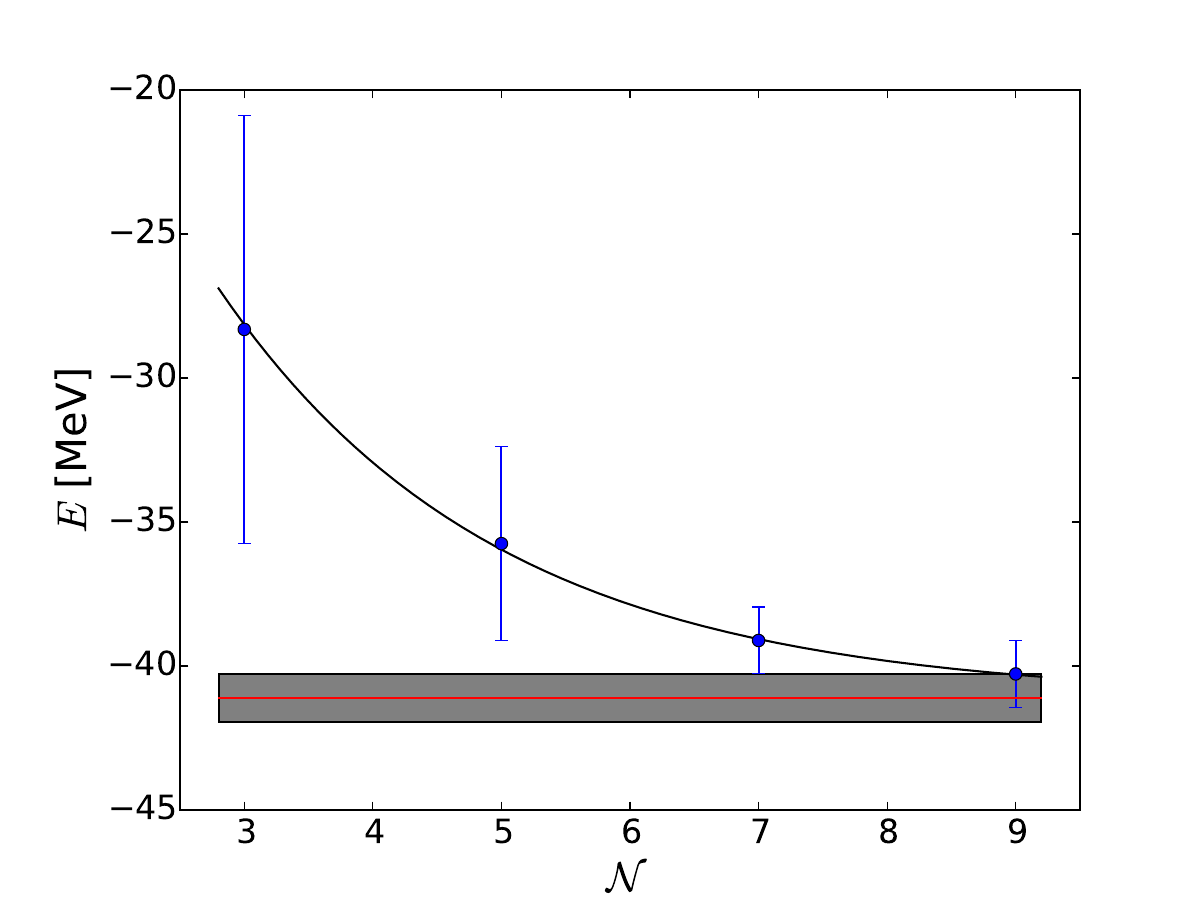}
    \includegraphics[scale=0.35]{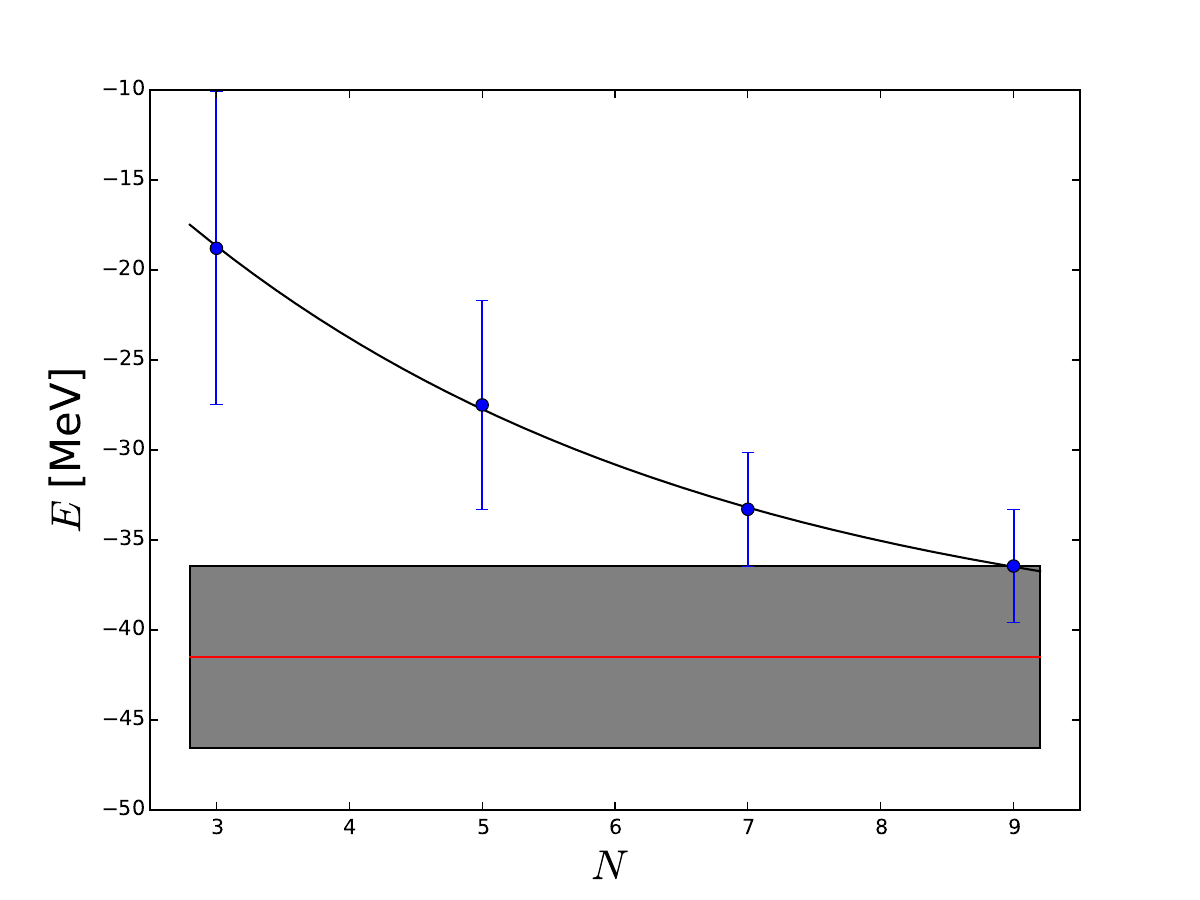}
 \end{center}   
    \caption{\label{fig:7Li3minusndep}$\cal N$-dependence of the ground state energy of $^7$Li for $\lambda=1.5$~fm$^{-1}$ (left) and 
      $\lambda=2.5$~fm$^{-1}$ (right). } 
\end{figure}

The change of the interaction leads to sufficient  softening so that the NCSM calculations converge 
significantly faster. As an example, we show in Fig.~\ref{fig:7Li3minusndep} the convergence 
with respect to the model space size for the ground state of $^7$Li for two different SRG parameters 
$\lambda_{SRG}=1.5$~fm$^{-1}$ and $2.5$~fm$^{-1}$. Clearly, the result for the larger 
$\lambda_{SRG}$ are not converged for the accessible model spaces. The uncertainty is about 5~MeV and therefore clearly larger than the expected chiral uncertainties. For the smaller $\lambda_{SRG}$, the 
uncertainty due to missing basis states is only about 1~MeV which is smaller 
than the uncertainty due to the truncation of the chiral expansion. Therefore, in this case, the accessible 
model spaces are sufficient to study effects of chiral interactions. In practice, many calculations 
are done with $\lambda_{SRG}=1.88$~fm$^{-1}$. This value is still large enough that induced many-body 
interactions are negligible and small enough to get converged results. In the following sections, we 
present some recent results for light hypernuclei exactly for this $\lambda_{SRG}$.

\subsection{CSB for \texorpdfstring{$p$}{p}-shell hypernuclei}
 \label{sec:csbpshell}

\begin{table} 
\centering
\renewcommand{\arraystretch}{1.5}
\begin{center}
\begin{tabular}{|l|l|r r|  rrr| r  | }
\hline
\multicolumn{2}{|c|}{} & $\Delta T$ & $\Delta V_\mathrm{NN}$ & \multicolumn{3}{c|}{$\Delta V_\mathrm{YN}$}  & $\Delta B_{\Lambda}$       \\[1pt]
\multicolumn{2}{|c|}{}  &  &  &  $^1S_0$  & $^3S_1$  &  total &  \\
\hline
\hline
 {$^7_\La$Be-$^7_\La$Li$^*$} &  NLO13-CSB & 8    & -24              & -49 & 26   & -24        & -40    \\
&  NLO19-CSB &6     & -41              & -43 & 42   & 0        & -35      \\
\cline{2-8}
& Hiyama~\cite{Hiyama:2009ki}  &      &   -70 &    &  &  200 & 150\\
& Gal~\cite{Gal:2015bfa} &  3   & -70  &  &  & 50  &  -17\\
\cline{2-8}
& experiment &  &  &  &  &   & $-145 \pm 107 $\\
\hline
\hline
 {$^7_\La$Li$^*$-$^7_\La$He} & NLO13-CSB &7     & -14               & -49 & 26    & -24         & -31        \\
 & NLO19-CSB & 5     & -21                 & -43 &  42    &  0     & -16        \\
\cline{2-8}
& Hiyama~\cite{Hiyama:2009ki}  &  & -80   &    &  & 200 & 130\\
& Gal~\cite{Gal:HYP2015} &  2   & -80  &  &  & 50  &  -28\\
\cline{2-8}
& experiment &  &  &  &  &  & $-279 \pm 141 $ \\
\hline
\hline
 {$^8_\La$Be-$^8_\La$Li}
& NLO13-CSB            & 12     & 7     &   100   &  56    &  159    &  178          \\
& NLO19-CSB            & 6      & -11       &   62     &  79  &  147    & 143      \\
\cline{2-8}
& Hiyama~\cite{Hiyama:2009ki}  &      & 40   &    &  & & 160\\
& Gal~\cite{Gal:2015bfa}  & 11    & -81  &  & &  119  & 49\\
\cline{2-8}
& experiment   &   &    &   &  &  & $51 \pm 80$ \\
\hline
\end{tabular}
\end{center}
\caption{ Contributions to CSB in the $A=7$ and $8$ isospin multiplets, 
based on the YN potentials NLO13(500) and NLO19(500) (including 
3N forces and SRG-induced YNN interactions). The results are for the original potentials 
(without CSB force) and for the scenario CSB1, see text. 
Results by Gal \cite{Gal:2015bfa} and by Hiyama et al.~\cite{Hiyama:2009ki} 
are included for the ease of comparison. 
All energies are in keV. The estimated uncertainties for $A=7$ and $8$ systems are  30 and 50~keV, respectively. 
}
\renewcommand{\arraystretch}{1.0}
\label{tab:csbsepener}
\end{table}

In Section~\ref{sec:csbyn}, we discussed the determination of the strength of CSB in YN interactions using the 
hypernuclear $A=4$ isospin multiplet $^4_\Lambda$He/$^4_\Lambda$H. Although the independence of the chiral realization
indicates that the approach is robust, cross checks with other available data are desirable. Such additional data can also serve to constrain the parameters more tightly in the future. 
Fortunately, data for more isospin multiplets are available (see Table~\ref{tab:csbsepener} and Refs.~\cite{Botta:2016kqd,Botta:2019has}). As one can see, the uncertainty of the data is still 
significant. The $A=8$ hypernuclei $^8_\Lambda$Be/$^8_\Lambda$Li  form an isospin doublet and the CSB of the 
separation energies is defined similarly to the one for $A=4$ hypernuclei in Eq.~(\ref{eq:csba4}). In this case, 
both separation energies are fairly well known, based on the current Mainz evaluation  \cite{HypernuclearDataBase},
the CSB is $51\pm 80$~keV. For $A=7$ hypernuclei, the ground states of  $^7_\Lambda$Be and $^7_\Lambda$He, and the excited isospin $T=1$ state of $^7_\Lambda$Li form an isospin triplet. In the latter case, the separation 
energy is calculated with respect to the first excited $T=1$ state in $^6$Li. Based on the excitation energy \cite{Tamura:2000ea} and the most recent determination of the ground state 
separation energy from the Mainz database \cite{HypernuclearDataBase}, one obtains $5.305 \pm 0.060$~MeV for the 
separation energy of this state. The current world average for the separation energy of $^7_\Lambda$Be 
is $5.160\pm 0.089$~MeV \cite{HypernuclearDataBase}. For $^7_\Lambda$He, the situation is more ambiguous 
since the old emulsion data and the new data from JLab are inconsistent with each other. 
Based on the 
new experiments at JLab \cite{HKSJLabE05-115:2016yge,HKSJLabE01-011:2012sgn}, one obtains  $5.584\pm 0.128$~MeV \cite{HypernuclearDataBase}. 
The experimental values for the CSB in the table are based on these values. 

The table shows our results for the CSB interactions of Sec.~\ref{sec:csbyn} 
and compares them to the experiment and the previous calculations of 
Refs.~\cite{Hiyama:2009ki,Gal:2015bfa}, performed within a cluster model and 
the shell model, respectively. Besides the total value $\Delta B_\Lambda$ 
for the CSB of the separation energies, also the different contributions 
from the kinetic energy $\Delta T$, the NN potential $\Delta V_{\rm NN}$
and the YN potential  $\Delta V_{\rm YN}$ are shown. For the latter one
also the singlet and triplet contributions are separately shown. The kinetic energy contribution is small in all cases. For the NN interaction, which 
also includes effects from the Coulomb force, our full calculation leads 
to much smaller values than the previous approximate ones. (Note that the 
values of Gal for this part have been taken over simply from the 
cluster calculations of Hiyama et al.). Also the contribution from 
the YN force is different in all of the approaches. In $A=7$, our 
full calculations predict generally smaller YN contributions 
sometimes even different in sign to the previous ones. Therefore, it is not 
surprising that the predicted CSB for the separation energies 
is smaller. Interestingly, we find that the sign of the contribution is 
different for $A=7$ and $A=8$. This is mainly caused by the opposite sign of 
the $^1{\rm S}_0$ contribution. This highlights that a proper understanding of the CSB of the light hypernuclei also requires a proper representation 
of the spin dependence. The heavier hypernuclei have recently been investigated within a 
mean field approach \cite{Sun:2025pfp}. In this case, the CSB is adjusted to $A=12$ mirror
hypernuclei and it was found that the spin dependence is small for heavier 
hypernuclei. Unfortunately, a direct comparison of both approaches for light 
hypernuclei and/or the resulting scattering lengths is not possible. 
It is also remarkable that the contribution 
from $^3{\rm S}_1$ is still comparable to the one of $^1{\rm S}_0$
although the affect on the corresponding scattering lengths is much larger 
for $^1{\rm S}_0$.

\begin{figure}
    \centering
    \includegraphics[width=0.5\linewidth]{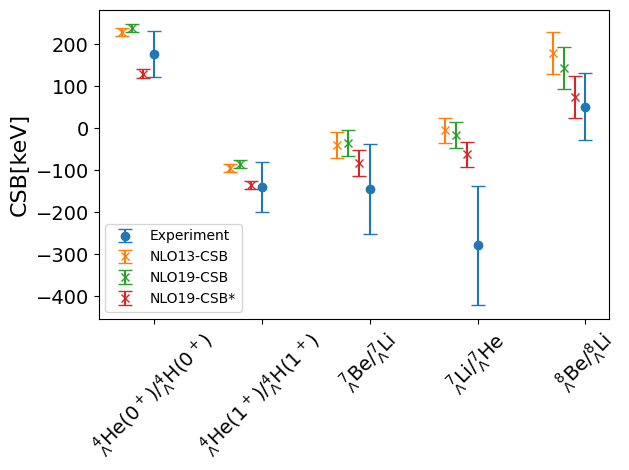}
    \caption{CSB of $\Lambda$ separation energies for $A=4$ to $8$ hypernuclei. Experimental values obtained as described in the text are compared to results for three CSB YN interactions: the scenario CSB1 for NLO13 (orange) and NLO19 (green)  of Ref.~\cite{Haidenbauer:2021wld} and 
    to the alternative scenario fitted to experimental values of the STAR 
    collaboration \cite{STAR:2022zrf} for NLO19 (red). Error bars on the theoretical values are 
    numerical uncertainties. }
    \label{fig:csbpshell}
\end{figure}

The total CSB of the separation energies is also presented in Fig.~\ref{fig:csbpshell}. 
The figure shows our results for the scenario CSB1 
of Ref.~\cite{Haidenbauer:2021wld} and an additional scenario (CSB*) fitted 
to a recent experiment of the STAR collaboration \cite{STAR:2022zrf}. 
It is reassuring to see that the predictions for NLO13 and NLO19 
are very similar to each other for the CSB1 scenario. This indicates 
that the theory uncertainty is small once the $A=4$ CSB is accurately 
determined. Using the STAR input (CSB*) for $A=4$, 
also the predictions for $A=7$ and $8$ shift visibly. 
Given the still quite larger experimental error bars, both 
scenarios are more or less consistent with the experimental data. 
The most problematic case is certainly 
$^7_\Lambda {\rm Li} /^7_\Lambda {\rm He} $. 
As discussed above, here the experimental situation is much less clear. 
Note also that the theoretical results indicate that the CSB 
for those two hypernuclei is similar to $^7_\Lambda {\rm Be} /^7_\Lambda {\rm Li} $
whereas the experiment suggests a larger value in the former systems. 
Also note that the CSB* scenario seems to be in better agreement with 
experiment and that both scenarios are able to reproduce the sign 
change of the CSB for $A=7$ and $8$. This could not be achieved within 
the cluster model \cite{Hiyama:2009ki}. 

In summary, our calculations can reproduce the 
non-trivial features for the CSB of $A=7$ and $A=8$ hypernuclei. This supports that 
the chiral interactions fitted to the $A=4$ data provide a realistic 
description for this aspect. In the future, more accurate data for 
separation energies of the light isomultiplets will provide firmer 
constraints for the strength of the $\Lambda n$ interaction. 

\subsection{Application of chiral YNN interactions}
 \label{sec:ncsmynn}

YN interactions based on phenomenological approaches result in very different 
predictions even for the light hypernuclei, and even when their description
of $\Lambda$N and $\Sigma$N data is quite similar \cite{Nogga:2001ef,Haidenbauer:2019boi}. 
It was therefore clear for many years that reliable predictions for the 
separation energies can only be obtained when including YNN interactions 
properly. With the chiral approach, the inclusions of consistent YN, NN, 3N and YNN 
forces becomes possible. The formulation of corresponding forces has already been discussed 
in Sec.~\ref{sec:BBB}. In this section, we discuss the first steps towards applying these interactions. 
The first applications of the $\Lambda$NN interaction have been done using momentum space 
in Refs.~\cite{Kamada:2023txx,Kohno:2023xvh}. In order to check the numerical implementation of 
the  partial waves decomposition, these results have been benchmarked in Ref.~\cite{Le:2024aox}. 

\begin{table}[t]
    \centering
    \begin{tabular}{|l  c   c   c l|}
\hline
$^A_\Lambda {\rm Z} (J^\pi,T)$ &   w/o YNN  & YNN(sat)   & YNN(sat+$C_2'$)  &  Expt.~\cite{HypernuclearDataBase}\\
 \hline
 \hline
{$^3_{\Lambda}\mathrm{H}(1/2^+,0)$} &  $0.121(4) $   & $0.125(4)$  & $0.155(3)$ & $0.164(43)  $  \\
\hline
$^4_{\Lambda}\mathrm{He}\,(0^+,1/2)$  &  $1.954(1)$   &   $ 2.027(3)$ &  $2.220(2)$ & $2.258(55) $  (average)   \\
&  &  &   & $2.169(42) $ $ (^4_{\Lambda}\mathrm{H})$ \\
&  &  &   & $2.347(36)$  $ (^4_{\Lambda}\mathrm{He})$ \\
\hline
$^4_{\Lambda}\mathrm{He}\,(1^+,1/2)$  &  $1.168(20)$   &   $ 1.010(11)$ & $0.984(12)$ &  $1.011(72) $  (average)  \\
&  &  &   & $1.081(46) $ $ (^4_{\Lambda}\mathrm{H})$ \\
&  &  &   &  $ 0.942(36)$ $ (^4_{\Lambda}\mathrm{He})  $\\
\hline
$^5_{\Lambda}\mathrm{He}(1/2^+,0)$  &  $ 3.518(20)$   &   $ 3.152(21)$ & $3.196(20)$ & $3.102(30)$\\
\hline
$^7_{\Lambda}\mathrm{Li}(1/2^+,0)$  & $5.719(56)$   &   $ 5.444(57)$ &  $5.623(52)$ & $5.619(60)$  \\
$^7_{\Lambda}\mathrm{Li}(3/2^+,0)$ & $5.522(70)$ &    $ 5.042(65)$ &  $5.040(57)$ & $4.927(60)$  \\
$^7_{\Lambda}\mathrm{Li}(5/2^+,0)$ & $3.440(66)$ &   $ 3.205(65)$ &  $3.356(60)$  & $3.568(60)$ \\
 \hline
  \end{tabular}
    \caption{Separation energies for $A=3-7$ hypernuclei with angular momentum and parity $J^\pi$ and isospin $T$
in MeV, calculated without and with inclusion of YNN 3BFs. 
The number in parenthesis indicates the estimated extrapolation uncertainties of the J-NCSM uncertainty 
and do not include the uncertainty due to higher order terms in the chiral expansion 
(see Fig.~\ref{fig:sepynn}). All results are based on the N4LO+(550) SMS NN 
             potential \cite{Reinert:2017usi}, the corresponding  N2LO(550) 3NF (see Table 1 of Ref.~\cite{Le:2023bfj}) and the N2LO(550) YN potential and include the SRG-evolution up to the 3B level for $\lambda_{SRG}=1.88$~fm$^{-1}$. In the third column (YNN(sat)) the YNN forces using properly adjusted parameters $H_1$ and $H_2$ are added. In the fourth column, the $C_2'$ term is added additionally. 
              }
    \label{tab:sepynn}
\end{table}

As described in Sec.~\ref{sec:BBB}, YNN forces depend on a larger number of LECs which 
essentially need to be determined by fits to hypernuclear data. Given the scarceness 
of the available data, 
this is impossible in the foreseeable future. The probably most viable path to a realistic 
inclusion of YNN force is therefore making use of the decuplet saturation as described in Sec.~\ref{subsec:BBBDec}. Using SU(6) symmetry, within decuplet saturation the 3BFs only 
depend on the two contact LECs $H_1$ and $H_2$. 
At the same time, the estimates discussed in Sec.~\ref{subsec:ynnestimate}
show that its contribution to $^3_\Lambda$H is likely negligible and that the $A=4$ hypernuclei are the 
lightest ones that can be used for a determination of the YNN LECs. Therefore,
in a na\"ive expectation one could determine $H_1$ and $H_2$ by a fit to the $0^+$ and $1^+$ 
states of the $A=4$
hypernuclei\footnote{The separation energy differences of $^4_\Lambda$H/$^4_\Lambda$He is due
to CSB and cannot be used to determine any parameter in the isospin conserving LO YNN forces}. 
Unfortunately, the explicit calculation showed that a consistent description of both 
states cannot be achieved with any combination of $H_1$ and $H_2$. In reality, the dependence 
of these separation energies on $H_1$ and $H_2$ is strongly correlated. These observations 
are easily explained when looking at the decuplet saturation for the most important 
$\Lambda$NN matrix elements (see Eq.~(\ref{eq:decsatlnn})). In turns out that these 
matrix elements 
only depend on the linear combination $H_1+3H_2$, i.e. essentially only on one LEC. 
This freedom in the parameter 
can be used to get a consistent description of the $1^+$ state in $A=4$ and 
of the $A=5$ separation energy. The $0^+$ state in $A=4$ is then not well described. 
In order to also improve the description of this state, it is necessary to add further terms 
of the YNN forces \cite{Le:2024rkd}. Since a full calculation without decuplet saturation 
involves too many unknown LECs, as already mentioned above, it was most practical to only 
select a simple term 
that promises to provide the necessary spin dependence. Such a term can be identified in the set 
of $\Lambda$NN contact terms of Eq.~(\ref{eq:lnncontact}). Decuplet saturation sets the spin dependent 
$C_2'$ term to zero. As shown in Ref.~\cite{Le:2024rkd}, by including this term with
a non-zero LEC a consistent description of $A=4$ and $A=5$ states is possible. In fact, 
the values of $H_1$ and $H_2$ were not even changed 
since the main contribution of this term is to the $0^+$ state. The resulting energies are shown in Tab.~\ref{tab:sepynn}. It is interesting to look at the contribution of the
individual 3BFs to $^3_\Lambda$H. 
The pure decuplet saturation (YNN(sat)) only contributes a tiny 4~keV to this energy. Only the additional $C_2'$ term 
adds about 30 keV additional binding to this hypernucleus. This is only slightly larger 
than the estimate of Sec.~\ref{subsec:ynnestimate} and still  below the experimental uncertainty 
of $40$~keV. The fit of the LECs was performed for the $A=4$ energies. Since the 
interaction does not include CSB terms, the averaged values for $^4_\Lambda$H and $^4_\Lambda$He 
were used. For YNN(sat) this is by construction in agreement with the experimental value for the $1^+$ state.
Adding the $C_2'$ term only led to changes within the uncertainties for this state, so that a refit 
of $H_1$ and $H_2$ was not necessary. 

\begin{figure}[t]
    \centering
    \includegraphics[width=0.7\linewidth]{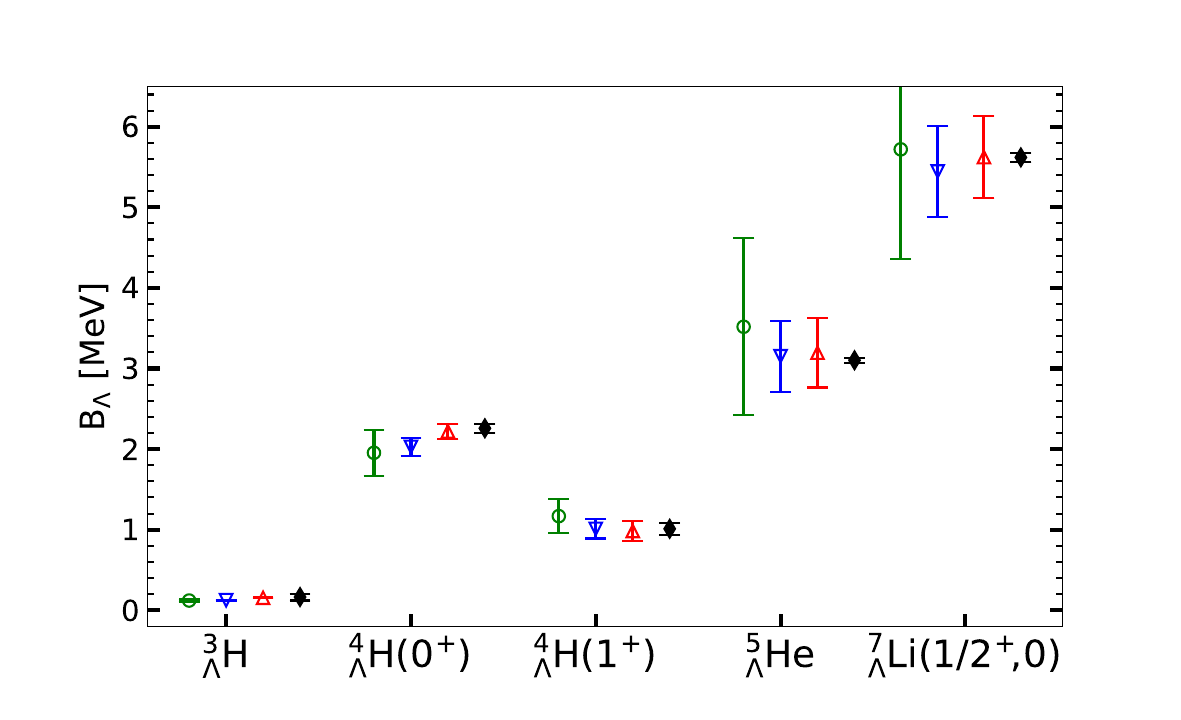}
    \caption{Separation energies of $A=3 \to 7$ hypernuclei based on the N4LO+(550) SMS NN 
             potential \cite{Reinert:2017usi}, the corresponding  N2LO(550) 3NF (see Table 1 of Ref.~\cite{Le:2023bfj}) and the N2LO(550) YN potential.
             The uncertainty is obtained as described in Sec.~\ref{subsec:ynnestimate}. 
             All interactions are evolved to $\lambda=1.88$~fm$^{-1}$ including the induced 3BFs.  
             The green circle is the result without chiral YNN force, and with the NLO uncertainty. The blue downward triangle is the result including the decuplet saturated YNN force and the red upward triangle also includes the $C_2'$ term.
             Here, the uncertainty is given by the N2LO uncertainty. The black diamond is the experimental result \cite{HypernuclearDataBase}. 
     }
    \label{fig:sepynn}
\end{figure}

In Fig.~\ref{fig:sepynn}, the energies are plotted together with uncertainties due to the 
truncation of the chiral expansion. Without YNN forces, we use the NLO uncertainty. 
When including the YNN force, we assume that we have included the dominant part of the 
YNN force and therefore use the N2LO uncertainty. The agreement of the experimental values 
is generally much better than expected from the uncertainty estimates which supports that 
our assumption is justified.  Clearly, the appropriate choice of the uncertainty estimate should 
be investigated more systematically in the future. 

It is remarkable that fixing the energy of the $1^+$ state 
also leads to a good description of the $^5_\Lambda$He ground state. The only 
visible difference of the YNN(sat) calculation and the one with $C_2'$ term appears 
for the $0^+$ state. It is also reassuring to observe that adding YNN forces improves the agreement 
with experiment for $^7_\Lambda$Li. Note that the YNN force acts repulsively for all of the examples 
except $^3_\Lambda$H and the $0^+$ state of $A=4$. 

Finally, in Fig.~\ref{fig:exynn}, the excitation energies for $^7_\Lambda$Li are shown. These energies 
are experimentally known with very high accuracy. Without YNN force, the doublet splitting for the lower 
states are too small and the ordering of states in the second doublet is wrong. For the lowest doublet, 
the uncertainties are quite small and indicate a deviation from experiment when no YNN force is 
considered. This improves when including the YNN forces. In the figure, we only show the calculation including 
the $C_2'$ term. Without this additional spin dependence, there is still an improvement but the increase 
of the splitting is somewhat smaller. The YNN force corrects the ordering 
within the upper doublet. For the isospin $1$ state, there is almost no effect 
from the YNN force. In summary, the YNN force that describes the $s$-shell hypernuclei 
fairly well also leads to an improved description for this $p$-shell hypernucleus. 

Note that our work on $p$-shell hypernuclei in general  
is still work in progress. It will be especially important to get more insight into the 
uncertainty estimates by, e.g. variation of the cutoff or by adding other YNN interaction
terms that are not contributing when one resorts to the decuplet saturation approximation. 
\begin{figure}[ht]
    \centering
    \includegraphics[width=0.7\linewidth]{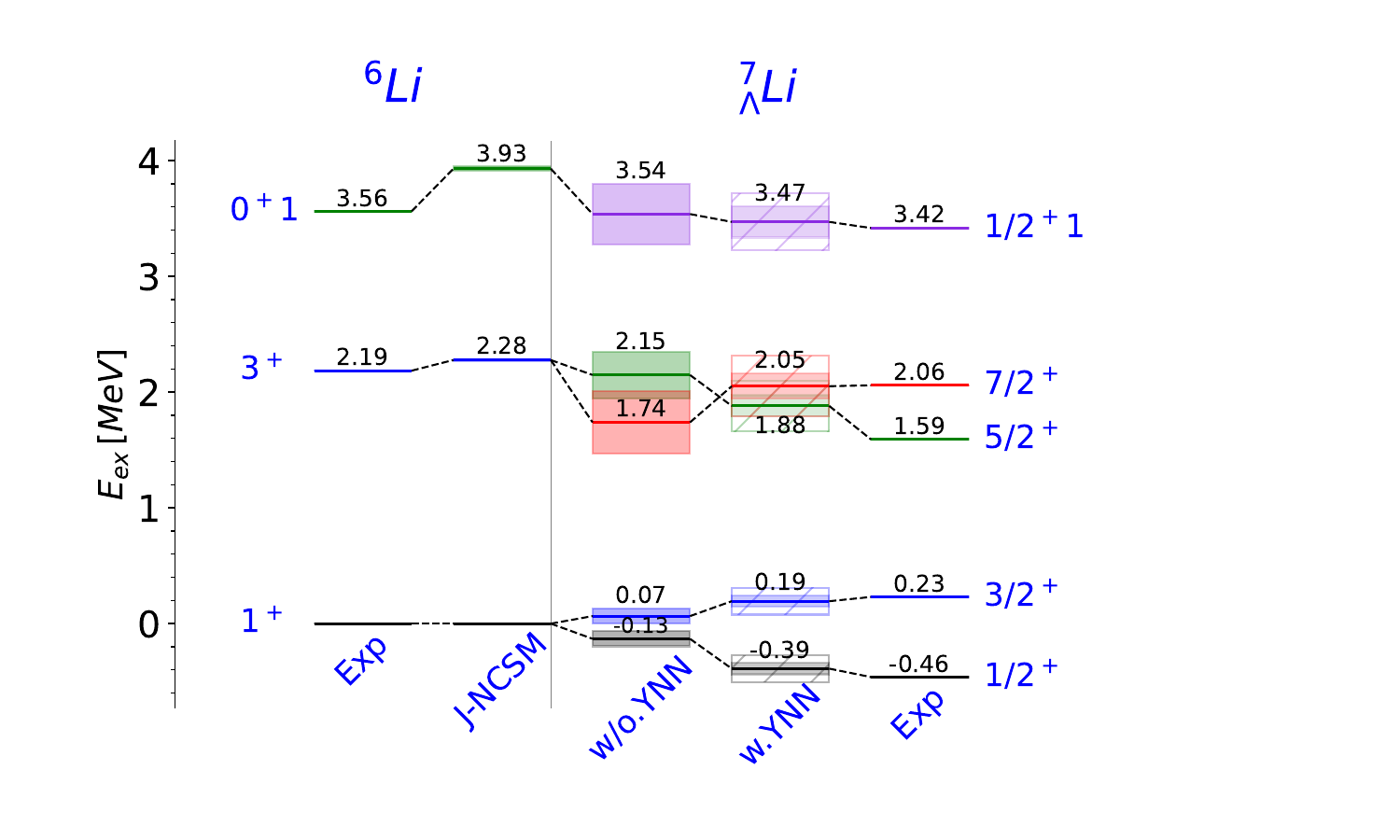}
    \caption{Spectrum of excitation energies of $^7_\Lambda$Li based on the N4LO+(550) SMS NN 
             potential \cite{Reinert:2017usi}, the corresponding  N2LO(550) 3NF (see Table 1 of Ref.~\cite{Le:2023bfj}) and the N2LO(550) YN potential.
             The uncertainty is obtained as described in Sec.~\ref{subsec:ynnestimate}. 
             All interactions are evolved to $\lambda=1.88$~fm$^{-1}$ including the induced 3BFs.
             Results are shown without chiral YNN forces (full band NLO uncertainty) and with chiral YNN forces (full band N2LO uncertainty, hatched band NLO uncertainty) including the $C_2'$ term. Experimental uncertainties are almost invisible.}
    \label{fig:exynn}
\end{figure}

\subsection{\texorpdfstring{$\Lambda \Lambda$}{Lambda-Lambda} hypernuclei and the strength of the \texorpdfstring{$S=-2$}{S=-2} interaction}
\label{sec:lamlamhyp}

As has been discussed in Sec.~\ref{subsec:LLXiN}, there is only very sparse direct scattering 
information on the $S=-2$ interaction. To determine the strength of the
$\Lambda \Lambda$ interaction, 
it is therefore even more important than in the YN sector to take information from hypernuclei
into account. But also experiments on $\Lambda \Lambda$ hypernuclei are difficult. Therefore, only a 
few events have been identified \cite{HypernuclearDataBase} for $A=6$ and $10$ to $13$ hypernuclei. 
Fortunately, the lightest system for which binding energies have been determined is $^{\ \ 6}_{\Lambda \Lambda}$He 
and is within the reach of J-NCSM calculations. Note that the most accepted experimental result 
for this hypernucleus \cite{Takahashi:2001nm} has been corrected in \cite{Nakazawa:2010zza} because 
the recommendation for the $\Xi$ mass was changed. 

Double $\Lambda$ hypernuclei have already been investigated within the stochastical variational 
approach \cite{Nemura:1999qp,Nemura:2004xb,Contessi:2019csf}, using for very light systems 
Faddeev-Yakubovsky equations \cite{Filikhin:2002wp}, 
and within cluster models \cite{Hiyama:1997ub,Hiyama:2018lgs,Filikhin:2003js,Filikhin:2002wm}. 
These calculations have mostly been performed with simplified interactions omitting 
$\Lambda \Lambda$-$\Sigma \Sigma$-$\Xi$N transitions. The common aim was to determine 
a reasonable strength of the underlying interactions such that all available 
hypernuclei could be described and use this to identify possible other bound states in $S=-2$. 
In Ref.~\cite{Contessi:2019csf}, a pionless EFT approach was used for the interaction 
and it was found that  $^{\ \ 5}_{\Lambda \Lambda}$He is most likely bound whereas  $^{\ \ 4}_{\Lambda \Lambda}$H 
is probably not particle stable against decay to $^3_{\Lambda}$H and $\Lambda$. Note that 
it is mostly the experimental binding energy of $^{\ \ 6}_{\Lambda \Lambda}$He that determines 
the strength of the $\Lambda \Lambda$ interaction in such a calculation, and that due to 
the Pauli principle only the spin singlet can contribute to the $S$-wave interaction. 
Ambiguities because of the spin dependence are therefore suppressed in this case. 

The first calculations based on realistic chiral interactions have been performed in 
Ref.~\cite{Le:2021wwz}. Commonly, one defines the so-called $\Lambda \Lambda$ excess binding energy
\begin{equation}
  \label{eq:lamlamexcess}
 \Delta B_{\Lambda \Lambda} (^{\ \ A}_{\Lambda \Lambda} {\rm X} )  =  B_{\Lambda \Lambda}(^{\ \ A}_{\Lambda \Lambda} {\rm X} ) - 2 B_{\Lambda}(^{A-1}_{\Lambda}{\rm X})\\
 = 2 E(^{A-1}_{\Lambda}{\rm X})  - E(^{\ \ A}_{\Lambda \Lambda} {\rm X})  - E(^{A-2}{\rm X})
 \end{equation}
as a difference of double $\Lambda$ separation energies $B_{\Lambda \Lambda}$ or binding energies 
$E(^{\ \ A}_{\Lambda \Lambda} {\rm X})$, the single  $\Lambda$ separation energies $B_{\Lambda}$ or binding energies 
$E(^{A-1}_{\Lambda}{\rm X}) $, and the energy of the core nucleus $E(^{A-2}{\rm X})$. It  
quantifies the contribution to the binding energy due to the $\Lambda \Lambda $ pair. 
In Ref.~\cite{Le:2021wwz}, the calculations have been performed neglecting all contributions of 3BFs. 
In order to still arrive at realistic values for the energies of these hypernuclei, the SRG evolution 
was done to specifical chosen $\lambda_{SRG}$ parameters that lead to a realistic description 
for the binding energy of the core nucleus and of the single $\Lambda$ core nuclei \cite{Le:2020zdu}. 
Specifically, for NN,
the SMS N4LO+(450) \cite{Reinert:2017usi} evolved to $\lambda_{\rm NN} = 1.6$~fm$^{-1}$ and, for YN, 
the NLO19(650) \cite{Haidenbauer:2019boi} evolved to $\lambda_{\rm YN} = 0.868$~fm$^{-1}$ was used.
For the $S=-2$ interaction, the LO(600) \cite{Polinder:2007mp}  and 
NLO(600) \cite{Haidenbauer:2015zqb} were employed. The parameters of these 
interactions are consistent with the scarce YY data that is available (see Sec.~\ref{subsec:LLXiN}).

\begin{figure}[t]
    \centering
    \includegraphics[width=0.45\linewidth]{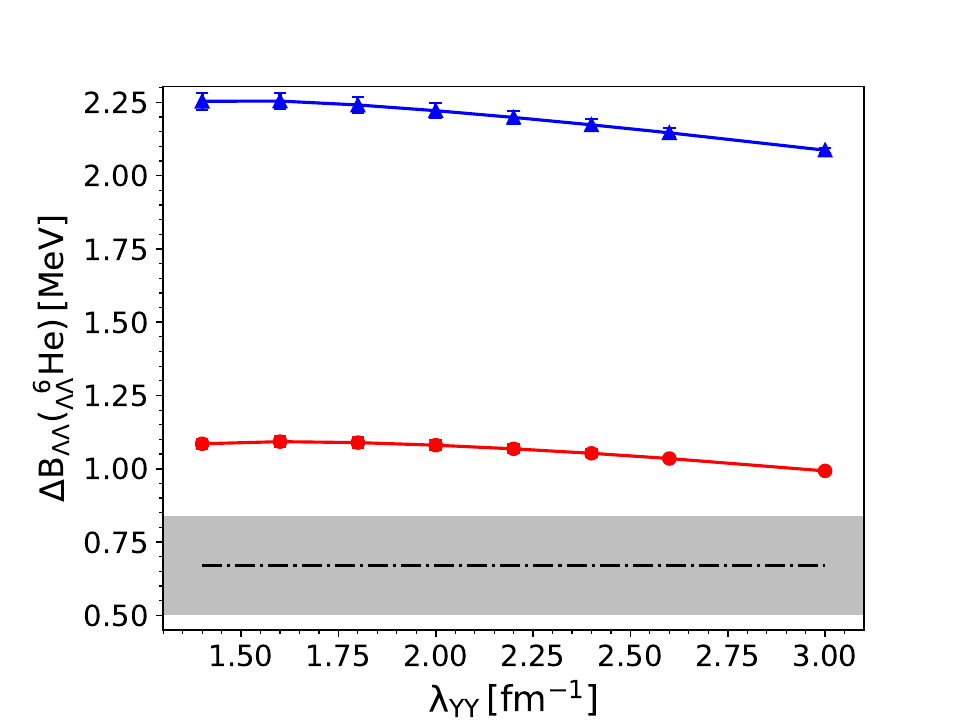} \hspace{0.5cm}
    \includegraphics[width=0.45\linewidth]{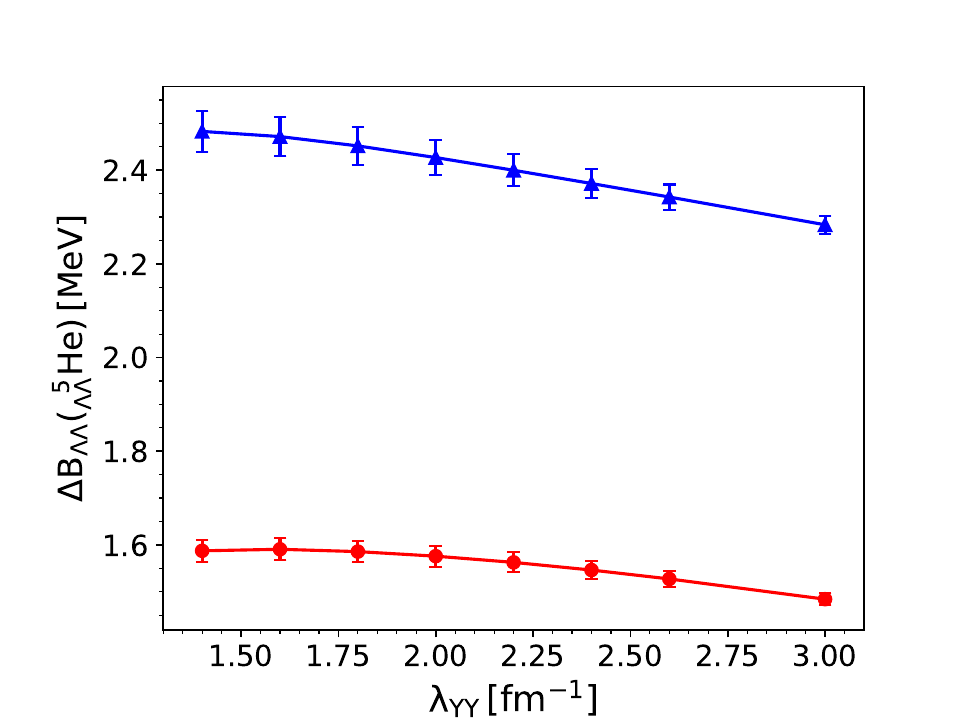}
    
    \caption{$\Lambda \Lambda$ excess energies $\Delta B_{\Lambda \Lambda}$ of $^{\ \ 6}_{\Lambda \Lambda}$He (left)  and $^{\ \ 5}_{\Lambda \Lambda}$He (right).
      depending on the SRG parameter of the YY interaction $\lambda_{YY}$. 
      Results for the LO(600) (blue triangles) and NLO(600) (red circles) YY interactions are compared. The experimental result for $^{\ \ 6}_{\Lambda \Lambda}$He is also shown (dashed dotted line and grey band) \cite{Takahashi:2001nm,Nakazawa:2010zza}.
      The SMS N4LO+(450) \cite{Reinert:2017usi} NN interaction evolved to $\lambda_{\rm NN} = 1.6$~fm$^{-1}$ and the  NLO19(650) \cite{Haidenbauer:2019boi} YN interaction evolved to $\lambda_{\rm YN} = 0.868$~fm$^{-1}$ was used. 
      Uncertainties of the theoretical calculations are due to extrapolation in model-space size.}
    \label{fig:excess56llHe}
\end{figure}

The left hand side of Fig.~\ref{fig:excess56llHe} shows the result for $^{\ \  6}_{\Lambda \Lambda}$He in comparison to 
the experimental result. Fortunately, the dependence on $\lambda_{YY}$ is small. 
LO and NLO energies are both more attractive than necessary to explain the data. Whereas the NLO 
one is still in fair agreement\footnote{Error bars are only due to the model-space extrapolation. The probably 
larger uncertainty due to the chiral expansion could not be quantified yet.}, 
the LO result is in clear deviation from the data. We stress that these results are pure predictions 
and that the energies of $^{\ \ 6}_{\Lambda \Lambda}$He have not been used to determine the YY force 
parameters.

Given that at least the NLO energy is quite realistic, it is now also interesting to predict 
energies of other $s$-shell double $\Lambda$ hypernuclei. The results for  $^{\ \ 5}_{\Lambda \Lambda}$He
are shown on the right hand side of Fig.~\ref{fig:excess56llHe}. Again the $\lambda_{YY}$ dependence is mild. 
First of all, the positive excess energy for LO and NLO indicate that this system is probably 
bound so that experimental searches seem to be promising. The results also show a significant increase 
of the excess energy from approximately $1$~MeV to $1.5$~MeV when going from $A=6$ to $A=5$ for the more realistic 
NLO interaction. It is conceivable 
that this increase is a signature of the transition to $\Xi$N which should be suppressed because of 
isospin conservation in $^{\ \ 6}_{\Lambda \Lambda}$He. An experimental determination of the energy for $A=5$
will therefore provide new and independent information on the YY interaction. 

\begin{figure}[t]
    \centering
    \includegraphics[width=0.45\linewidth]{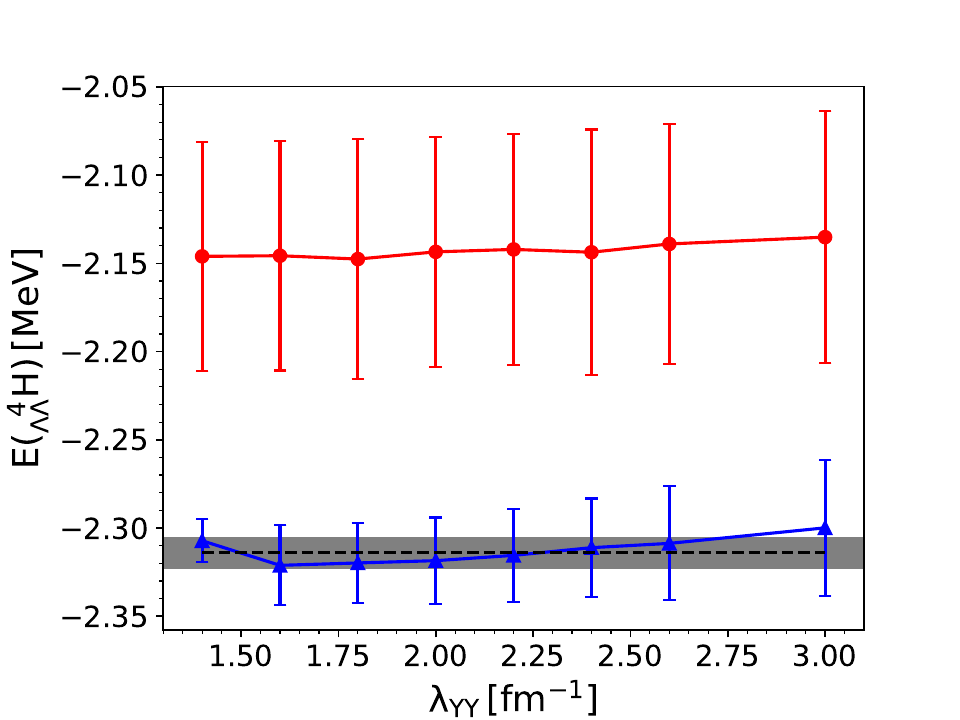} \hspace{0.5cm}
    \includegraphics[width=0.45\linewidth]{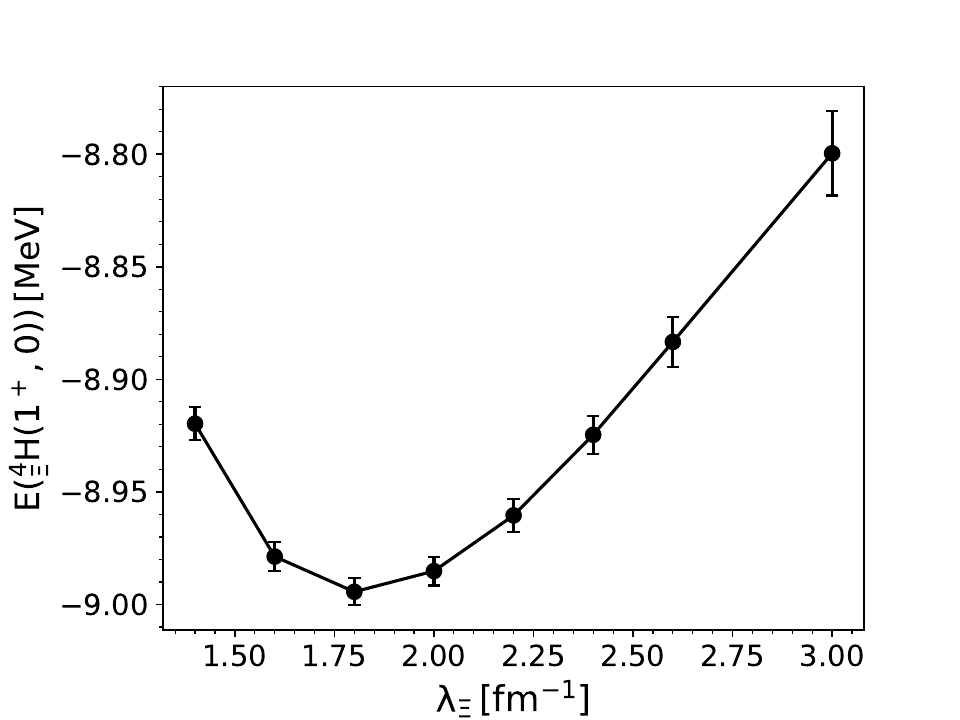}
    
    \caption{Binding energies $E$  of $^{\ \ 4}_{\Lambda \Lambda}$H (left) and $^{4}_{\Xi}\mathrm{H}(1^+,0))$ (right)
      depending on the SRG parameter of the YY ($\Xi$N) interaction $\lambda_{YY}$ ($\lambda_{\Xi}$). The SMS N4LO+(450) \cite{Reinert:2017usi} 
      NN interaction evolved to $\lambda_{\rm NN} = 1.6$~fm$^{-1}$ and the  NLO19(650) \cite{Haidenbauer:2019boi} YN 
      interaction evolved to $\lambda_{\rm YN} = 0.868$~fm$^{-1}$ was used. On the left, results for the LO(600) (blue triangles) 
      and NLO(600) (red circles) YY interactions are compared to the experimental binding energy 
      of $^3_\Lambda$H (dashed  line and grey band) \cite{Juric:1973zq}. On the right, the 
      results for the NLO $\Xi$N interaction was used (black circles and line). 
      Uncertainties of the theoretical calculations are due to extrapolation in model-space size.}
    \label{fig:srgdepLLXiN}
\end{figure}

Going further, we show energy results for $^{\ \ 4}_{\Lambda \Lambda}$H on the left-hand side of  Fig.~\ref{fig:srgdepLLXiN}. 
Energies below the hypertriton energy shown as a grey band imply a bound $A=4$ double 
$\Lambda$ hypernucleus. The LO result is close to this threshold, however, the more realistic NLO 
results are clearly above the threshold making a bound $A=4$ system unlikely. 

In summary, we find that realistic chiral $S=-2$ interactions lead to a reasonable description of 
the $A=6$ system and predict that an $A=5$ double $\Lambda$ hypernucleus likely exists. An experimental 
confirmation of this expectation will be most welcome in view of the possible insight into the 
importance for $\Lambda \Lambda$-$\Sigma \Sigma$-$\Xi$N conversion.

\subsection{\texorpdfstring{$\Xi$}{Cascade}-hypernuclei as a probe of the \texorpdfstring{$\Xi$N}{Cascade-N} interaction}
\label{sec:xihyp}

The $\Xi$N interaction is strongly linked to the $\Lambda \Lambda$ interaction 
because of the importance of $\Lambda \Lambda$-$\Sigma \Sigma$-$\Xi$N conversion. 
In the last section, we have therefore considered it as a byproduct of the $\Lambda\Lambda$
case. 
But the interaction itself is also interesting because several different $S$-wave 
channels contribute. We will see below that 
the different states of the light $\Xi$-hypernuclei reflect the freedom of 
spin and isospin in the $\Xi$N channels. 
Moreover, the $\Xi$N interaction could have direct impact on the EOS and the 
properties of neutron stars \cite{Schaffner-Bielich:2020psc}. While in some
scenarios $\Xi$ hyperons appear only at fairly large density \cite{Vidana:2024ngv}, 
others suggest a very early onset of the appearance of 
$\Xi$´s (specifically of the $\Xi^-$) in 
nuclear matter \cite{Weissenborn:2011ut,Oertel:2014qza,Ofengeim:2019fjy}. 
As discussed in Sec.~\ref{subsec:LLXiN}, 
only very scarce experimental information on the strength of the interaction is available 
and the lattice QCD simulations for this interaction still bear large uncertainties. 
In fact, the original version of the chiral $S=-2$ potential \cite{Haidenbauer:2015zqb} leads to a repulsive 
interaction in the $^{(2I+1)(2S+1)}L_J=^{\, 33\!\!  }{\rm S}_{1}$ partial wave and was too repulsive 
to explain observations of $\Xi$-hypernuclei. In Ref.~\cite{Haidenbauer:2018gvg}, the potential 
was therefore appropriately readjusted while keeping it consistent with the available 
empirical constraints on the $\Xi$N interaction. Our calculations below use this updated potential.

In order to further constrain the LECs, more detailed experimental information on $\Xi$-hypernuclei are 
highly desirable. 
Due to the strong conversion process to $\Lambda \Lambda$, their identification 
is not so easy since a decay to $\Lambda \Lambda$ can be expected. Older experimental searches 
were inconclusive only providing hints \cite{E224:1998uzu,AGSE885:1999erv}, but in recent years 
some evidence for a $^{13}_{\, \Xi}$B bound state 
was found \cite{KEKE176:2009jzw} and, in the reaction $^{12}{\rm C}(K^-,K^+)$, also for $^{12}_{\, \Xi}$Be \cite{Ichikawa:2024fjf}.
In emulsion experiments several 
events of a possible $^{15}_{\, \Xi}$C bound state have been identified 
\cite{Nakazawa:2015joa,J-PARCE07:2020xbm,Yoshimoto:2021ljs} which 
seem to stem from different states of this hypernucleus. 
Unfortunately, some of the events found are not
fully consistent with each other, see the overviews in 
Refs.~\cite{Yoshimoto:2021ljs,Nakazawa:2023xad}, 
triggering a discussion on the correct assignment 
and its impact on the strength of the interaction \cite{Friedman:2022huy}.

Theoretically, the $\Xi$-hypernuclei have been studied within mean field approaches 
\cite{Hiyama:2018lgs,Jin:2019sqc,Friedman:2022huy,Tanimura:2022ahi,Ding:2024gdv} 
and a cluster approximation \cite{Hiyama:2008fq}. 
Such calculations help to clarify the assignment of the states although an 
unambiguous result has not been reached yet. Unfortunately, a direct calculation within the J-NCSM 
is not possible at this time because the interesting $\Xi$-hypernucleus state 
will be converted into a $\Lambda \Lambda$ state so that the calculations would then
correspond 
to the related double $\Lambda$ states. Fortunately, it turns out that the 
$\Xi$N-$\Lambda \Lambda$ transition is accidentally weak for chiral interactions \cite{Haidenbauer:2015zqb,Haidenbauer:2018gvg} 
and in lattice QCD simulations \cite{HALQCD:2019wsz}. This justifies removing the $\Lambda \Lambda$ channel and 
readjusting the $\Xi$N interaction so that all observables not involving  the $\Lambda \Lambda$ states 
are well reproduced. The transition matrix elements 
of the interaction can then be included perturbatively for a width estimate \cite{Hiyama:2008fq,Hiyama:2019kpw}. 

$\Xi$-nucleus bound states have already been studied based on the ESC08c \cite{Nagels:2015slg} potential. 
For this interaction several $A=4$ bound states have been found. Based on the HAL 
QCD potential only the state with angular momentum, parity and isospin $(J^\pi,T)=(1^+,0)$ 
could be confirmed \cite{Hiyama:2019kpw}. The $\Xi$NN system was also investigated for chiral interactions
and it was found that the $A=3$ $\Xi$-hypernuclei are not bound \cite{Miyagawa:2021krh}. 

\begin{figure}[t]
    \centering
    \includegraphics[width=0.7\linewidth]{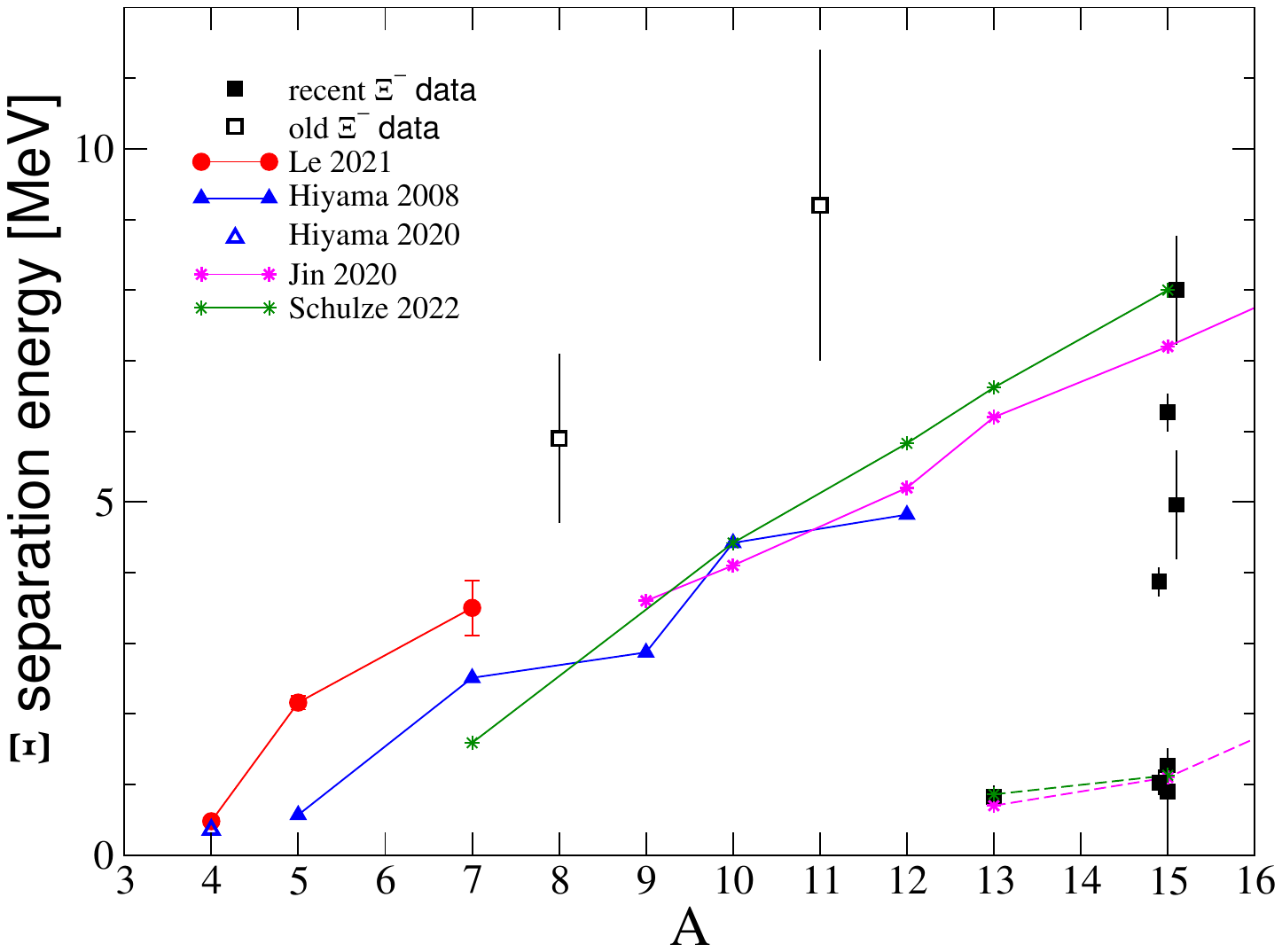}
    \caption{$\Xi$ hypernuclei: past and present experiments, predictions from theory.  
Shown are theoretical results by Jin et al.~\cite{Jin:2019sqc} and Schulze et al.
\cite{Schulze:2022zaj} from a Skyrme model (stars),
by Hiyama et al., obtained in a cluster model
\cite{Hiyama:2008fq} and from a variational calculation \cite{Hiyama:2019kpw} (filled and open
triangles), and the NCSM results by Le et al.~\cite{Le:2021gxa} (circles). 
Recent data for $^{15}_{\Xi}$C~\cite{J-PARCE07:2020xbm,Yoshimoto:2021ljs} and 
$^{13}_{\Xi}$B~\cite{KEKE176:2009jzw} are indicated by filled squares. Older but 
inconclusive results from the past \cite{Dover:1982ng} are shown as open squares.
}
    \label{fig:Xi}
\end{figure}

In Ref.~\cite{Le:2021gxa}, $A=4-7$ $\Xi$ hypernuclei have been investigated based on the modified 
chiral interactions. For the J-NCSM calculation, the transition to the 
$\Lambda \Lambda$ channels has been removed and the LECs of the $\Xi$N $^{11}{\rm S}_0$ 
partial wave have been readjusted so that the $\Xi$N data is described as in the original 
interaction. The conversion to the $\Sigma \Sigma$ channel has not been changed since the system cannot 
decay into this channel. As usual, the interactions are SRG-evolved in order to improve the 
convergence. On the right hand side of Fig.~\ref{fig:srgdepLLXiN}, the $\Xi$ separation energy 
$B_{\Xi}( ^{4}_{\Xi}\mathrm{H}(1^+,0))$ of the $(J^\pi,T)=(1^+,0)$ state is shown for given 
SRG parameters of the $NN$ and $YN$ interaction. Compared to $\Lambda \Lambda$ states, 
the SRG-dependence is more visible but it is still small enough to gain insight into possibly 
bound states. For a detailed quantitative analysis in the future, the evolution probably needs to be 
performed up to the $A=3$ level. Based on a specific choice of the SRG parameters 
($\lambda_{\rm NN}=\lambda_{YY}=1.6$~fm$^{-1}$), the explicit calculations confirm that it is likely 
that several $\Xi$NNN bound states exist and that their width is often small. Ref.~\cite{Le:2021gxa}
also gave explicit analytic expressions for the contribution of $s$-waves channels to the 
different states in an $s$-wave approximation. Clearly, the different $\Xi$N partial wave contribute very 
differently to the various $s$-shell $\Xi$-hypernuclei. The pattern for $\Xi$NNN bound states is therefore 
valuable to disentangle the spin dependence of the $\Xi$N interaction. Ref.~\cite{Le:2021gxa} also 
gives the contribution of the different $S$-wave partial waves to the binding energy of the 
various $A=4$ states and $^5_{\Xi}$H and $^7_{\Xi}$H. Also for these light systems, an important 
contribution to the binding is coming from the  $^{33}{\rm S}_1$ partial wave. 

It is interesting that the chiral 
interactions also led to several $A=4$ bound states although its strength is often very similar to the 
HAL QCD potential. The explicit calculation also showed that $^5_{\Xi}$H is bound and its width is rather 
small. This nucleus cannot provide much insight into the spin dependence since the quantum 
numbers of the $\alpha$ particle core imply that there is likely only one state 
bound. However, the large density of the $\alpha$ particle is a good example to $\Xi$ binding 
with a denser environment. Finally, Ref.~\cite{Le:2021gxa} also provides a prediction of the 
binding energy of $^7_{\Xi}$H. This hypernucleus is of special interest  because one expect soon 
some experimental results \cite{Fujioka:2019prop} and also because it promises 
to test the strength of the neutron-$\Xi$ interaction given the large number of neutrons. 

Clearly, one can expect several light $\Xi$-hypernuclear bound states. Experimental data 
of these states can provide valuable information on the spin and isospin dependence of the 
$\Xi$N interaction. 


In order to provide a general overview of the situation regarding light $\Xi$ hypernuclei,
in Fig.~\ref{fig:Xi} we summarize data from past and present experiments, and 
results from different theoretical approaches. 
Besides the predictions from Ref.~\cite{Le:2021gxa}, indicated by circles, we show the result of 
the variational calculation of Hiyama et al. \cite{Hiyama:2019kpw} (open triangle).  
The latter is based on an interaction derived 
by the HAL QCD collaboration from lattice simulations
close to the physical point \cite{HALQCD:2019wsz}.
However, in the actual calculations the various 
YY channels of the original HAL QCD potential
are renormalized into an effective $\Xi$N interaction and the latter is then treated within the so-called Gaussian expansion method.
 Earlier results for $\Xi$ hypernuclei with $A=5-12$ by Hiyama et al. were obtained within a cluster model \cite{Hiyama:2008fq} (filled triangles). 
 In this case the underlying effective $\Xi$N 
 interaction was fixed to the single-particle
 potential of $U_\Xi = -14$~MeV, as suggested by the
 analysis in Ref.~\cite{AGSE885:1999erv}. 
Regarding mean-field calculations we show exemplary two Skyrme model results by 
Jin et al.~\cite{Jin:2019sqc} and Schulze et al. \cite{Schulze:2022zaj} (stars).
The former calculation (model SLX3) is rooted on the value of 
$B_\Xi = 1.11 \pm 0.25$ MeV, interpreted as $p$-state
of the hypernucleus $^{15}_{\Xi}$C, 
the so-called Kiso event \cite{Nakazawa:2015joa}. In the latter case the
model parameters have been fixed to produce the $\Xi^-$ removal energies for the
$s$- and $p$-states of $^{15}_{\Xi}$C of $8$ and $1.13$~MeV, respectively. 
Note that similar predictions were reported 
by Tanimura et al. within a relativistic mean field approach \cite{Tanimura:2022ahi} and 
for the density-dependent relativistic mean-field model of Ding et al.~\cite{Ding:2024gdv}. 
As far as experimental results are concerned we include here 
values for $^{15}_{\Xi}$C reported in Refs.~\cite{Yoshimoto:2021ljs}
and \cite{J-PARCE07:2020xbm},
and for $^{13}_{\Xi}$B from Ref.~\cite{KEKE176:2009jzw}
(filled squares). 
For purely illustrative reasons we show also binding energies of some 
inconclusive $\Xi$ hypernuclei reported in the past (open squares) \cite{Dover:1982ng}.

One can see from Fig.~\ref{fig:Xi} that the
overall trend of 
the predictions based on the chiral $S=-2$ 
interaction, obtained within a microscopic approach,
is roughly in line with results for heavier 
$\Xi$ hypernuclei calculated in a phenomenological
way. Nonetheless, one should keep in mind that there
might be certainly some overestimation of the separation 
energies, given that induced three-body forces
that result from the SRG transformation have not
yet been properly implemented. 


\section{Nuclear lattice effective field theory}\label{sec:fifth}

So far, we have considered various continuum approaches combined with forces from chiral EFT.
An alternative is given by Nuclear Lattice Effective Field Theory (NLEFT), where the
Euclidean space-time is discretized and the world is represented by a finite box,
see Refs.~\cite{Lee:2008fa,Lahde:2019npb} for detailed expositions. Recent
results are summarized in~\cite{Lee:2025req}. Here, we only discuss the basic
features of NLEFT and focus on its extension to hypernuclei.

\subsection{Basics of NLEFT}

\begin{figure}[ht]
\centering
\includegraphics[scale=0.35]{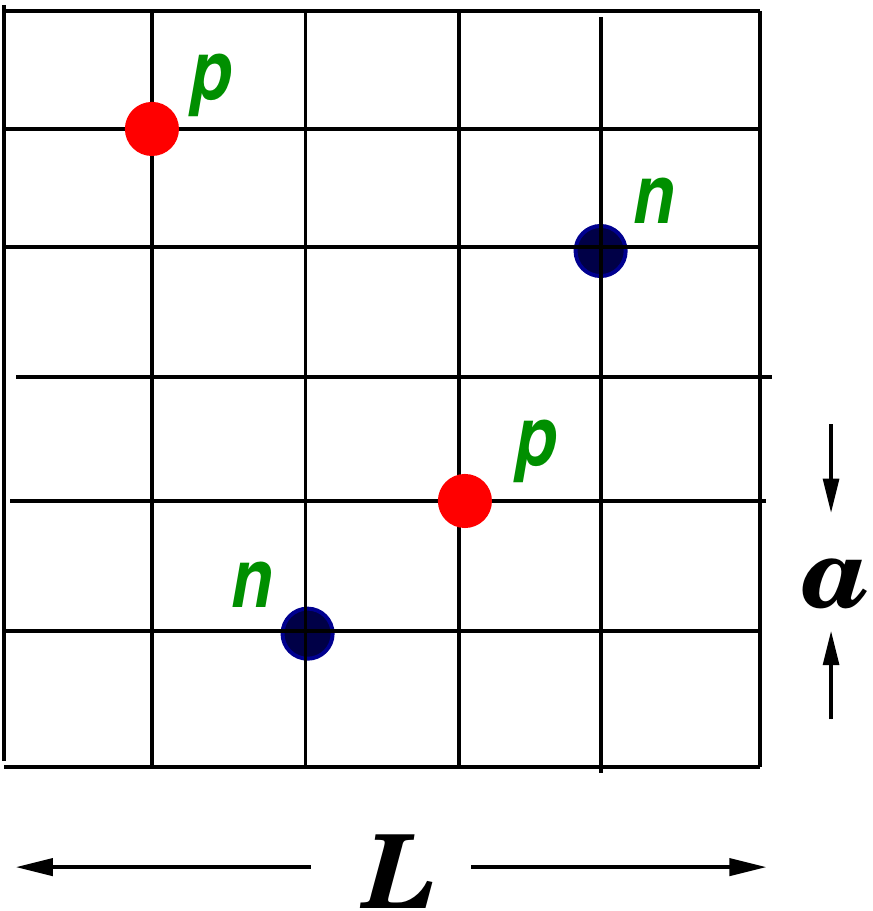}
\caption{
Sketch of the lattice used in NLEFT. For details, see the text.
\label{fig:NuclLatt}
}
\end{figure}

Let us briefly summarize the main ingredients of NLEFT as a novel method to tackle the
nuclear few- as well as the many-body problem:
\begin{itemize}
\item Space-time is discretized on a hypercubic lattice with the volume
  $V =  L \times L  \times L \times L_t$, where $L$ is the extension in the spatial
  directions and $L_t$ the extension in Euclidean time, see Fig.~\ref{fig:NuclLatt}.
  $L$ is usually taken sufficiently large so as to avoid finite volume corrections,
  say $L =10$ (in lattice units given by the lattice spacing $a$),  whereas $L_t$ in principle tends to
  infinity but in practice is taken large.
\item The box is further characterized by finite lattice spacings $a$ and $a_t$ in the
  spatial and temporal directions, respectively. In contrast to lattice QCD,
  $a$ is kept finite and is varied between 1 and 2~fm, corresponding to a fine and
  a coarse lattice, respectively. A finite $a$ entails a maximum momentum,
  \begin{equation}
   p_{\rm max} = \frac{\pi}{a} \simeq 315-630\,{\rm MeV}~,
  \end{equation}
  for $a= 2-1$~fm. Thus,  $1/a$ serves as an UV cutoff and defines the EFT on the lattice. 
  Quite differently, $a_t$ is taken very small to
  ensure convergence as $L_t \to \infty$, typically $a_t= 0.001$~MeV$^{-1}$.
\item The lattice spacing $a$ is bounded from below by the nucleon size, as the EFT can not
  resolve the nucleon structure. Also, it is bounded from above by not exceeding too much
  the average distance of nucleons in nuclei. In nuclear matter ($\rho = \rho_0 = 0.17$~fm$^{-3}$), 
  the average distance between nucleons is $d\simeq 1.8\,$fm. Still, physics should be independent of
  the choice of $a$, which is indeed the case if one goes to sufficiently high orders
  in the chiral expansion~\cite{Alarcon:2017zcv,Klein:2018lqz}.  
\item Nucleons are placed on the sites as depicted in Fig.~\ref{fig:NuclLatt}. Their
  interactions are given by a discretized version of the chiral NN and 3N potentials
  and the Coulomb interaction between protons can easily be included~\cite{Epelbaum:2010xt}.
  Note further that the masses of the nucleons (and of the pions)  are given by their physical values.
\item Simulations of many-baryon systems are usually plagued by sign oscillations,
  induced by the finite density (chemical potential). These pose a severe problem to lattice
  QCD calculations of the phase diagram of strongly interacting matter or 
  calculations of atomic nuclei. In the
  case of NLEFT, these sign oscillations are strongly suppressed due to the approximate
  Wigner SU(4) (spin-isospin)  symmetry of the nuclear interactions, see~\cite{Wigner:1936dx,Mehen:1999qs,Chen:2004rq}.
  In fact, Wigner's SU(4) symmetry can be exploited to construct a minimal nuclear interaction,
  that can reproduce the ground state properties of light nuclei, medium-mass nuclei, and neutron matter
  simultaneously with no more than a few percent error in the energies and charge radii,
  requiring only four parameters~\cite{Lu:2018bat}. Later, we will see how such a
  framework can be exploited in hypernuclear calculations.
\item The central object of NLEFT is the 
   correlation function for $A$ nucleons,
  \begin{equation}
   Z_A(\tau) = \langle \Psi_A | \exp(-\tau H)|\Psi_A\rangle~,
  \end{equation}
  with $\Psi_A$ a Slater determinant for $A$ free nucleons or a more sophisticated (correlated)
  initial/final state with given spin and parity, like e.g. $J^P=0^+$ for the $^{12}$C ground state (g.s.).
  Here, $\tau$ is the Euclidean time and $H$ is the Hamiltonian which is expanded in terms of small momenta and pion masses (chiral expansion, with the small parameters collectively denoted as $Q$).  From this, one calculates the transient energy $E_A(\tau) = -{d}\ln Z_A (\tau)/d\tau$,
  which allows one to extract the g.s. energy via:
  \begin{equation}
   E_A^0 = \lim\limits_{\tau \to \infty} E_A(\tau)~.
  \end{equation}
  In practical calculations, $\tau$ is limited due to sign oscillations, but for large enough
  values of $\tau$ one can fit to asymptotic formulas. Excited states can also be extracted,
  in this case $Z_A$ becomes a matrix, which upon diagonalization delivers the first, second,
  third, ... state with given quantum numbers of spin and parity, like the tower of
  $0^+$ states in the  $^{12}$C spectrum. The expectation value of any normal-ordered operator
  can be calculated in a similar fashion by inserting the operator at mid-times.

\item Smearing is an important tool to enhance the simulation signals. In NLEFT, both local
  and non-local smearings, corresponding to velocity-independent and velocity-dependent
  interactions, are employed. Local smearing is performed on the level of the nucleon
  densities, such as for the spin- and isospin-independent density $\rho(\vec{n})$,
  \begin{equation}
  {\rho}(\vec{n})= \sum_{i,j=0,1}
  {a}^{\dagger}_{i,j}(\vec{n})  \, {a}^{\,}_{i,j^{\prime}}(\vec{n})
  + s_{\rm L} \sum_{|\vec{n}-\vec{n}^{\prime}|^2 = 1} \,
  \sum_{i,j=0,1}  {a}^{\dagger}_{i,j}(\vec{n}^{\prime}) \, {a}^{\,}_{i,j}(\vec{n}^{\prime})\,,
  \end{equation}
  with $s_{\rm L}$ the local smearing parameter, $\vec{n}$, $\vec{n}^{\prime}$ represent points on the
  lattice with  spin $i = 0, 1$ (up, down) and isospin $j = 0, 1$ (proton, neutron) indices.
  The non-local smearing is done on the level of the nucleon creation/annihilation
  operators,
  \begin{equation}
  \tilde{a}_{i,j}^{(\dagger)}(\vec{n})=a_{i,j}^{(\dagger)}(\vec{n})+s_{\rm NL}\sum_{|\vec{n}^{\prime}-\vec{n}|=1}a_{i,j}^{(\dagger)}(\vec{n}^{\prime})\,,
  \end{equation}
  with $s_{\rm NL}$ the non-local smearing parameter. In few-body systems, these two types of smearing
  can not be disentangled, quite differently to medium-mass and heavy nuclei (as discussed below). Note further
  that through this smearing, one effectively accounts for a number of higher-order operators, as distances can be translated into derivatives on the lattice.
\item An important ingredient in NLEFT simulations is the representation of the interactions in
  terms of auxiliary fields, such as
  \begin{equation}
    \exp\left[-\frac{C}{2}\left(N^\dagger N\right)^2 \right]
    = \sqrt{\frac{1}{2\pi}}\,\int^{+\infty}_{-\infty} ds \exp \left[-\frac{s^2}{2} + \sqrt{C} \, \,s \left(N^\dagger N\right)\right]~,
  \end{equation}
  with $s$ an auxiliary field that couples to the nucleon density, $\rho = N^\dagger N$.
  Similarly, spin- and isospin-dependent
  multi-nucleon operators and pion exchanges can be represented by labelled auxiliary fields. 
  This makes NLEFT optimally suited for  parallel computing as shown in Fig.~\ref{fig:auxfield}.
  \begin{figure}[h]
  \centering
  \includegraphics[scale=0.5]{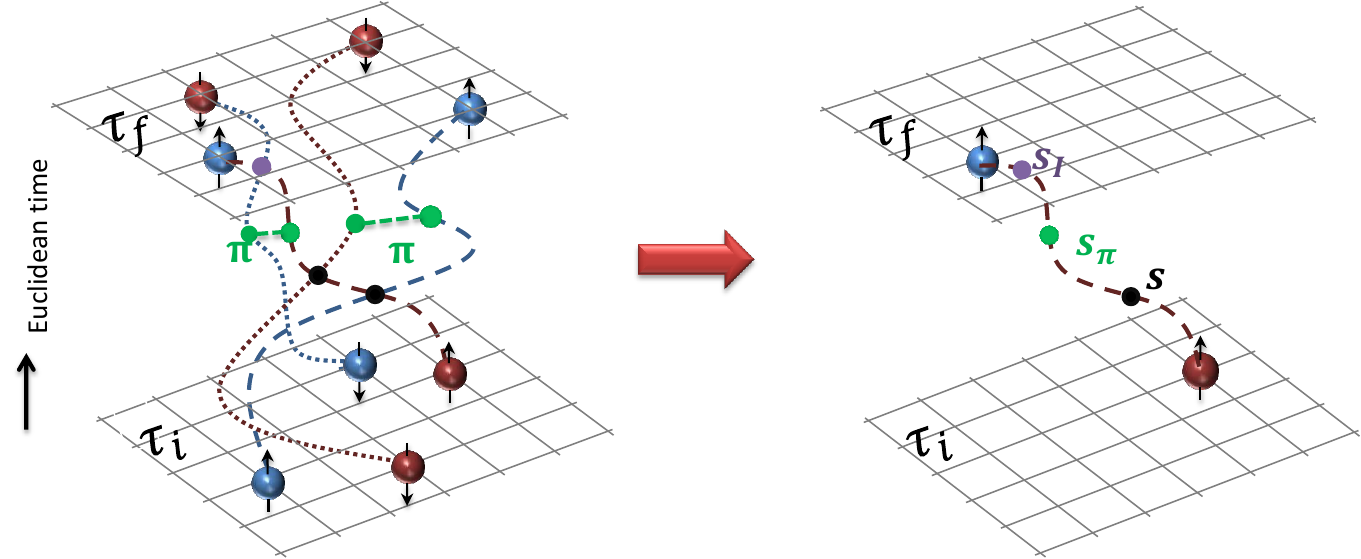}
  \caption{
    Simulation of a $^4$He nucleus (left panel) that decomposes into four independent worldline simulations
    due to the use of auxiliary fields (right panel). In the left panel, explicit interactions (pion exchanges
    and contact terms) are shown. In the right panel, these interactions are represented by auxiliary field insertions
    on the nucleon worldline.
    Figure courtesy of Serdar Elhatisari.
    \label{fig:auxfield}}
  \end{figure}
\item By construction, all possible initial/final states are considered in NLEFT. Often, one starts
  with free waves subject to the lattice boundary conditions and consistent with the quantum
  numbers of the state under consideration. However, different initial states can be constructed.
  To be specific, consider the $^4$He nucleus. These initial states can consist
  of four nucleons on different lattice sites, but also doublets, triplets and even quartets, the
  latter consisting of two protons and two neutrons, each pair with one spin up and one spin down
  as allowed by the Pauli principle. Thus,  NLEFT has the advantage of capturing  the
  full set of many-body correlations, or stated differently, clustering emerges naturally.
  Furthermore, one can also work in a basis given by shell  model states, as detailed in~\cite{Shen:2021kqr,Shen:2022bak}.
\item Another important ingredient to achieve the required high precision is the recently developed
  method called {\em wavefunction matching}~\cite{Elhatisari:2022zrb}. Despite the approximate
  Wigner SU(4) symmetry, NLEFT suffers from
  severe sign problems in its application to high-fidelity chiral forces (beyond N2LO) described by the chiral
  Hamiltonian $H_\chi$. In the wavefunction matching method, the unitarily transformed $H_\chi' = UH_\chi^{} U$ is mapped onto
  a simple Hamiltonian $H_S$, where the $H_S$ is largely free of sign oscillations. This simple Hamiltonian
  consists of smeared two-nucleon SU(4) symmetric contact interactions as well as regularized one-pion exchange. If $H_\chi'$
  is sufficiently close to  $H_S$, first-order perturbation theory in $H_\chi'-H_S$ can be used efficiently to calculate the
  higher-order chiral forces up-to-and-including N3LO. Furthermore, smeared 3NFs are then fitted to masses
  of selected nuclei ranging from $A=3$ to $A=58$. Consequently, nuclear charge radii as well the EOS
  for neutron matter and for nuclear matter can be predicted, showing good agreement with the data. 
\end{itemize}  

\subsection{Lattice Hamiltonian}

Here, we briefly discuss the state-of-the-art lattice Hamiltonian for NN and 3N forces, that
forms the basis of the calculations of the nuclear core of the hypernuclei. This is based on the
high-fidelity chiral interaction at N3LO and  the quantum many-body approach, called wavefunction matching,
developed in Ref.~\cite{Elhatisari:2022zrb}. The Hamiltonian $H$ is,
\begin{equation}
H = T + V_{\rm OPE} + V_{\rm C} + V_{\rm 3N}^{\rm Q^3} + V_{\rm 2N}^{\rm Q^4} + W_{\rm 2N}^{\rm Q^4}\,,
\label{eq:H-N3LO}
\end{equation}
where $T$ is the kinetic energy term constructed using fast Fourier transforms to produce the exact dispersion relation
$E_N = p^2/(2m_{N})$, and $m_{N}=938.92$~MeV is the nucleon mass. The one-pion-exchange potential $V_{\rm OPE}$ is given in terms
of the regularization method from Ref.~\cite{Reinert:2017usi},
\begin{align}
V_{\rm OPE}  = - &   \frac{g_A^2}{8F^2_{\pi}}\ \, \sum_{{\bf n',n},S',S,I}
:\rho_{S',I\rm }^{(0)}(\vec{n}')f_{S',S}(\vec{n}'-\vec{n})   \rho_{S,I}^{(0)}(\vec{n}):   -C_{\pi} \, \frac{g_A^2}{8f^2_{\pi}} \sum_{{\bf n',n},S,I}
:\rho_{S,I}^{(0)}(\vec{n}')
f^{\pi}(\vec{n}'-\vec{n})
  \rho_{S,I}^{(0)}(\vec{n}):\,,
\label{eq:OPEP-full}
\end{align}
where $g_{A}=1.287$ the axial-vector coupling constant (adjusted to account for the Goldberger-Treiman
discrepancy~\cite{Fettes:1998ud}), $F_{\pi}=92.2$~MeV the pion decay constant, and $\rho_{SI}(\vec{n})$ is the spin-
and isospin-dependent  density operator,
\begin{align}
\rho^{(d)}_{S,I}(\vec{n}) = & \sum_{i,j,i^{\prime},j^{\prime}=0,1} a^{\dagger}_{i,j}(\vec{n}) \, [\sigma_{S}]_{ii^{\prime}} \,
[\tau_{I}]_{jj^{\prime}} \, a^{\,}_{i^{\prime},j^{\prime}}(\vec{n})
 + s_{\rm L}
\sum_{|\vec{n}-\vec{n}^{\,\prime}|^2 = 1}^d \,
\sum_{i,j,i^{\prime},j^{\prime}=0,1} a^{\dagger}_{i,j}(\vec{n}^{\,\prime}) \, [\sigma_{S}]_{ii^{\prime}} \, [\tau_{I}]_{jj^{\prime}} \, a^{\,}_{i^{\prime},j^{\prime}}(\vec{n}^{\,\prime}) \,,
\label{eqn:appx--005}
\end{align} with $\vec{\tau}, \vec{\sigma}$ the Pauli-(iso)spin matrices and annihilation (creation) operators $a^{}$ ($a^{\dagger}$).
Further, $f_{S',S}$ is the locally-regulated pion correlation function,
\begin{align}
f_{S',S}(\vec{n}'-\vec{n}) = &\frac{1}{L^3}\sum_{\vec{q}}
\frac{q_{S'}q_{S} \, e^{-i\vec{q}\cdot(\vec{n}'-\vec{n})-(\vec{q}^2+M^2_{\pi})/\Lambda_{\pi}^2}}{\vec{q}^2 + M_{\pi}^2} \,,
\end{align}
where $f^{\pi}$ is a local regulator defined in momentum space,
\begin{align}
f^{\pi}(\vec{n}'-\vec{n})
= &
\frac{1}{L^3}
\sum_{\vec{q}}
e^{-i\vec{q}\cdot(\vec{n}'-\vec{n})-(\vec{q}^2+M^2_{\pi})/\Lambda_{\pi}^2}\,,
\end{align}
with $\vec{q} = \vec{p}- \vec{p}^{\prime}$ the momentum transfer ($\vec{p}$ and $\vec{p}^{\prime}$ are the relative incoming
and outgoing momenta). In addition, $C_{\pi}$ is the coupling constant of the OPE counter term given by,
\begin{align}
C_{\pi} = & -
\frac{\Lambda_{\pi} (\Lambda_{\pi}^2-2M_{\pi}^{2}) + 2\sqrt{\pi} M_{\pi}^3\exp(M_{\pi}^2/\Lambda_{\pi}^{2}){\rm erfc}(M_{\pi}/\Lambda_{\pi})}
{3 \Lambda_{\pi}^3}\,,
\end{align}
with
\begin{equation}
  {\rm erfc}(x) = \frac{2}{\sqrt{\pi}}\,\int_x^\infty e^{-t^2} \, dt
\end{equation}  
the complementary error function, 
$\Lambda_{\pi}=300$~MeV the regulator parameter  and $M_{\pi}=134.98$~MeV the pion mass. Furthermore, $V_{\rm C}$
is the Coulomb interaction, $V_{\rm 3N}^{\rm Q^3}$ denotes the 3N potential, and $V_{\rm 2N}^{\rm Q^4}$ is the 2N short-range
interaction at N3LO. In addition,   $W_{\rm 2N}^{\rm Q^4}$ is the required 2N Galilean invariance restoration (GIR) interaction
at N3LO. For more details of the Coulomb interaction and the 2N short-range interactions, see Ref.~\cite{Li:2018ymw}.
The 3N interactions at ${\rm Q^3}$ consist of three different topologies. These are the  locally smeared contact interactions,
the OPE interaction with the locally smeared two-nucleon contact terms, and the two-pion exchange
potential~\cite{Friar:1998zt,Epelbaum:2002vt,Epelbaum:2009zsa}. Further, we have two additional SU(4) symmetric 3N potentials
denoted by $V_{c_{E}}^{(l)}$ and $V_{c_{E}}^{(t)}$. Therefore, the 3N interaction at ${\rm Q^3}$ is given by
\begin{align}
V_{\rm 3N}^{\rm Q^3} = V_{c_{E}}^{(0)} + V_{c_{E}}^{(1)} + V_{c_{E}}^{(2)} +   V_{c_{E}}^{(l)} + V_{c_{E}}^{(t)} +
V_{c_{D}}^{(0)} + V_{c_{D}}^{(1)} + V_{c_{D}}^{(2)} + V_{\rm 3N}^{\rm (TPE)}
\,.
\label{eqn:V_NNLO^3N--001}
\end{align}
The two-pion exchange potential  can be separated into  three parts,
\begin{align}
V_{\rm 3N}^{\rm (TPE1)}
= & \frac{c_{3}}{f_{\pi}^{2}}\,
\frac{g_{A}^{2}}{4 f_{\pi}^{2}}
\, \sum_{S,S^{\prime},S^{\prime\prime},I}
\sum_{\vec{n},\vec{n}^{\,\prime},\vec{n}^{\,\prime\prime}}	
\, : \, \rho_{S^{\prime},I}^{(0)}(\vec{n}^{\,\prime}) \,
f_{S^{\prime},S}(\vec{n}^{\,\prime}-\vec{n})
f_{S^{\prime\prime},S}(\vec{n}^{\,\prime\prime}-\vec{n})
\rho_{S^{\prime\prime},I}^{(0)}(\vec{n}^{\,\prime\prime}) \,
\rho^{(0)}(\vec{n})
\, : \,
\label{eqn:V_TPE1^3N--001}
\end{align}
\begin{align}
V_{\rm 3N}^{\rm (TPE2)}
= &
-\frac{2c_{1}}{f_{\pi}^{2}}\,
\frac{g_{A}^{2} \, M_{\pi}^{2}}{4 f_{\pi}^{2}}
\, \sum_{S,S^{\prime},I}
\sum_{\vec{n},\vec{n}^{\,\prime},\vec{n}^{\,\prime\prime}}	
\, : \, \rho_{S^{\prime},I}^{(0)}(\vec{n}^{\,\prime}) \,
f_{S^{\prime}}^{\pi\pi}(\vec{n}^{\,\prime}-\vec{n})
f_{S}^{\pi\pi}(\vec{n}^{\,\prime\prime}-\vec{n})
\rho_{S,I}^{(0)}(\vec{n}^{\,\prime\prime}) \,
\rho^{(0)}(\vec{n})
\, : \, \,,
\label{eqn:V_TPE2^3N--001}
\end{align}
\begin{align}
V_{\rm 3N}^{\rm (TPE3)}
=   \frac{c_{4}}{2f_{\pi}^{2}}
&
\left( \frac{g_{A}}{2 f_{\pi}}\right)^{2}
\sum_{S_{1},S_{2},S_{3}}
\sum_{I_{1},I_{2},I_{3}}
\sum_{S^{\prime},S^{\prime\prime}}
\sum_{\vec{n},\vec{n}^{\,\prime},\vec{n}^{\,\prime\prime}}
\varepsilon_{S_1,S_2,S_3}
\varepsilon_{I_1,I_2,I_3}
\nonumber \\
&
\times	
\, : \, \rho_{S^{\prime},I_{1}}^{(0)}(\vec{n}^{\,\prime}) \,
f_{S^{\prime},S_{1}}(\vec{n}^{\,\prime}-\vec{n})
f_{S^{\prime\prime},S_{2}}(\vec{n}^{\,\prime\prime}-\vec{n})
\rho_{S^{\prime\prime},I_{2}}^{(0)}(\vec{n}^{\,\prime\prime}) \,
\rho_{S_{3},I_{3}}^{(0)}(\vec{n})
\, : \, \,,
\label{eqn:V_TPE3^3N--001}
\end{align}
where the locally smeared spin-isospin symmetric density operator is defined as, \begin{align}
\rho^{(d)}(\vec{n}) = \sum_{i,j=0,1} a^{\dagger}_{i,j}(\vec{n}) \, a^{\,}_{i,j}(\vec{n})
+
s_{\rm L}
\sum_{|\vec{n}-\vec{n}^{\,\prime}|^2 = 1}^d \,
\sum_{i,j=0,1} a^{\dagger}_{i,j}(\vec{n}^{\,\prime}) \, a^{\,}_{i,j}(\vec{n}^{\,\prime})
\,,
\label{eqn:appx--001}
\end{align}
and the LECs of two-pion exchange potentials are fixed from pion--nucleon scattering data,
$c_{1}=-1.10(3)$, $c_{3}=-5.54(6)$ and $c_{4}=4.17(4)$ all in GeV$^{-1}$~\cite{Hoferichter:2015tha}. Next, the OPE
interaction with the locally smeared 2N contact terms is given by,
\begin{align}
V_{c_{D}}^{(d)}
= -\frac{c_{D}^{(d)} \, g_{A}}{4f_{\pi}^{4} \Lambda_{\chi}}\,  \sum_{\vec{n},S,I}
\sum_{\vec{n}^{\,\prime},S^{\prime}}
\, : \,
\rho^{(0)}_{S^{\prime},I}(\vec{n}^{\,\prime})
f_{S^{\prime},S}(\vec{n}^{\,\prime}-\vec{n})
\rho^{(d)}_{S,I}(\vec{n})
\rho^{(d)}(\vec{n})
\, : \,
\,.
\label{eqn:V_cD-sL-001}
\end{align}
A graphical representation of the first three terms with $d=0,1,2$ is given in Fig.~\ref{fig:cDsmear}.
\begin{figure}[htb!]
\centering
\includegraphics[width=0.2\textwidth]{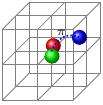}\quad\quad
\includegraphics[width=0.2\textwidth]{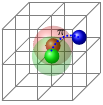}\quad\quad
\includegraphics[width=0.2\textwidth]{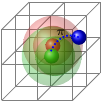}
\caption{
  Smearing of the LEC $c_D$. {\bf Left panel:} Non-smeared case $d=0$.
  {\bf Middle panel:} Smearing with $d=1$. {\bf Right panel:} Smearing with $d=2$.
  Note that $d=0,1,2$ refers to the same, the neighbouring and the next-to-neighbouring 
  lattice point, in order.
\label{fig:cDsmear}
}
\end{figure}

Furthermore, the locally smeared contact interactions take the form,
\begin{align}
V_{c_{E}}^{(d)}= \frac{c_{E}^{(d)}}{6}
\, \sum_{\vec{n},\vec{n}^{\,\prime},\vec{n}^{\,\prime\prime}}
\,
\left[\rho^{(d)}(\vec{n}) \right] ^3\,,
\label{eqn:V_cE-sL-001}      \end{align}
and finally two additional SU(4) symmetric potentials denoted by $V_{c_{E}}^{(l)}$ and $V_{c_{E}}^{(t)}$ are defined as,
\begin{align}
V_{c_{E}}^{(l)} = c_{E}^{(l)} \, \sum_{\vec{n},\vec{n}^{\,\prime},\vec{n}^{\,\prime\prime}}
\rho^{(d)}(\vec{n})\,
\rho^{(d)}(\vec{n}^{\,\prime}) \, \rho^{(d)}(\vec{n}^{\,\prime\prime}) \delta_{|\vec{n}-\vec{n}^{\,\prime}|^2,1} \, \,
\delta_{|\vec{n}-\vec{n}^{\,\prime\prime}|^2,1} \, \,  \delta_{|\vec{n}^{\,\prime}-\vec{n}^{\,\prime\prime}|^2,4},  \label{eqn:V_cE-l-001}
\end{align}
\begin{align}
V_{c_{E}}^{(t)} = c_{E}^{(t)}
\, \sum_{\vec{n},\vec{n}^{\,\prime},\vec{n}^{\,\prime\prime}}
\rho^{(d)}(\vec{n})\,
\rho^{(d)}(\vec{n}^{\,\prime}) \, \rho^{(d)}(\vec{n}^{\,\prime\prime}) \delta_{|\vec{n}-\vec{n}^{\,\prime}|^2,2} \, \,  \delta_{|\vec{n}-\vec{n}^{\,\prime\prime}|^2,2} \, \,  \delta_{|\vec{n}^{\,\prime}-\vec{n}^{\,\prime\prime}|^2,2}\,.
\label{eqn:V_cE-t-001}      \end{align}
It is important to stress that in contrast to the
continuum case, where we just have two LECs, namely $c_E$ and $c_D$, these are smeared here over
neighbouring lattice sites and appear with independent LECs $c_{D,E}^{(0)}, c_{D,E}^{(1)}, c_{D,E}^{(2)},...\,$. As noted before, this smearing effectively emulates a number of
higher-order interactions involving derivatives.
In lattice units, these LECs take the values
\begin{eqnarray}
c_{D}^{(0)} &=& -1.2787~,~~
c_{D}^{(1)} = -2.5665~,~~
c_{D}^{(2)} = -0.2578~,\nonumber\\
c_{E}^{(0)} &=& \phantom{-}3.3724~,~~
c_{E}^{(1)} = \phantom{-}4.9896~,~~
c_{E}^{(2)} = -1.0876~,~~ \nonumber\\
c_{E}^{(l)} &=&  -0.4991~,~~
c_{E}^{(t)} = \phantom{-}0.06575~.
\end{eqnarray}
Finally, we note that the pion-nucleon vertices are not smeared, so that we can take the
values of the dimension-two LECs $c_i$ from Ref.~\cite{Hoferichter:2015tha}.

\subsection{Lattice Hamiltonian with hyperons}

Consider now interactions containing additional hyperons. At present, two different NLEFT approaches
to hypernuclei exist. The first combines the high-fidelity chiral NN and 3N forces with
leading order YN and YNN forces, all smeared locally and non-locally. In this framework,
hypernuclei up to $^{16}_\Lambda$O have been investigated. Second, an extension of the
minimal nuclear into into the strangeness $S=-1,2$ sectors also has been constructed. In this
framework, some single and double $\Lambda$ hypernuclei have been studied together with
the EoS of nuclear matter and of (hyper)neutron matter. Clearly, these studies require
improvements in the years to come.

All NLEFT studies have so far been using LO  S-wave $\Lambda N$  contact interactions:
\begin{equation}
V_{Y N} = \frac{1}{4}C_{YN}^S(\mathbbm{1}-\boldsymbol{\sigma_1}\cdot\boldsymbol{\sigma_2})
+\frac{1}{4}C_{YN}^T(3+\boldsymbol{\sigma_1}\cdot\boldsymbol{\sigma_2})
\label{eq:V-lambdaN}
\end{equation}
where $\boldsymbol{\tau},\boldsymbol{\sigma}$ are Pauli-(iso)spin matrices and $C_{YN}^{S,T}$
are the respective LECs. In addition, contact three-body $\Lambda NN$ and $N\Lambda\Lambda$
forces as derived in Ref.~\cite{Petschauer:2015elq} have been utilized,
\begin{eqnarray}
V_{YNN} &=& C_1(\mathbbm{1}-\boldsymbol{\sigma}_2\cdot\boldsymbol{\sigma}_3)(3+\boldsymbol{\tau}_2\cdot\boldsymbol{\tau}_3)
+ C_2 \,\boldsymbol{\sigma}_1\cdot(\boldsymbol{\sigma}_2+\boldsymbol{\sigma}_3)(\mathbbm{1}-\boldsymbol{\tau}_2\cdot\boldsymbol{\tau}_3)
+ C_3 (3+\boldsymbol{\sigma}_2\cdot\boldsymbol{\sigma}_3)(\mathbbm{1}-\boldsymbol{\tau}_2\cdot\boldsymbol{\tau}_3)\,,\nonumber\\
V_{YYN} &=& D_1 \, \mathbbm{1} \, \mathbbm{1}~,
\label{eq:V-LambdaNN}
\end{eqnarray}  
in terms of 3 LECs for the $\Lambda NN$ and 1 LEC for the $\Lambda\Lambda N$ interaction (without smearing). Note that 
the hyperon in the YNN interaction is always counted as the first particle. 
As noted before, 3BFs appear at N2LO in the 
chiral power counting \`a la Weinberg. In pionless EFT the three-body forces, however, 
would be LO~\cite{Bedaque:1998kg}. Since in these calculations the explicit two-pion exchange interactions
where not considered, but effectively simulated by the smearing discussed below, 
it is legitimate to promote the 3BFs in the $S=-1$ and the $S=-2$ sectors to LO
(or consider the extention of the minimal interaction to include hyperons discussed below).
Note further that the one-pion exchange is suppressed for the $\Lambda N$ 
interaction due to isospin symmetry and hence not part of this effective leading 
order interaction. Finally, due to the fact that high-fidelity chiral interactions between 
nucleons are well tested, 
those set the smearing parameters as well as the lattice spacing for the hypernuclear 
interactions.  Therefore, the lattice Hamiltonian is defined as,
\begin{align}
H = H_{\rm N3LO} + T_{Y} + V_{Y N} + V_{YNN} + V_{YYN}\,,
\label{eq:H-001}
\end{align}
where $H_{\rm N3LO}$ is the high-fidelity Hamiltonian for nucleons~\cite{Elhatisari:2022zrb}, 
$T_{Y}$ is the kinetic energy term for $\Lambda$ 
hyperons defined by using fast Fourier transforms to produce the exact dispersion 
relations $E_\Lambda=p^2/(2m_{\Lambda})$ with hyperon mass 
$m_{\Lambda}=1115.68$~MeV, $V_{Y N}$ and $V_{YNN}$, $V_{YYN}$ are the hyperon-nucleon 
and hyperon-nucleon-nucleon, hyperon-hyperon-nucleon interactions given in Eqs.~(\ref{eq:V-lambdaN}) 
and (\ref{eq:V-LambdaNN}), respectively. Note that in these calculations, no other
hyperons have been considered, and thus the $\Sigma$-$\Lambda$ conversion is subsumed in the
fitted LECs. As so far in this approach only the $S=-1$ hypernuclear sector has been investigated,
let us discuss the smearing of the $\Lambda NN$ 3BFs. The locally smeared forms of the interactions given 
in Eq.~(\ref{eq:V-LambdaNN}) are given by,
\begin{eqnarray}
\label{eqn:V_c3-sL-001}
  V_{1}^{(d)} & = & 3 \,
\sum_{\boldsymbol{n}}
 : \left\{[\rho^{(d)}(\boldsymbol{n})]^2 \,
    - \sum_{S} \, [\rho_{S}^{(d)}(\boldsymbol{n})]^2
    \right\}
    \xi^{(d)}(\boldsymbol{n}):
   + \sum_{\boldsymbol{n},I} 
  : \left\{ [\rho_{I}^{(d)}(\boldsymbol{n})]^2 
  - \sum_{S} \, [\rho_{SI}^{(d)}(\boldsymbol{n})]^2
  \right\}
    \xi^{(d)}(\boldsymbol{n}) : \,,
\nonumber\\
    V_{2}^{(d)} & = & 
    2 \,
    \sum_{\boldsymbol{n}}:
    \rho^{(d)}(\boldsymbol{n})
    \sum_{S} \rho_{S}^{(d)}(\boldsymbol{n})
     \xi_{S}^{(d)}(\boldsymbol{n}) :
       -2 \, \sum_{\boldsymbol{n},S,I}:
        \rho_{I}^{(d)}(\boldsymbol{n})
        \rho_{SI}^{(d)}(\boldsymbol{n})
         \xi_{S}^{(d)}(\boldsymbol{n}): \,,
\\
    V_{3}^{(d)} & = &  3 \,
\sum_{\boldsymbol{n}}
: \left\{[\rho^{(d)}(\boldsymbol{n})]^2 \,
    - \sum_{I} \, [\rho_{I}^{(d)}(\boldsymbol{n})]^2 \right\}
    \xi^{(d)}(\boldsymbol{n}) :
         + \sum_{\boldsymbol{n},S} 
        : \left\{ [\rho_{S}^{(d)}(\boldsymbol{n})]^2 
        - \sum_{I} \, [\rho_{SI}^{(d)}(\boldsymbol{n})]^2
        \right\}
          \xi^{(d)}(\boldsymbol{n}) : \,. \nonumber
\end{eqnarray}
in terms of the various smeared nucleon densities and $\xi^{(d)}(\boldsymbol{n}), \xi_{S}^{(d)}(\boldsymbol{n})$
are the locally smeared hyperon densities,
\begin{eqnarray}
 \xi^{(d)}(\boldsymbol{n}) &=& \sum_{i=0,1}  
         b^{\dagger}_{i}(\boldsymbol{n}) \, b^{\,}_{i}(\boldsymbol{n})   
                  + s_{\rm L}
          \sum_{|\boldsymbol{n}-\boldsymbol{n}^{\prime}|^2 = 1}^d 
          \,
          \sum_{i,j=0,1} 
          b^{\dagger}_{i}(\boldsymbol{n}^{\prime}) \, b^{\,}_{i}(\boldsymbol{n}^{\prime})\,,\\
  \xi^{(d)}_{S}(\boldsymbol{n}) &=& \sum_{i,i^{\prime}=0,1}  
         b^{\dagger}_{i}(\boldsymbol{n}) \, [\sigma_{S}]_{i,i^{\prime}} \, b^{\,}_{i^{\prime}}(\boldsymbol{n})
          + s_{\rm L}
          \sum_{|\boldsymbol{n}-\boldsymbol{n}^{\prime}|^2 = 1}^d 
          \,
           \sum_{i,i^{\prime}=0,1}  
            b^{\dagger}_{i}(\boldsymbol{n}^{\prime}) \, [\sigma_{S}]_{i,i^{\prime}} 
              \, b^{\,}_{i^{\prime}}(\boldsymbol{n}^{\prime}) \,.         
\end{eqnarray}  
Here, the $b_i^{}, b_i^\dagger$ are hyperon annihilation, creation operators and 
the superscript $d$ describes the range of the local smearing. We consider 
different choices up to $d = 3$ corresponding to 2.28~fm. In addition, 
for these interactions we set $s_{\rm L} = 0.5$. Therefore, the locally smeared 
three-body interactions are labelled as $V_{C_k}^{(d=0)}$, $V_{C_k}^{(d=1)}$, $V_{C_k}^{(d=2)}$ and 
$V_{C_k}^{(d=3)}$ with $k = 1,2,3$. We also define the non-locally smeared forms of 
the interactions given  in Eq.~(\ref{eq:V-LambdaNN}),
\begin{eqnarray}
  \label{eqn:V_c1-sL-001}
  V_{1}^{s_{\rm NL}} &= & 
    3 \,
    \sum_{\boldsymbol{n}}
     : \left\{ [\hat{\rho}(\boldsymbol{n})]^2 \,
        - \sum_{S} \, [\hat{\rho}_{S}(\boldsymbol{n})]^2 \right\}
        \hat{\xi}(\boldsymbol{n}) :
      + \sum_{\boldsymbol{n},I} 
      : \left\{ [\hat{\rho}_{I}(\boldsymbol{n})]^2 
      - \sum_{S} \, [\hat{\rho}_{SI}(\boldsymbol{n})]^2
      \right\}
      \hat{\xi}(\boldsymbol{n}) : \,,
      \nonumber\\
        V_{2}^{s_{\rm NL}} &= & 
        2 \,
        \sum_{\boldsymbol{n}}:
        \hat{\rho}(\boldsymbol{n})
        \sum_{S} \hat{\rho}_{S}(\boldsymbol{n})
        \hat{\xi}_{S}(\boldsymbol{n}): 
         -2 \, \sum_{\boldsymbol{n},S,I}:
            \hat{\rho}_{I}(\boldsymbol{n})
            \hat{\rho}_{SI}(\boldsymbol{n})
            \hat{\xi}_{S}(\boldsymbol{n}): \,,  \\    
        V_{3}^{s_{\rm NL}} &= & 
        3 \,
    \sum_{\boldsymbol{n}}
     : \left\{[\hat{\rho}(\boldsymbol{n})]^2 \,
        - \sum_{I} \, [\hat{\rho}_{I}(\boldsymbol{n})]^2\right\}
        \hat{\xi}(\boldsymbol{n}) : 
            + \sum_{\boldsymbol{n},S} 
            : \left\lbrace [\hat{\rho}_{S}(\boldsymbol{n})]^2 
            - \sum_{I} \, [\hat{\rho}_{SI}(\boldsymbol{n})]^2
            \right\rbrace
              \hat{\xi}(\boldsymbol{n}) : \,.  \nonumber 
\end{eqnarray}
in terms of the non-locally smeared hyperon densities
\begin{eqnarray}
 \hat{\xi}(\boldsymbol{n}) &=& \sum_{i=0,1} \tilde{b}_{i}^\dagger (\boldsymbol{n})
                                     \tilde{b}_{i}^{} (\boldsymbol{n})\,,\quad
     \hat{\xi}_{S}(\boldsymbol{n}) = \sum_{i,i^{\prime}=0,1} \tilde{b}_{i}^\dagger (\boldsymbol{n})
                                [\boldsymbol{\sigma}_{S}]_{i,i^{\prime}}
                                \tilde{b}_{i^{\prime}}^{} (\boldsymbol{n})\,, \nonumber\\
   \tilde{b}_{i}(\boldsymbol{n})&=& b_{i}(\boldsymbol{n}) 
                             + s_{\rm NL}\sum_{|\boldsymbol{n}^{\prime}-\boldsymbol{n}|=1}
                               b_{i}(\boldsymbol{n}^{\prime})\,,                             
\end{eqnarray}  
in terms of the strength of the non-locality  of the interactions, ${s_{\rm NL}}$. One considers three
different values of this parameter, ${s_{\rm NL}}=0.1, 0.2, 0.3$, labeled
 $V_{C_k}^{0.1}$, $V_{C_k}^{0.2}$ and $V_{C_k}^{0.3}$, respectively,  with $k = 1,2,3$. 
Now one  considers all possible versions of the $\Lambda NN$ 
interaction, constructed from the combinations of these smeared versions 
of $V_{C_1}$, $V_{C_2}$, and $V_{C_3}$, which leads to a total set 
of $343$ combinations. By systematically analyzing each 
combination using hypernuclei from light to medium mass, we determine 
the optimal configuration for the $\Lambda NN$ 
interaction which give a good description for hypernuclei.
Such an analysis is not yet available. In the results presented later,
a limited set of hypernuclei is included in the fitting procedure
and one then considers the combination with the lowest 
 Root Mean Square Deviation (RMSD) defined as follows
\begin{equation}\label{eq:RMSD}
\text{RMSD}(S)=\sqrt{\frac{1}{M_S}\sum_{i\in S}
\left(\frac{^{i}B^c_\Lambda-{}^{i}B^{\exp}_\Lambda}{^{i}B^{\exp}_\Lambda}\right)^2},
\end{equation}
where $^i B_\Lambda^c$ is the evaluated $\Lambda$ separation energy and 
$^i B^{\exp}_\Lambda$ is the experimental separation energy for each hypernucleus within the set $S$.
The size of the set of hypernuclei is given by $M_S$ of the set $S$, which contains
well measured hypernuclei from the light and medium mass region, starting from the four-body systems.
The corresponding $^i B^{\exp}_\Lambda$ are taken from Ref.~\cite{HypernuclearDataBase}.

There exists also a so-called minimal hypernuclear interaction that builds
upon the significant achievements of the EFT within Wigner's SU(4) spin-isospin symmetry,
which is referred to as the minimal nuclear interaction. An extension of this is pionless EFT at LO
for nucleons, see also~\cite{Konig:2016utl}), derived from this minimal nuclear interaction. This
approach allows us to make use of the well-established theoretical framework by the minimal nuclear
interaction, providing a solid basis for  calculations of hypernuclei and the equation of state of
neutron  matter including hyperons. This was developed in Ref.~\cite{Tong:2024jvs}.
As before, we only  consider  $\Lambda$ hyperons, so that the
$\Lambda-\Sigma^0$ transition induces three-body forces which are effectively represented by $\Lambda NN$
forces. For the YN and YY interactions, we also utilize minimal interactions assuming that these
interactions are spin symmetric. Therefore, the minimal Hamiltonian for hypernuclear physics is defined as,
\begin{align}
H= & \, T +\frac{c_{NN}}{2}\sum_{\vec{n}}
\,:\,\left[
\tilde{\rho}(\vec{n})
\right]^2
\,:\,
+\frac{c_{NN}^{T}}{2}\sum_{I,\vec{n}}
\,:\,
\left[
\tilde{\rho}_{I}(\vec{n})
\right]^2
\,:\,
+ c_{N\Lambda}\sum_{\vec{n}}
\,:\,
\tilde{\rho}(\vec{n})
\tilde{\xi}(\vec{n})
\,:\,
+ \frac{c_{\Lambda\Lambda}}{2}\sum_{\vec{n}}
\,:\,
\left[
\tilde{\xi}(\vec{n})
\right]^2
\,:\,
\nonumber\\
 &
+V^{\rm GIR}_{NN}
+V^{\rm GIR}_{N\Lambda}
+V^{\rm GIR}_{\Lambda\Lambda}
+V_{\rm C}
+V_{NNN}
+V_{NN\Lambda}
+V_{N\Lambda\Lambda}
\,,
\label{eq:H-min}
\end{align}
Here, $T$ is the kinetic energy term defined by using fast Fourier transforms to produce the
exact dispersion relations $E_N = p^2/(2m_{N})$ and $E_\Lambda =p^2/(2m_{\Lambda})$ with nucleon mass $m_{N}=938.92$~MeV
and hyperon mass $m_{\Lambda}=1115.68$~MeV, $c_{NN}$ is the coupling constant of the SU(4) symmetric short-range 2N
interaction, $c_{NN}^{T}$ is the coupling constant of the isospin-dependent short-range 2N interaction,
that breaks SU(4) symmetry (see the discussion below), $c_{N\Lambda}$ ($c_{\Lambda\Lambda}$) is the LEC
of the spin-symmetric short-ranged YN (YY) interaction, and $\tilde{\xi}$ is  the hyperon density operator,
that is smeared both locally and non-locally, as given above.
Further, $V_{\text{C}}$ represents the Coulomb interaction, see Ref.~\cite{Li:2018ymw} for details.
The nonlocal smearing applied on the lattice introduces an explicit dependence on the center-of-mass momentum,
thereby breaking Galilean invariance. Consequently, in Eq.~(\ref{eq:H-001}) we introduce $V^{\text{GIR}}_{NN}$,
$V^{\text{GIR}}_{N\Lambda}$, and $V^{\text{GIR}}_{\Lambda\Lambda}$, which denote the Galilean invariance restoration
(GIR) interactions for the nucleon-nucleon, nucleon-hyperon, and hyperon-hyperon interactions, respectively.
We refer the reader to Ref.~\cite{Li:2019ldq} for further details.
Finally, in this approach, the three-baryon interactions $V_{NNN}$, $V_{NN\Lambda}$, and $V_{N\Lambda\Lambda}$,
are also given in Eq.~(\ref{eq:H-min}).  The 3N forces take the form
\begin{align}\label{eq:VNNN}
V_{NNN}
=
\sum_{i = 1,2}
\frac{c_{NNN}^{(d_i)}}{6}
\,
\sum_{\vec{n}}
\,:\,
\left[
\rho^{(d_i)}(\vec{n})
\right]^3
\,:\,,
\end{align}
where the parameter $d_i$ denotes the range of local smearing with $0 \leq d_1 < d_2 \leq 3$ (in lattice units). 
In this case, the 3BFs between two nucleons and one hyperon
are defined with two different choices of local smearing 
\begin{align}
V_{NN\Lambda}
=
\sum_{i = 1,2}
\frac{c_{NN\Lambda}^{(d_i)}}{2}
\,
\sum_{\vec{n}}
\,:\,
\left[
\rho^{(d_i)}(\vec{n})
\right]^2 \xi^{(d_i)}(\vec{n})
\,:\,,
\label{eqn:V-NNY}
\end{align}
and similarly for the interactions involving one nucleon and two hyperons,
\begin{align}
V_{N\Lambda\Lambda}
=
\sum_{i = 1,2}
\frac{c_{N\Lambda\Lambda}^{(d_i)}}{2}
\,
\sum_{\vec{n}} \,
 \,:\,
 \rho^{(d_i)}(\vec{n})  \,
\left[
\xi^{(d_i)}(\vec{n})
\right]^2
\,:\,,
\label{eqn:V-NYY}
\end{align}
where ${\rho}$ (${\xi}$) is then purely locally smeared nucleon (hyperon) density operator with annihilation and
creation operators, ${a}$ (${b}$) and ${a}^{\dagger}$ (${b}^{\dagger}$) for nucleons (hyperons),
\begin{align}
{\rho}^{(d)}(\vec{n}) = \sum_{i,j=0,1}
{a}^{\dagger}_{i,j}(\vec{n}) \, {a}^{\,}_{i,j}(\vec{n})
+
s^{\rm 3B}_{\rm L}
 \sum_{|\vec{n}-\vec{n}^{\prime}|^2 = 1}^{d}
 \,
 \sum_{i,j=0,1}
{a}^{\dagger}_{i,j}(\vec{n}^{\prime}) \, {a}^{\,}_{i,j}(\vec{n}^{\prime})
\,,
\label{eqn:rho-local}
\end{align}
\begin{align}
{\xi}^{(d)}(\vec{n}) = \sum_{i=0,1}
{b}^{\dagger}_{i}(\vec{n}) \, {b}^{\,}_{i}(\vec{n})
+
s^{\rm 3B}_{\rm L}
 \sum_{|\vec{n}-\vec{n}^{\prime}|^2 = 1}^{d}
 \,
 \sum_{i=0,1}
{b}^{\dagger}_{i}(\vec{n}^{\prime}) \, {b}^{\,}_{i}(\vec{n}^{\prime})
\,.
\label{eqn:xi-local}
\end{align}
Here, the parameter $d$ gives the range of local smearing with $0\leq d \leq 3$ and $s^{\rm 3B}_{\rm L}$
defines the strength of the local smearing. Note that here no nonlocal smearing is applied to the
3BFs. This is related to the fact that this approach was applied simultaneously to hypernuclei as
as well as dense hypernuclear matter, where  it is found that for generating a stiff neutron matter EoS,
only local smearing should be used. 

Next, we have to discuss how to treat the $\Lambda$ hyperons in the simulations.
At present, there exist two different methods that we will discuss separately.

\subsubsection{Impurity lattice MC formalism}

The impurity lattice Monte Carlo (ILMC) method has been introduced in Ref.~\cite{Elhatisari:2014lka}
in the context of a Hamiltonian  theory of spin-up and spin-down fermions, and applied to the intrinsically
non-perturbative physics of Fermi polarons in two dimensions in Ref.~\cite{Bour:2014bxa}.
The ILMC method is particularly useful for the case where only one (or two) fermion(s)
(of either species) is immersed in a ``sea'' of the other species. Within the standard auxiliary
field Monte Carlo method, such an extreme imbalance would often lead to unacceptable sign oscillations
in the Monte Carlo probability weight (except by using a certain trick, as discussed later).
In the ILMC method, the minority particle is ``integrated out'',
resulting in a formalism where only the majority species fermions appear as explicit degrees of freedom,
while the minority fermion is represented by a ``worldline'' in Euclidean projection time.
The spatial position of this worldline is updated using Metropolis moves, while the interactions between
the majority fermions are described by the usual auxiliary field formalism~\cite{Lahde:2019npb}.

The ILMC formalism for hypernuclei with one $\Lambda$ was developed in~\cite{Frame:2020mvv}.
One starts from the  partition function via the Grassmann path integral
\begin{align}
\mathcal{Z} = \int 
\Bigg[\prod_{\substack{\vec n, n_t \\ s=N,Y}}
d\zeta_s^{}(\vec n, n_t^{}) d\zeta_s^*(\vec n, n_t^{}) \Bigg]
\exp(-S[\zeta, \zeta^*]),
\label{Z_pi}
\end{align}
where the subscripts $N$ and $Y$ refer to all nucleon and hyperon degrees of freedom, respectively.
For simplicity, let us assume that the YN and NN interactions are spin-independent and we neglect
Coulomb interactions. Trotterizing the the Euclidean action in Eq.~(\ref{Z_pi}) gives
\begin{align}
& S[\zeta, \zeta^*] \equiv \sum_{n_t} \bigg\{
S_t^{}[\zeta, \zeta^*,n_t^{}] +
S_Y^{}[\zeta^{}, \zeta^*,n_t^{}] 
+ S_N^{}[\zeta^{}, \zeta^*,n_t^{}] 
+ S_{YN}^{}[\zeta, \zeta^*,n_t^{}] + S_{NN}^{}[\zeta, \zeta^*,n_t^{}]
\bigg\},
\end{align}
where the component due to the time derivative is
\begin{align}
S_t^{}[\zeta, \zeta^*,n_t^{}]  \equiv 
\!\!\! \sum_{\vec n,s = N,Y} 
\zeta_s^*(\vec n,n_t^{})
 \bigg[ \zeta_s^{}(\vec n,n_t^{}+1) - \zeta_s^{}(\vec n,n_t^{}) \bigg],
\end{align}
while $S_Y^{}$ and $S_N^{}$ describe the kinetic energies of the hyperons and nucleons, respectively.
Further, $S_{YN}$ provides the YN interaction, and $S_{NN}$
the NN interaction. Note that  actual NLEFT calculations are performed using the transfer matrix MC method.
As noted in Ref.~\cite{Elhatisari:2014lka}, the Grassmann and transfer matrix formulations are related by
\begin{align}
& \mathrm{Tr} \big\{ : f_{N_t-1}^{}[a_s^{}(\vec n),a_{s^\prime}^\dagger(\vec n^\prime)] : 
\cdots : f_0^{}[a_s^{}(\vec n),a_{s^\prime}^\dagger(\vec n^\prime)] : \big\} =
\nonumber \\
& \int \Bigg[
\prod_{\substack{\vec n, n_t \\ s=N,Y}}
d\zeta_s^{}(\vec n, n_t^{}) d\zeta_s^*(\vec n, n_t^{}) \Bigg]
\exp\left(-\sum_{n_t} S_t^{}[\zeta, \zeta^*, n_t^{}]\right)
\, \prod_{n_t=0}^{N_t-1}
f_{n_t}^{}\big[\zeta_s^{}(\vec n, n_t^{}), \zeta_{s^\prime}^*(\vec n^\prime, n_t^{})\big],
\label{relation_3}
\end{align}
where $f$ is an arbitrary function, $a_s^\dagger$ and $a_s^{}$ denote creation and annihilation operators for
the fermion degrees of freedom, and colons
represent normal ordering. We shall now consider the explicit forms of the YN and NN interactions,
and use Eq.~(\ref{relation_3}) to relate expressions in the Grassmann and transfer matrix formulations.
To be specific, consider now the $\Lambda$ hyperons. The most simple form of the action reads
\begin{eqnarray}
S &=& S_Y + S_{YN}~, \nonumber\\
S_Y &=& h \sum_{\vec n} 
\zeta_Y^*(\vec n,n_t^{}) \zeta_Y^{}(\vec n,n_t^{})
 - h \sum_{\vec n} \sum_{l = 1}^3 \: \zeta_Y^*(\vec n,n_t^{})
 \bigg[ \zeta_Y^{}(\vec n+\hat e_l^{},n_t^{}) + \zeta_Y^{}(\vec n-\hat e_l^{},n_t^{}) \bigg]~,\nonumber\\
S_{YN} &=& \alpha_t^{} C_{YN}^{} \sum_{\vec n}\rho_N^{}(\vec n, n_t^{}) \rho_Y^{}(\vec n, n_t^{})~,
\end{eqnarray}  
where $m_Y$ is the hyperon mass, $h = {\alpha_t^{}}/(2m_Y^{})$,
and $\alpha_t^{} = a_t^{}/a$ is the ratio of temporal and spatial lattice spacings.
The LEC $C_{YN}^{}$ may be fitted, for instance, to the empirical hypertriton binding energy.
Using Eq.~(\ref{relation_3}), the hyperon contributions are described by the
transfer matrix operator 
\begin{equation}
M = \, : \exp\left(
- \alpha_t^{} T - \alpha_t^{} C_{YN}^{} \sum_{\vec n}  \rho_N^{}(\vec n)  \rho_Y^{}(\vec n)\right) :~,
\label{ham1}
\end{equation}
where 
\begin{equation}
\rho_N^{}(\vec n)  \equiv \sum_{i,j} \rho_{i,j}^{}(\vec n)
\equiv \sum_{i,j} a^{\dagger}_{i,j}(\vec n) a^{}_{i,j}(\vec n)~,
\quad  \rho_Y^{}(\vec n) \equiv a_Y^\dagger(\vec n) a_Y^{}(\vec n)~,
\end{equation}
are density operators for nucleons and hyperons, respectively.
Note that this is a simplified version of the pionless
EFT calculation of Ref.~\cite{Hammer:2001ng}, 
which also included a three-body interaction at LO. Treating the
NN interactions in the usual fashion, we are now in the position
to integrate out the hyperon degrees of freedom and
derive a ``reduced'' transfer matrix, which refers to the nucleon degrees of freedom only.
For simplicity (and without loss of generality), we shall neglect the NN interaction term for the purpose
of the derivation,
and consider the case of a single hyperon Y and nucleon N (which can be thought of as representing any one of the
spin-isospin combinations $i,j$ of the full theory). Next, we write the transfer matrix element between time slices
$n_t$ and $n_t+1$ in terms of 
\begin{equation}
| \chi_{n_t^{}}^N, \chi_{n_t^{}}^Y \rangle \equiv \prod_{\vec n}
\left\{
\left[ a_N^\dagger (\vec n) \right]^{\chi^N_{n_t^{}}(\vec n)}
\left[ a_Y^\dagger (\vec n) \right]^{\chi^Y_{n_t^{}}(\vec n)}
\right\} | 0 \rangle,
\end{equation}
where the $\chi^s_{n_t^{}}(\vec n)$ count the occupation numbers for nucleons and hyperons on time
slice $n_t$ and  spatial lattice site $\vec n$. Following the relations established in
Ref.~\cite{Elhatisari:2014lka}, we express the transfer matrix element as
\begin{align}
& \langle \chi_{n_t^{}+1}^N, \chi_{n_t^{}+1}^Y | \hat M | \chi_{n_t^{}}^N, \chi_{n_t^{}}^Y \rangle =
\nonumber \\
& \quad \prod_{\vec n} \left\{
\left[\frac{\overrightarrow{\partial\,}}{\partial\zeta_N^*(\vec n,n_t^{})}\right]^a
\left[\frac{\overrightarrow{\partial\,}}{\partial\zeta_Y^*(\vec n,n_t^{})}\right]^b 
\right\}
X(n_t^{}) M(n_t^{})
\, \left. \times \: \prod_{\vec n^{\prime}} \left\{
\left[\frac{\overleftarrow{\partial\,}}{\partial\zeta_N^*(\vec n^{\prime},n_t^{})}\right]^c 
\left[\frac{\overleftarrow{\partial\,}}{\partial\zeta_Y^*(\vec n^{\prime},n_t^{})}\right]^d
\right\} \right |_{\substack{\zeta_N^* = \zeta_N^{} = 0 \\ \zeta_Y^* = \zeta_Y^{} = 0}},
\end{align}
where
\begin{equation}
a = \chi^N_{n_t^{}+1}(\vec n)~, \quad
b = \chi^Y_{n_t^{}+1}(\vec n)~, \quad
c = \chi^N_{n_t^{}}(\vec n^{\prime})~, \quad
d = \chi^Y_{n_t^{}}(\vec n^{\prime})~,
\end{equation}
are integers which are either $0$ or~$1$. Further,
\begin{align}
X(n_t^{}) & \equiv \prod_{\vec n}
\exp(\zeta_N^*(\vec n, n_t^{}) \zeta_N^{}(\vec n, n_t^{}))
\, \times \exp(\zeta_Y^*(\vec n, n_t^{}) \zeta_Y^{}(\vec n, n_t^{})),
\end{align}
and
\begin{equation}
M(n_t^{}) \equiv \exp(-S_\mathrm{kin}^{}[\zeta, \zeta^*,n_t^{}])
\exp(-S_\mathrm{int}^{}[\zeta, \zeta^*,n_t^{}]),
\end{equation}
are Grassmann functions (to be defined below). The impurity worldline is considered static for the purposes of
this derivation, although it will be updated by the Metropolis algorithm in
the actual MC simulations. From one time slice to the next, the impurity may either
remain on the same lattice site, or hop to a nearest-neighbor site. For the case where the impurity remains on a given
lattice site $\vec n^{\prime\prime}$, we have
\begin{align}
& \langle \chi_{n_t^{}+1}^N, \chi_{n_t^{}+1}^Y |  M | \chi_{n_t^{}}^N, \chi_{n_t^{}}^Y \rangle =
\nonumber \\
& \quad \prod_{\vec n} \left\{
\left[\frac{\overrightarrow{\partial}}{\partial\zeta_N^*(\vec n,n_t^{})}\right]^{\chi^N_{n_t^{}+1}(\vec n)} \right\}
{\slashed X}(n_t^{}) {\slashed M}_{\vec n^{\prime\prime}, \vec n^{\prime\prime}}^{}(n_t^{})
\, \left. \prod_{\vec n^{\prime}} \left\{
\left[\frac{\overleftarrow{\partial}}{\partial\zeta_N^*(\vec n^{\prime},n_t^{})}\right]^{\chi^N_{n_t^{}}(\vec n^{\prime})} 
\right\} \right |_{\zeta_N^* = \zeta_N^{} = 0},
\end{align}
with
\begin{equation}
\chi^Y_{n_t^{}}(\vec n^{\prime\prime}) = 1, \quad
\chi^Y_{n_t^{}+1}(\vec n^{\prime\prime}) = 1,
\end{equation}
and thus
\begin{align}
& {\slashed X}(n_t^{}) {\slashed M}_{\vec n^{\prime\prime}, \vec n^{\prime\prime}}^{}(n_t^{}) =
 \frac{\overrightarrow{\partial\,}}{\partial\zeta_Y^*(\vec n^{\prime\prime},n_t^{})}
X(n_t^{}) \exp(-S_\mathrm{kin}^{}[\zeta, \zeta^*,n_t^{}]) 
\left. \times \: \exp(-S_\mathrm{int}^{}[\zeta, \zeta^*,n_t^{}]) 
\frac{\overleftarrow{\partial}}{\partial\zeta_Y^{}(\vec n^{\prime\prime},n_t^{})}
\right |_{\zeta_Y^* = \zeta_Y^{} = 0},
\label{derivatives}
\end{align}
where
\begin{align}
{\slashed X}(n_t^{}) \equiv \prod_{\vec n}
\exp(\zeta_N^*(\vec n, n_t^{}) \zeta_N^{}(\vec n, n_t^{})).
\end{align}
and we define ${\slashed M}(n_t^{})$ as the ``reduced'' transfer matrix. 
Specifically, we take $S_\mathrm{kin}^{}[\zeta, \zeta^*,n_t^{}] = S_N^{}[\zeta, \zeta^*,n_t^{}] + 
S_Y^{}[\zeta^{},\zeta^*,n_t^{}]$, 
%
with the nearest-neighbor expression 
for the hyperon kinetic term and also 
$S_\mathrm{int}^{}[\zeta, \zeta^*,n_t^{}] = S_{YN}^{}[\zeta,\zeta^*,n_t^{}]$
as the YN interaction.
By evaluating the derivatives in Eq.~(\ref{derivatives}), we find
\begin{align}
{\slashed M}_{\vec n^{\prime\prime}, \vec n^{\prime\prime}}^{}(n_t^{}) & = 
\exp(-S_N^{}[\zeta,\zeta^*,n_t^{}])
\, \left(1-6h-\alpha_t^{} C_{YN}^{} \rho_N^{}(\vec n^{\prime\prime},n_t^{})\right),
\end{align}
which can be  written as
\begin{align}
& {\slashed M}_{\vec n^{\prime\prime}, \vec n^{\prime\prime}}^{}(n_t^{}) \simeq (1-6h)
\, \exp\left(-S_N^{}[\zeta,\zeta^*,n_t^{}]
-\frac{\alpha_t^{} C_{YN}^{}}{1-6h} \rho_N^{}(\vec n^{\prime\prime},n_t^{}) \right),
\label{Mgrass_stat}
\end{align}
where the last factor, which encodes the interaction between the nucleons and the single hyperon impurity,
has been exponentiated. Thus, Eq.~(\ref{Mgrass_stat}) is the reduced Grassmann transfer matrix for the case 
where the impurity worldline remains stationary. For the case of a long-range YN interaction,
Eq.~(\ref{Mgrass_stat}) should be replaced by an expression of the form
\begin{align}
{\slashed M}_{\vec n^{\prime\prime}, \vec n^{\prime\prime}}^{}(n_t^{}) 
\simeq & \: (1-6h) \, \exp\bigg(-S_N^{}[\zeta,\zeta^*,n_t^{}]
- \frac{\alpha_t^{}}{1-6h} \sum_{\vec n^\prime} G(\vec n^\prime - \vec n^{\prime\prime}) 
\rho_N^{}(\vec n^{\prime},n_t^{}) \bigg),
\end{align}
whereby the hyperon impurity now also interacts with nucleons not on the same spatial lattice site,
and $G(\vec n^\prime - \vec n^{\prime\prime})$ is the nucleon Greens function (propagtor).
As a check, setting
$G(\vec n^\prime - \vec n) = C_{YN}^{} \delta(\vec n^\prime - \vec n)$
then the expression  for a contact interaction is recovered. 
Another possibility permitted by the nearest-neighbor YN kinetic term is
\begin{equation}
\chi^Y_{n_t^{}}(\vec n^{\prime\prime}) = 1, \quad
\chi^Y_{n_t^{}+1}(\vec n^{\prime\prime} \pm \hat e_l^{}) = 1,
\end{equation}
such that
\begin{align}
& {\slashed X}(n_t^{}) {\slashed M}_{\vec n^{\prime\prime} \pm \hat e_l^{}, \vec n^{\prime\prime}}(n_t^{}) = 
\nonumber \\
& \quad \frac{\overrightarrow{\partial}}{\partial\zeta_Y^*(\vec n^{\prime\prime} \pm \hat e_l^{},n_t^{})}
X(n_t^{}) \exp(-S_\mathrm{kin}^{}[\zeta, \zeta^*,n_t^{}]) 
\, \left. \times \: \exp(-S_\mathrm{int}^{}[\zeta, \zeta^*,n_t^{}]) 
\frac{\overleftarrow{\partial}}{\partial\zeta_Y^{}(\vec n^{\prime\prime},n_t^{})}
\right |_{\zeta_Y^* = \zeta_Y^{} = 0},
\label{derivatives_hop}
\end{align}
which gives
\begin{equation}
{\slashed M}_{\vec n^{\prime\prime} \pm \hat e_l^{}, \vec n^{\prime\prime}}^{}(n_t^{}) =
h \exp\left(-S_N^{}[\zeta, \zeta^*,n_t^{}]\right),
\label{Mgrass_hop}
\end{equation}
for the reduced Grassmann transfer matrix,
when the impurity hops to a neighboring lattice site. 
Having determined the form of the reduced Grassmann transfer matrices, 
these can be translated to the transfer matrix formulation.
The corresponding operators are
\begin{equation}
{\slashed M}_{\vec n^{\prime\prime}, \vec n^{\prime\prime}}^{}
= (1 - 6h) : \exp\left(
- \alpha_t^{}T_N - \frac{\alpha_t^{} C_{YN}^{}}{1 - 6h} 
\hat \rho_N^{}(\vec n^{\prime\prime})
\right) :,
\label{Mslash_stat}
\end{equation}
from Eq.~(\ref{Mgrass_stat}), and
\begin{equation}
{\slashed M}_{\vec n^{\prime\prime} \pm \hat e_l^{}, \vec n^{\prime\prime}}^{}
= h : \exp(-\alpha_t^{} T_N) :,
\label{Mslash_hop}
\end{equation}
from Eq.~(\ref{Mgrass_hop}), with  $T_N$ the nucleon kinetic energy.
A few comments about this implementation of the ILMC formalism are in order.
In the MC codes, Eq.~(\ref{Mslash_stat}) is evaluated as
\begin{equation}
{\slashed M}_{\vec n^{\prime\prime}, \vec n^{\prime\prime}}^{}
\sim \left(1 - \alpha_t^{}T_N - \frac{\alpha_t^{} C_{YN}^{}}{1 - 6h} 
\hat \rho_N^{}(\vec n^{\prime\prime}) - V_{\phi N}\right),
\label{Mslash_stat_code}
\end{equation}
which  also includes the NN interaction through term $V_{\phi N}$ that gives
the auxiliary field coupling(s) to the nucleon density.
For ILMC, the nucleons are treated as distinguishable particles, and the hyperon as a classical
worldline during the Euclidean time evolution, as depicted in Fig.~\ref{fig_worldline}.
\begin{figure}[htb!]
\begin{center}
\includegraphics[width = .35\columnwidth]{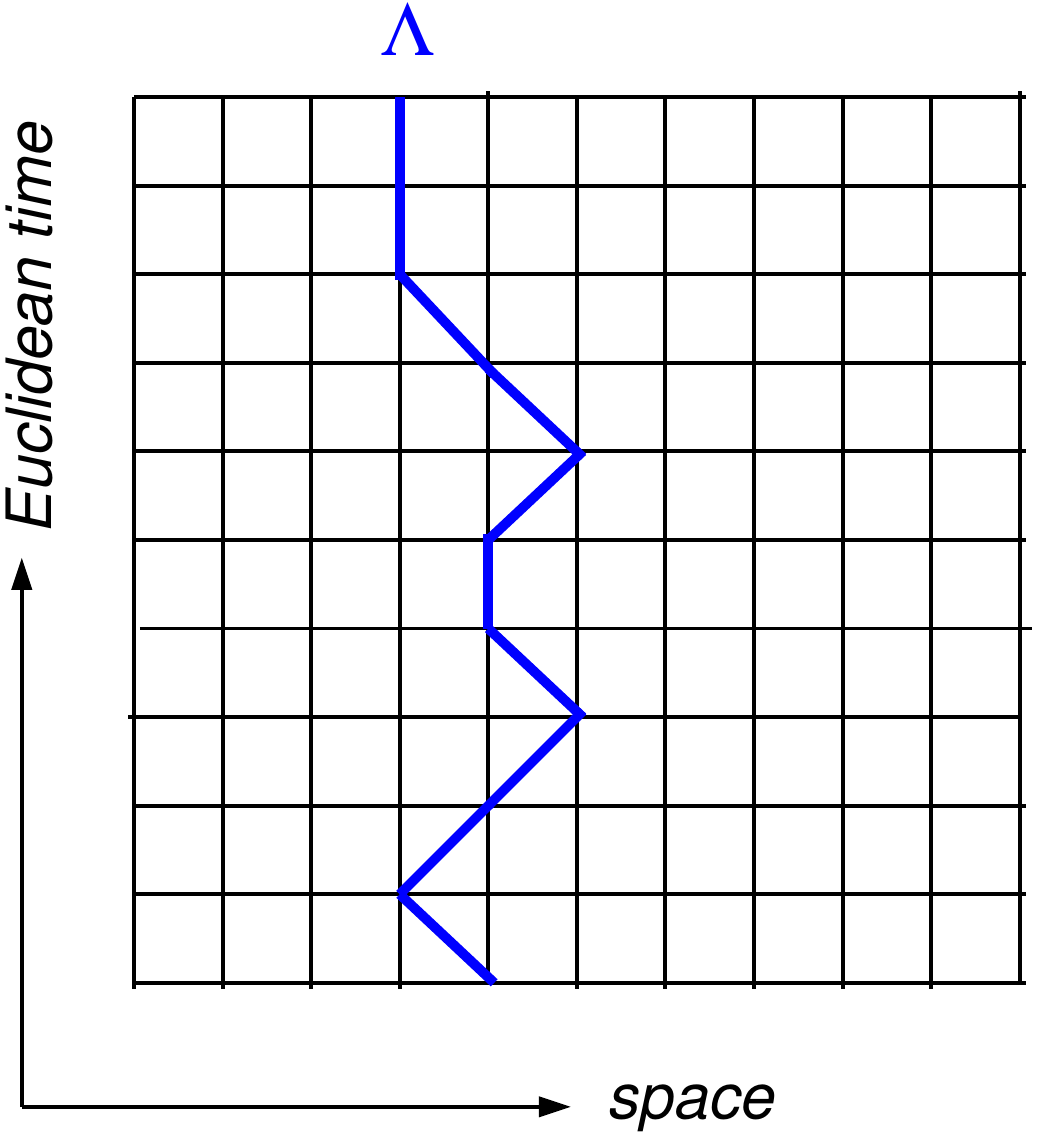}
\caption{Illustration of the hyperon worldline. In the reduced transfer matrix formalism, the hyperon (here, the $\Lambda$) has been ``integrated out'', and the interaction
between the hyperon and the nucleons is mediated by an effective ``background field'' generated by the hyperon worldline.
\label{fig_worldline}}
\end{center}
\end{figure}
This induces a three-body interaction when two nucleons and
the hyperon occupy the same site, which is absent in Eq.~(\ref{ham1}).
To be able to benchmark the  ILMC codes against exact Euclidean time projection calculations of Eq.~(\ref{ham1}), 
the induced interaction
\begin{equation}
H_{YNN}^{} = - \frac{\alpha_t^{} C_{YN}^2}{2(1-6h)} \sum_{\vec n} 
\rho_N^{}(\vec n) \rho_N^{}(\vec n) \rho_Y^{}(\vec n),
\label{H_spurious}
\end{equation}
has to be added to the original transfer matrix~(\ref{ham1}).
This induced 3BF is a lattice artifact which disappears when $\alpha_t^{} \to 0$.
Let us now discuss how ILMC calculations are performed using the Projection Monte Carlo (PMC) method.
Assume first that the impurity has been fixed at a given spatial lattice site, and that no
``hopping'' of the impurity occurs during the Euclidean time evolution. We shall then relax this constraint, and
discuss a practical algorithm for updating the configuration of the hyperon worldline.

\noindent
{\bf Stationary impurity:}~For a stationary hyperon impurity, the reduced transfer matrix is given by
Eq.~(\ref{Mslash_stat}), and for the purposes of the PMC calculation, we define the Euclidean projection amplitude
\begin{equation}
Z_{jk}^{}(N_t^{}) \equiv \langle \psi_j^{} | {\slashed M}^{N_t^{}} | \psi_k^{} \rangle,
\end{equation}
where $j$ and $k$ denote different initial cluster states.
As usual, this is expressed as a determinant of single-particle amplitudes, which gives
\begin{equation}
Z_{jk}^{}(N_t^{}) = \mathrm{det} \, M_{p\times p}^{jk},
\label{proj_ampl}
\end{equation}
where
\begin{equation}
M_{p\times p}^{jk} = \left(
\begin{array}{c c c}
\vspace{.1cm}
\langle \phi_{0,j}^{} | {\slashed M}^{N_t^{}} | \phi_{0,k}^{} \rangle & 
\langle \phi_{0,j}^{} | {\slashed M}^{N_t^{}} | \phi_{1,k}^{} \rangle & \cdots \\
\vspace{.1cm}
\langle \phi_{1,j}^{} | {\slashed M}^{N_t^{}} | \phi_{0,k}^{} \rangle &
\langle \phi_{1,j}^{} | {\slashed M}^{N_t^{}} | \phi_{1,k}^{} \rangle & \cdots \\
\vdots & \vdots & \ddots
\end{array}
\right),
\end{equation}
for $p$ nucleons. By means of the projection amplitudes~(\ref{proj_ampl}), we construct
\begin{equation}
[M^a_{}(N_t^{})]_{qq^\prime}^{} \equiv \sum_{q^{\prime\prime}}Z_{qq^{\prime\prime}}^{-1}(N_t^{})
Z_{q^{\prime\prime}q^\prime}^{}(N_t^{}+1),
\label{M_adiab}
\end{equation}
which is known as the adiabatic transfer matrix. If we denote the eigenvalues of~(\ref{M_adiab}) by 
$\lambda_i(N_t)$, we find
\begin{equation}
\lambda_i^{}(N_t^{}) = \exp(-\alpha_t^{}E_i^{}(N_t^{}+1/2)),
\end{equation}
such that the low-energy spectrum is given by the transient energies
\begin{equation}
E_i^{}(N_t^{}+1/2) = -\frac{\log(\lambda_i^{}(N_t^{}))}{\alpha_t^{}},
\end{equation}
at finite temporal lattice spacing $a_t$. For the case of a single trial cluster state with $p$ nucleons, 
Eq.~(\ref{proj_ampl}) reduces to 
\begin{equation}
Z(N_t^{}) = \mathrm{det} \, M_{p\times p}^{00}~.
\end{equation}
The ground-state energy is obtained from
\begin{equation}
E_0^{}(N_t^{}+1/2) = -\frac{\log(Z(N_t^{}+1)/Z(N_t^{}))}{\alpha_t^{}},
\end{equation}
in the limit $N_t \to \infty$, where
the exact low-energy spectrum of the transfer matrix will be recovered.
Note that the argument $N_t+1/2$ is conventionally
assigned to the transient energy computed from the ratio of projection amplitudes evaluated at
Euclidean time steps $N_t+1$ and $N_t$. 
For the hypertriton, we have $p = 2$ nucleons after the impurity hyperon has been integrated out.
We start the Euclidean time projection with a single initial trial cluster state ($j = k = 0$)
consisting of a spin-up proton, and a spin-up neutron. 
As there are no terms that mix spin or isospin, the other components of each single-particle state
are set to zero, and  remain so during the PMC calculation. For the spatial parts of the nucleon wave functions,
we may choose, for example, the zero-momentum state
\begin{equation}
| \phi_{0,0}^{} \rangle = | \phi_{1,0}^{} \rangle = \langle 0,0,0 \rangle, 
\end{equation}
in the notation of Ref.~\cite{Elhatisari:2014lka}, which denotes 
plane-wave orbitals in a cubic box. In principle, we may also choose any other plane-wave state with 
non-zero momentum (see Table~1 of Ref.~\cite{Elhatisari:2014lka}), or any other more complicated trial state.
For the heavier nuclei, it is indeed better to choose an initial state where the nucleons are clustered together.
In this case we sum over all possible translations of the cluster in order construct an initial state with zero total momentum.

\noindent
{\bf Hopping impurity:}~If the hyperon impurity is allowed to hop between nearest-neighbor sites
(from one Euclidean time slice to the next), the Euclidean projection amplitude becomes a sum over hyperon
worldline configurations. This gives
\begin{equation}
Z_{jk}^{}(N_t^{}) \equiv \sum_{\vec n_0^{}, \ldots, \vec n_{N_t}^{}}
\langle \psi_j^{} |{\slashed M}^{N_t^{}}_{\{\vec n_j\}} | \psi_k^{} \rangle,
\end{equation}
where the product
\begin{equation}
{\slashed M}^{N_t^{}}_{\{\vec n_j\}} \equiv
{\slashed M}_{\vec n_{N_t}^{}, \vec n_{N_t-1}^{}}^{}
{\slashed M}_{\vec n_{N_t-1}^{}, \vec n_{N_t-2}^{}}^{}
\ldots
{\slashed M}_{\vec n_2^{}, \vec n_1^{}}^{}
{\slashed M}_{\vec n_1^{}, \vec n_0^{}}^{},
\end{equation}
is expressed in terms of the reduced transfer matrices~(\ref{Mslash_stat}) and~(\ref{Mslash_hop}).
Here, $\vec n_j$ denotes the spatial position of the hyperon impurity 
on time slice $j$. The expressions for the projection amplitude and determinant are generalized to
\begin{equation}
Z_{jk}^{}(N_t^{}) = \sum_{\vec n_0^{}, \ldots, \vec n_{N_t}^{}} \mathrm{det} \, M_{p\times p}^{jk},
\end{equation}
where
\begin{equation}
M_{p\times p}^{jk} = \left(
\begin{array}{c c c} 
\vspace{.1cm}
\langle \phi_{0,j}^{} | {\slashed M}^{N_t^{}}_{\{\vec n_j\}} | \phi_{0,k}^{} \rangle & 
\langle \phi_{0,j}^{} | {\slashed M}^{N_t^{}}_{\{\vec n_j\}} | \phi_{1,k}^{} \rangle & \cdots \\
\vspace{.1cm}
\langle \phi_{1,j}^{} | {\slashed M}^{N_t^{}}_{\{\vec n_j\}} | \phi_{0,k}^{} \rangle &
\langle \phi_{1,j}^{} | {\slashed M}^{N_t^{}}_{\{\vec n_j\}} | \phi_{1,k}^{} \rangle & \cdots \\
\vdots & \vdots & \ddots
\end{array}
\right),
\end{equation}
such that the determinant is now to be computed over all possible hyperon wordline configurations.
We note that the worldline configuration is to be updated stochastically using a Metropolis-type algorithm.
Thus, the proposed changes in the impurity worldline are accepted or rejected by importance sampling
with $|Z_{jj}(N_t)|$ as the probability weight function. Here, $j$ denotes one of the initial trial
nucleon cluster states.

\noindent{\bf Worldline updates:}~The updating of the impurity worldline is handled in two steps:
The generation of a new proposed worldline, and a Metropolis accept/reject 
step to determine whether to use the generated worldline. Here, the worldline $W(\vec n, n_t^{})$ is a function
of only the lattice  site $\vec n$ and the Euclidean time step $n_t^{}$, and is equal to 1 where the impurity is
present, and 0 at all other lattice points.  From the expressions of the reduced transfer matrices, 
the worldline at two adjacent time steps, $W(\vec n^\prime,n_t^{})$ and $W^\prime(\vec n^\prime, n_t^{}+1)$ 
must obey the relation $|\vec n - \vec n^\prime | \leq 1$. For an illustration of the impurity (hyperon)
worldline, see Fig.~\ref{fig_worldline}.
For the non-interacting worldline, we can generate new configurations from the free probabilities, determined
from the reduced  transfer matrices. In this case, $P_h = h$ is the hopping probability, and $P_s = (1 - 6h)$ is
the probability to remain stationary. When initializing the worldline at the beginning of the MC simulation, we may
start from a configuration  where the worldline is completely stationary (``cold start''), a configuration where the
worldline either hops or remains  stationary with equal probabilities (``hot start''), or one where the worldline either
hops or remains stationary according to $P_h$ and $P_s$ (``warm start'').
At the beginning of every sweep through the lattice, we propose a new worldline to use for that sweep.
This is done  by taking the previous worldline and choosing a random time at which we cut the worldline and
regenerating it either in the forwards and backwards time direction.  The new worldline is then accepted or rejected
using a Metropolis accept or reject condition to preserve detailed balance associated with the absolute value of the amplitude.

In Table~\ref{hypertriton_benchmark} we present benchmark
calculations of the IMLC results for $^3_\Lambda$H in comparison with exact
transfer matrix calculations.  We present the energy versus Euclidean time.
\begin{table}
\begin{center}
\caption{Impurity lattice Monte Carlo results for the energy of $^3_\Lambda$H versus Euclidean time in
    comparison with exact transfer matrix results for periodic box length 15.8~fm.
\label{hypertriton_benchmark}}
\bgroup
\def\arraystretch{1.2}
\begin{tabular}{|c|c|c|c|}
\hline
$L_t$ & $t$ (MeV$^{-1}$) & ILMC (MeV) & Exact (MeV) \\
\hline
50 & 0.1667 & $-$1.0878(6) & $-$1.0878 \\
100 & 0.3333 & $-$1.4598(9) & $-$1.4590 \\
150 & 0.5000 &$-$1.6778(11) & $-$1.6760 \\
200 & 0.6667 &$-$1.7975(13) & $-$1.7966 \\
250 & 0.8333 &$-$1.8630(17) & $-$1.8614 \\
300 & 1.0000 & $-$1.8971(18) & $-$1.8954 \\
\hline
\end{tabular}
\egroup
\end{center}
\end{table}
We see that the agreement is quite good.  The initial nucleon states for these calculations are
chosen to be constant functions corresponding to zero momentum wave functions.  The hyperon initial
wave function is also taken be a constant function.  These exact transfer matrix calculations include
the induced three-baryon interaction described in Eq.~(\ref{H_spurious}).

In Ref.~\cite{Frame:2020mvv}, results for light hypernuclei where 
presented using spin-independent nucleon-nucleon and hyperon-nucleon interactions to test the computational power of the method. The resulting hypertriton binding energy comes out as $B_{\Lambda}^{} = 0.22~\text{MeV}$,
which is reasonable given the simplified nature of the calculation. In 
Fig.~\ref{4H5He_Lambda} we show the ILMC results for the $^4_\Lambda$He/$^4_\Lambda$H and the $^5_\Lambda$He hypernuclei, which lead to the binding energies
of $0.46(5)$~MeV and $3.96(6)$~MeV, where the error is mostly due to the
Euclidean time extrapolation but no effort has been made to address the
issue of higher orders. These values are again comparable to the experimental
ones, showing that this approach is indeed capable of capturing the main
features of light hypernuclei.

\begin{figure}[t]
        \begin{center}
        \includegraphics[width=0.30\textwidth,angle=-90]{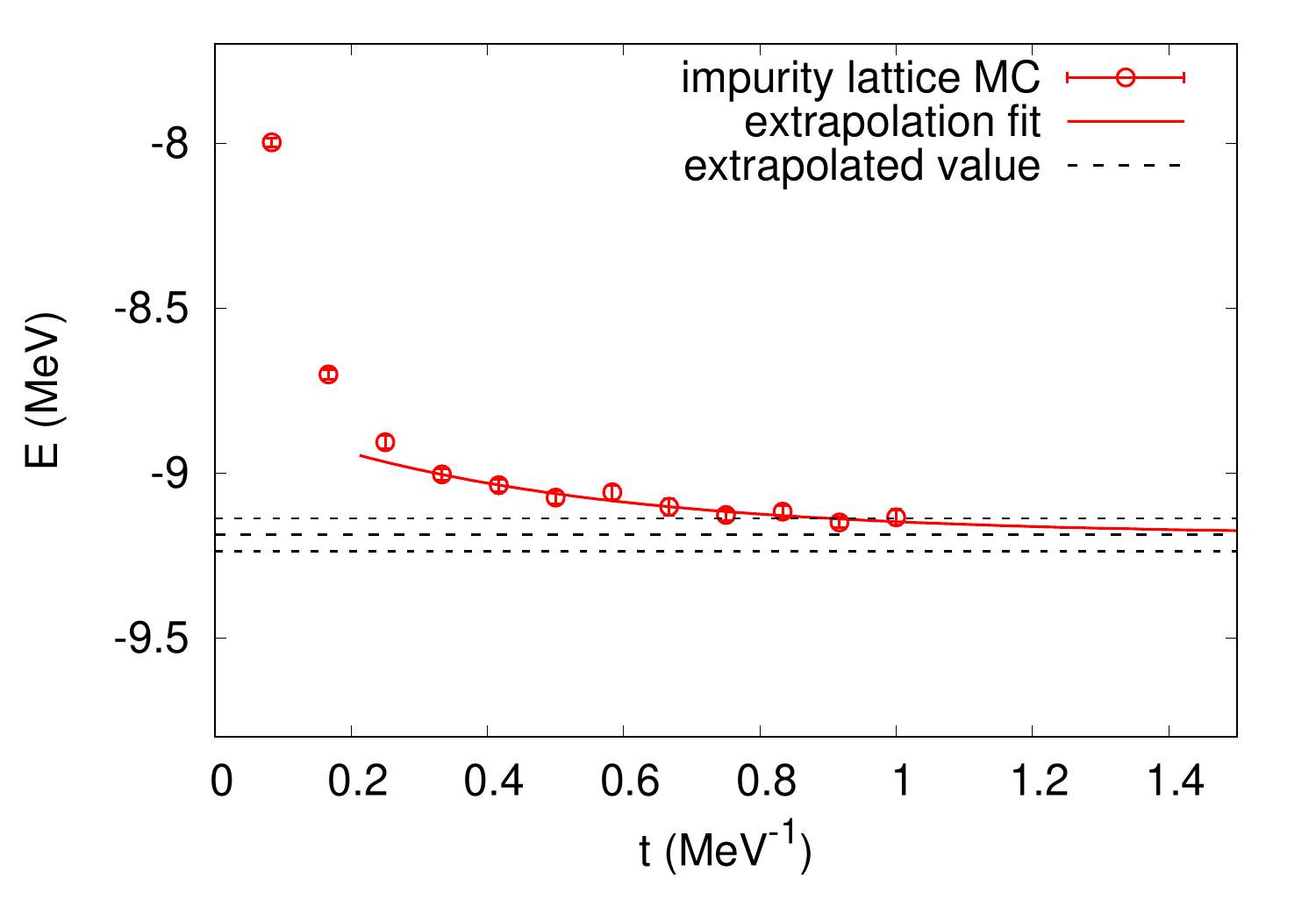}~~
        \includegraphics[width=0.30\textwidth,angle=-90]{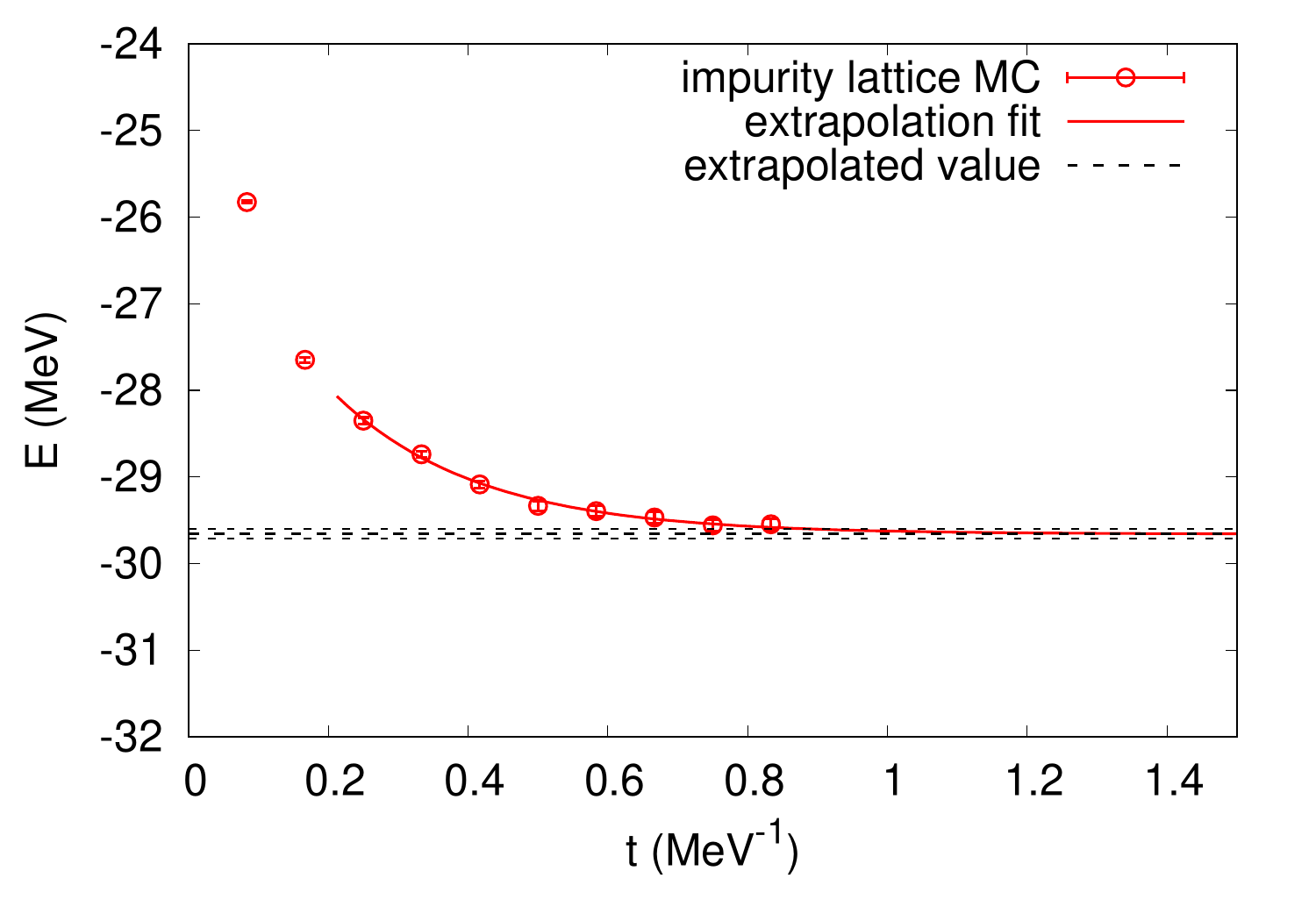}
        \caption{{\bf Left panel:} ILMC results for the $^4_{\Lambda}$H/$^4_{\Lambda}$He energy versus Euclidean projection time in a periodic box size of $L = 15.8$~fm.  
         {\bf Right panel:}  ILMC results for the $^5_\Lambda$He energy versus Euclidean time
          in a periodic box size of $L = 9.9$~fm.  In both cases the ground state energy is extracted using an exponential ansatz for the asymptotic time dependence.
          \label{4H5He_Lambda}}
        \end{center}
\end{figure}

In Ref.~\cite{Hildenbrand:2022imw}, the ILMC formalism for the case of two distinguishable impurities in a bath
of polarized fermions was developed. As before, the majority particles are treated as explicit degrees of freedom,
while the impurities are described by worldlines. The latter serve as localized auxiliary fields, which affect the
majority particles.  The method was applied  to non-relativistic three-dimensional systems of two impurities and a
number of majority particles where both the impurity-impurity interaction and the impurity-majority 
interaction have zero range. For the  case of an attractive impurity-majority interaction the
formation and disintegration of bound states as a function of the 
impurity-impurity interaction strength was studied. This formalism could be applied to $\Lambda\Lambda$ hypernuclei, but
this has so far not been done. An unsolved problem in the ILMC formalism is the smearing of the worldline, which has so
far prevented further detailed applications of this approach.

\subsubsection{Auxiliary field formalism}

In NLEFT,  the Auxiliary Field Quantum Monte Carlo (AFQMC) method is used as it leads to a significant
suppression of the sign oscillations~\cite{Lahde:2019npb}. AFQMC represents a powerful computational
framework within quantum many-body physics, particularly tailored for investigating strongly correlated systems.
This method addresses the challenge of solving the full $A$-body Schrödinger equation by introducing
a Hubbard-Stratonovich transformation. This transformation incorporates auxiliary fields to
decouple particle densities, thereby enhancing the applicability of MC techniques.
In essence, within the AFQMC formalism, individual nucleons evolve as if they are single particles
in a fluctuating background of auxiliary fields. As has been shown recently~\cite{Tong:2024jvs,Hildenbrand:2024ypw}, 
$\Lambda$ hyperons can
also be included in this formalism, which is a major step forward in the study of hypernuclei and
hyperons in dense neutron matter.

To be concrete, let us consider a discrete auxiliary field formulation for the SU(4) symmetric short-ranged NN
interaction given in Eq.~(\ref{eq:H-min}),
\begin{align}
   : \exp \left( -\frac{a_{t} \, c_{NN}}{2} \, \tilde{\rho}^2
          \right) :
=
\sum_{k = 1}^{3} \, w_{k} \,
: \exp \left( \sqrt{-a_{t} \, c_{NN}}  \, s_{k} \, \tilde{\rho} \right) \, :~.
\label{eqn:AFQMC-NN}
\end{align}
From a Taylor expansion of Eq.~(\ref{eqn:AFQMC-NN}) one determines the constants $s_{k}$ and $w_k$
as $s_{1} = -s_{3}=\sqrt{3}$, $s_{2} = 0$, $w_{1} = w_3 = 1/6$ and $w_2 = 2/3$.
This NN interaction obeys the Wigner SU(4) symmetry.
Since we use minimal forces for the YN  and YY interactions, it is possible to derive an
auxiliary field formulation for systems including neutrons, protons and $\Lambda$ hyperons.
This derivation involves replacing the isospin SU$_T(2)$ with flavor SU$_F(3)$ within Wigner's SU(4)
symmetry framework, and the combined spin ($S$) and flavor ($F$) invariance ultimately leads to
the SU(6) symmetry~\cite{Gursey:1964htz}. However, the fact that the strengths of the NN and YN interactions
are different breaks this SU(6) symmetry, and there is no longer an approximate symmetry similar to
Wigner's SU(4) symmetry used in Eq.~(\ref{eqn:AFQMC-NN}). Nevertheless, one can exploit the fact
that $|c_{NN}|>|c_{N\Lambda}|>|c_{\Lambda\Lambda}|$, allowing us to introduce an auxiliary field formulation with
an approximate SU(6) symmetry that protects our simulations including $\Lambda$ hyperons against
strong sign oscillations. For that, express
the spin and isospin independent two-baryon interactions in Eq.~(\ref{eq:H-min}) as,
\begin{align}
V_{\rm 2B} = \frac{c_{NN}}{2}\sum_{\vec{n}}
\,:\,
\left[
\tilde{\rho}(\vec{n})
\right]^2
\,:\,
+ c_{N\Lambda}\sum_{\vec{n}}
\,:\,
\tilde{\rho}(\vec{n})
\tilde{\xi}(\vec{n})
\,:\,
+ \frac{c_{\Lambda\Lambda}}{2}\sum_{\vec{n}}
\,:\,
\left[
\tilde{\xi}(\vec{n})
\right]^2
\,:\,
\,,
\label{eqn:NY-Potential-000}
\end{align}
and this potential (\ref{eqn:NY-Potential-000}) can be rewritten in the following form,
\begin{align}
V_{\rm 2B} = \frac{c_{NN}}{2}\sum_{\vec{n}}
\,:\,
\left[
\tilde{\slashed{\rho}}(\vec{n})
\right]^2
\,:\,
+
\frac{1}{2}
\left(
c_{\Lambda\Lambda}
-\frac{c_{N\Lambda}^2}{c_{NN}}
\right)
\sum_{\vec{n}}
\,:\,
\left[
\tilde{\xi}(\vec{n})
\right]^2
\,:\,
\,,
\label{eqn:NY-Potential-010}
\end{align}
where $\tilde{\slashed{\rho}}$ is defined as,
\begin{align}
\tilde{\slashed{\rho}} =
\tilde{\rho} +
\frac{c_{ N\Lambda}}{c_{NN}} \,  \tilde{\xi}
\,.
\label{eqn:Density-rho-bar}
\end{align}
In Eq.~(\ref{eqn:NY-Potential-010}) the leading contribution comes from the first term on the right-hand side. It
is treated non-perturbatively, while the remaining term is computed using first-order perturbation theory.
Hence, we define a new Hubbard-Stratonovich transformation for the first term in Eq.~(\ref{eqn:NY-Potential-010}),
enabling the simulations of systems consisting of both arbitrary number of nucleons and  arbitrary
number of $\Lambda$ hyperons with a single auxiliary field,
\begin{align}
   : \exp \left( -\frac{a_{t} \, c_{NN}}{2} \, \tilde{\slashed{\rho}}^2\right) :
=
\sum_{k = 1}^{3} \, w_{k} \,
: \exp \left( \sqrt{-a_{t} \, c_{NN}}  \, s_{k} \,  \tilde{\slashed{\rho}} \right) :\, .
\label{eqn:AFQMC-NY}
\end{align}
It is evident that the solution for the auxiliary field variables $s_{k}$ and weights $w_{k}$ is consistent
with systems containing only nucleons.
This extension of the  AFQMC method introduced here broadens hypernuclear calculations by enabling
simulations with any number of hyperons. In addition, the approach can be effectively applied to a wide
range of systems, see e.g.~\cite{Sedrakian:2005zj}. For understanding that, consider two distinct families of
particles and call them $A$ and $B$, and assume that all interactions are attractive.
When the square of the interaction strength between particle types $A$ and $B$, denoted as $c_{AB}^2$,
is of comparable magnitude to the product of the interaction strengths within the same particle types,
$c_{AA}^{}c_{BB}^{}$, the overall coupling of the second term in Eq.~(\ref{eqn:NY-Potential-010}) becomes
very small, enabling a perturbative treatment and calculations with a single auxiliary field. Furthermore,
when $c_{AA}^{} c_{BB}^{} \geq c_{AB}^2$, the second term's overall coupling is attractive, so that calculations still can be
performed with two auxiliary fields. However, only in the case of $c_{AA}^{} c_{BB}^{} \ll c_{AB}^2$, the
overall coupling of the second term becomes repulsive which leads to significant sign problems.

Finally, we discuss the NN interaction $\sim c_{NN}^T$,
known to break SU(4) symmetry and to induce significant sign oscillations,
which was previously disregarded in minimal nuclear interaction studies.
Since we aim at constraining nuclear forces by using the ground state energies of finite hypernuclei and
the saturation properties of symmetric nuclear matter,
this isospin interaction is treated non-perturbatively. We employ a Hubbard-Stratonovich transformation and
introduce a discrete auxiliary field defined as,
\begin{align}
    : \exp \left( -\frac{a_{t} \, c_{NN}^T}{2} \, \sum_{I} \, \tilde{\rho}_{I}^2
           \right) :
 =
 \sum_{k = 1}^{3} \, w_{k} \,
 : \exp \left( \sqrt{-a_{t} \, c_{NN}^{T}}  \, \sum_{I} \,  s_{k,I} \, \tilde{\rho}_{I} \right):\, .
 \label{eqn:AFQMC-NN-I}
 \end{align}
To minimize the occuring sign oscillations in finite nuclei, we focus on systems with equal numbers of protons
and neutrons. Furthermore, in the simulations  of pure neutron matter and hyper-neutron matter, this term can
be omitted due to the absence of particles breaking isospin symmetry,  allowing for sign oscillation-free simulations.
Similarly, in the studies based on the Hamiltonian given in Eq.~\eqref{eq:H-001}, the same trick is
employed to include the $\Lambda$ hyperon in the simulations.

\subsection{Results from the minimal hypernuclear interaction}

Here, we consider the results obtained from the minimal hypernuclear model. We note that in the original work
on that topic~\cite{Tong:2024jvs},
hypernuclei where considered together with the equation of state of nuclear as well as (hyper)neutron matter,
that is neutron matter in the  absence (presence) of hyperons. In this review, we focus  mostly on the
results for hypernuclei and refer to a detailed discussion of the EoS of hyper-neutron matter to Ref.~\cite{Tong:2024jvs}.
First, one has to give the lattice parameters. A spatial lattice spacing of $a = 1.1$~fm and a temporal lattice spacing of
$a_t = 0.2$~fm are used. The local and nonlocal smearing parameters are $s_{\rm L}=0.06$ and $s_{\rm NL} = 0.6$, respectively.
For the three-baryon interaction, the local smearing parameter is set  to $s^{\rm 3B}_{\rm L} = 0.06$. To compute the g.s.
energies of finite nuclei and hypernuclei, various periodic cubic lattices ranging in length from $13.2$~fm to $19.7$~fm
are used. Furthermore, for the computation of pure neutron matter and hyper-neutron matter energies one uses
lattices with a length of $6.6$~fm and imposes  average twisted boundary conditions to efficiently eliminate
finite volume effects, see Ref.~\cite{Lu:2019nbg}.
Second, one must fix parameters in the nucleonic sector, i.e. the LECs $c_{NN}^{}, c_{NN}^T$ and $c_{NNN}^{(d_i)}$,
see Eqs.~\eqref{eq:H-min},\eqref{eq:VNNN}. This are determined from a fit to the $^1S_0$ and $^3S_1$ np phase
shifts and to the  to the saturation properties of symmetric nuclear matter considering all possible combinations
of $d_1$ and $d_2$ with $0 \leq d_1 < d_2 \leq 3$. Through this process, one arrives at six distinct interactions,
enabling one to quantify the theoretical uncertainty.  The resulting EoSs for nuclear and neutron matter are
shown in Fig.~\ref{figS1}. Together with the binding energy of some typical nuclei, namely $E(^4{\rm He}) = 29.4(5)$~MeV,
$E(^8{\rm Be} )= 58.4(9)$~MeV,  $E(^{12}{\rm C}) = 87.1(10)$~MeV, and $E(^{16}{\rm 0}) = 121.8(7)$~MeV, this shows that
this minimal model captures the basic features of nuclear binding. 
 \begin{figure}[htbp]
~~~  \includegraphics[width=0.45\textwidth]{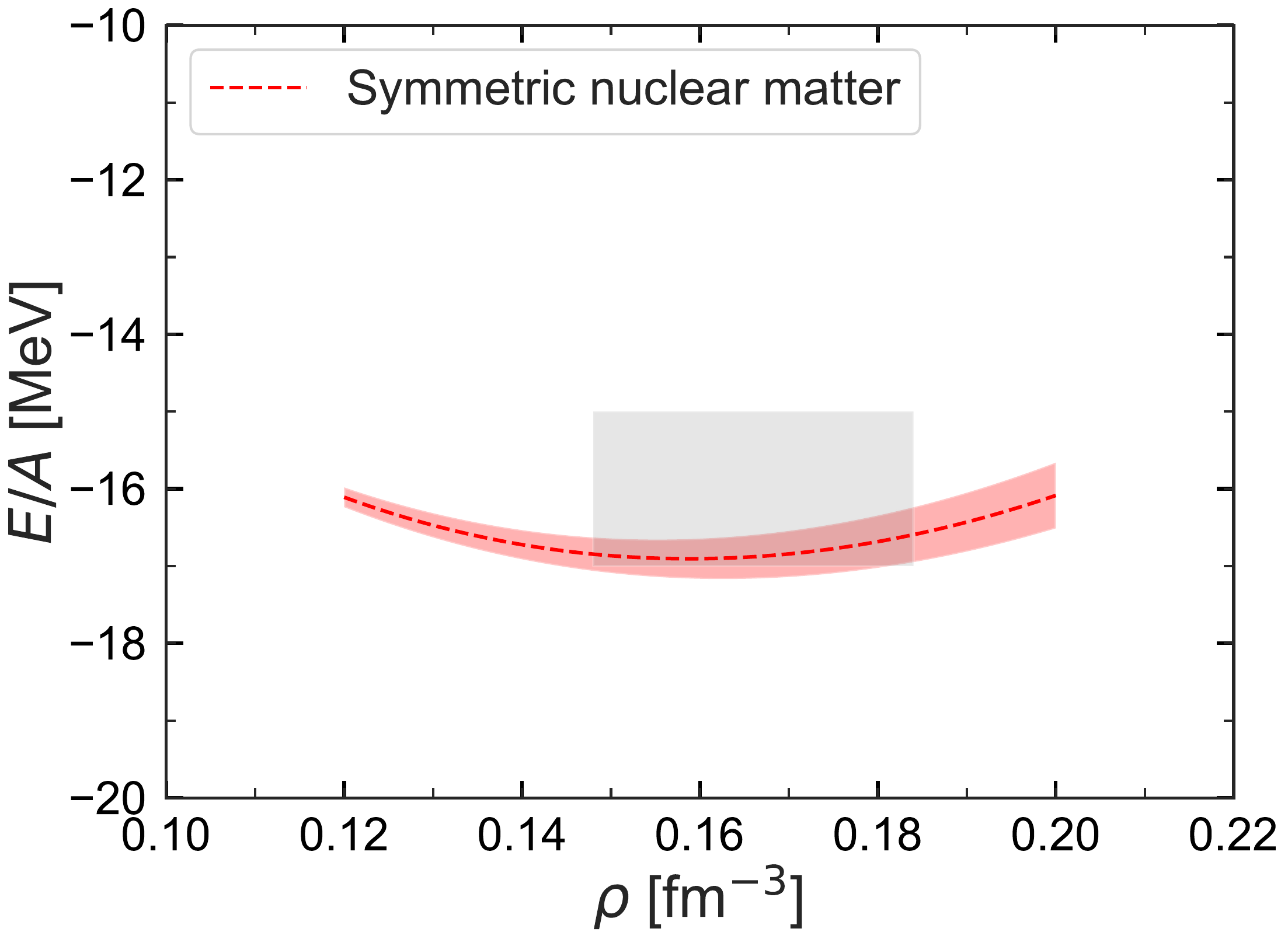}
~~~  \includegraphics[width=0.45\textwidth]{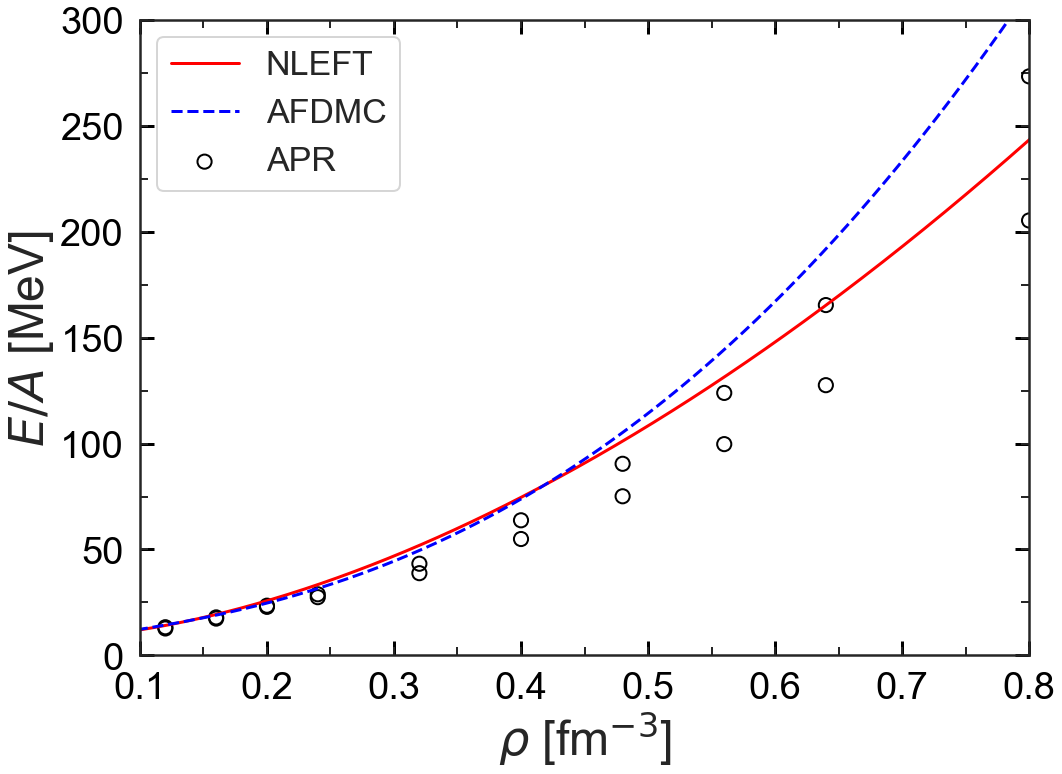}
\caption{
{\bf Left Panel:} Energy per nucleon as a function of density for symmetric nuclear matter from the NLEFT calculations.
The gray shaded area indicates the empirical values.
{\bf Right Panel:} Energy per nucleon as a function of density for symmetric nuclear matter from the NLEFT calculations (red
line) in comparison to the APR~\cite{Akmal:1998cf} (black circles) and  AFDMC~\cite{Gandolfi:2013baa} (blue dashed line)
calculations.
\label{figS1}}
\end{figure}
Third, we must fix the LECs in the $S=-1,-2$ sectors, which are $c_{N\Lambda}, c_{\Lambda\Lambda}$ in the two-particle
and $c_{NN\Lambda}^{(d_i)}, c_{N\Lambda\Lambda}^{(d_i)}$ in the three-particle sector. These parameters are
determined as follows: For the $\Lambda N$ interaction, one fits the  experimental total cross-section data
$\sigma_{\rm tot}(\Lambda N\to \Lambda N)$ for laboratory momenta below 600~MeV and in the absence of cross section data
in the $=-2$ sector, the $^1{\rm S}_0$ phase shift derived from chiral EFT at NLO~\cite{Haidenbauer:2015zqb} is
also fitted. To pin down the parameters of the 3BFs, the separation energies for single-$\Lambda$ and double-$\Lambda$
hypernuclei are fitted as collected in Table~\ref{tab:hypern}, where also
a few predictions are given. As in the nucleonic sector, one obtains an overall satisfactory description of hypernuclei.
For the discussion of the neutron matter EoS in the presence of hyperons and the resulting neutron star properties, we refer to Refs.~\cite{Tong:2024jvs,Tong:2025sui}.

\begin{table}[htbp]
\centering
\caption{$\Lambda$ separation energies for single-$\Lambda$ and 
double-$\Lambda$ hypernuclei  in the minimal interaction model (in MeV). The first error is the statistical one
whereas the second error is the systematic one (due to the 
three-baryon forces). The $^*$ marks a prediction. Note that the experiemntal values for the
two Be double-$\Lambda$ hypernuclei are contested.}\label{tab:hypern}
\begin{tabular}{|c||c|c|c|c|c|c|c|}
\hline
Nucleus & $_\Lambda^5$He &  $_\Lambda^9$Be & $_\Lambda^{13}$C & $_{\Lambda\Lambda}^6$He & $_{\Lambda\Lambda}^{10}$Be & $_{\Lambda\Lambda}^{12}$Be \\
\hline
NLEFT$_{\rm min}$ & $3.40(1)(1)$ & $5.72(5)(4)$ & $10.54(17)(29)^*$ & $6.91(1)(1)$ & $13.30(7)(12)^*$ & $21.22(56)(21)^*$ \\
Exp.            & $3.10(3)$    & $6.61(7)$    & $11.80(16)$       & $6.91(16)$   & $14.70(40)$ & $ 21.48(121)$\\
\hline
\end{tabular}
\end{table}

\subsection{Results based on high-fidelity chiral forces}

In Ref.~\cite{Hildenbrand:2024ypw} $\Lambda$-hypernuclei where considered based on the Hamiltonian
Eq.~\eqref{eq:H-001} with $V_{YYN}=0$ and the locally and non-locally smeared 3BFs given in
Eq.~\eqref{eqn:V_c3-sL-001} and Eq.~\eqref{eqn:V_c1-sL-001}, respectively.
A spacial lattice spacing of $a=1.32$~fm and a temporal lattice spacing of $a_t=1/1000\,$MeV$^{-1}$ were used.
All interactions share a similar set of local and non-local smearing parameters of $s_{\rm L}=0.07$ and $s_{\rm NL}=0.5$,
the same parameters are also used in Ref.~\cite{Elhatisari:2022zrb} for the NN interaction. The presence of
non-local smearing results in an  explicit dependence on the center-of-mass momentum, thus breaking Galilean invariance.
As in the case of the  nucleon-nucleon interactions in Ref.~\cite{Elhatisari:2022zrb}, 
in~\cite{Hildenbrand:2024ypw}  the nucleon-hyperon GIR interaction was included. For a detailed discussion of
GIR interactions on the lattice, see Ref.~\cite{Li:2019ldq}. Having fixed the lattice parameters, the LECs of the YN
interaction ($C_{YN}^{S,T}$) are determined from the unpolarized $\Lambda N\to \Lambda N$ cross section while
keeping consistency with the scattering parameters of the best continuum interaction~\cite{Haidenbauer:2023qhf}.
Furthermore, the hypertriton is described as a shallow bound state to be able to properly split the
two  S-wave channels. Then, the smeared $\Lambda NN$ forces are fitted to a number of hypernuclei, more precisely,
two scenarios are considered. In scenario~$1$, the 3BFs are constrained only by the light 
$A=4$ and $A=5$ systems, while in scenario~$2$  hypernuclei  up to $A=16$ are used. Note also
that no CSB breaking effects are considered, so in the four-body system the average of $^4_\Lambda$H and $^4_\Lambda$He is taken.

The results of this study can be summarized as follows. As shown in Fig.~\ref{fig:hyperhifi}, considering
2BFs alone reproduces already the observed trends, however, there is clearly room left for 3BFs. The results
when including these are also shown in that figure for scenario~1.  A rather good description
of the g.s. energies of ${}^9_\Lambda\text{Be}^{}$,  ${}^{13}_{\hphantom{1}\Lambda}\text{C}^{}$ and
${}^{16}_{\hphantom{1}\Lambda}\text{O}^{}$ is found. However,  the
${}^7_\Lambda\text{Li}^{\frac{1}{2}^+}$-${}^7_\Lambda\text{Li}^{\frac{3}{2}^+}$ splitting is not well reproduced.  Within scenario~2, one finds that the
improvement in the final results is minimal, as the data is already well-described within uncertainties,
see Table~\ref{tab:hyperres}. For more details on the determination of the 3BFs, the reader is referred
to Ref.~\cite{Hildenbrand:2024ypw}.
However, the inclusion of heavier hypernuclei may be necessary in the future to address mid- to heavy-mass hypernuclei.
Furthermore, in~\cite{Hildenbrand:2024ypw} it was shown that decuplet saturation~\cite{Petschauer:2020urh}, which means 
$C_3=C_1$ and $C_2=0$, gives a fair description of the hypernuclei considered, but can certainly not be
considered a precision tool.  Note that similar problems for decuplet saturation were also observed in \cite{Le:2024rkd} and fixed  by choosing $C_2 \ne 0$ (see Sec.~\ref{sec:ncsmynn}). 
\begin{table}[htb]
\centering
\begin{tabular}{|c|c|c|c|}
\hline
Nucleus & Scenario~1 [MeV] & Scenario~2 [MeV] & Experiment [MeV] \\
\hline
${}^3_\Lambda\text{H}$        & $0.12\pm 0.06$ & $0.12\pm 0.06$ & $0.16\pm 0.04$\\
${}^4_\Lambda\text{H}^{0^+}$   & $2.26\pm 0.19$ & $2.27\pm 0.19$ & $2.25\pm 0.04$\\
${}^4_\Lambda\text{H}^{1^+}$   & $0.97\pm 0.19$ & $1.02\pm 0.19$ & $1.01\pm 0.05$\\
${}^5_\Lambda\text{He}$       & $3.10\pm 0.13$ & $3.11\pm 0.13$ & $3.10\pm 0.03$\\
${}^7_\Lambda\text{Li}^{\frac{1}{2}^+}$ & $5.52\pm 0.97$ & $5.51\pm 0.97$ & $5.62\pm0.06$\\
${}^7_\Lambda\text{Li}^{\frac{3}{2}^+}$ & $5.36\pm 0.97$ & $5.37\pm 0.97$ & $4.93\pm0.06$\\
${}^9_\Lambda\text{Be}^{}$    & $6.72\pm 0.55$ & $  6.73\pm 0.56$ & $6.61\pm 0.07$\\   
${}^{13}_{\hphantom{1}\Lambda}\text{C}^{}$ & $11.44\pm 0.84$ & $11.47\pm 0.85$ & $11.96\pm 0.07$\\
${}^{16}_{\hphantom{1}\Lambda}\text{O}^{}$ & $12.72\pm 1.61$ & $13.00\pm 1.61$ & $13.00\pm 0.06$\\
\hline
\end{tabular}
\caption{$\Lambda$ separation energies in the two different scenarios of fitting the
$YNN$ forces. The experimental values  are taken from \cite{HypernuclearDataBase}, where  
the four-body systems were averaged.}
\label{tab:hyperres}
\end{table}
\begin{figure}[ht]
\centering 
\includegraphics[scale=0.25]{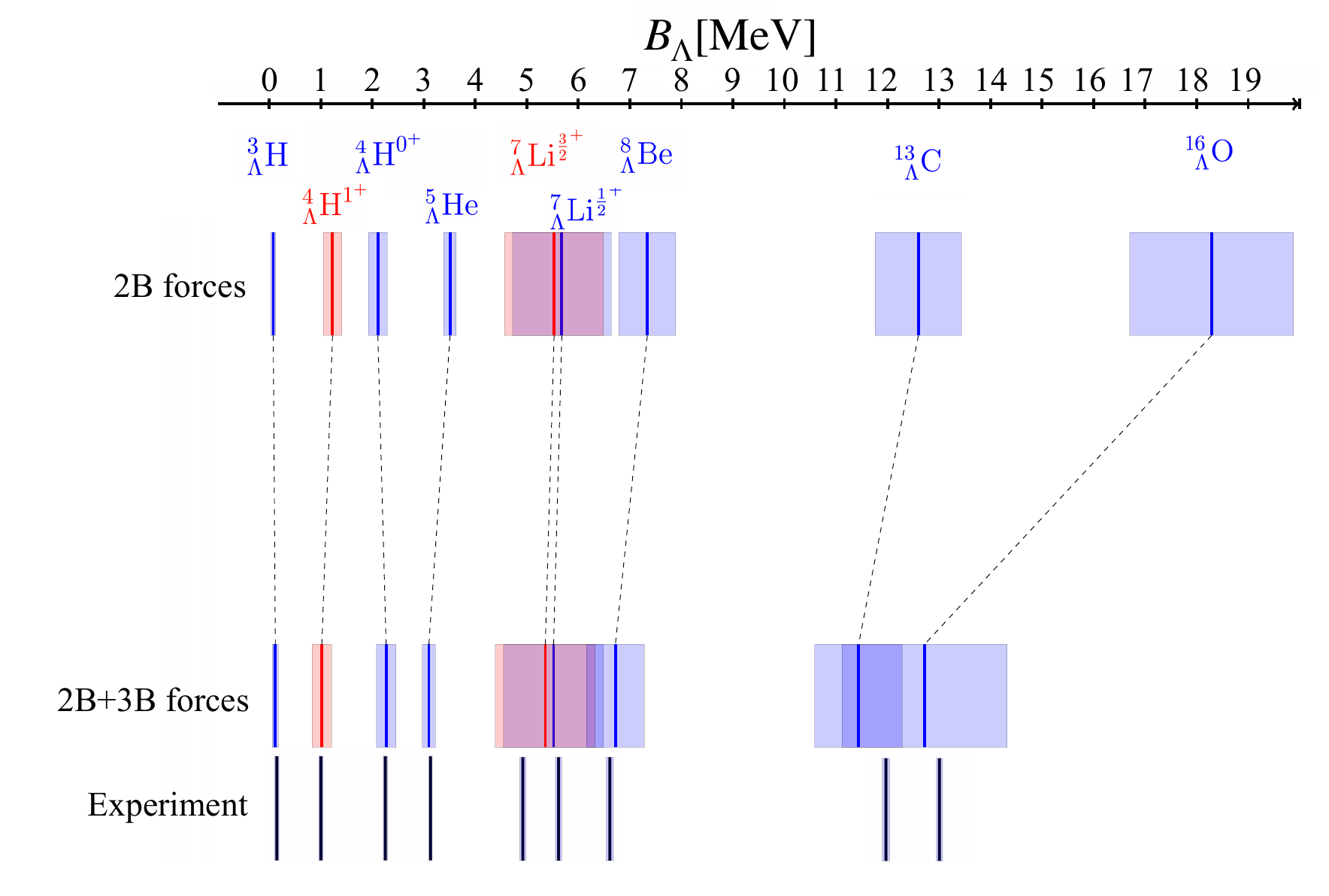}
\caption{
$\Lambda$ separation energies the $YNN$ forces in scenario~1.
The experimental values are taken from  \cite{HypernuclearDataBase}, where
the four-body systems were averaged. Ground states are depicted in blue, excited states in 
    red. The uncertainties are indicated by the shaded areas.
\label{fig:hyperhifi}
}
\end{figure}

Clearly, a number of improvements of this type of calculation are necessary,
such as the inclusion of the explicit two-pion exchange in the NN interaction, the inclusion of higher
order terms in the hyperon-nucleon interaction as well  the $\Lambda$-$\Sigma$  conversion, which
could be done perturbatively as suggested in Ref.~\cite{Beane:2003yx}.



    \section{Summary and outlook}\label{sec:sum}

Over the past 10 years, there has been considerable progress in the treament
and in our understanding of hypernuclear physics. It could be achieved, 
on the one hand side, by adopting chiral effective field theory, an approach
that has been rather successfully employed in studies of nuclear forces, 
as basis for deriving the interaction of hyperons too. 
Thereby, the principal merits of chiral EFT, 
(i) an organizational scheme (``power counting'') that allows to 
quantify the uncertainty of predictions, and
(ii) a consistent treatment of two- and many-body forces, 
are likewise incorporated into hypernuclear studies.
On the other hand, and again following the development in the nuclear
sector, rigorous calculations with so-called {\it ab initio} methods 
have become feasible for hypernuclei as well.
For example, within the no-core shell model (NCSM) computations 
of hypernuclei up to A=10 and beyond have been reported, 
based on state-of-the-art YN and YNN potentials. 

In this work we have reviewed the present status of {\it ab initio} 
calculations of light hypernuclei with interactions based on chiral 
effective field theory.   
To begin with, we have summarized the basics to derive the forces between 
octet baryons (N, $\Lambda$, $\Sigma$, $\Xi$) up to next-to-next-to-leading 
order in SU(3) chiral EFT.
The effective baryon-baryon potentials include contributions from pure four-baryon contact terms, one-meson-exchange diagrams, and two-meson-exchange diagrams 
involving leading and subleading meson-baryon vertices.
The leading three-baryon forces, which formally start to contribute at 
N2LO, consist of a three-baryon contact interaction, a one-meson exchange 
and a two-meson exchange component.
For illustration explicit expressions of the potentials for the $\Lambda$NN interaction in the spin and isospin basis have been discussed and compared 
to those for the NNN system. It has also been shown how  
the emerging low-energy constants can be estimated via decuplet saturation,
which leads to a promotion of some parts of the three-baryon forces to NLO.

Regarding the BB interaction, selective results for cross sections in the 
$\La p$, $\Si^-p$, $\Si^+p$, and $\Xi^-p$ channels have been provided. 
Those testify that an excellent description of the low-energy scattering 
cross sections could be achieved.

In the second part of this review, we have presented the formalisms 
of {\it ab initio} methods for studies of light hypernuclei. This concerns
Faddeev- and Faddeev-Yakubovsky equations on one hand side and
the no-core shell model on the other side. In addition, explicit
results for separation energies of light hypernuclei based on 
chiral NN, YN and YY potentials have been reported. 
In some calculations effects from charge-symmetry breaking in the
$\Lambda$N interaction have been taken into account, while in 
others chiral three-body
forces in the NNN and YNN systems have been included.  
Another promising approach discussed here is NLEFT, where nuclear
properties can be worked out to high precision, so far up to $A=58$
for conventional nuclei. In this review, we have concentrated on the
first hypernuclear calculations in NLEFT, that offer an alternative to
the NCSM.

Overall the results for $\Lambda$ hypernuclei obtained with the {\it ab initio} 
approaches considered in the review are very encouraging. Obviously,
despite of the incomplete knowledge of the underlying
$\Lambda$N interaction and the sizeable uncertainties of the 
available scattering data, the spectrum of light hypernuclei is 
qualitatively well reproduced, up to A=7 within the NCSM and even up to
A=16 with the NLEFT method. This signals that the bulk properties
of the $\Lambda$N and $\Sigma$N interactions are reasonably well reflected 
by the scarce two-body data and also that those properties are appropriately 
represented by the chiral potentials which are the starting point 
of the few-body calculations. Remarkably,
once three-body forces are included in the calculations, in line  
with and of an order of magnitude that is consistent with the employed power 
counting, a good quantitative agreement with the empirical separation 
energies is achieved for essentially all considered hypernuclei. 
Furthermore, charge symmetry breaking in the $\Lambda$N interaction, 
as evidenced by the measured level splittings of several light mirror hypernuclei, 
can be well accounted for when, in addition to the long-established 
pion-exchange contribution associated with $\Lambda$-$\Sigma^0$ mixing,
appropriate contact interactions are included, which arise in a proper and 
consistent treatment within chiral EFT. 

As next step the more subtle and detailed properties of the YN interaction have
to be explored and established. On the experimental side work in this direction has
already started. Here measurements of the $\Sigma^+$p and $\Sigma^-$p differential
cross sections have been performed and reported. Corresponding measurements for 
$\Lambda$p scattering, which are more directly relevant for studies of 
$\Lambda$ hypernuclei, are planned for the future \cite{Miwa:2022coz,Miwa:2025adw}.  
Also, as already pointed out,
the actual experimental values of some of the energy levels of light $\Lambda$ 
hypernuclei are still controversial. In particular, the uncertainty for the 
lightest system, the hypertriton is still uncomfortably large. Any more
accurate determination if its value \cite{Chen:2025eeb} would provide a very strong 
constraint on the $\Lambda$N interaction, specifically the $^1S_0$ scattering length. 

The next and also straightforward step is certainly to take into
account CSB in the $\La$N channel of the SMS YN potentials. 
With CSB forces included a direct comparison of the predicted
separation energies with the measured splittings of the 
$^4_\Lambda$H/$^4_\Lambda$He hypernuclei 
in the $0^+$ and $1^+$ states will be possible. 

It is also desirable to extend the application of chiral potentials 
to heavier $\La$ hypernuclei, i.e. at least up to A=10 or so.  
Among other things, this would allow to investigate the possible 
role of $\Lambda$N P-waves on the pertinent separation energies. 

The so far reported $\Xi$ hypernuclei $^{13}_\Xi$B and 
$^{15}_\Xi$C are much heavier than those which have been 
predicated/calculated within the NCSM up to now. An extension of 
the calculations to those systems is desirable but computationally
quite challenging. Anyway, as a first step one would need to include
the SRG induced YNN 3BFs in the evaluation of the separation energies
of those $\Xi$ hypernuclei that have been already studied in the past.
Though no essential change in the separation energies is expected, 
it is still interesting to establish reliably the predictions for the lightest bound $\Xi$ hypernuclei and to pin down the expected energies
more quantitatively. 

Furthermore, 
it would be also interesting to explore the existence of possible 
light $\Sigma$ hypernuclei based on the chiral YN potentials. 
Indeed, so far there is only undisputed evidence for a single nucleus, namely a $^4_\Si$He $I = 1/2$ 
quasibound state, established in the experiments by Hayano et al.~\cite{Hayano:1988pn} and Nagae et al.~\cite{Nagae:1998tj},
see also the review \cite{Nagae:2022rvm}, 
and investigated theoretically by Harada et al.~\cite{Harada:1998fj}.  
Recently a possible $\Si^0nn$ state has been reported by the HALL~A
Collaboration at JLab~\cite{HallA:2022qqj}. Its location is
$3.14 \pm 0.84$~MeV below the $\Si^0nn$ threshold and the
suggested quantum numbers are $I=1$, $J^P=1/2^+$. A corresponding 
theoretical study has been presented in Ref.~\cite{Garcilazo:2022pgt}.

A first Faddeev calculation of $\Lambda d$ scattering based on 
the chiral YN potentials NLO13 and NLO19 has been published 
recently by Kohno and Kamada~\cite{Kohno:2024tkq}. 
There are ongoing experiments for determining $\Lambda$d scattering
observables by the CLAS Collaboration at JLab~\cite{Miwa:2025}. 
However, the expected $\Lambda$ momentum range of $0.5$ to 
$1.5$~GeV/c is significantly higher than what has been considered 
in Ref.~\cite{Kohno:2024tkq}. 
Regarding the $\Lambda d$ properties at low energies, measurements
of correlation functions by the STAR Collaboration~\cite{Hu:2023iib},
and similar efforts by the HADES \cite{HADES:2024} and
ALICE \cite{ALICE:2025} collaborations,
could provide some insight \cite{Haidenbauer:2020uew,Kohno:2025zgb}. 
Of course, in this context one has to 
keep in mind the caveats pointed out in 
Ref.~\cite{Epelbaum:2025aan}. 

Finally, we note that NLEFT allows to investigate much
heavier hypernuclei due to the mild scaling in $A$.
However, before performing such calculations, a more precise representation of the two- and three-baryon
forces consistent with the high-fidelity two- and three-nucleons forces has to be implemented. Such work is underway.


    \section*{Acknowledgements}
It is a great pleasure to thank Shahin Bour,  Serdar Elhatisari, Dillon Frame, Hans-Werner Hammer, Fabian Hildenbrand, 
Norbert Kaiser, Hiroyuki Kamada, Michio Kohno, Timo L\"ahde, Hoai Le, Dean Lee, Susanna Liebig, Kazuya Miyagawa. Stefan Petschauer, Henk Polinder, 
Zhengxue Ren, Xiang-Xiang Sun, Hui Tong, Isaac Vida\~na and Wolfram Weise 
for sharing their insights into the topics discussed here. 
This work was supported in part by the European
Research Council (ERC) under the European Union's Horizon 2020 research
and innovation programme (grant agreement No. 101018170) and by the MKW NRW
under the funding code NW21-024-A.
The work of UGM was also supported in part by the CAS President's International
Fellowship Initiative (PIFI) (Grant No.~2025PD0022).


	\appendix
	\renewcommand*{\thesection}{\Alph{section}}
	

	\bibliography{extra,inspire}

	



	
\end{document}